\documentclass[useAMS,usenatbib]{mn2e}

\usepackage{graphicx}
\usepackage{natbib}
\usepackage{subfigure}
   \usepackage{amsmath}
   \usepackage{amsfonts}   
   \usepackage{amssymb}    
\usepackage{longtable,lscape}
\def\aap{A\&A}
\def\apjl{ApJL}
\def\apjs{ApJS}

\title[Lyman break and UV-selected galaxies at $z \sim 1$]{Lyman break and UV-selected galaxies at $z \sim 1$ \\ II. PACS-100$\mu$m/160$\mu$m FIR detections\thanks{{\it Herschel} is an ESA space observatory with science instruments provided by European-led Principal Investigator consortia and with important participation from NASA.}}
\author[I. Oteo et al.]
{\parbox{\textwidth}{I. Oteo,$^{1,2}$\thanks{E-mail: \texttt{ioteo@iac.es}},
G. Magdis$^{3}$, 
\'A. Bongiovanni$^{1,2,4}$,
A.M. P\'erez-Garc\'ia$^{1,2,4}$, 
J. Cepa$^{1,2}$, 
B. Cedr\'es$^{1,2}$,
A. Ederoclite$^{5}$, 
M. S\'anchez-Portal$^{4,6}$, 
J. A. L. Aguerri$^{1,2}$, 
E. J. Alfaro$^{8}$, 
B. Altieri$^{9}$, 
P. Andreani$^{10,11}$, 
T. Aparicio-Villegas $^{8,25}$, 
H. Aussel$^{12}$, 
N. Ben\' itez$^{8}$, 
S. Berta$^{13}$, 
T. Broadhurst $^{14}$, 
J. Cabrera-Ca\~no$^{15}$, 
F. J. Castander$^{16}$, 
M. Cervi\~no$^{1,8}$, 
A. Cimatti$^{17}$, 
D. Cristobal-Hornillos$^{5,8}$, 
E. Daddi$^{18}$, 
D. Elbaz$^{18}$, 
A. Fernandez-Soto$^{19,20}$, 
N. F\"orster Schreiber$^{13}$, 
R. Genzel$^{13}$, 
R. M. Gonzalez-Delgado$^{8}$, 
C., Husillos$^{8}$, 
L. Infante$^{21}$, 
E. Le Floc'h$^{18}$, 
D. Lutz$^{13}$, 
B. Magnelli$^{13}$, 
R. Maiolino$^{22}$, 
I. M\' arquez$^{8}$, 
V. J. Mart\' inez$^{23,24}$, 
J. Masegosa$^{8}$, 
I. Matute$^{8}$, 
M. Moles$^{5}$, 
A. Molino$^{8}$, 
A. del Olmo$^{8}$, 
J. Perea$^{8}$, 
R. P\' erez-Mart\' inez$^{7}$, 
I. Pintos-Castro$^{1,2,26}$, 
A. Poglitsch$^{13}$, 
J. Polednikova$^{1,2}$, 
P. Popesso$^{13}$, 
M. Povi\'c$^{8}$, 
F. Pozzi$^{17}$, 
F. Prada $^{8}$, 
J. M. Quintana$^{8}$, 
L. Riguccini$^{18}$, 
E. Sturm$^{13}$, 
L. Tacconi$^{13}$, 
I. Valtchanov$^{6}$, and
K. Viironen$^{5}$
}\vspace{0.4cm}\\
$^{1}$Instituto de Astrof{\'i}sica de Canarias (IAC), E-38200 La Laguna, Tenerife, Spain\\
$^{2}$Departamento de Astrof{\'i}sica, Universidad de La Laguna (ULL), E-38205 La Laguna, Tenerife, Spain\\
$^{3}$Department of Physics, University of Oxford, Keble Road, Oxford OX1 3RH\\
$^{4}$Asociaci\' on ASPID. Apartado de Correos 412, La Laguna, Tenerife, Spain\\
$^{5}$Centro de Estudios de F\'isica del Cosmos de Arag\' on, Plaza San Juan 1, Planta 2, Teruel, 44001, Spain\\
$^{6}$Herschel Science Centre (ESAC). Villafranca del Castillo, Spain\\
$^{7}$XMM/Newton Science Operations Centre (ESAC). Villafranca del Castillo. Spain\\
$^{8}$Instituto de Astrof\'isica de Andaluc\' ia (CSIC), Glorieta de la Astronom\'ia s/n, EÐ18008 Granada, Spain\\
$^{9}$Herschel Science Centre, European Space Astronomy Centre, Villanueva de la Ca÷nada, 28691 Madrid, Spain\\
$^{10}$ESO, Karl-Schwarzschild-Str. 2, 85748 Garching, Germany\\
$^{11}$INAF Ð Osservatorio Astronomico di Trieste, Via Tiepolo 11, 34143 Trieste, Italy\\
$^{12}$Laboratoire AIM-Paris-Saclay, CEA/DSM/Irfu - CNRS - Universit«e Paris Diderot, CE-Saclay, F-91191 Gif-sur-Yvette, France\\
$^{13}$Max-Planck-Institut f$\ddot{u}$r extraterrestrische Physik, Postfach 1312, Giessenbachstra\ss e 1, 85741 Garching, Germany\\
$^{14}$School of Physics and Astronomy, Tel Aviv University, Israel\\
$^{15}$Facultad de F\'isica. Departamento de F\'isica At\'omica, Molecular y Nuclear. Universidad de Sevilla, Sevilla, Spain\\
$^{16}$Institut de Ciencies de lÕEspai, IEEC-CSIC, Barcelona, Spain\\
$^{17}$Dipartimento di Astronomia, Universitˆ di Bologna, Via Ranzani 1, 40127 Bologna, Italy\\
$^{18}$Laboratoire AIM, CEA/DSM-CNRS-UniversitŽ Paris Diderot, IRFU/Service dÕ Astrophysique, B‰t. 709, CEA-Saclay, 91191 Gif-sur-Yvette Cedex, France\\
$^{19}$Instituto de F\' isica de Cantabria (CSIC-UC), EÐ39005, Santander, Spain\\
$^{20}$Unidad Asociada Observatorio Astron—mico (Universitat de Valncia / IFCA-CSIC), Parc Cient'fic UV, 46980 Paterna, Spain\\
$^{21}$Departamento de Astronom\'ia, PontiÞcia Universidad Catolica, Santiago, Chile\\
$^{22}$Cavendish Laboratory, University of Cambridge, 19 J. J. Thomson Avenue, Cambridge, CB3 0HE, UK\\
$^{23}$Observatori Astron˜mic de la Universitat de Valncia, ÊParc Cient'fic UV, 46980 Paterna, Spain\\
$^{24}$Departament d«Astronomia i Astrof'sica, Universitat de Valncia, 46100 Burjassot, Spain\\
$^{25}$Observat—rio Nacional-MCT, Rua JosŽ Cristino, 77. CEP 20921-400, Rio de Janeiro-RJ, Brazil\\
$^{26}$Centro de Astrobiolog\'{i}a, INTA-CSIC, P.O. Box - Apdo. de correos 78, Villanueva de la Ca\~nada Madrid 28691, Spain
}

\begin{document}

\date{Accepted ??. Received ??; in original form ??}

\pagerange{\pageref{firstpage}--\pageref{lastpage}} \pubyear{2002}

\maketitle

\clearpage

\label{firstpage}

\begin{abstract}
In this work we report the PACS-100$\mu$m/160$\mu$m detections of a sample of \textrm{42} GALEX-selected and FIR-detected Lyman break galaxies (LBGs) at $z \sim 1$ located in the COSMOS field and analyze their ultra-violet (UV) to far-infrared (FIR) properties. The detection of these LBGs in the FIR indicates that they have a dust content high enough so that its emission can be directly detected. According to a spectral energy distribution (SED) fitting with stellar population templates to their UV-to-near-IR observed photometry, PACS-detected LBGs tend to be bigger ($R_{\rm eff} \sim 4.1$ kpc), more massive ($\log{\left( M_{*}/M_\odot \right)} \sim 10.7$), dustier ($E_s(B-V) \sim 0.40$), redder in the UV continuum ($\beta \sim -0.60$), and UV-brighter ($\log{\left( L_{\rm UV}/L_\odot \right)} \sim 10.1$) than PACS-undetected LBGs. PACS-detected LBGs at $z \sim 1$ are mostly disk-like galaxies and are located over the green-valley and red sequence of the color-magnitude diagram of galaxies at their redshift. By using their UV and IR emission, we find that PACS-detected LBGs tend to be less dusty and have slightly higher total star-formation rates (SFRs) than other PACS-detected UV-selected galaxies within their same redshift range. As a consequence of the selection effect due to the depth of the FIR observations employed, all our PACS-detected LBGs have total IR luminosities, $L_{\rm IR}$, higher than $10^{11}L_\odot$ and thus are luminous IR galaxies (LIRGs). However, none of the PACS-detected LBGs are in the ultra-luminous IR galaxies (ULIRGs) regime, $L_{\rm IR} \geq 10^{12}L_\odot$, where the FIR observations are complete. The finding of ULIRGs-LBGs at higher redshifts ($z \sim 3$) suggests an evolution of the FIR emission of LBGs with cosmic time. In an IRX-$\beta$ diagram, PACS-detected LBGs at $z \sim 1$ tend to be located around the relation for local starburst similarly to other UV-selected PACS-detected galaxies at their same redshift. Consequently, the dust-correction factors obtained with their UV continuum slope allow to determine their total SFR, unlike at higher redshifts. However, the dust attenuation derived from UV to NIR SED fitting overestimates the total SFR for most of our PACS-detected LBGs in age-dependent way: the overestimation factor is higher in younger galaxies. This is likely due to the typical degeneracy between dust attenuation and age in the SED-fitting with synthetic templates and highlights the importance of the FIR measurements in the analysis of star-forming galaxies at intermediate redshifts.

\end{abstract}

                


\section{Introduction}\label{intro}

There are some methods which have been traditionally used to look for star-forming (SF) high-redshift galaxies. Among them, one of the most employed and successful is the Lyman-break or dropout technique, that segregates the so-called Lyman break galaxies (LBGs). The technique is based on sampling the Lyman break of galaxies with two broad-band filters close in wavelength, each one located on one side of the break. Since the location in wavelength of the Lyman break depends upon redshift, different combination of filters pick up galaxies at different redshifts. Many samples of LBGs have been found following that procedure, mainly at $z \gtrsim 3$ by employing optical filters in ground-based telescopes \citep{Madau1996,Steidel1996,Steidel1999,Steidel2003,Verma2007,Iwata2007,Yabe2009}, but also at $z \sim 2$ \citep{Hathi2010,Ly2009,Ly2011,Haberzettl2012,Basu2011}, and at lower redshifts \citep{Burgarella2006,Burgarella2007,Barger2008,Chen2013,Oteo2013}.

In order to accurately study the evolution and the bolometric emission of LBGs it is essential to have information of the light that is absorbed by dust in the UV and re-emited in-turn in the far-infrared (FIR). With a careful analysis of the UV slope, coupled with appropriate dust attenuation models, \cite{Vijh2003} identify some of the most heavily attenuated specimens in LBGs at $z \sim 2-4$. Furthermore, \cite{Finkelstein2009a} find that LBGs at $z \sim 4$ can be detectable with ALMA after a few hours of integration. However, until recently, few high-redshift LBGs had been individually detected in FIR \citep{Chapman2000,Chapman2009,Siana2009,Rigopoulou2010} and, therefore, a complete and accurate study at different redshifts has not been possible yet. With the advent of the ESA's \emph{Herschel} space observatory \citep{Pilbratt2010} and its Photodetector Array Camera and Spectrometer (PACS) \citep{Poglitsch2010} and Spectral and Photometric Imaging REceiver (SPIRE) \citep{Griffin2010} instruments we are in possession of deep FIR data that enable us to study the FIR spectral energy distribution (SED) around the dust emission peak of LBGs as never before. \cite{Magdis2010LBGs} studied for the first time using PACS data the FIR SED of infrared luminous LBGs at $z \sim 3$. Although none of their galaxies are individually detected with \emph{Herschel}, a stacking analysis suggests a median IR luminosity of $L_{\rm IR} = 1.6 \times 10^{12} L_\odot$. Additionally, complementing their study with multi-wavelength data they obtained that for their IR luminosity, MIPS-LBGs are warmer than sub-millimeter luminous galaxies while they fall in the locus of the $L_{\rm IR} - T_{\rm d}$ relation of the local ultra-luminous IR galaxies (ULIRGs). Also with a stacking analysis, \cite{Lee2012} study the FIR emission of LBGs at $3.3 \lesssim z \lesssim 4.3$ obtaining that their IR-to-UV luminosity ratio ($L_{\rm IR} / L_{\rm UV}$) is low  compared to that observed for $z \sim 2$ LBGs. Furthermore, they obtain that the correlation between the UV continuum slope and the $L_{\rm IR} / L_{\rm UV}$ ratio suggest dust properties similar to those of local starbursts except for the galaxies with the highest luminosities. \cite{Davies2013} employed stacked detections in SPIRE bands of LBGs at $z \sim 3$ to find that a significant fraction of the star formation in these systems is obscured and that the extinction-corrected SFR does not exhibit the large values that would be predicted by their UV continuum slope. The latter indicates that the method of assuming an intrinsic UV-slope and correcting for dust attenuation may be invalid for this sample and that these are not in fact the most actively SF systems. \cite{Burgarella2011} report the first SPIRE-250$\mu$m and SPIRE-350$\mu$m FIR detections of LBGs by employing data from the Herschel Multi-tiered Extragalactic Survey (HerMES) project \citep{Oliver2010}. They found SPIRE detections for 12 LBGs at $0.7 \lesssim z \lesssim 1.6$ and one at $z \sim 2.0$. All these galaxies are high mass, luminous IR galaxies, and have redder NUV-$U$ and $U$-$R$ color than other SPIRE-undetected galaxies. \cite{Oteo2013_LBGsz3} find 16 LBGs at $z \sim 3$ which are individually detected in PACS under the depth of the observations carried out in the frame of the GOODS-\emph{Herschel} survey \citep{Elbaz2011}. They found that PACS-detected LBGs at $z \sim 3$ are ULIRGs or hyper-luminous IR galaxies (HyLIRGs) and that the dust attenuation factor obtained from the SED-derived dust attenuation with \cite{Bruzual2003} templates or from the UV continuum slope cannot reproduce the total SFR of these galaxies.

In this study, we continue the work started in \cite{Oteo2013} about the physical properties of LBGs and other UV-selected galaxies at $0.8 \leq z \leq 1.2$ in the Cosmic Evolution Survey \citep[COSMOS,][]{Scoville2007} field. In that work, we analyze the stellar populations of a sample of GALEX-selected LBGs by combining UV GALEX measurements and the optical to near-IR data coming from the Advanced Large, Homogeneous Area Medium Band Redshift Astronomical (ALHAMBRA) survey \citep{Moles2008}. This combination provides an excellent photometric coverage of the UV-to-near-IR SED of those galaxies. As result, it was found that LBGs at $z \sim 1$ are mostly young galaxies, with a median age of 341 Myr, and have intermediate dust attenuation, $E_s (B-V) \sim 0.2$. Due to their selection criteria they are UV-bright galaxies and have high star-formation rates (SFRs). The median value of the stellar mass of the LBGs in the sample is $\log{M_*/M_\odot} = 9.74$. Now, we enlarge the photometric coverage of the SEDs of those galaxies by employing IRAC-5.8$\mu$m, IRAC-8.0$\mu$m, MIPS-24$\mu$m, and PACS-100$\mu$m/PACS-160$\mu$m data. Actually, one of the advantages of dealing with galaxies at $z \sim 1$ is that their typical observed FIR fluxes make a significant number of  objects detectable under the depth of PACS observations used in this work (see Section \ref{PACS_counterparts}). This allow us to carry out the study of their physical properties avoiding the uncertainties of the optical-based methods that do not take into account the dust emission in the FIR, like the SED-fitting with \cite{Bruzual2003} templates. The combination of GALEX, ALHAMBRA, IRAC, MIPS, and PACS data provides a photometric coverage of the SEDs of galaxies which has not been reached with previous data sets. The paper is structured as follows: in Section \ref{PACS_counterparts} we define the selection criteria employed to look for our SF galaxies at $z \sim 1$, explain how their SEDs were fitted with synthetic templates, and report their FIR counterparts. In Section \ref{dust_attenuation} we study the dust attenuation of our PACS-detected galaxies as a function of the rest-frame UV, total IR, and bolometric luminosity and UV continuum slope. Section \ref{starformation} shows the study of the SFR and the dust correction factors. We analyze the relation between stellar mass and total and specific SFR in Section \ref{sfr_mass_plane}. The location of our PACS-detected galaxies in a color-magnitud diagram is presented in Section \ref{CCMMDD}, and their morphology is examined in Section \ref{morfo_ACS}. The radio emission of our SF galaxies at $z \sim 1$ is shown in Section \ref{radio_VLA}. Finally, we summarize the main conclusions of the work in Section \ref{conclusions}.

Throughout this paper we assume a flat universe with $(\Omega_m, \Omega_\Lambda, h_0)=(0.3, 0.7, 0.7)$, and all magnitudes are listed in the AB system \citep{Oke1983}.

\section{UV-selected galaxies and their FIR detections}\label{PACS_counterparts}

In \cite{Oteo2013} we selected those LBGs whose Lyman break is located between the GALEX FUV and NUV filters, which are centered at the effective wavelengths 1528\AA\ and 2271\AA\ and have bandwidths of 1344-1786 \AA\ and 1771-2831 \AA, respectively. The LBGs were selected by imposing the color criterion:

\begin{equation}\label{criterion}
FUV-NUV>1.5 
\end{equation}

Additionally, and in order of ruling out as much interlopers as possible, we limit the photometric redshift range of the sources to $0.8 \leq z \leq 1.2$. It should be noted that the combination of GALEX and ALHAMBRA data provides accurate photometric redshifts at $z \sim 1$ \citep{Oteo2013}.

Those galaxies that are undetected in the FUV channel due to the presence of a strong Lyman break were also included in the sample. FUV observations in the COSMOS field have a limiting magnitude of $FUV \sim 27 \,{\rm mag}$ \citep{Oteo2013}. Then, galaxies brighter than that value in the wavelength range covered by the FUV filters should be detected. Since we select galaxies with a $FUV-NUV$ color cut of 1.5, that limiting magnitude would imply a limit of 25.5 mag in the NUV channel. To be conservative, we choose a limit in the NUV channel of 25 mag and include in the previous sample of LBGs those galaxies that are brighter than 25 mag in the NUV channel, are at $0.8 \leq z \leq 1.2$, and are undetected in the FUV channel. The non detection in the FUV channel under those restrictions is likely due to a strong Lyman break. AGN contamination was ruled out by discarding galaxies with X-ray detections \citep{Elvis2009,Salvato2011}. In total, 1225 GALEX-selected LBGs are segregated, with a median redshift of 0.93.

The photometric redshift and other properties such as age, dust attenuation, stellar mass, or UV continuum slope were derived by fitting \cite{Bruzual2003} (hereafter BC03) templates to their observed GALEX+ALHAMBRA fluxes. These templates were built by assuming a \cite{Salpeter1955} initial mass function distributing stars from $0.1 M_\odot$ to $100 M_\odot$, a fixed value of the metallicity ($Z=0.2Z_\odot$) and a time-independent star formation history (SFH). We considered a fixed value of metallicity \citep[see also][]{Gonzalez2011,Stark2013} because it is a parameter difficult to constrain with an SED-fitting procedure \citep[see for example]{Debarros2012,Oteo2012a}. Actually, by running ZEBRA with BC03 templates associated to different metallicities we  obtained in \cite{Oteo2013} very similar values of the $\chi^2_r$ for individual galaxies, preventing us from constraining the metallicity of our galaxies. The changes in the SED-derived parameters associated to fits with different metallicities are not strongly significant, but they are within their typical uncertainties. The choice of the constant SFR is justified for the comparison with the results obtained with the \cite{Kennicutt1998} calibrations. Therefore, the SED-derived parameters with BC03 templates presented in this work should be understood as those obtained with those assumptions of metallicity and SFH. We considered values of age from 1 Myr to 7 Gyr in steps of 10 Myr from 1 Myr to 1 Gyr and in steps of 100 Myr from 1 Gyr to 7 Gyr. Dust attenuation is included in the templates via the \cite{Calzetti2000} law and parametrized through the color excess in the stellar continuum, $E_s(B-V)$, for which we select values ranging from 0 to 0.7 in steps of 0.05. We also include intergalactic medium absorption adopting the prescription of \cite{Madau1995}. The fits are carried out by using the Zurich Extragalactic Bayesian Redshift Analyzer \citep[ZEBRA,][]{Feldmann2006} code which, in its maximum-likelihood mode, employs a $\chi^2$ minimization algorithm over the templates to find the one which fits the observed SED of each input object best. We have not included the contribution of emission lines in the SED fits \citep{Zackrisson2008,Schaerer2009,Schaerer2010,Schaerer2011,Atek2011,Schaerer2013,deBarros2012} since we will deal with dusty FIR-detected galaxies and, according to \cite{deBarros2012}, the contribution of emission lines in their SED-derived parameters is not expected to be significant. Each fitted template has a characteristic value of the $\chi^2$. We define the reduced $\chi^2$, $\chi^2_r$, as the ratio between the $\chi^2$ associated to each fitted template and the numbers of filters minus one employed in the SED-fitting procedure, $\chi^2_r=\chi^2/(N-1)$ \citep[see for example][]{Debarros2012}. Only galaxies with $\chi^2_r < 10$ where included in the sample. The uncertainties in the SED-derived parameters were derived as the difference between the best-fitted value and its corresponding weighted average, defined as  $WA=\sum_i^N P_i f_i/N$, where $P_i$ is the probability that a given template, $i$, can represent the observed SED of a given galaxy, $f_i$ is the value of one of the physical properties associated to the $i-th$ template, and $N$ is the number of templates (see \cite{Oteo2013} for more details). In the case of the UV continuum slope, the uncertainties considered here are those raising from the fit of the UV continuum to the exponential law.

In \cite{Oteo2013} we obtained that the photometric redshifts derived with the combination of GALEX and ALHAMBRA data are very accurate, $\sigma_z = (|z_{\rm phot} - z_{\rm spec}|)/(1+z_{\rm spec}) < 0.05$, for most of our galaxies. As in \cite{Oteo2013}, we adopt the spectroscopic redshifts for those galaxies with available optical spectroscopy in the zCOSMOS survey \citep{Lilly2007} to redo the SED fits. For those galaxies without zCOSMOS counterpart, we employ the photometric redshift and their associated SED-fitting results. Once each SED is fitted with a BC03 template, the UV continuum slope is obtained for each galaxy directly from its best-fit template by fitting a power law in the form $f_\lambda \sim \lambda^\beta$ \citep{Calzetti1994} to their UV continuum. In this process we employ the photometric points corresponding to the rest-frame wavelength range between $1300\AA \lesssim \lambda \lesssim 3000\AA$. This approach has the advantage of using all the available fluxes of each source, resulting in higher S/N determinations. The rest-frame UV luminosity, defined in a $\nu L_\nu$ way, for each source is obtained from its best fitted template shifted to the rest-frame and by convolving it with a top-hat filter, 300\AA\ width, centered in 1500 \AA.

With the aim of placing LBGs in a general scenario of FIR-detected SF galaxies at $0.8 < z < 1.2$ we select photometrically a sample of UV-detected galaxies in the same redshift range than the GALEX-selected LBGs: we consider all the galaxies detected in GALEX (at least in the NUV band) that have an ALHAMBRA counterpart, are detected in PACS-100$\mu$m or PACS-160$\mu$m, and their photometric redshifts are within $0.8 \leq z \leq 1.2$ It should be pointed out that LBGs are contained in this general sample of UV-selected galaxies. The subsample of UV-selected and PACS-detected galaxies which are not LBGs will be called \emph{UV-faint galaxies}. We emphasize that we consider LBGs and UV-faint galaxies independently because the main goal of this work is analyzing the FIR SED of the intermediate-redshift analogs of the color-selected high-redshift LBGs.

Due to the selection criteria, the main difference between LBGs and UV-selected galaxies is their rest-frame UV luminosity. The selection of LBGs in the FUV channel or the condition $NUV < 25$ for those FUV-undetected implies that GALEX-selected LBGs are intrinsically brighter in the rest-frame UV. The general sample of UV-selected galaxies contains UV-fainter galaxies than the LBGs because their only restriction is the photometric redshift and the detection in the NUV and FIR (regardless the FUV and NUV magnitude). 

\begin{figure*}
\centering
\includegraphics[width=0.3\textwidth]{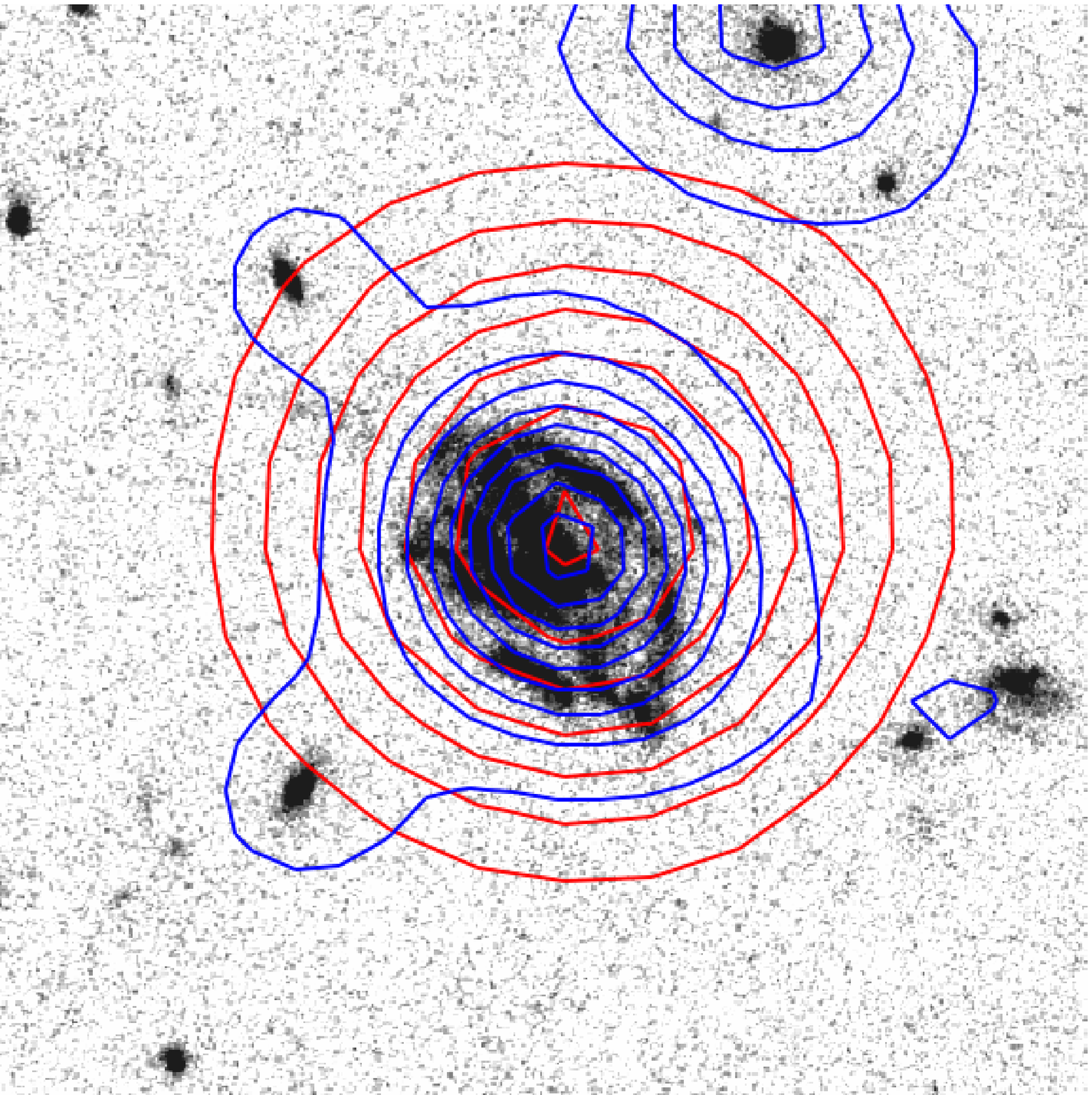}
\includegraphics[width=0.3\textwidth]{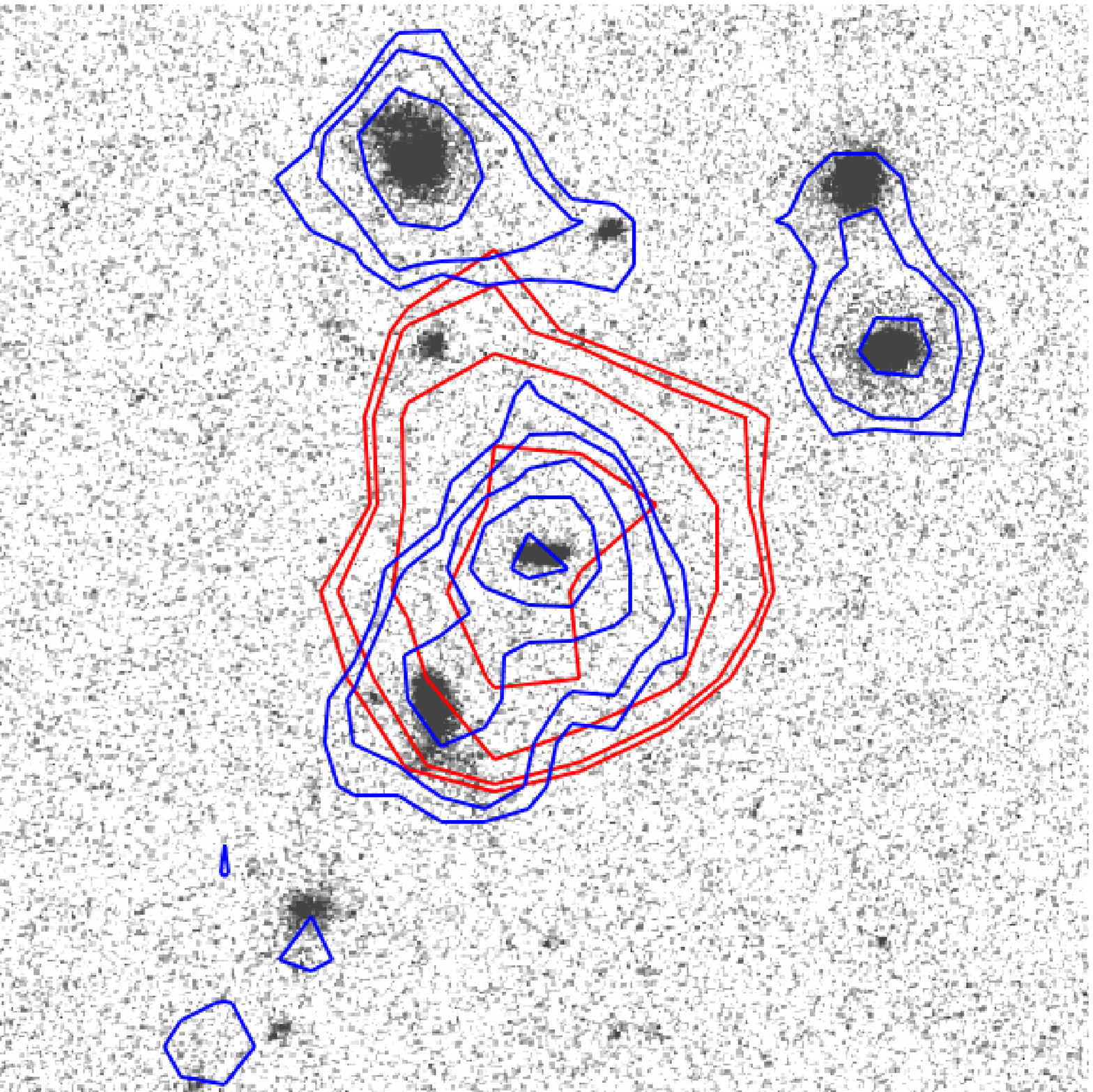}
\caption{ACS I-band images along with the IRAC-3.6$\mu$m (blue) and MIPS-24$\mu$m (red) contours for two PACS-detected LBGs. The PACS-detected LBGs are the galaxies located in the center of each image. To show how good are MIPS coordinates with respect to optical data we have selected for the plot one of the best cases and one of the worst cases of FIR identifications. \emph{Left}: Favorable case where the PACS-detected LBG is a big galaxy without close neighbors. \emph{Right}: Unfavorable case where the PACS-detected galaxy is small and it is surrounded by some close neighbors. In the two cases, it is clear that the MIPS-24$\mu$m emission (and therefore the PACS emission) truly comes from the PACS-detected LBG.             
}
\label{nnnnn}
\end{figure*}

The PACS FIR data employed in this work are taken within the framework of the PACS Evolutionary Probe (PEP) Herschel guaranteed time key project, optimized for FIR studies of galaxy evolution \citep{Lutz2011}. As part of the PEP project, COSMOS field was targeted in PACS-100$\mu$m and PACS-160$\mu$m. The PACS fluxes used in this study were extracted by using MIPS-24$\mu$m priors \citep{Magnelli2009,LeFloch2009}. The usage of MIPS-24$\mu$m priors is justified because the 24$\mu$m emission is strongly correlated to the FIR emission and, therefore, the catalogues created will not contain a large excess of sources without FIR counterparts. This substantially avoids deblending FIR sources into several unrealistic counterparts, as it might have occur if, for examples, optical based catalog with very high source density, had been employed. The catalogues were extracted by using a PSF-fitting method. The PSFs were directly derived from the science maps when performing a blind source extraction with the code Starfinder \citep{Diolaiti2000,Diolaiti2000b} and were cut to a 3 pixel radius to extract the catalogues in order to minimize filtering effects on the wings of bright sources. The aperture corrections were determined by independent observations of very bright objects, such as the asteroid Vesta, by simply accounting for the limited radius in the observed PSFs. More details of the reduction and source extraction can be consulted in \cite{Lutz2011} and references therein. The limiting $3\sigma$ fluxes are 5.0 and 11.0 mJy in the PACS-100$\mu$m and PACS-160$\mu$m bands, respectively. 

\begin{figure}
\centering
\includegraphics[width=0.45\textwidth]{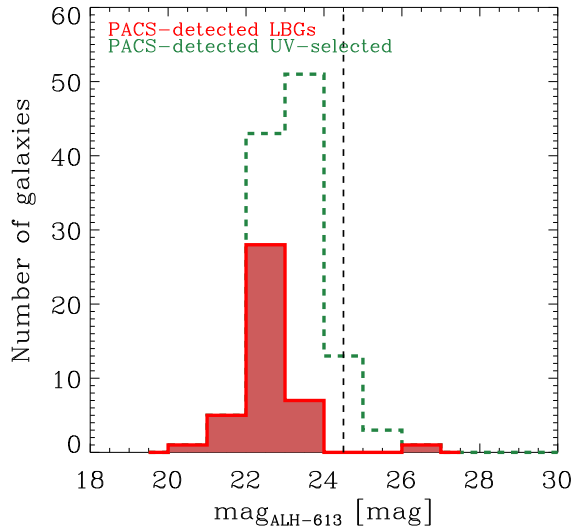}
\caption{Distribution of our PACS-detected LBGs (red filled histogram) and PACS-detected UV-selected galaxies (green dashed histogram) magnitude in the ALH-613 band. The vertical dashed line represents a magnitude of 24.5 mag and indicates the approximate minimum brightness limit down to which the SED-fitting in the UV, optical, and near-IR with ALHAMBRA data can be considered reliable.     
}
\label{mag_613_PACS}
\end{figure}

We look for the possible PACS detections of our UV-selected galaxies by using their optical ALHAMBRA positions as reference and considering a matching radius of 2''. It should be noted that since we are using PACS catalogs built with MIPS-24$\mu$m priors, the coordinates of the sources detected by PACS are those corresponding to the MIPS-24$\mu$m observations. We find that 42 LBGs and 65 UV-faint galaxies are individually detected either in PACS-100$\mu$m or PACS-160$\mu$m. We call PACS-detected galaxies to those galaxies which are detected in, at least, one PACS band. Table \ref{table_detections} summarizes the detections in each PACS band for each kind of PACS-detected galaxy. By inspecting the ACS I-band (F814W) images of the PACS-detected galaxies (see Section \ref{morfo_ACS}) we do not find evidence that the FIR fluxes of our sources are contaminated by close neighbors. In Figure \ref{nnnnn} we show the optical ACS I-band cutouts along with the IRAC-3.6$\mu$m and MIPS-24$\mu$m contours of two PACS-detected LBGs. With the aim of studying the possible effect of source confusion and analyzing how good are the MIPS-based coordinates (employed as priors) and the optical ones we represent one of the best favorable (big galaxy with no close companions) and the least favorable cases (where it can be thought that there is source confusion with close neighbors). It can be seen that in the two cases, the MIPS-24$\mu$m emission (and therefore the PACS emission) truly comes from the PACS-detected LBG and, thus, there is no evidence of source confusion. In \cite{Oteo2013} we found that the SED-fitting with ALHAMBRA data in the optical and near-IR could be consider reliable only for galaxies brighter than 24.5 mag in the ALH-613 band. Fainter galaxies in the optical have low values $\chi^2_r$ likely due due to high photometric uncertainties rather than to truly good fitting. In Figure \ref{mag_613_PACS} we represent the optical ALH-613 magnitude of our PACS-detected galaxies. It can be seen that most of the galaxies are brighter than the previous value and, therefore, the values of $\chi^2_r$ are likely due to a good SED-fitting rather than high photometric uncertainties. Additionally, Figure \ref{mag_613_PACS} indicates that PACS-detected LBGs are typically brighter ($m_{\rm ALH-613} \sim 22.4$) than UV-faint galaxies ($m_{\rm ALH-613} \sim 23.2$). This is a direct consequence of their selection criteria. Furthermore, PACS-detected LBGs are among the optical brightest galaxies among the whole sample of GALEX-selected LBGs presented in \cite{Oteo2013}.

\begin{table}
\caption{\label{table_detections}Summary of the stacking results for the different bins of rest-frame UV luminosity, stellar mass, and age (see text for more details).}
\centering
\begin{tabular}{cccc}
\hline\hline
 Galaxy & PACS-100$\mu$m & PACS-160$\mu$m & Both PACS bands \\
 LBGs	&	13	&	11	&	18	\\
 UV-faint	&	23	&	14	&	28	\\
\hline
\hline
\end{tabular}
\end{table}

\begin{figure*}
\centering
\includegraphics[width=0.33\textwidth]{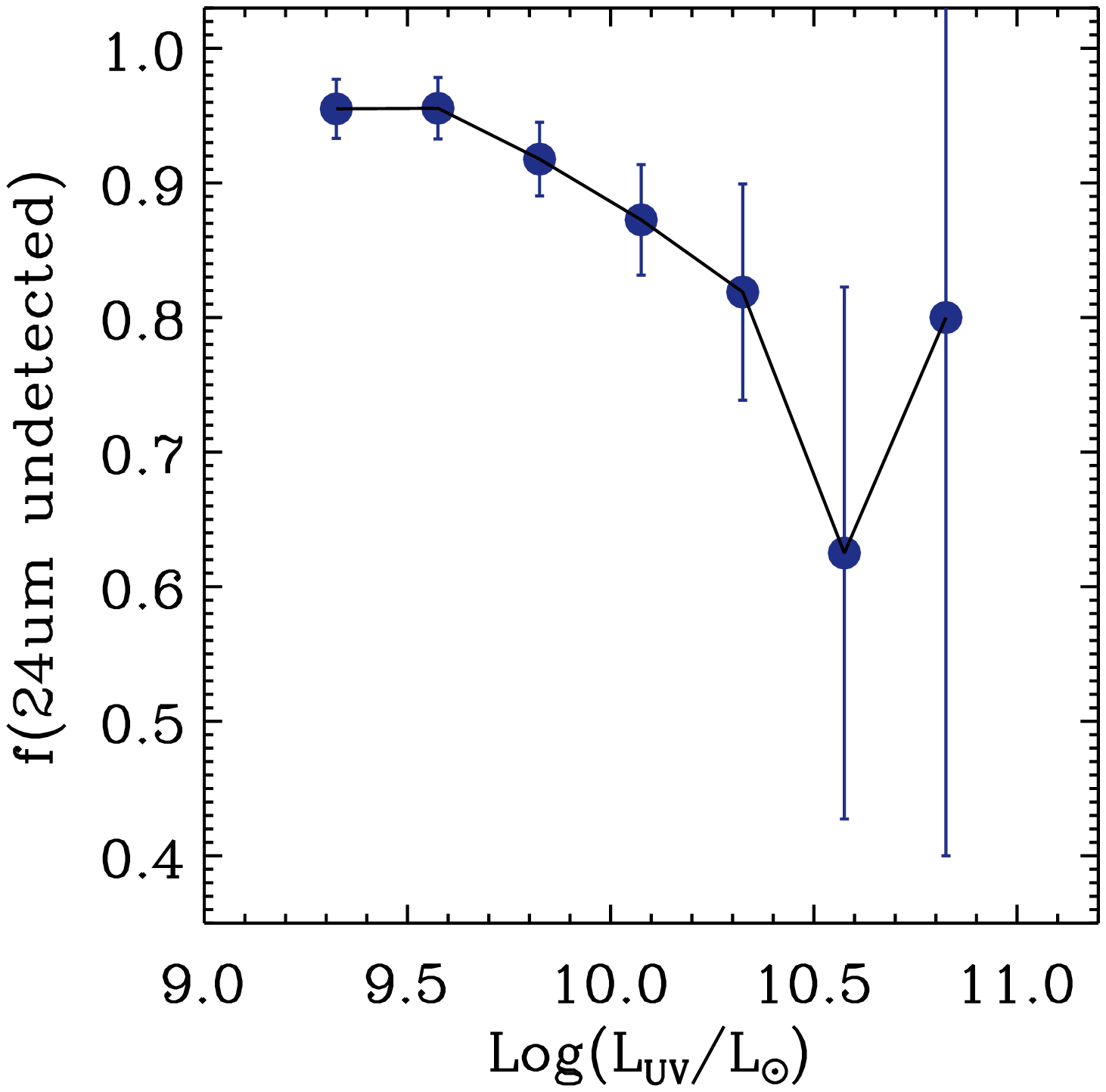}
\includegraphics[width=0.33\textwidth]{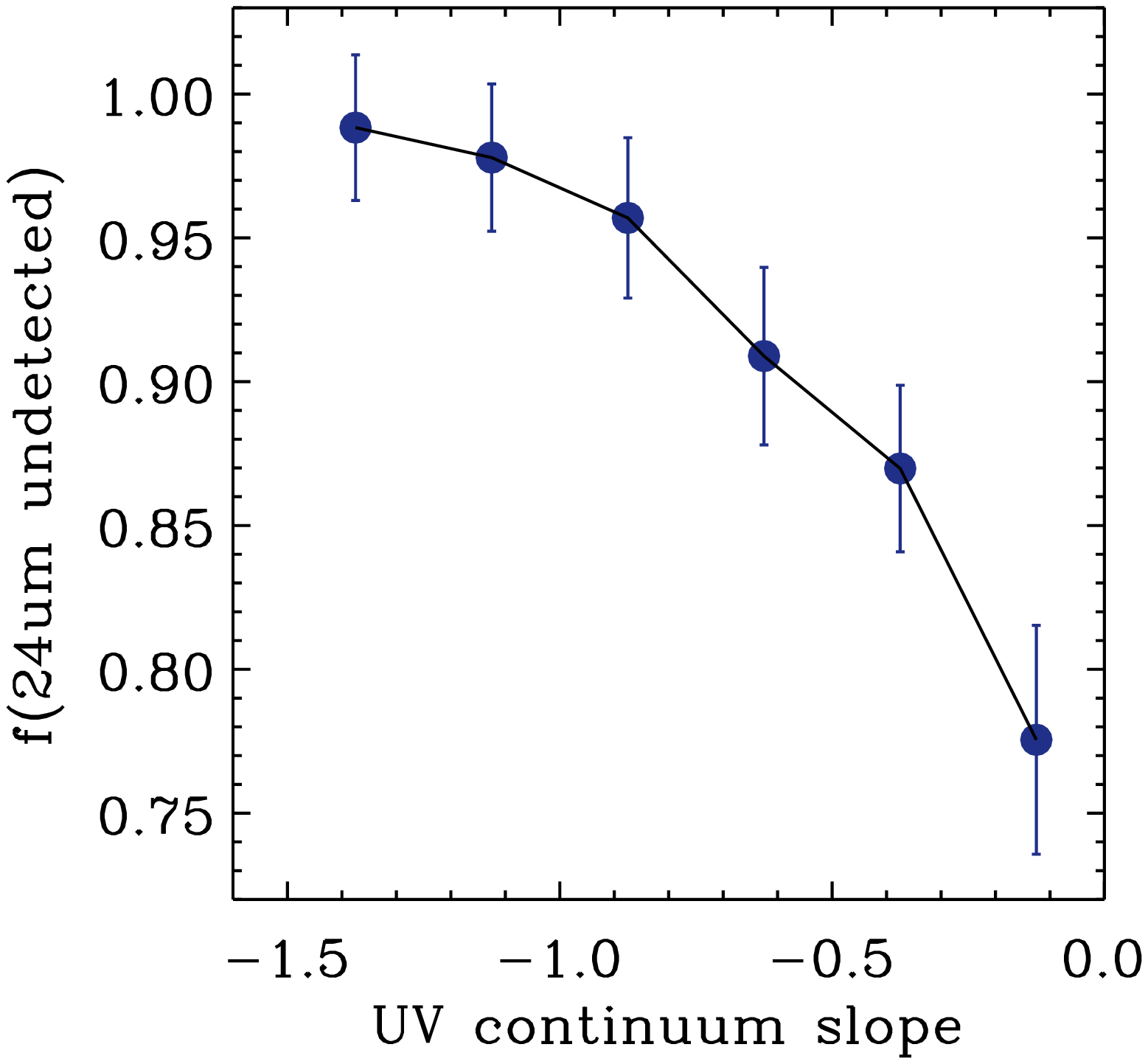} \\
\includegraphics[width=0.33\textwidth]{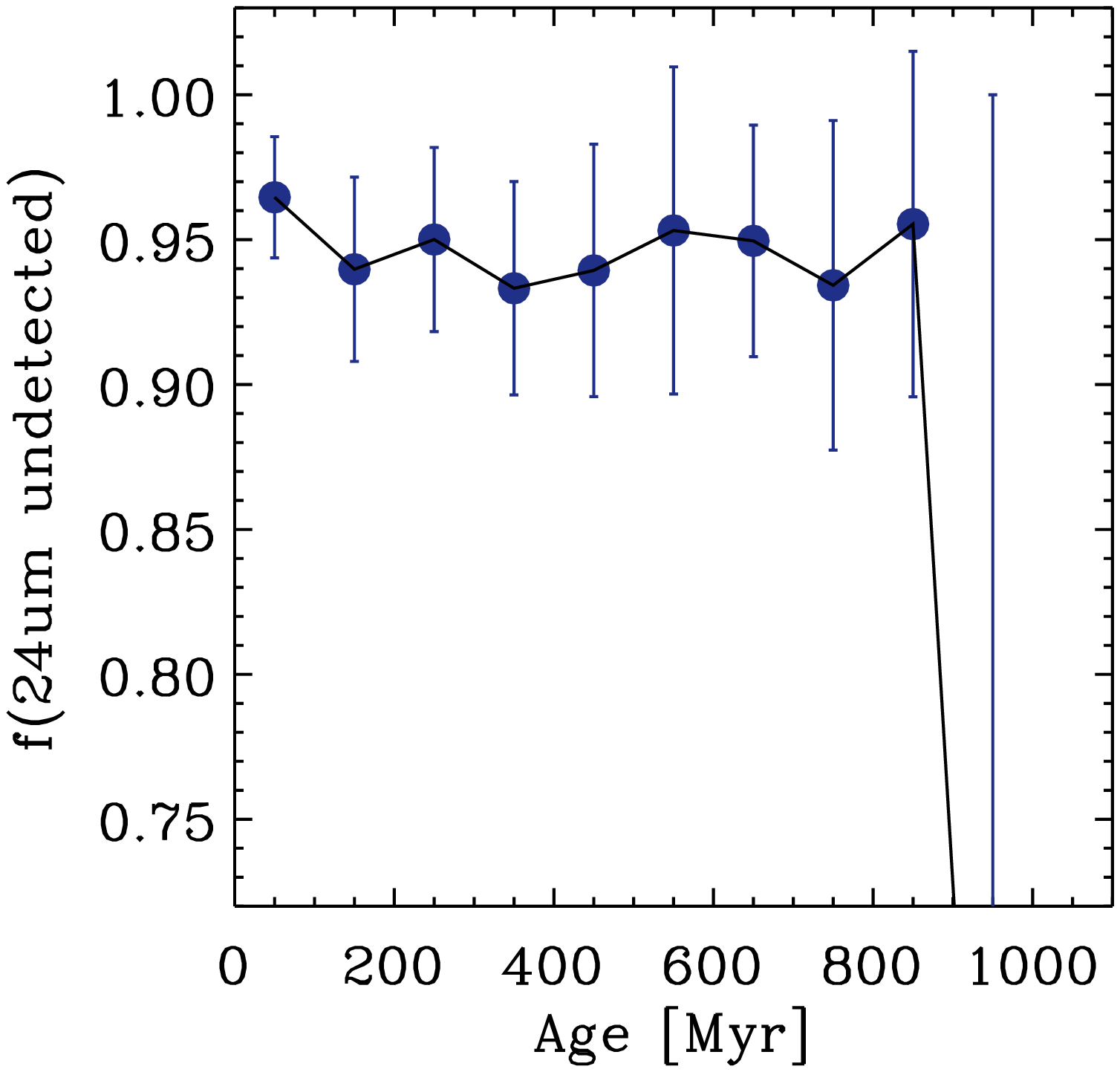}
\includegraphics[width=0.33\textwidth]{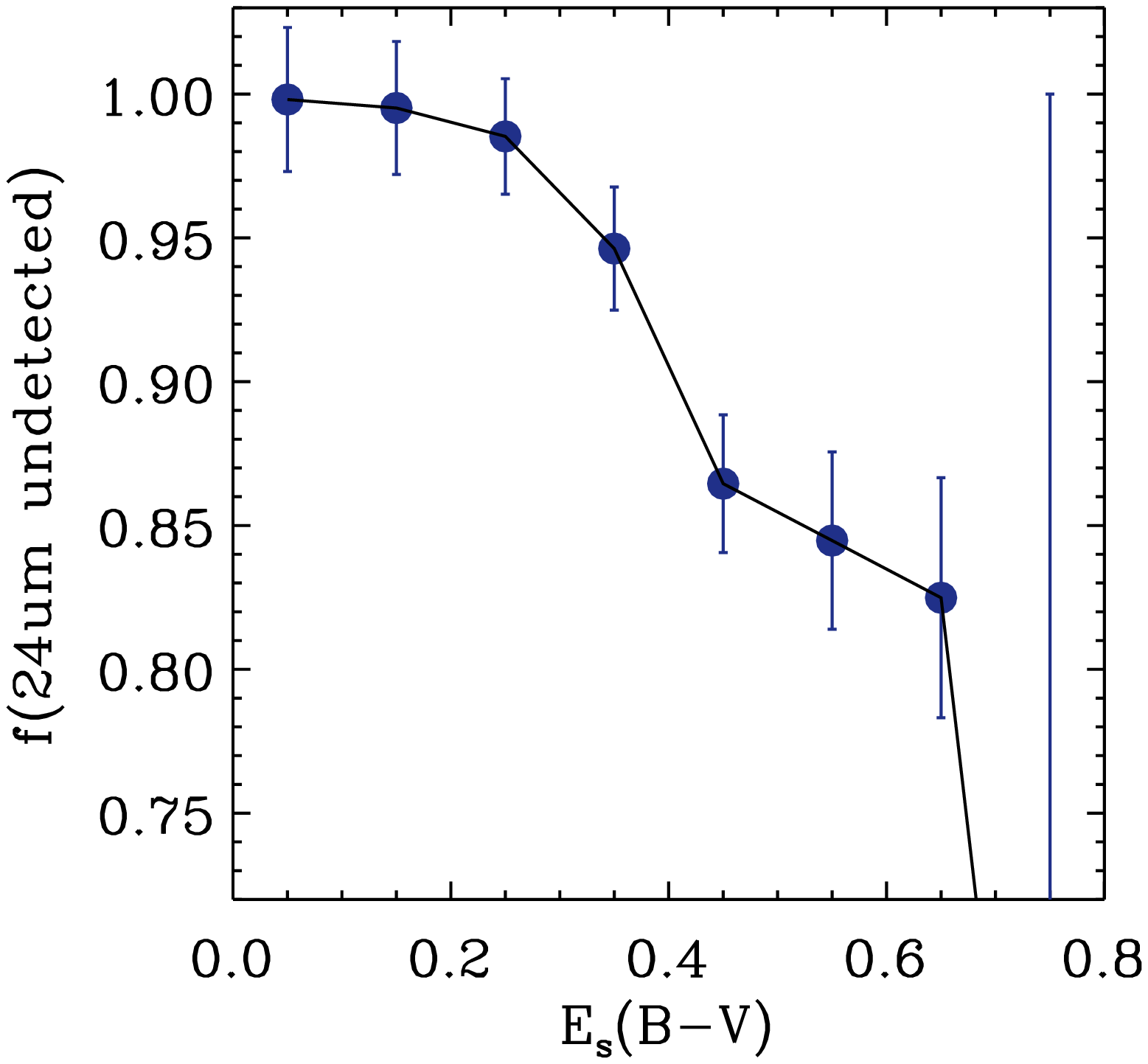}
\includegraphics[width=0.33\textwidth]{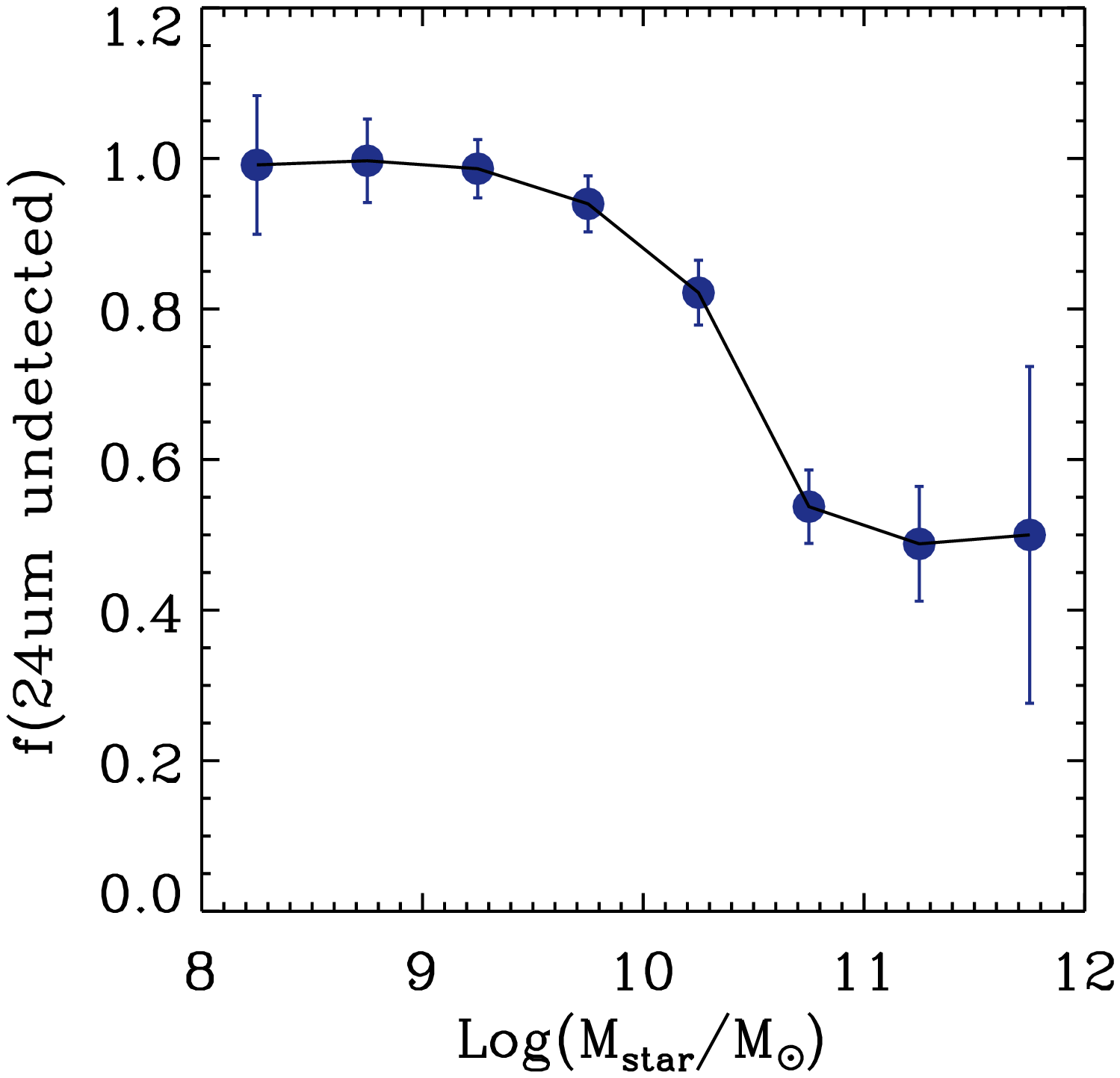}
\caption{Undetected fraction in MIPS-24$\mu$m of the whole population of PACS-detected LBGs and UV-faint galaxies at $z \sim 1$ as a function of their rest-frame UV luminosity, UV continuum slope, age, dust attenuation and stellar mass. Error bars have been obtained assuming Poisson statistic. The wider error bars in the points associated to the UV-brightest, oldest, dustiest, and most massive galaxies are due to a low number of galaxies in those specific bins.
              }
\label{mipsdetections}
\end{figure*}

Summarizing, we define the samples of PACS-detected LBGs, UV-selected, and UV-faint galaxies as:

\begin{eqnarray}
\{ {\rm LBGs:} \} = \{ {\rm GALEX+ALHAMBRA:} \nonumber \\ 
(i) \quad FUV-NUV > 1.5 \quad {\rm or} \nonumber \\ NUV < 25 \quad {\rm and} \quad FUV-undetected \nonumber \\
(ii) \quad 0.8 < z < 1.2 \nonumber \\
(iii) \quad {\rm non-AGNs} \nonumber \\
(iv) \quad {\rm PACS-detected} \} \nonumber
\end{eqnarray}

\begin{eqnarray}
\{ {\rm UV-selected:} \} = \{ {\rm GALEX+ALHAMBRA:} \nonumber \\ 
(i) \textrm{NUV-detected with no restriction on FUV} \nonumber \\
(ii) 0.8 < z < 1.2 \nonumber \\
(iii) {\rm non-AGNs} \nonumber \\
(iv) {\rm PACS-detected} \} \nonumber
\end{eqnarray}

\begin{eqnarray}
\{ {\rm UV-faint} \} = \{ {\rm UV-selected} \} - \{ {\rm LBGs} \} \nonumber
\end{eqnarray}

\begin{figure*}
\centering
\includegraphics[width=0.3\textwidth]{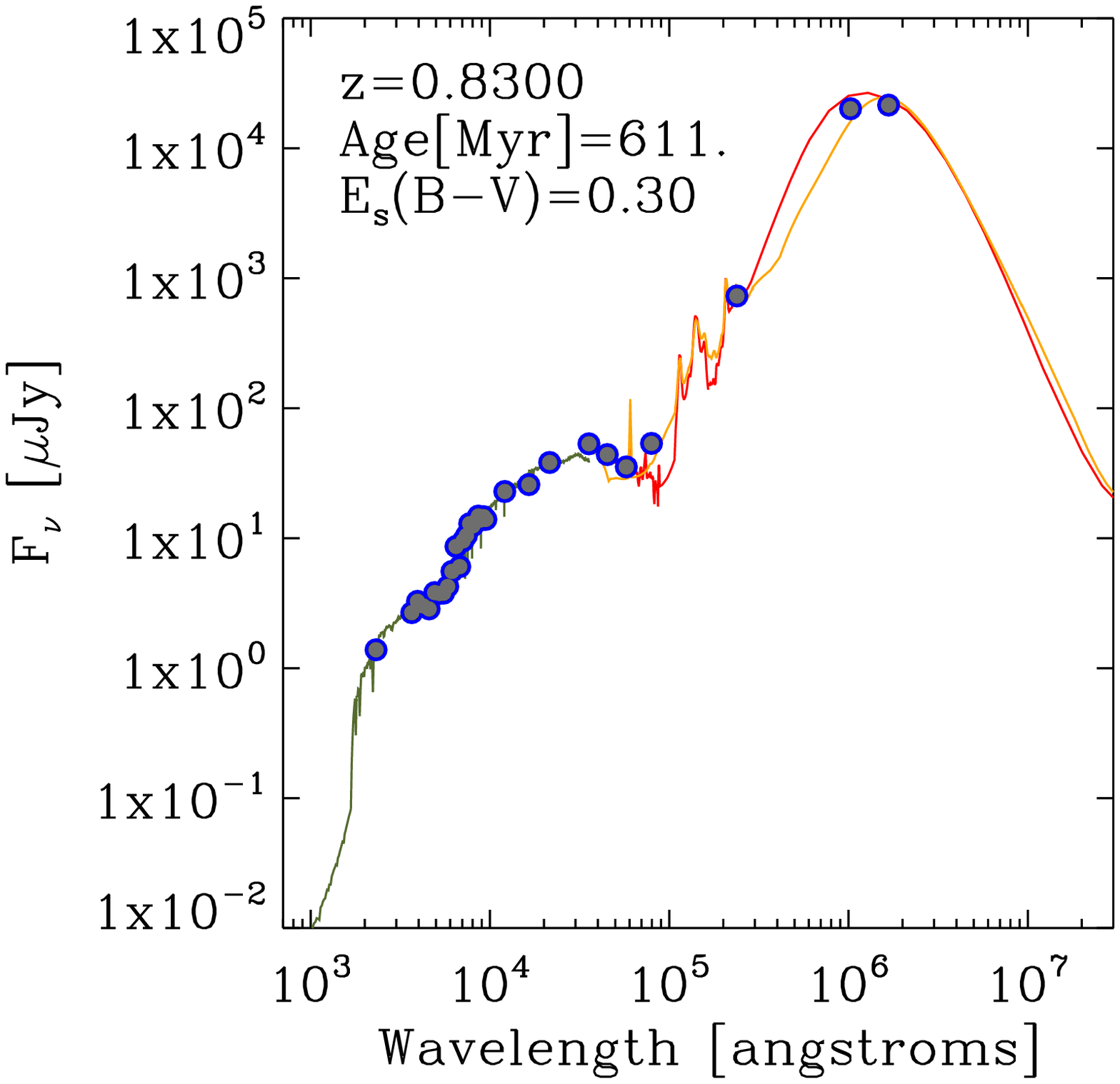}
\includegraphics[width=0.3\textwidth]{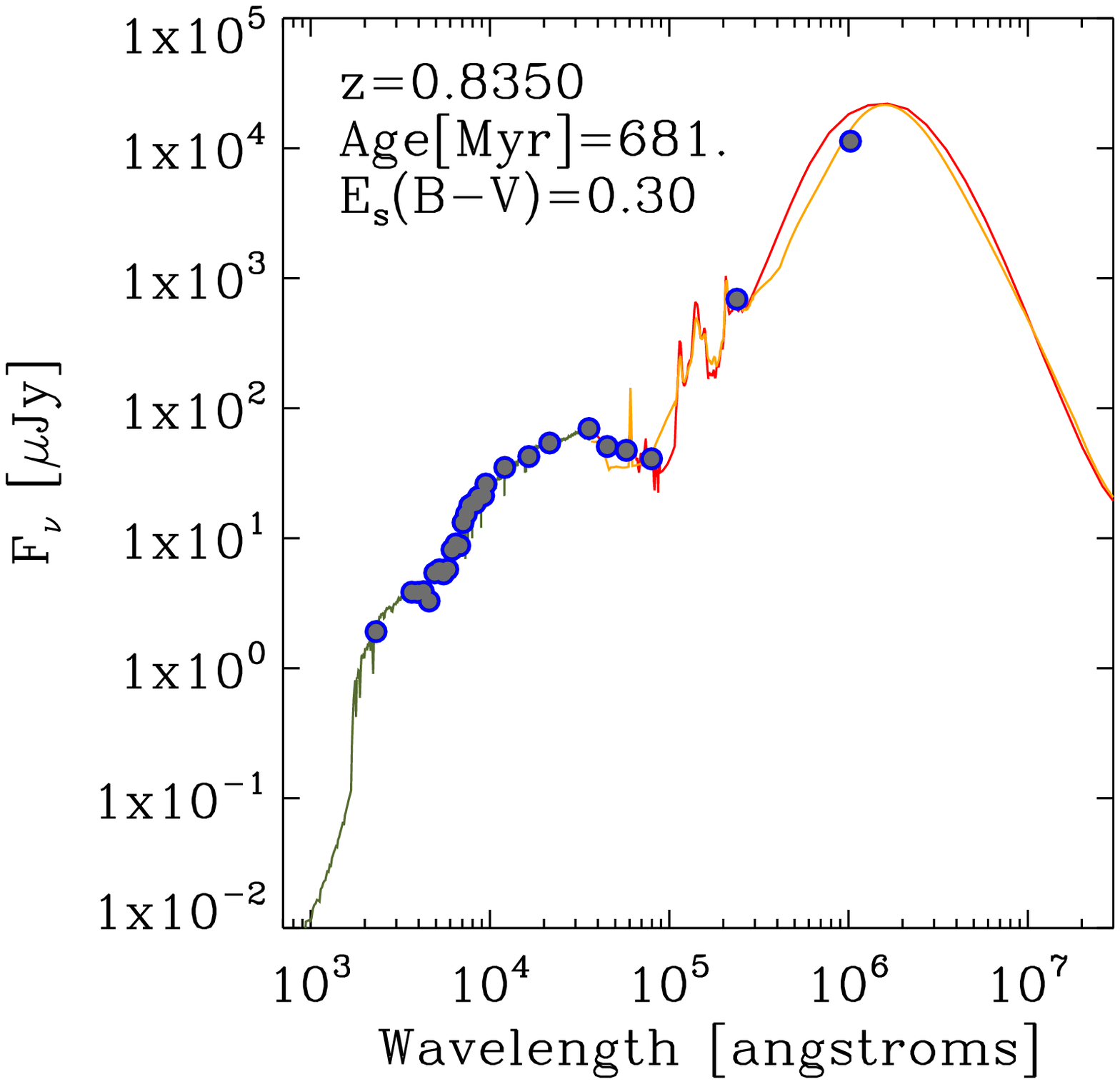}
\includegraphics[width=0.3\textwidth]{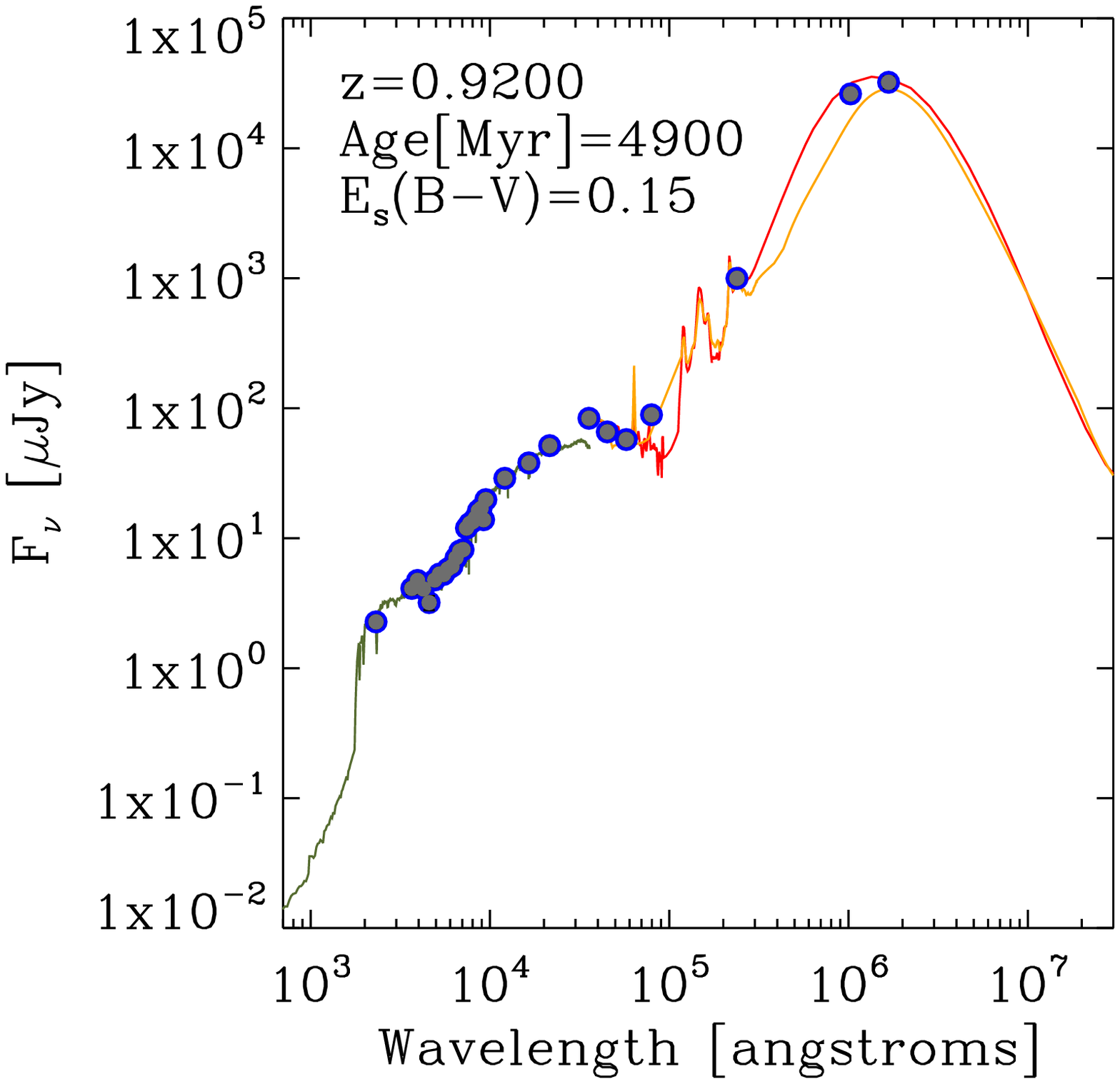}

\includegraphics[width=0.3\textwidth]{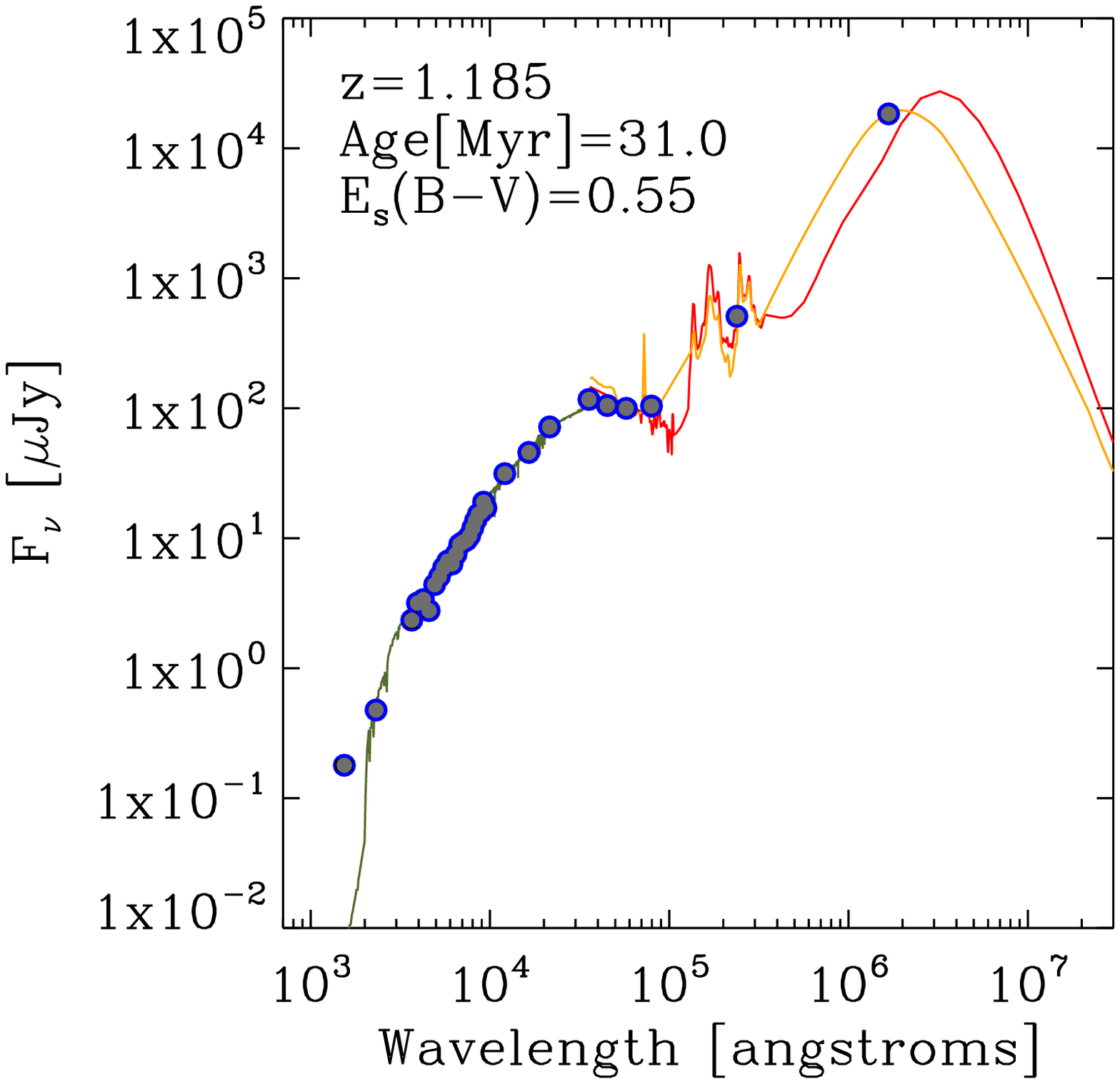}
\includegraphics[width=0.3\textwidth]{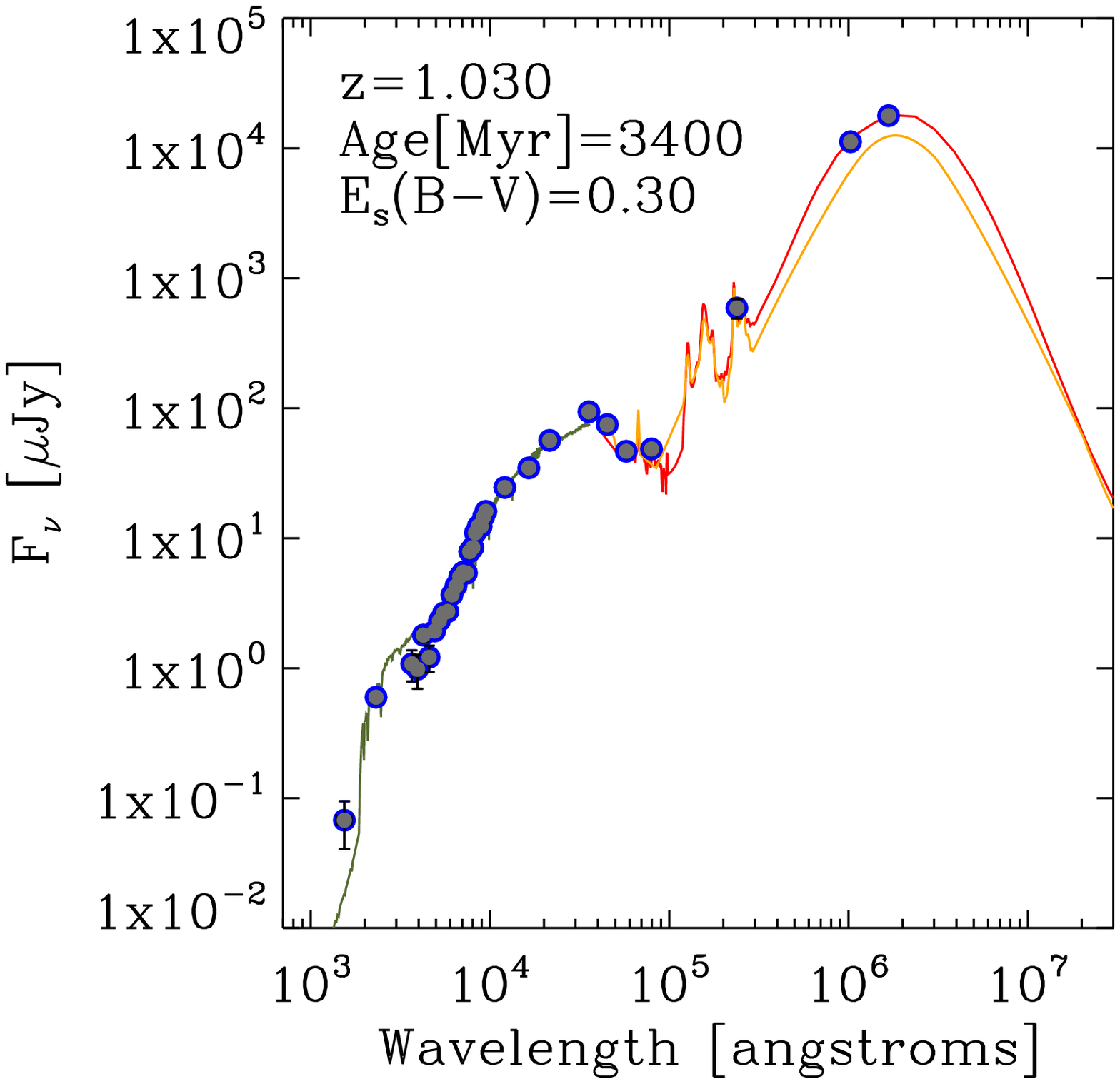}
\includegraphics[width=0.3\textwidth]{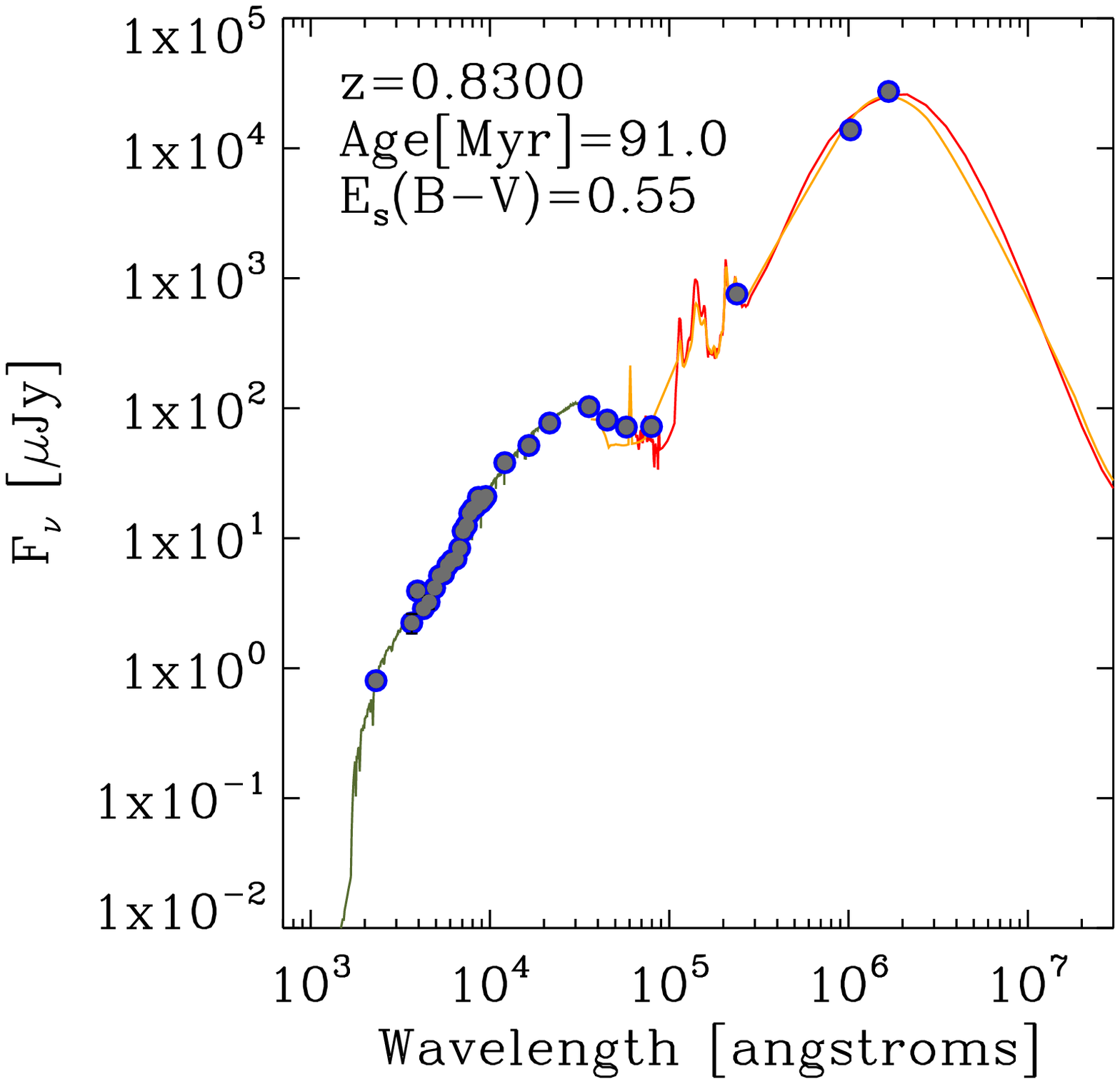}

\includegraphics[width=0.3\textwidth]{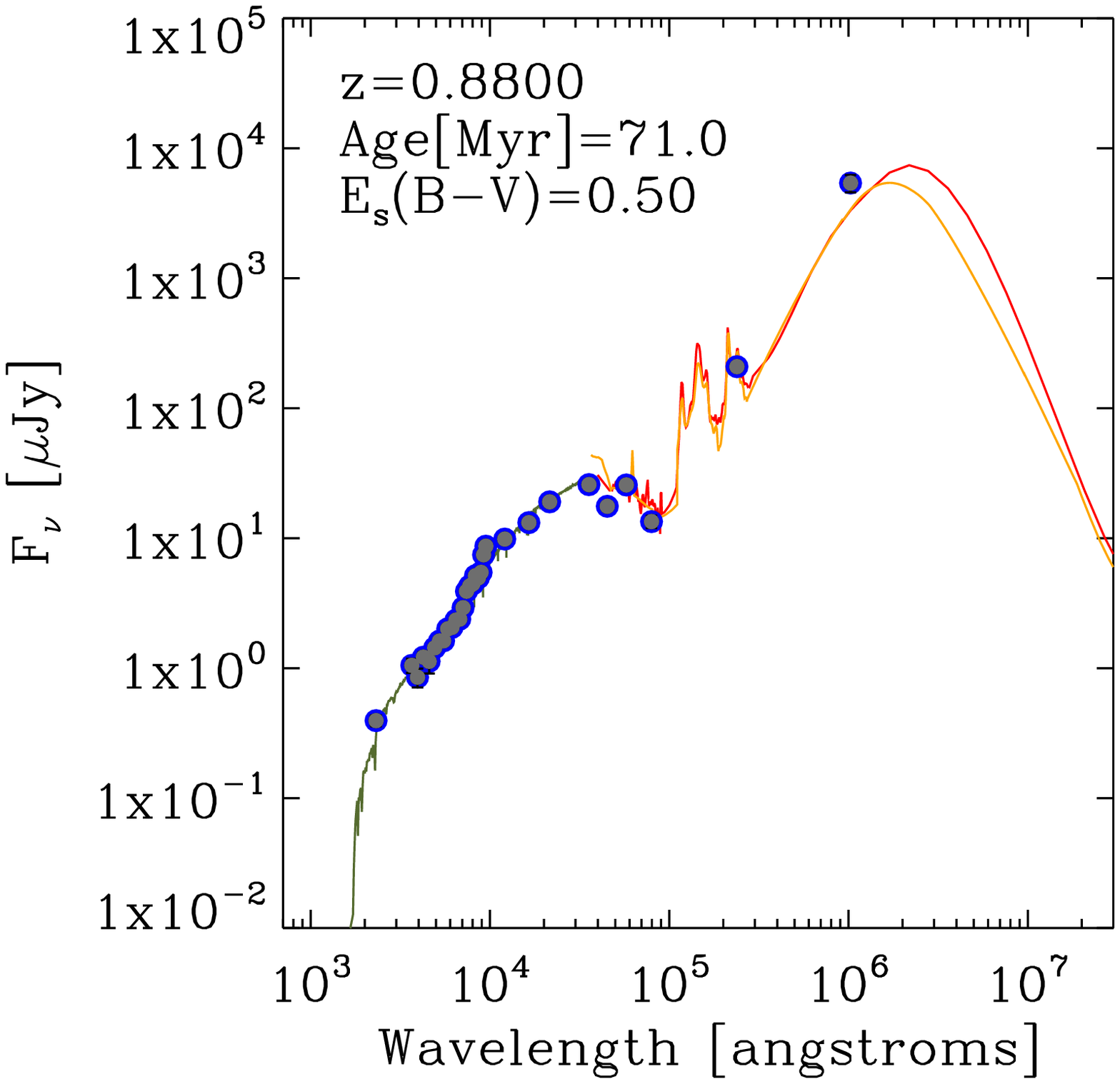}
\includegraphics[width=0.3\textwidth]{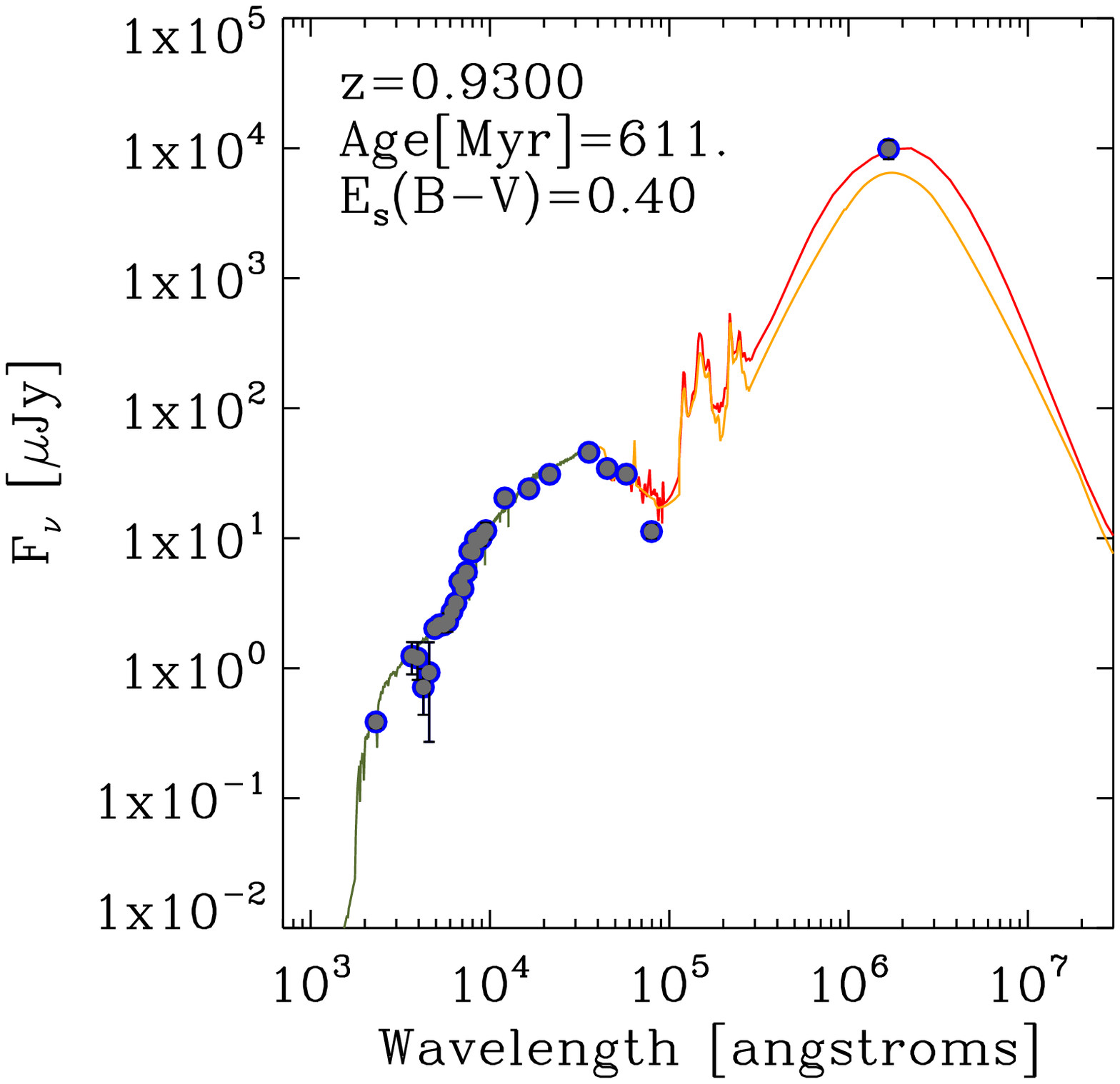}
\includegraphics[width=0.3\textwidth]{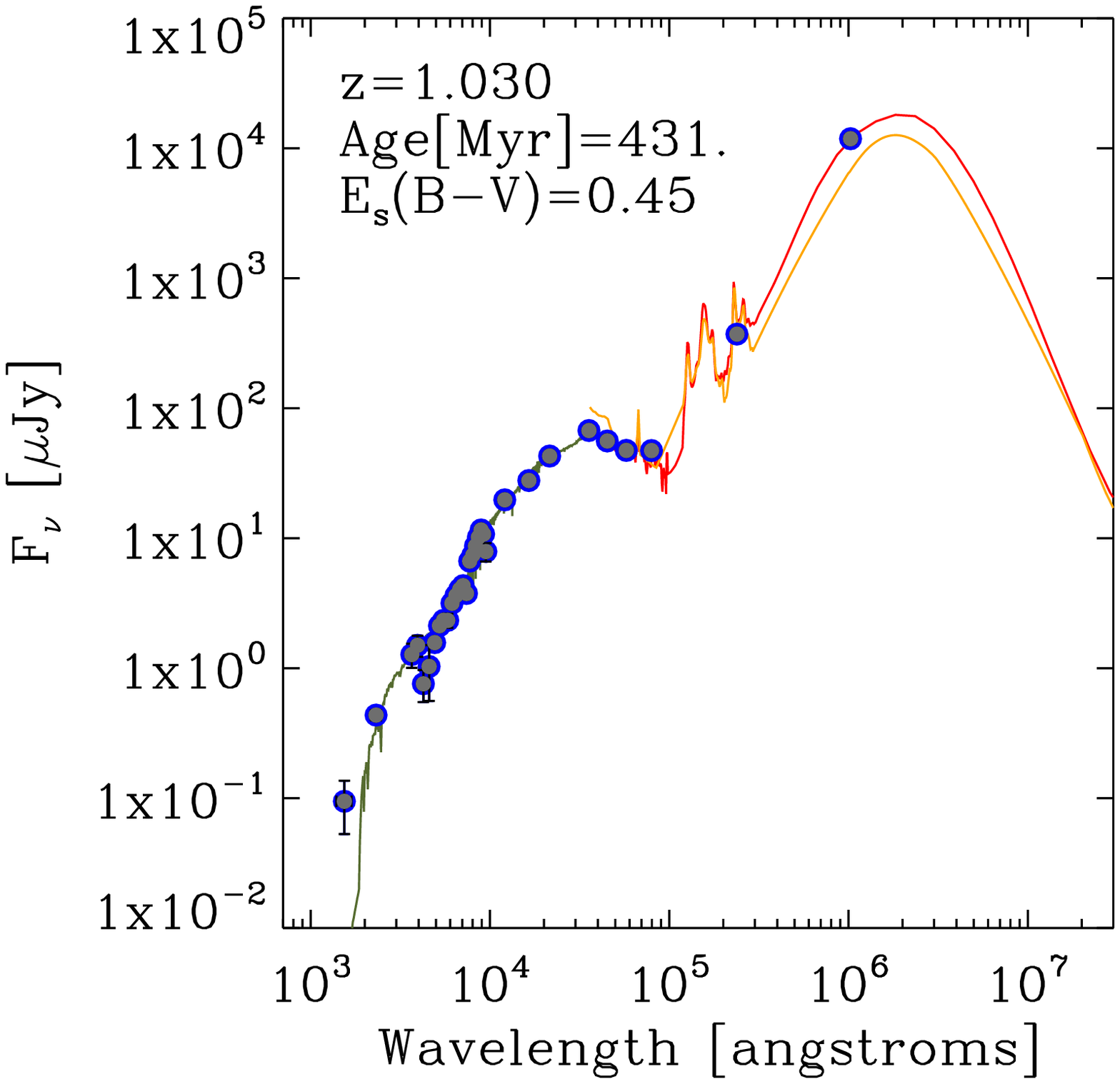}
\caption{Rest-frame UV-to-FIR SED of nine randomly selected PACS-detected LBGs. These examples are representative of the SED-fitting results for our whole sample of our PACS-detected galaxies. The grey points are the observed UV to FIR fluxes of the galaxies. The green curves are the best-fitted \citet{Bruzual2003} templates to the UV to IRAC-4.5$\mu$m fluxes of each galaxy. The \citet{Bruzual2003} templates considered in the representations and fits are associated to time-independent SFH and a fixed metallicity of $Z=0.2Z_\odot$ (see \citet{Oteo2013} for more details). SED-derived redshift, age and dust attenuation associated to the represented \citet{Bruzual2003} templates are also indicated. The orange and red curves are the best-fitted \citet{Chary2001} and \citet{Dale2002} templates, respectively, to the IRAC-5.8$\mu$m to PACS fluxes of each galaxy. 
              }
\label{SEDs}
\end{figure*}

Figure \ref{SEDs} shows the UV-to-FIR SED of nine PACS-detected LBGs that represent the typical SED fitting results for our galaxies. It can be seen that the combination of GALEX, ALHAMBRA, IRAC, MIPS, and PACS data provides an excellent photometric coverage of the SEDs of the studied galaxies. 

Mid- and far-infrared detection rates of UV-selected galaxies vary with basic properties of the galaxies. Because of the larger detection rate in the mid-IR compared to the far-IR we discuss mid-IR detection rates, but trends in the FIR are consistent. Figure \ref{mipsdetections} gives the detection rate in MIPS-24$\mu$m of the whole sample of UV-selected galaxies at $0.8 < z < 1.2$ built in \cite{Oteo2013}, as a function of their SED-derived rest-frame UV luminosity, UV continuum slope, age, dust attenuation, and stellar mass. MIPS-24$\mu$m detections were searched by using a matching radius of 2'' and using the data coming from the S-COSMOS survey \citep{Sanders2007}. 


First, it can be seen that the detection rate in MIPS-24$\mu$m increases with the rest-frame UV luminosity, i.e. UV-bright galaxies are more likely detected in the FIR. This trend has been also reported in other works which analyze the IR properties of UV-selected galaxies at lower \citep[see for example][]{Oteo2012b} and higher redshifts \citep[see for example][]{Reddy2010}. Furthermore, the SPIRE detected LBGs reported in \cite{Burgarella2011} are UV-bright galaxies, with an average value of $\log{\langle L_{UV}/L_\odot \rangle}$=10.7, being most of them UV-luminous galaxies. Other clear tendencies between the detection rate in the FIR and the SED-derived physical properties of galaxies are those related to the UV continuum slope, dust attenuation, and stellar mass. Figure \ref{mipsdetections} indicates that galaxies which have redder UV continuum and are dustier and more massive are more likely detected in the FIR. We do not find any significant relation between the MIPS-24$\mu$m detection rate and the age of the galaxies.

\subsection{Rest-frame UV, IR, and bolometric luminosities}\label{lumi_IR}

\begin{figure*}
\centering
\includegraphics[width=0.45\textwidth]{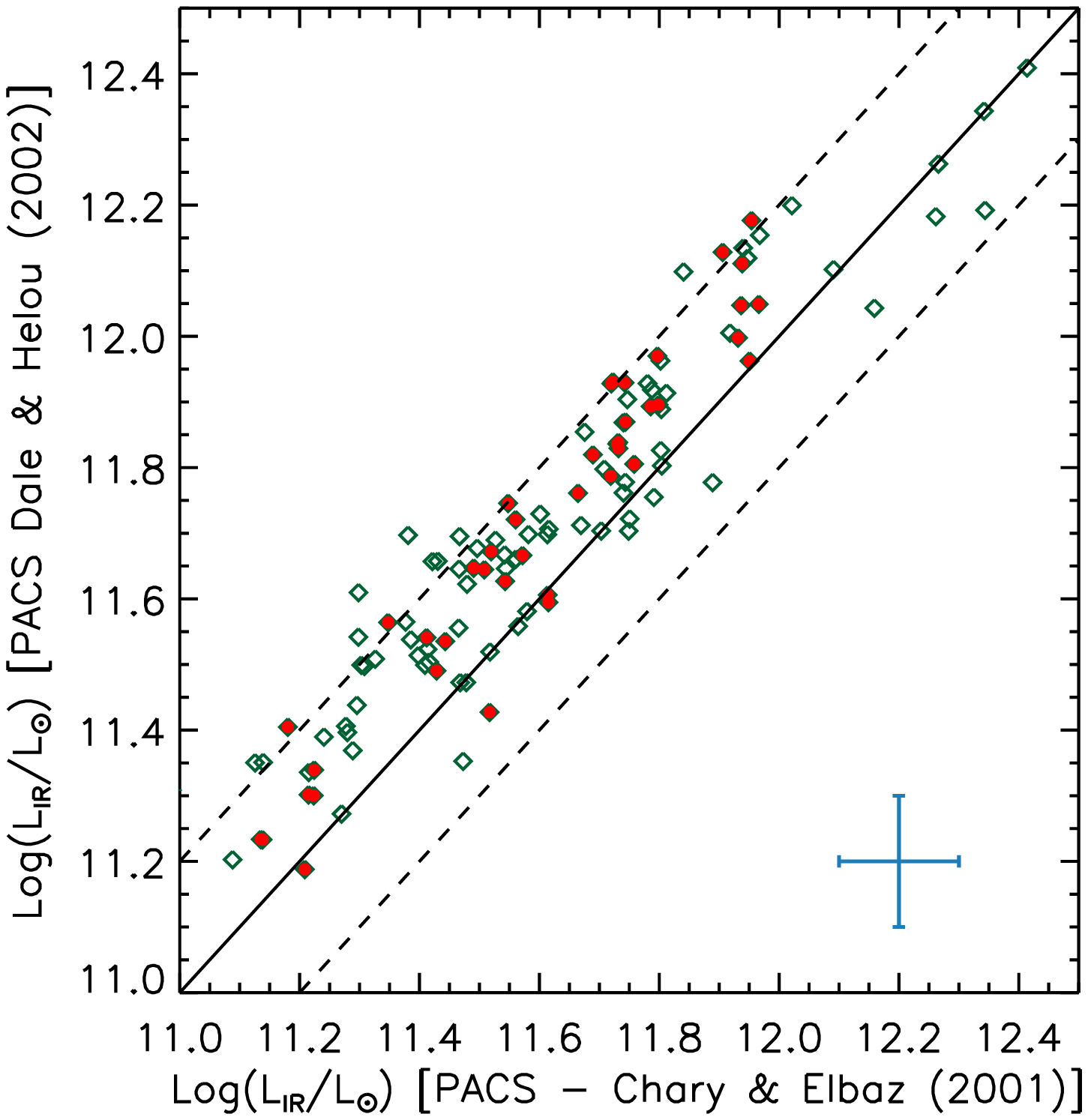}\\
\includegraphics[width=0.45\textwidth]{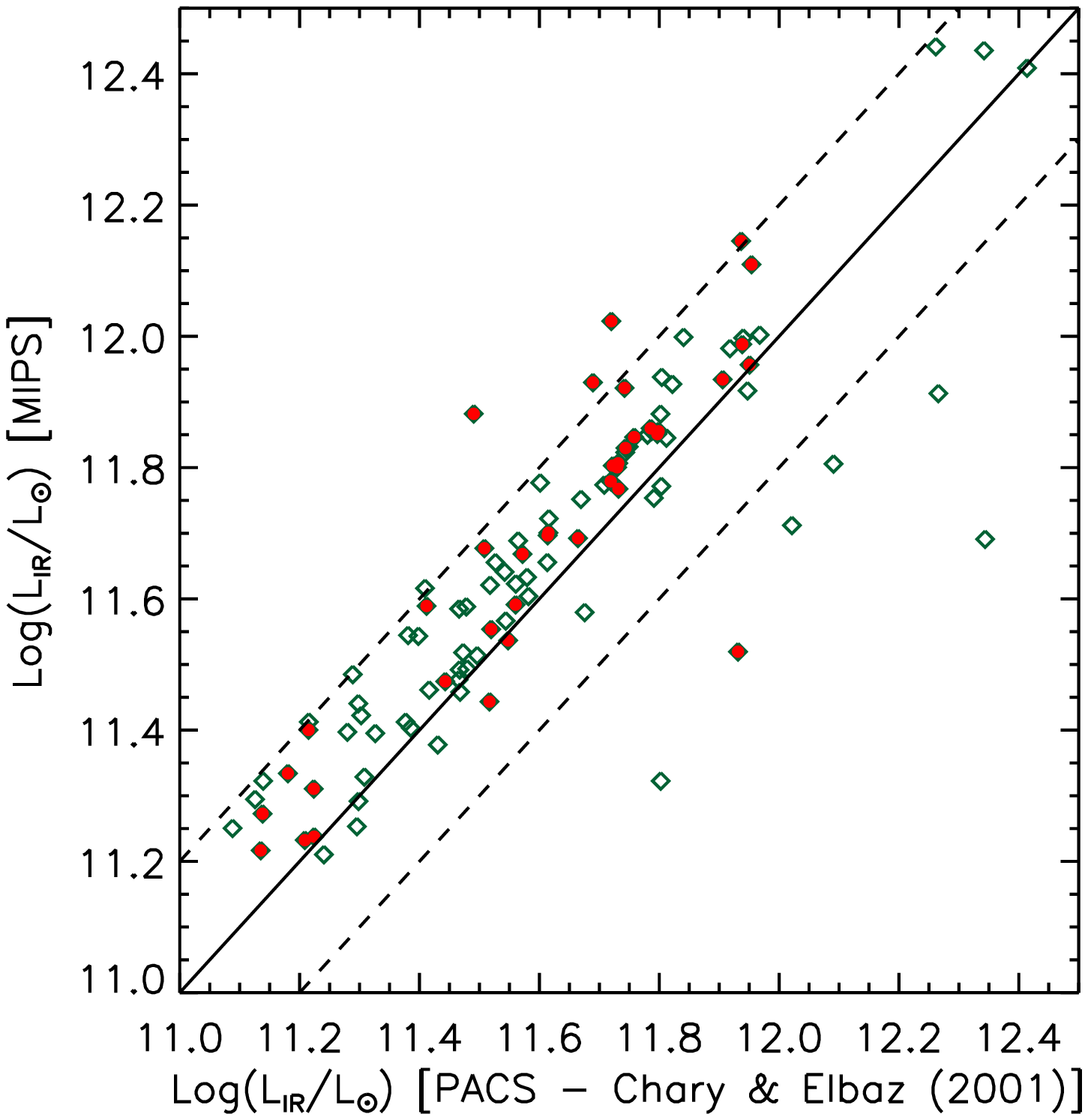}
\includegraphics[width=0.45\textwidth]{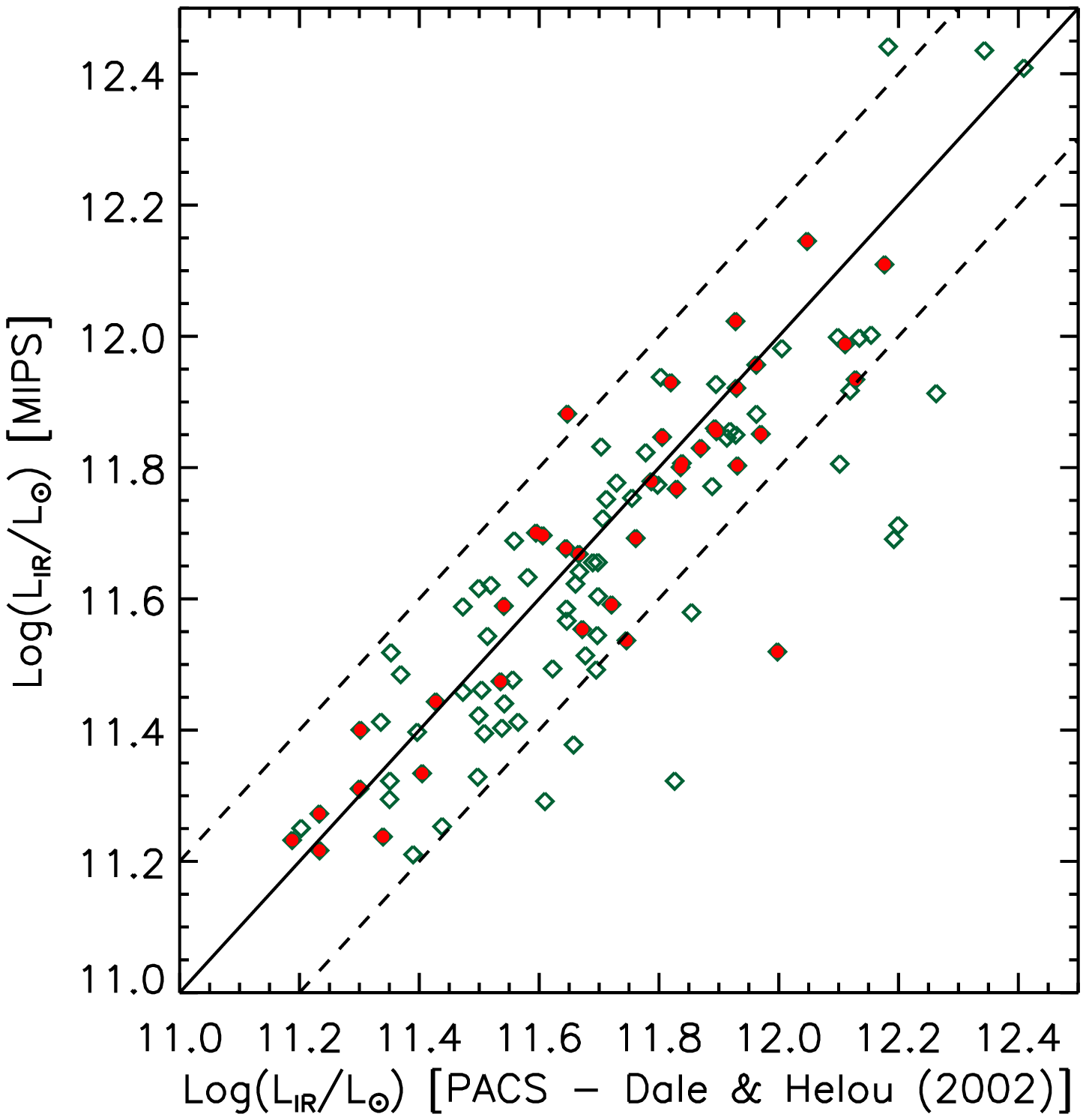}
\caption{\emph{Upper panel}: Comparison between the total IR luminosity for our PACS-detected LBGs (red dots) and UV-faint galaxies (green diamonds) with \citet{Chary2001} and \citet{Dale2002} templates. The blue bars indicate the typical uncertainties in the determinations of the total IR luminosity. \emph{Bottom panels}: Comparison between the MIPS-24$\mu$m-derived and PACS-derived total IR luminosities for our PACS-detected LBGs (red dots) and UV-faint galaxies (green diamonds) with \citet{Chary2001} (left panel) and \citet{Dale2002} (right panel) templates. Only galaxies with available MIPS and PACS fluxes are included. The MIPS-based $L_{\rm IR}$ are obtained by using single band extrapolations from the observed MIPS-24$\mu$m to $L_{\rm IR}$ employing the CE01 templates. The PACS-based total IR luminosities are derived by fitting CE01 templates to the observed IRAC-8.0$\mu$m, MIPS-24$\mu$m, PACS-100$\mu$m, and PACS-160$\mu$m fluxes of the sources. Straight line indicates where both determinations would agree and dashed lines indicate deviations of $\pm 0.2$ dex around the one-to-one relation.
              }
\label{mips}
\end{figure*}

\begin{figure*}
\centering
\includegraphics[width=0.9\textwidth]{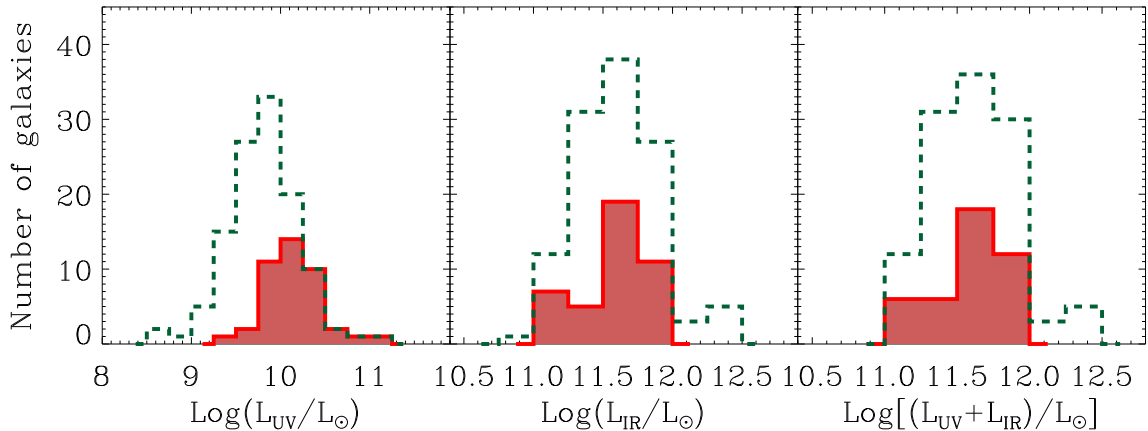}
\caption{Distribution of the rest-frame UV (\emph{left}), IR (\emph{middle}) and bolometric (\emph{right}) luminosities for our PACS-detected LBGs (red shaded histograms) and the whole sample of LBGs and UV-faint galaxies (green dashed histograms).
              }
\label{lumibol}
\end{figure*}

As indicated before, the rest-frame UV luminosities, $L_{\rm UV}$, for the galaxies studied in this work are obtained from the BC03 templates which fit the observed UV-to-near-IR SED of each galaxy best by shifting them to the rest-frame and convolving them with a top-hat filter centered in 1500 \AA\ with 100\AA\ width. Throughout this study, we consider that the rest-frame UV luminosities indicated as $L_{\rm UV}$ are expressed in $\nu L_{\nu}$ units, whereas the rest-frame UV luminosities indicated as $L_{1500}$, are expressed in $L_{\nu}$ units. With the aim of obtaining the total IR luminosities, $L_{\rm IR} [8-1000\, \mu {\rm m}]$, of our PACS-detected galaxies we carry out SED fits with ZEBRA to their observed IRAC-8.0$\mu$m, MIPS-24$\mu$m, and PACS-100$\mu$m/PACS-160$\mu$m fluxes with \cite{Chary2001} templates. Once the best templates are found they are shifted to the rest-frame and then integrated between 8 and 1000$\mu$m to obtain $L_{\rm IR}$. This way we avoid applying the relation between $L_{\rm IR}$ and SED shape, known to fail at high redshift \citep{Elbaz2010,Elbaz2011}. We have computed the uncertainties in the total IR luminosities by modifying the IRAC/MIPS/PACS fluxes randomly within their uncertainties and obtaining the new $L_{\rm IR}$ with the same procedure above. We repeat this 1000 times and find that the typical deviations in $\log{\left(L_{\rm IR}/L_\odot\right)}$ are within 0.1 dex. For comparison, we also fit the FIR SED of our PACS-detected galaxies (in the same bands) with \cite{Dale2002} (hereafter DH02) templates. The typical uncertainties for the total IR luminosity obtained with DH02 templates are obtained in the same way than for those obtained with CE01 templates and have also a value of $\log{\left(L_{\rm IR}/L_\odot\right)}$=0.1 dex. The upper panel of Figure \ref{mips} shows the relation between the total IR luminosities derived with CE01 and DH02 templates. It can be seen that DH02 templates tend to give higher values for our PACS-detected galaxies, although many of the points are within the uncertainties. In Figure \ref{SEDs} nine examples of IR SED fittings can be found (red curves).

In the bottom panels of Figure \ref{mips} we compare the total IR luminosities derived with PACS (with both CE01 and DH02 templates) data with those obtained from single band extrapolations with MIPS-24$\mu$m data only \citep[see][]{Elbaz2010}. It can be seen that MIPS-24$\mu$m alone tends to slightly overestimate the total IR luminosities of our PACS-detected sources in only 0.1 dex typically when employing CE01 templates. For DH02 templates, the points are nearer the one-to-one relation. For both kinds of FIR templates and for most of the galaxies, the deviations are lower than 0.2 dex. Therefore, there is a good agreement between those quantities within the uncertainties of the total IR luminosities. This is only true for FIR-bright galaxies with PACS detections, but it does not have to be true for FIR fainter sources detected in MIPS-24$\mu$m and PACS-undetected. There is no reason to choose between one of the two sets of FIR templates. It should be remarked that the fact that the points are located closer to the one-to-one relation when using DH02 templates does not mean that these templates provide better determinations of the total IR luminosity since the MIPS-24$\mu$m single band extrapolations does not have to apply for our galaxies. From now on, we decide to report the results obtained with CE01 templates.

Finally, we define the bolometric luminosity, $L_{\rm bol}$, of our galaxies as the sum of the UV and IR luminosities: $L_{\rm bol} = L_{\rm UV} + L_{\rm IR}$ \citep[see for example][]{Reddy2006}. 

Figure \ref{lumibol} shows the distribution of $L_{\rm UV}$, $L_{\rm IR}$, and $L_{\rm bol}$ for our PACS-detected galaxies. As expected from their selection criteria and supported by the results of a KS test, LBGs tend to be UV-brighter than UV-faint galaxies. However, despite their difference in the UV selection criteria, the $L_{\rm IR}$ span within a similar range. Their total IR luminosities place PACS-detected LBGs and UV-faint galaxies in the luminous IR galaxies (LIRGs) regime, i.e. they have $10^{11} < L_{\rm IR}/L_\odot < 10^{12}$. The median value of $L_{\rm IR}$ for PACS-detected LBGs and UV-faint galaxies are $\log{\left( L_{\rm IR}/L_\odot \right)} = 11.7$ and  $\log{\left( L_{\rm IR}/L_\odot \right)} = 11.6$, respectively. It should be noted that the fact that all our PACS-detected galaxies have $L_{\rm IR} > 10^{11}L_\odot$ is a consequence of the depths of the PACS observations in the COSMOS field, since they are not deep enough to recover galaxies with $L_{\rm IR} < 10^{11}L_\odot$. However, we do not find any PACS-detected LBG at $0.8 \lesssim z \lesssim 1.2$ in the ULIRG regime ($10^{12} < L_{\rm IR}/L_\odot < 10^{13}$) or with higher total IR luminosities, where the observations are complete. A population of ULIRG-LBG do appear at higher redshifts, suggesting evidence for an evolution of the FIR emission of LBGs \citep{Oteo2013_LBGsz3}. A similar behavior has been suggested for the other classical population of high-redshift SF galaxies: Ly$\alpha$ emitters \citep{Oteo2011,Oteo2012a,Oteo2012b}. The distributions of the bolometric luminosities are similar to the IR ones, since the UV contribution to the bolometric luminosity is low for LIRGs.


\subsection{Stacking analysis}\label{stack}


\begin{table}
\caption{\label{table_stacking}Summary of the stacking results for the different bins of rest-frame UV luminosity, stellar mass, and age considered in this work (see text for more details).}
\centering
\begin{tabular}{ccc}
\hline\hline
 		Bin											&		$N$		&		$f_{160\mu{\rm m}}$ [mJy]	\\
$9.75 \leq \log{\left( L_{\rm UV}/L_\odot \right)} \leq 10.0$ 			&		435		&		0.81	$\pm$ 0.21			\\	
$10.00 < \log{\left( L_{\rm UV}/L_\odot \right)} \leq 10.25$	 		&		338		&		0.59	$\pm$ 0.24			\\			 
$10.25 < \log{\left( L_{\rm UV}/L_\odot \right)} \leq 10.50$ 		&		108		&		0.78	$\pm$ 0.40			\\
$9 \leq \log{\left( M_*/M_\odot \right)} \leq 10.0$  				&		612		&		0.48	$\pm$ 0.13			\\
$10.0 \leq \log{\left( M_*/M_\odot \right)} \leq 11.0$ 				&		352		&		0.98	$\pm$ 0.23			\\
Age $<$ 100 Myr 										&		238		&		0.45	$\pm$ 0.27			\\
\hline
\hline
\end{tabular}
\end{table}

Many of our GALEX-selected LBGs are undetected in PACS-100$\mu$m/PACS-160$\mu$m. In order to gain insight about the FIR properties of these PACS-undetected LBGs we perform a stacking analysis in the PACS-160$\mu$m band. We choose this band for the stacking because it is the closest one to the dust emission peak and, therefore, the determination of the FIR properties will be more accurate than with the PACS-100$\mu$m band. We apply here the traditional stacking procedure previously adopted in some other works based on \emph{Herschel} data \citep{Rodighiero2010,Magdis2010LBGs,Reddy2012}. We extract 60''x60'' images centered in the optical position of the PACS-undetected galaxies and then these cutouts are combined by using a median algorithm. We use the median image instead of the mean to ensure that the result is not very dependent on the presence of bright outliers. In the stacking we use the cutouts extracted from residual PACS-160$\mu$m images, which are obtained from the science images by removing all the PACS-160$\mu$m detected sources in the catalog with $SNR > 3$.


\begin{figure}
\centering
\includegraphics[width=0.2\textwidth]{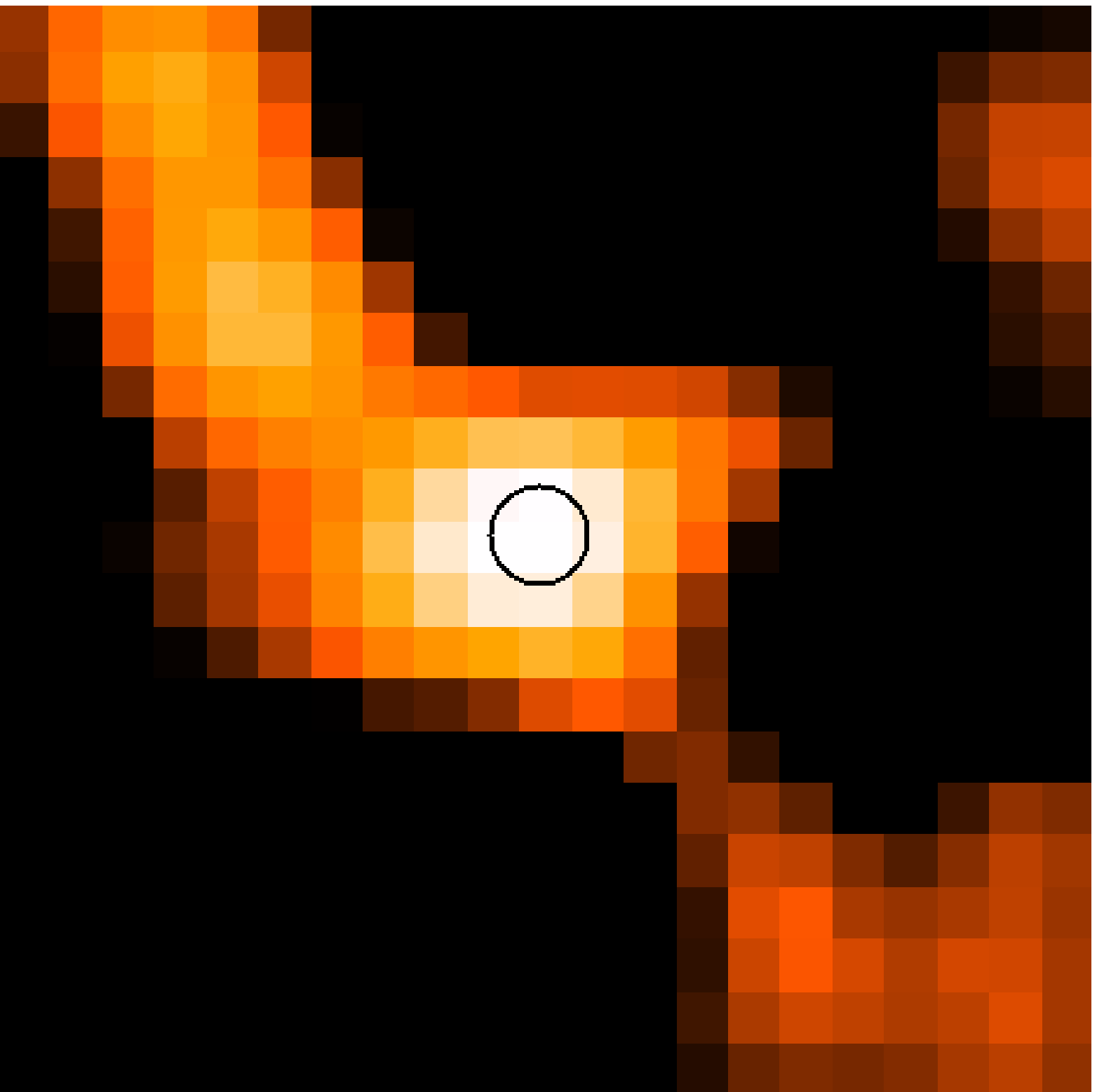}
\includegraphics[width=0.2\textwidth]{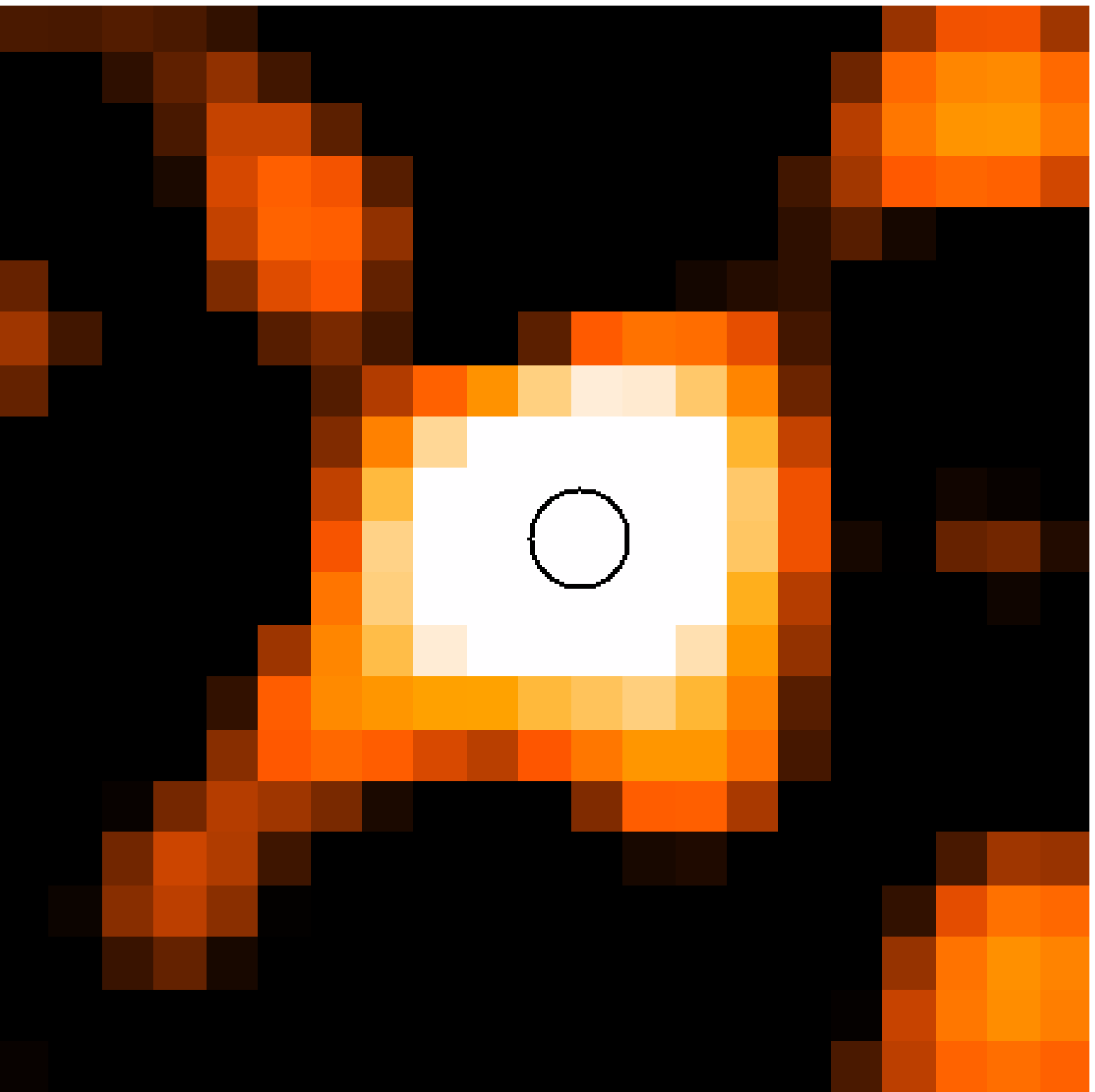}
\caption{PACS-160$\mu$m stacked images (60'' $\times$ 60'') in the two bins where a stacked detection is recovered (see text for more details). \emph{Left}: stacking for galaxies with $9.75 \leq \log{\left( L_{\rm UV}/L_\odot \right)} \leq 10.0$. \emph{Right}: stacking for galaxies with $10.0 \leq \log{\left( M_*/M_\odot \right)} \leq 11.0$              
}
\label{stackk}
\end{figure}

We consider different samples to stack among the whole sample of PACS-undetected LBGs related to their different properties that will be studied in this work in the next sections. In this way, we arbitrarily split our sample of PACS-undetected LBGs in bins of rest-frame UV luminosities, stellar mass, and SED-derived age, as indicated in Table \ref{table_stacking}. We distinguish only one range of age with the aim of studying if the youngest galaxies in our sample behaves differently from the rest of the population. The number of galaxies in each bin is also indicated in Table \ref{table_stacking}.

In order to measure the uncertainties in the stacked images we follow a similar procedure than those employed, for example, in \cite{Reddy2012} or \cite{Magdis2010LBGs}. We perform stacking in the PACS-160$\mu$m residual map in $N$ random positions, where $N$ is the number of galaxies in each bin of rest-frame UV luminosity, stellar mass, and age. We then measure the resultant fluxes in the stacked images, and repeat this process 1000 times. It should be noted that the stacking in this case should be done with the residual images since otherwise there would be a high contribution of FIR emission of real PACS-detected sources. The width of the distribution of the stacked fluxes in each bin is considered as the uncertainty in the stacked PACS-160$\mu$m flux in each case. The lower number of galaxies in each bin, the higher the uncertainty in the stacked flux is. These uncertainties span from $\Delta f_{160\mu \rm{m}}= 0.40\, {\rm mJy}$ to $\Delta f_{160\mu \rm{m}}= 0.13\, {\rm mJy}$. The uncertainties in each bin are shown in Table \ref{table_stacking}. In addition of giving a estimation of the uncertainty of the stacking procedure, this process also provides information about the probability of obtaining significant stacked fluxes when stacking in random positions along the images and, therefore, representing fluxes not related to real sources. If the width of the distribution was as high as the stacked flux obtained in the real optical position of the LBGs there would be a non-negligible probability of a chance measurement of stacked flux in a position which do not correspond to real sources and, therefore, the stacking procedure would not be reliable. Following this idea, we will consider that we get a stack detection if each stacked PACS-160$\mu$m flux is higher than 3*$\Delta f_{160\mu {\rm m}}$.

As a first step, and in order to study the reliability of the stacking analysis, we perform a stacking in the PACS-160$\mu$m band of all the GALEX-selected sources which are detected in MIPS-24$\mu$m but undetected in PACS. At the redshift range of our galaxies, MIPS-24$\mu$m single band extrapolations to the total IR luminosities give reasonably good results \citep{Elbaz2010,Elbaz2011}. Therefore, in this case we can compare the MIPS-derived total IR luminosities with those obtained from the stacked PACS-160$\mu$m flux and study the agreement between the results. The PACS-160$\mu$m stacked flux obtained for the MIPS-detected LBGs with the cutouts taken from the PACS-160$\mu$m residual image is $f_{160{\mu\rm m}} = 2.7 \pm 0.4\, {\rm mJy}$. In order to obtain its associated total IR luminosity we employ a single band extrapolation method from PACS-160$\mu$m flux to the total IR luminosity, obtaining a value of $\log{\left( L_{\rm IR}/L_\odot\right)} = 11.0$. The median value of the MIPS-24$\mu$m flux for the MIPS-detected/PACS-undetected LBGs is $f_{24\mu{\rm m}} = 0.2 \pm 0.1$, where the indicated uncertainty correspond to the standard deviation of the distribution of the MIPS-24$\mu$m fluxes. Applying again a single band extrapolation, we obtain that its associated total IR luminosity is $\log{\left( L_{\rm IR}/L_\odot\right)} = 11.3$. This value is slightly higher than that obtained from stacking in PACS-160$\mu$m but still compatible with the typical uncertainties of the single band extrapolations for obtaining the total IR luminosity and the uncertainties of the MIPS-24$\mu$m and PACS-160$\mu$m fluxes of the stacked sources.
 
For each of the bins we applied the stacking procedure outlined above with the PACS-160 residual images. We only find stacked detections in the bin associated to the lowest rest-frame UV luminosity and the bin associated to the highest stellar mass. Figure \ref{stackk} shows the stacked PACS-160$\mu$m images in those two ranges where we recover a stacked detection. The stacked fluxes are shown in Table \ref{table_stacking}. The total IR luminosities associated to the stacked PACS-160$\mu$m fluxes are obtained by using single band extrapolations \citep{Elbaz2010}. In the following analysis, the values of the rest-frame UV luminosities, stellar mass, and age for each bin are represented by the median values for all the galaxies belonging to each one.

\subsection{SED-derived properties of PACS-detected galaxies}\label{properties_PACS_detected}

In \cite{Oteo2013} we analyzed the physical properties of GALEX-selected LBGs at $0.8 \lesssim z \lesssim 1.2$ by employing an SED fitting procedure with BC03 templates. This allowed us to study their rest-frame UV luminosity, UV continuum slope, age, dust attenuation, dust-corrected total SFR, and stellar mass. Furthermore, we have discussed before the trends between the MIPS-24$\mu$m detections and the SED-derived physical properties, finding that those objects more likely detected in the FIR are those more massive, dustier, redder, and UV-brighter. In this section we examine the SED-derived physical properties of our PACS-detected LBGs and study the differential properties between PACS-detected and PACS-undetected ones. 

\begin{figure*}
\centering
\includegraphics[width=0.25\textwidth]{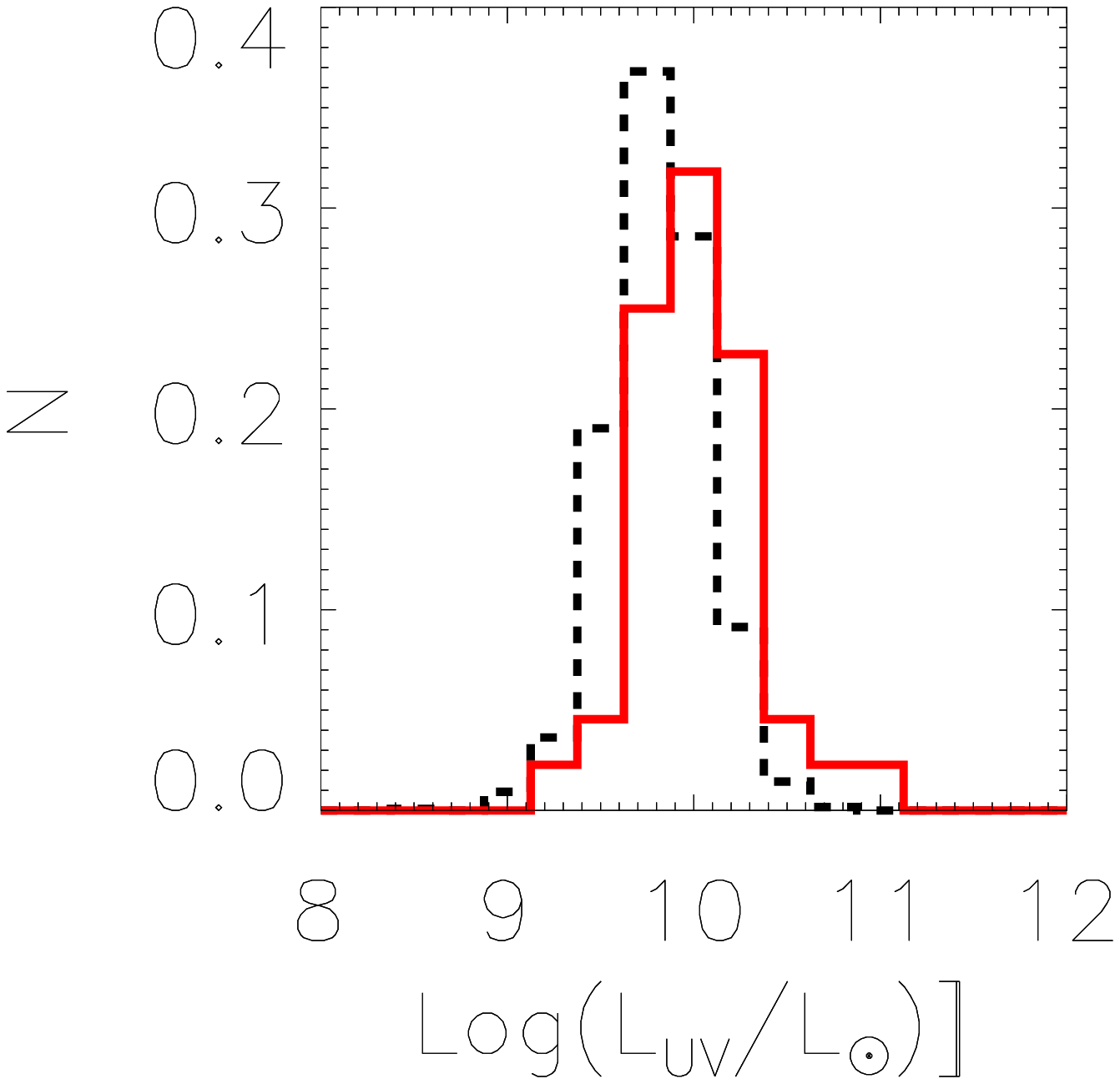}
\includegraphics[width=0.25\textwidth]{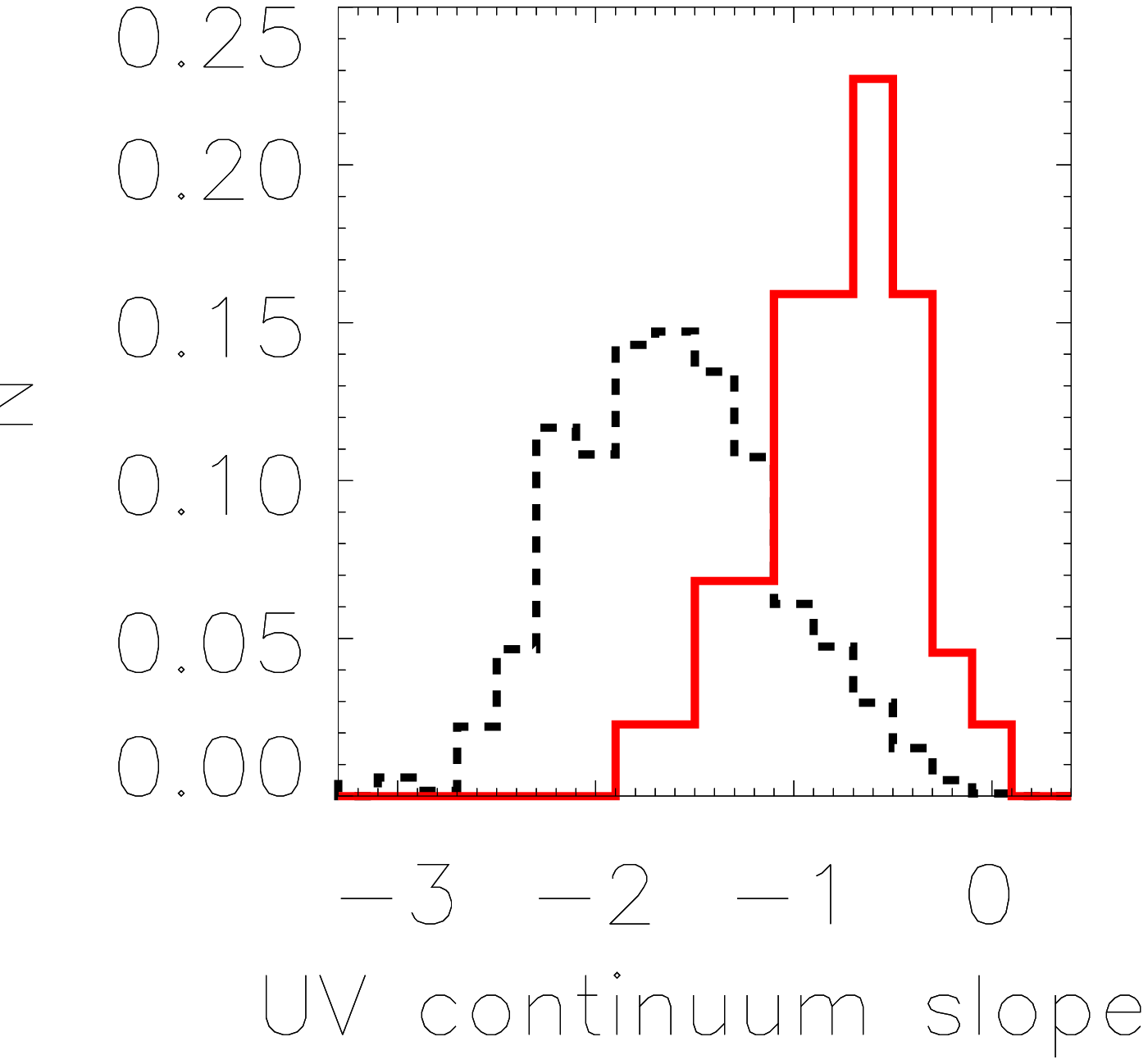} 
\includegraphics[width=0.25\textwidth]{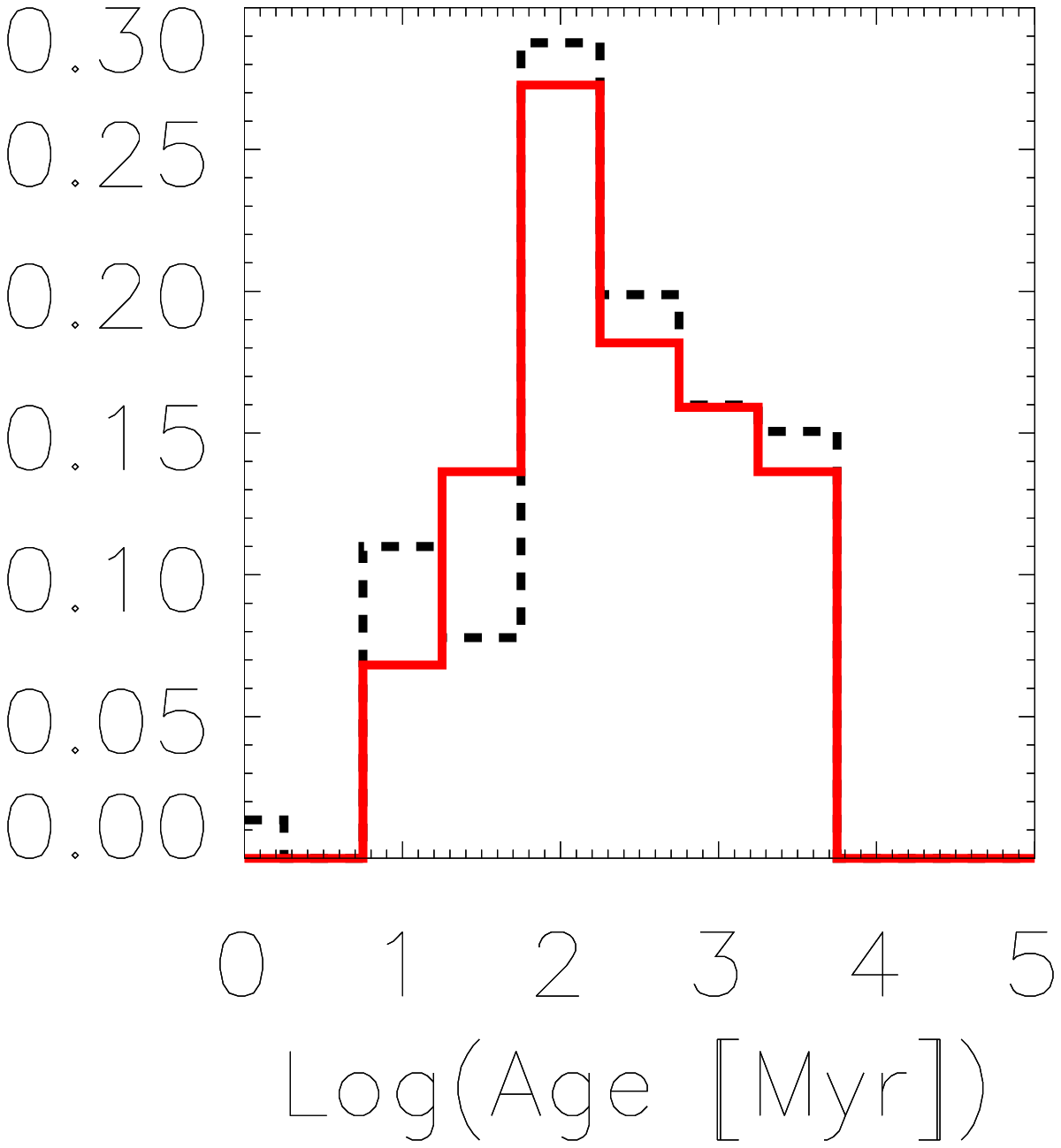} \\ 
\includegraphics[width=0.25\textwidth]{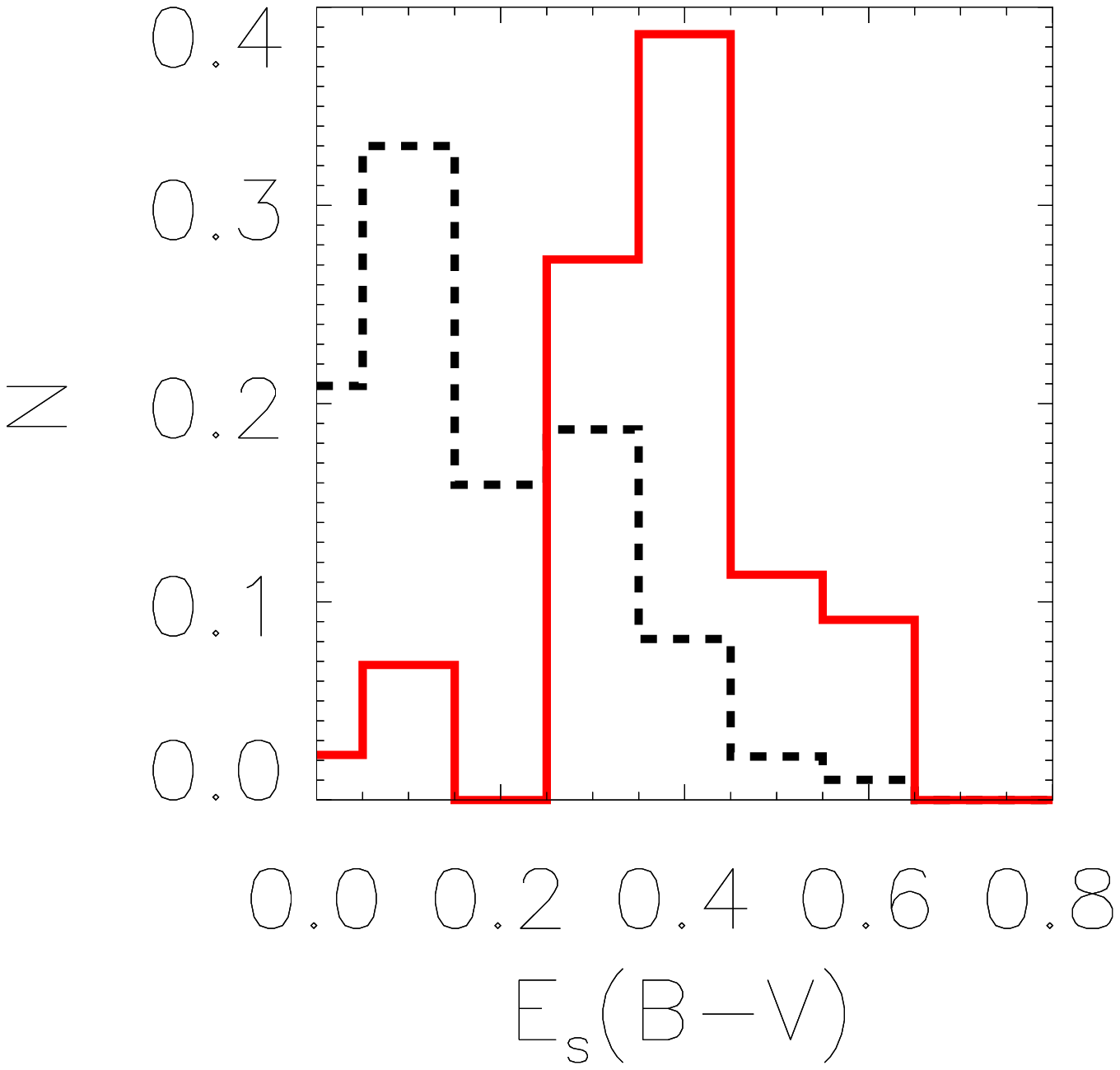}
\includegraphics[width=0.25\textwidth]{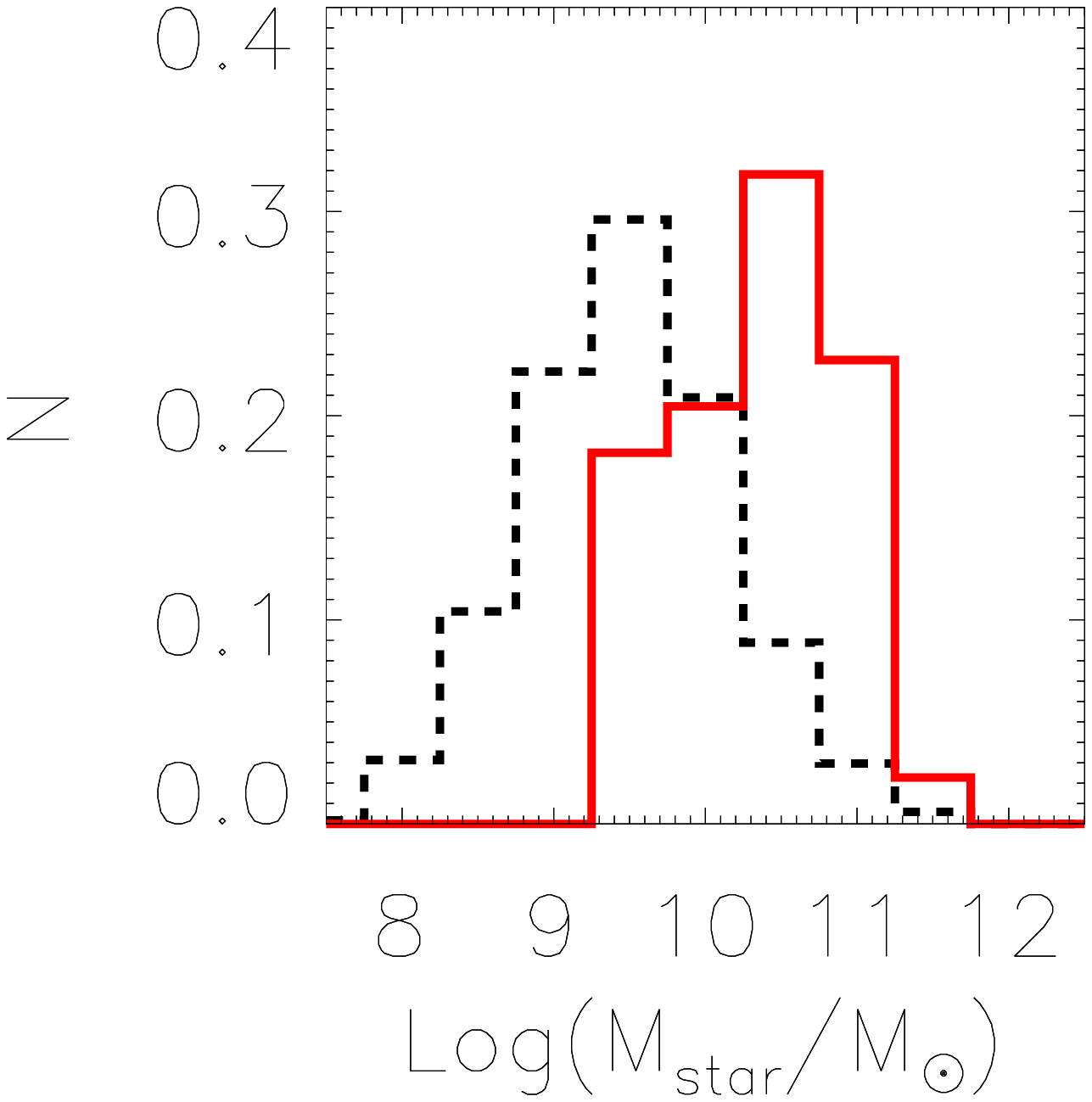}
\includegraphics[width=0.25\textwidth]{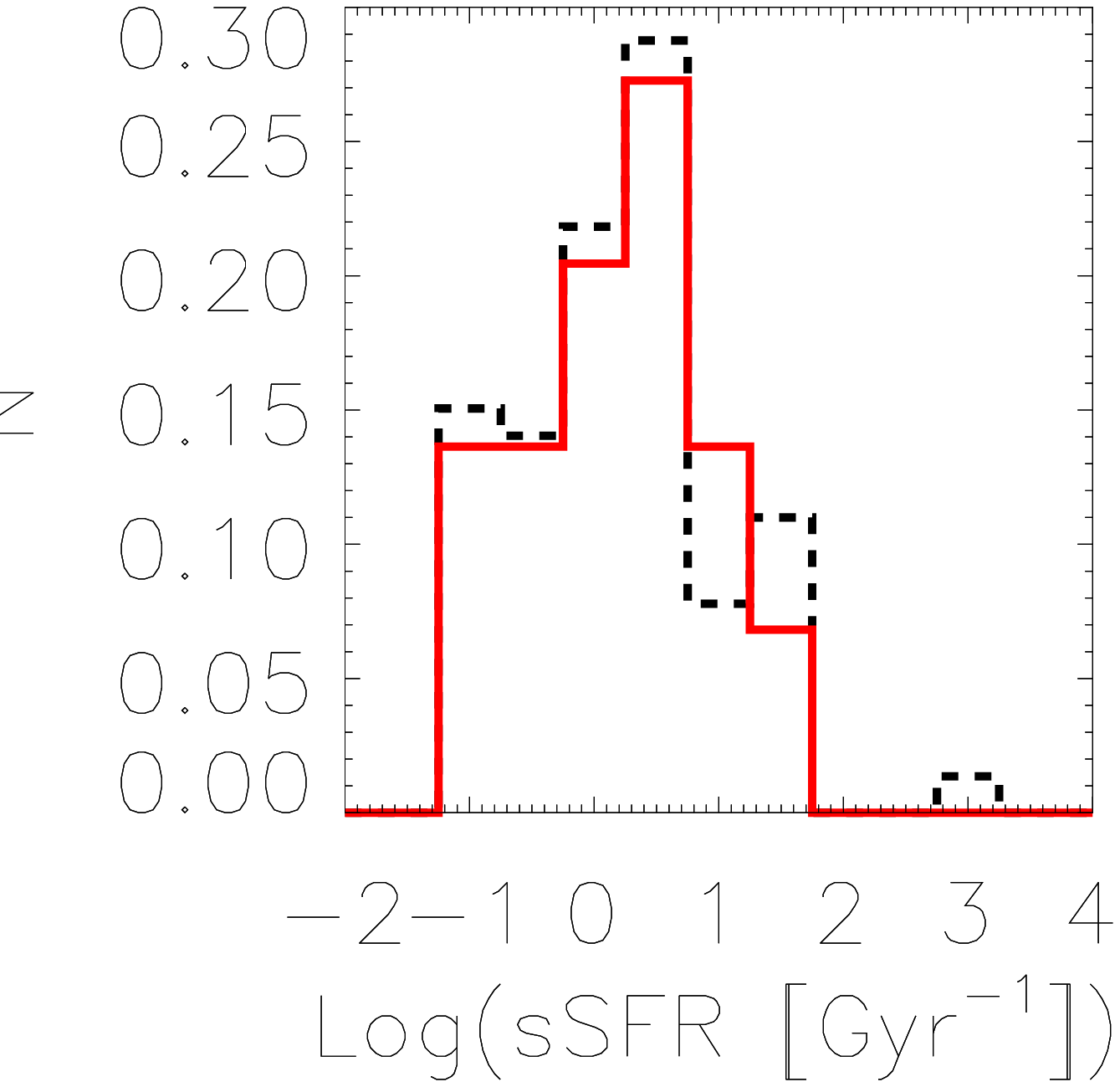}
\caption{Distribution of the SED-derived rest-frame UV luminosity (\emph{upper left}), UV continuum slope (\emph{upper middle}), age (\emph{bottom right}), dust attenuation (\emph{bottom left}), stellar mass (\emph{bottom middle}), and specific SFR (\emph{bottom right}) for our PACS-detected LBGs (red histograms) and PACS-undetected LBGs (black dashed histograms). Histograms have been normalized to their area in order to clarify the representations.
              }
\label{properties_PACS}
\end{figure*}

\emph{Properties of PACS-detected LBGs}: We focus first on the SED-derived physical properties of the PACS-detected LBGs (Figure \ref{properties_PACS}). Their stellar populations have a median age of 341 Myr, rest-frame UV luminosity of $\log{\left( L_{\rm UV}/L_\odot \right)}$=10.1, dust attenuation of $E_s(B-V) = 0.4$, UV continuum slope of $\beta = -0.6$, stellar mass of $\log{\left( M_{*}/M_\odot \right)}$ = 10.1, and specific SFR $\log{\left( sSFR [{\rm Gyr}^{-1}] \right) } = 0.51$. The median value of the stellar mass of our PACS-detected LBGs is slightly lower than that reported in \cite{Burgarella2011} for their SPIRE-detected LBGs. This difference is likely due to their different IR-selection criterion: the selection in SPIRE bands segregates the most massive galaxies. These values of the stellar mass of our PACS-detected LBGs are similar to the stellar masses of LBGs at $z \sim 3$ detected in IRAC-8$\mu$m and MIPS-24$\mu$m \citep{Magdis2010_IRAC}.

\emph{Differences between PACS-detected and PACS-undetected LBGs}: In order to examine the differential properties between PACS-detected and PACS-undetected LBGs, we plot in Figure \ref{properties_PACS} the distribution of UV luminosities, UV continuum slope, age, dust attenuation, stellar mass, and specific SFR ($sSFR = SFR/M_*$) for the PACS-undetected LBGs with black dashed histograms. It can be seen that, as it could be expected from Figure \ref{mipsdetections}, PACS-undetected LBGs are intrinsically UV-brighter and less attenuated, have lower stellar mass, and present bluer UV continuum slopes than their PACS-detected analogues. However, there is no remarkable difference in age or  sSFR. The no correlation with age is in agreement with Figure \ref{mipsdetections}. The median values of the distributions of PACS-undetected LBGs are: $\log{\left( L_{\rm UV}/L_\odot \right)} = 9.72$, $\beta = -1.55$, Age = 341 Myr, $E_s(B-V) = 0.2$, $\log{\left( M_*/M_\odot \right)} = 9.91$, and $\log{\left( sSFR [{\rm Gyr}^{-1}] \right) } = 0.46$.

\section{Dust attenuation}\label{dust_attenuation}

\begin{figure*}
\centering
\includegraphics[width=0.49\textwidth]{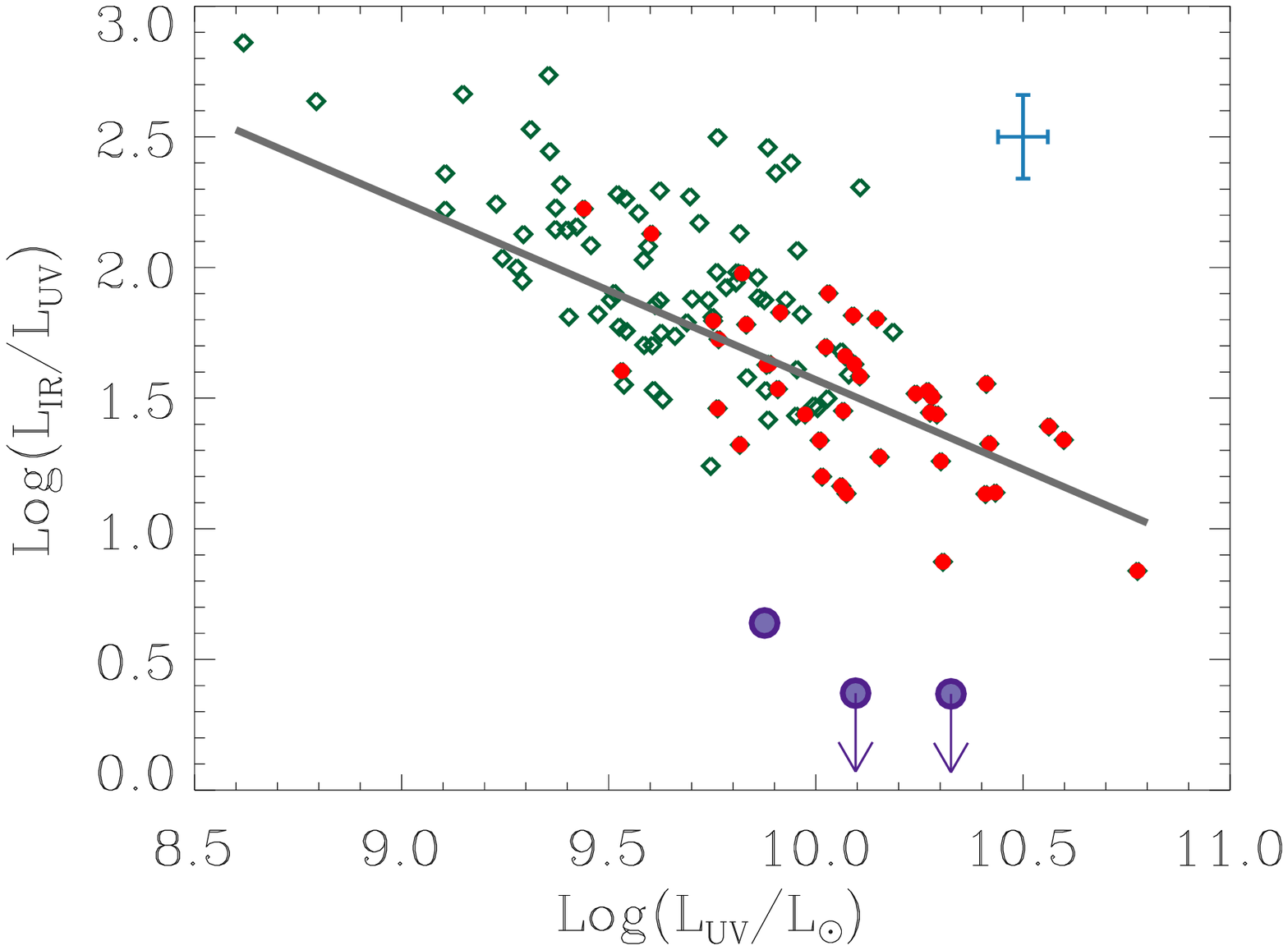}
\includegraphics[width=0.49\textwidth]{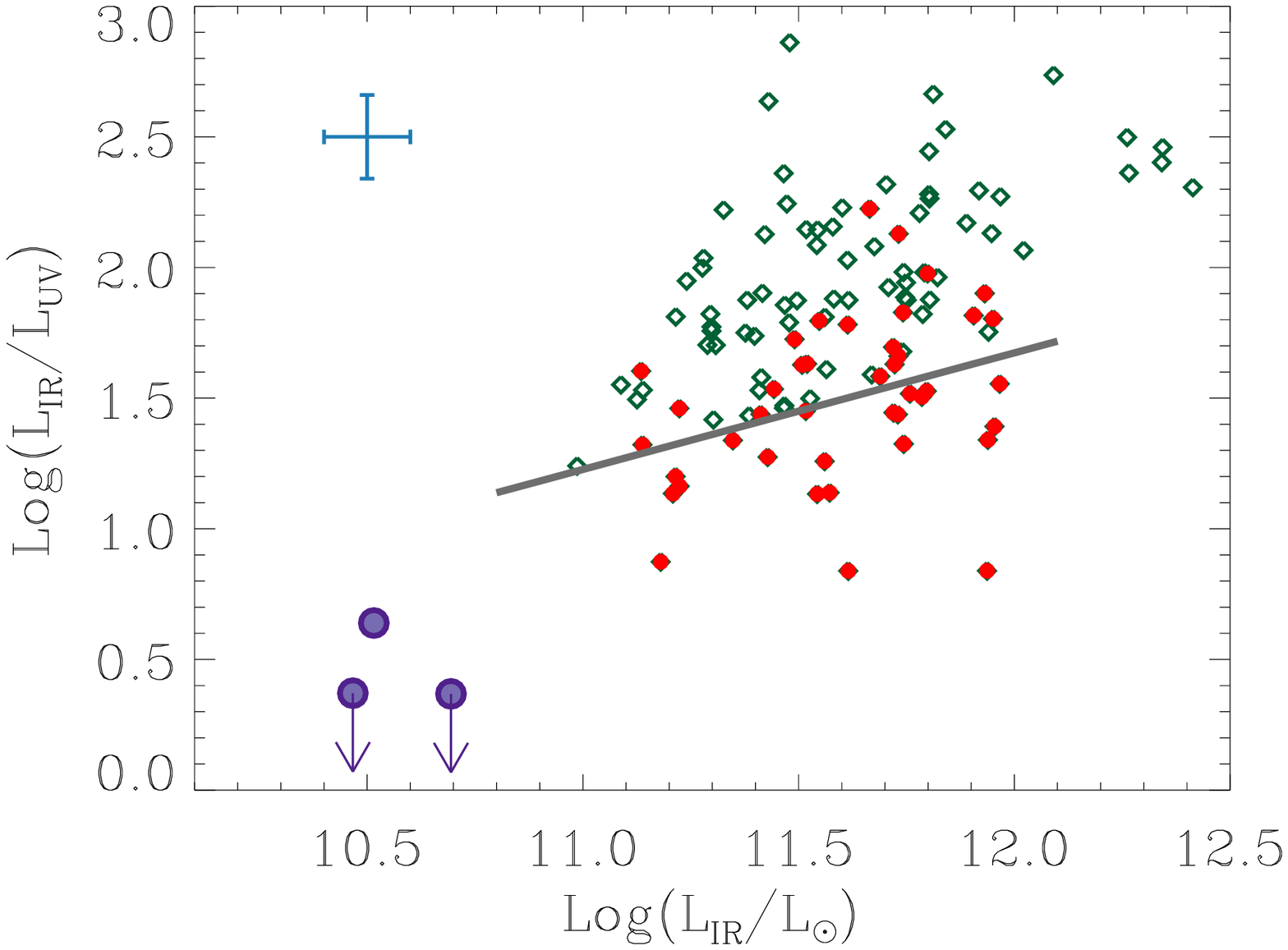}
\caption{\emph{Left:} IR/UV ratio (related to dust attenuation through Eqn. \ref{A_NUV}) as a function of the rest-frame UV luminosity for PACS-detected LBGs (red filled dots) and the general population of UV-selected galaxies (green open diamonds). \emph{Right:} IR/UV ratio as a function of the total IR luminosity for PACS-detected LBGs (red filled dots) and UV-faint galaxies (green open diamonds). In both panels, the grey solid lines represent a linear fit to the PACS-detected LBGs. PACS-stacked LBGs in the three bins of rest-frame UV luminosity are represented by the filled purple dots. The arrows indicate the two bins of rest-frame UV luminosity where no stacked detections are recovered. The blue error bars indicate the typical uncertainties of the parameters shown in the horizontal and vertical axes.
}
\label{dust}
\end{figure*}

Part of the light emitted by galaxies in the UV is absorbed by dust and re-emitted in-turn in the FIR. Therefore, the most accurate way to obtain dust attenuation of galaxies is comparing the total IR and rest-frame UV luminosities. In fact, calibrations between dust attenuation and the $L_{IR}/L_{UV}$ ratio have been already built. To obtain the dust attenuation of our PACS-detected galaxies we use the calibration of \cite{Buat2005}:

\begin{equation}\label{A_NUV}
A_{FUV} = -0.0333x^3 + 0.3522x^2 + 1.1960x + 0.4967
\end{equation}

\noindent where A$_{FUV}$ is the dust attenuation in the FUV band and $x=\log{\left(L_{IR}/L_{UV}\right)}$. The conversion from A$_{FUV}$ to the dust attenuation in 1200\AA, $A_{1200\AA}$, is made by using \cite{Calzetti2000} reddening law. Figure \ref{dust} shows the dust attenuation of our PACS-detected LBGs (red dots) and UV-faint galaxies (green open triangles) as parametrized by their $L_{\rm IR}/L_{\rm UV}$ ratio \citep{Buat2005} as a function of their $L_{\rm UV}$ and $L_{\rm IR}$. PACS-detected LBGs have dust attenuation spanning $2 \lesssim A_{\rm 1200\AA} {\rm [mag]} \lesssim 5$, with a median value of 3.6 mag. In the comparison between LBGs and UV-faint galaxies, Figure \ref{dust} indicates that LBGs tend to be among the less dusty objects in our sample of PACS-detected galaxies. The median value of dust attenuation of our PACS-detected UV-faint galaxies is $A_{1200\AA} = 4.3$ mag, respectively. The difference in dust attenuation is a consequence of the UV and IR luminosities: LBGs are UV-brighter and have similar IR luminosities than UV-faint galaxies, implying that they are also less dustier according to the \cite{Buat2005} calibration.

There seems to be a tight trend between dust attenuation and the rest-frame UV luminosity: the dust attenuation decreases when the rest-frame UV luminosity increases (see left panel of Figure \ref{dust}). A similar result was reported by \cite{Burgarella2011} for their SPIRE-detected LBGs and by \cite{Buat2012} at higher redshifts. We have fitted a linear relation to the dust attenuation and rest-frame UV luminosity of PACS-detected LBGs, shown in Figure \ref{dust} with a grey straight line. This relation, although fitted for PACS-detected LBGs only, is also followed by the PACS-detected UV-faint galaxies. This tendency should be understood as valid for the envelope corresponding to highest values of dust attenuation, since we are only considering FIR-detected galaxies and this selection tends to segregate the dustiest galaxies at $z \sim 1$.  We also show in the left panel of Figure \ref{dust} the three points corresponding to the stacking of the galaxies in three bins of rest-frame UV luminosities. The stacked points suggest that when introducing FIR-fainter galaxies the relation between dust attenuation and rest-frame UV luminosity has a  much higher spread, and, therefore, that relation is a direct consequence of the FIR selection effect.

On the IR side, there also seems to be a trend between dust attenuation and total IR luminosity in the sense that more IR-luminous PACS-detected LBGs are dustier (see right panel of Figure \ref{dust}). We have also fitted here a linear relation for the PACS-detected LBGs, shown with the grey straight line. However, in this case PACS-detected UV-faint galaxies do not follow that relation but they tend to be located  above it and, therefore, for the same IR luminosity, PACS-detected UV-faint galaxies tend to have higher values of dust attenuation than PACS-detected LBGs. This is again a direct consequence of the similarity in the IR luminosities and the difference in the rest-frame UV brightness between the two kinds of galaxies. Interestingly, in this case the stacked points associated to the three bins of rest-frame UV luminosity do tend to follow the linear relation fitted for PACS-detected LBGs although with some deviation toward lower values of dust attenuation. This might indicate that when including FIR-fainter galaxies, the correlation between dust attenuation and total IR luminosity still holds. We show with a dashed grey straight line a linear fit to the points corresponding to both the PACS-detected LBGs and the PACS-stacked ones.

\subsection{Bolometric luminosity and dust attenuation}\label{bolometric_dust}

\begin{figure*}
\centering
\includegraphics[width=0.8\textwidth]{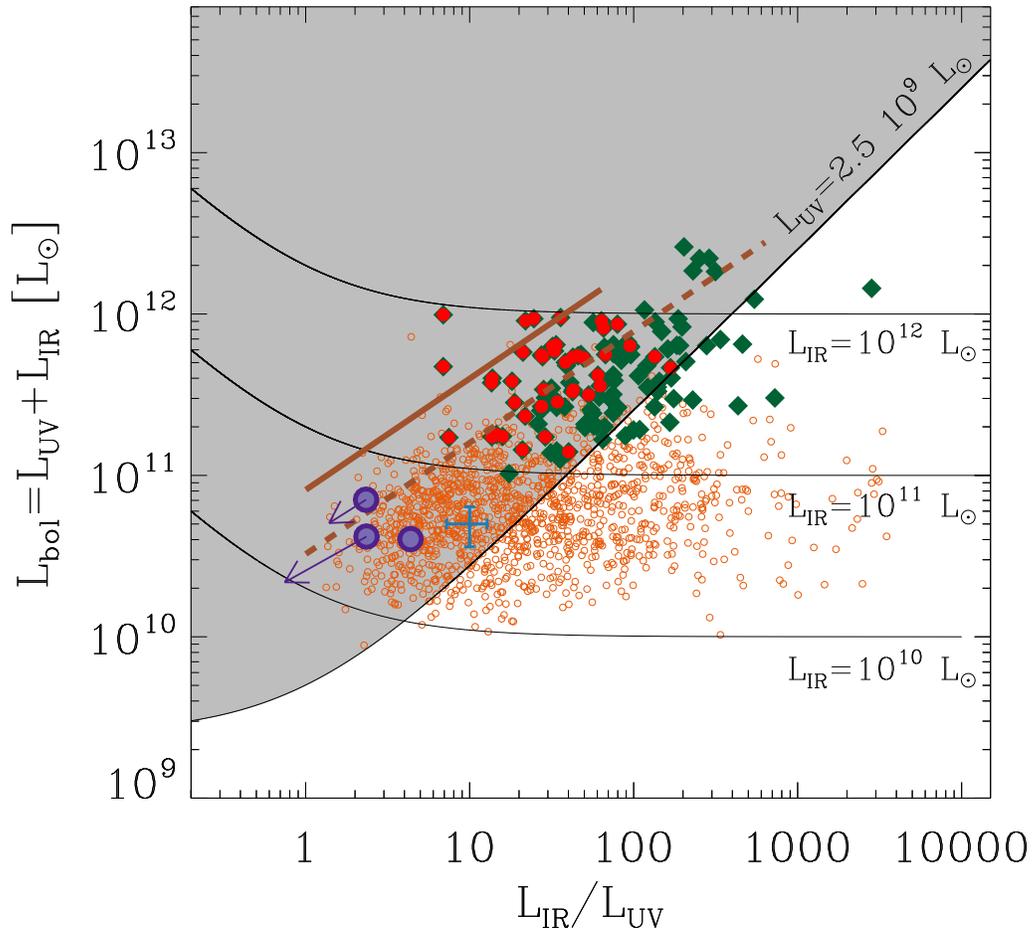}
\caption{Bolometric luminosity as a function of the dust attenuation parametrized by the IR/UV ratio. PACS-detected LBGs and the general population of UV-selected galaxies are represented by red dots and green diamonds, respectively. A sample of UV-selected galaxies at $z \sim 0.3$ with UV and PACS measurement is also represented with yellow open dots. Black curves indicate different constant values of the rest-frame UV and total IR luminosities. The value chosen for the rest-frame UV luminosity is $\log{\left( L_{\rm UV}/L_\odot \right)} = 9.2$, which is the minimum typical value for our PACS-detected galaxies at $z \sim 1$ (see text for more details). Grey shaded zone represents the location of the galaxies with rest-frame UV luminosity above that threshold. The values chosen for the total IR luminosity are those differentiating the three main kinds of FIR-selected galaxies: normal SF galaxies, LIRGs, and ULIRGs: $L_{\rm IR} = 10^{10}$, $10^{11}$, and $10^{12}L_\odot$, respectively. We also include in the plot the bolometric luminosities and IR/UV ratios associated to the stacked points in the three bins of rest-frame UV luminosity with purple filled dots. The arrows indicate upper limits for the two ranges where no stacked detections are recovered. The solid brown line is the \citet{Reddy2010} relation between bolometric luminosity and dust attenuation for galaxies at $z \sim 2$. The dashed brown line is plotted to guide the eye about the relation between bolometric luminosity and dust attenuation for our PACS-detected galaxies. It has the same slope that the relation of \citet{Reddy2010} but the zero-point is 2.5 times lower. The blue error bars indicate the typical uncertainties of the parameters shown in the horizontal and vertical axes.
}     
\label{sum_coc}
\end{figure*}

Figure \ref{sum_coc} shows the bolometric luminosity against dust attenuation for our PACS-detected LBGs and UV-faint galaxies at $z \sim 1$. In order to compare the behavior of UV-selected galaxies at different redshifts, we also consider a sample of PACS-detected galaxies at $0.2 \lesssim z \lesssim 0.4$. This sample is built by considering all the galaxies in the COSMOS field which are detected both in GALEX and PACS (100$\mu$m or 160$\mu$m) and whose photometric redshifts taken from COSMOS survey \citep{Capak2007} indicate that they are at $0.2 \lesssim z \lesssim 0.4$. This sample is included in Figure \ref{sum_coc} with small yellow open dots. The dissimilar range of rest-frame UV luminosities between the samples of UV-selected galaxies at different redshifts is due to their selection criteria: at higher redshifts only the brightest galaxies are selected. Therefore, with the aim of comparing samples at different redshifts, their rest-frame UV luminosities must be limited to the same range to avoid any UV selection bias. In our case, the sample at $z \sim 0.3$ contains galaxies with lower rest-frame UV luminosities than the sample at $z \sim 1$. Therefore, we have to limit the sample at $z \sim 0.3$ to the same rest-frame UV luminosity range than the galaxies at $z \sim 1$. It can be seen in Figure \ref{lumibol} that the number of PACS-detected UV-selected galaxies at $z \sim 1$ significantly drops for $\log{\left(L_{\rm UV}/L_\odot\right)} \lesssim 9.4$. Thus, we only take into consideration in this analysis those galaxies both at $z \sim 1$ and $z \sim 0.3$ whose rest-frame UV luminosities are above that value. The zone in Figure \ref{sum_coc} which satisfies that condition is shaded in grey.

Focusing first on the galaxies at $z \sim 1$, Figure \ref{sum_coc} indicates that there might be a relation between the bolometric luminosity and the dust attenuation for galaxies with $\log{\left(L_{\rm UV}/L_\odot\right)} \geq 9.4$ in the sense that higher values of dust attenuation correspond to higher values of the bolometric luminosity. However, the scatter is significant, about 0.45 dex. Such a correlation between those variables has also been reported in other works at similar \citep{Buat2009} and higher redshifts \citep{Reddy2010}. At $z \sim 2.0$, \cite{Reddy2010} found a strong relation between $L_{\rm bol}$ and dust attenuation with a scatter of 0.23 dex about a linear fit, which is very similar to the trend found in \cite{Reddy2006}. This relation is plotted in Figure \ref{sum_coc} with a continuous brown line. It should be pointed out that in the derivation of such a relation, they included both MIPS-detected and MIPS-undetected galaxies and, given that they work with galaxies at $z \sim 2$, their UV luminosities tend to be higher those for our PACS-detected galaxies and therefore their galaxies are bolometrically more luminous than ours. Despite these differences, our points tend to follow a linear relation with a similar slope than that found in \cite{Reddy2010} but with a lower zero-point due to the difference in the rest-frame UV luminosity between the two samples. Only the least dusty and most bolometrically luminous PACS-detected LBGs at $z \sim 1$ are in agreement with the \cite{Reddy2010} relation. We plot in Figure \ref{sum_coc} with a brown dashed line a linear relation with the same slope than that in \cite{Reddy2010} but with the bolometric luminosities divided by a factor of 2.5. It can be seen that this relation is well followed by the PACS-detected LBGs and also by the the stacked LBGs in PACS-160$\mu$m associated to the three ranges of rest-frame UV luminosity. Furthermore, the PACS-detected galaxies at $z \sim 0.3$ with $\log{\left(L_{\rm UV}/L_\odot\right)} \geq 9.4$ tend to follow the same relation that our LBGs at $z \sim 1$, filling in the $L_{\rm bol} - L_{\rm IR}/L_{\rm UV}$ diagram the gap between our PACS-detected and PACS-stacked LBGs.

Considering the whole sample of UV-and-FIR-detected galaxies at $z \sim 0.3$ regardless the rest-frame UV luminosities, it can be seen that there is a large spread in the relation between the bolometric luminosity and the dust attenuation of these galaxies. That spread prevent us from formulating any relation between bolometric luminosity and dust attenuation. In the analysis at $z \sim 1$, we cut the rest-frame UV luminosity of the galaxies to $\log{\left( L_{\rm UV}/L_\odot \right)} \geq 9.4$ and obtain that, with that limitation, there is a relation between $L_{\rm bol}$ and $L_{\rm IR}/L_{\rm UV}$ although with same scatter. The large scatter at $z \sim 0.3$ makes us to speculate that the relation found at $z \sim 1$ is only a consequence of the cut in the rest-frame UV luminosity and also in the total IR luminosity (consequence of the PACS limiting fluxes in the COSMOS field). If we could go deeper in the rest-frame UV and IR luminosity functions, both at $z \sim 1$ and at higher redshifts, the relation between bolometric luminosity and dust attenuation would not be so clear.

\subsection{Dust attenuation and UV continuum slope}\label{dust_beta}

Part of the light emitted by galaxies in the UV is absorbed by dust. This absorption takes place in a wavelength-dependent way: bluer wavelengths are more attenuated than redder ones. This selective attenuation implies a change in the UV continuum slope with dust attenuation: for a given age, SFH, metallicity, and when the dust attenuation increases, bluer zones of the UV continuum are more attenuated that redder zones, producing an increasing of the UV continuum slope, $\beta$, from negative to less negative or even positive values. Furthermore, as it has been commented before, the absorbed light by dust in the UV is in turn re-emitted in the FIR and the ratio between the total IR and rest-frame UV luminosities is the best tracer of dust attenuation. Therefore, the UV continuum slope and dust attenuation could be correlated: higher (redder) values of the UV continuum slope could be due to higher values of dust attenuation. However, it should be also considered that the relation between UV continuum and dust attenuation could be partially hidden by the presence of evolved stellar populations, which also tend to redden the UV continuum of galaxies. 

The UV continuum slope is a quantity relatively easy to determine since it only requires a good sampling of the rest-frame UV continuum, which is usually achieved for galaxies in a wide range of redshifts \citep{Bouwens2006,Hathi2008,Bouwens2009,Bouwens2011,Wilkins2011,Finkelstein2012,Hathi2012}. However, the accurate measurement of dust attenuation, mostly at high redshift, is very challenging since it requires information of the dust emission in the FIR. In the local, low and intermediate-redshift universe, a relatively high percentage of galaxies exhibit observed FIR fluxes high enough to be detected with current instrumentation. However, in the high-redshift universe, a very low percentage of galaxies are individually detected in the FIR and, therefore, the accurate determination of dust attenuation is only possible for some the IR-brightest sources. This is one of the main reasons why many authors try to recover the dust attenuation from the values of the UV continuum slope by using some relations between the UV slope and dust attenuation, the so-called IRX-$\beta$ relations \citep[see for example][]{Meurer1999,Boissier2007,Boquien2009,Boquien2012,Overzier2011,Murphy2011,Buat2012,Heinis2013}\footnote{IRX refers to $\log{\left( L_{\rm IR} / L_{\rm UV}\right)}$.}. The IRX-$\beta$ relations that are found in the literature have been defined for specific kind of galaxies at specific redshifts. This is the case, for example, of local star-burst (SB) \citep[][hereafter M99]{Meurer1999} or local SF galaxies \citep[][hereafter B07]{Boissier2007}.

\begin{figure*}
\centering
\includegraphics[width=0.8\textwidth]{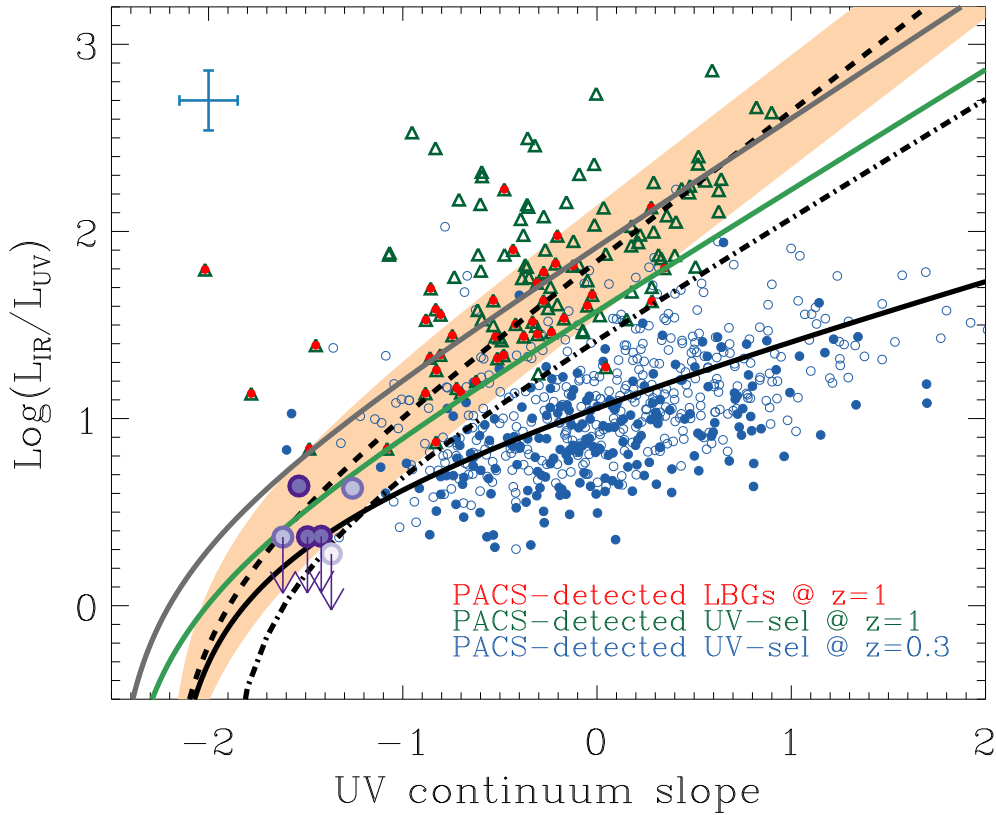}
\caption{Dust attenuation parametrized by the IR/UV ratio (IRX) as a function of the UV continuum slope for our PACS-detected LBGs and UV-selected galaxies, represented by red filled dots and green open triangles, respectively. For comparison, we also represent a sample of UV-selected PACS-detected galaxies at $z \sim 0.3$ with blue filled dots for galaxies with $\log{\left(L_{\rm UV} / L_\odot\right)} \geq 9.8$ and with blue open dots for galaxies whose $\log{\left(L_{\rm UV} /L_\odot \right)} < 9.8$. Dashed curve is the law for local star-bursts of \citet{Meurer1999} (M99), dashed dotted curve is the correction of \citet{Takeuchi2012} to the M99 law, and continuous line is the curve of \citet{Boissier2007} for normal local star-forming galaxies. The shaded region represents the dispersion of IRX around the M99 law. The IRX-$\beta$ relations of \citet{Buat2012} and \citet{Heinis2013} are over-plotted with a grey and green solid curves, respectively. We also plot with purple filled dots the points corresponding to the staked PACS-160$\mu$m fluxes for the six bins considered in the stacking procedure (see text for more details). The arrows indicate upper limits for the bins where no stacked detections are recovered. The blue error bars indicate the typical uncertainties (taken from \citet{Oteo2013}) of the parameters shown in the horizontal and vertical axes.
              }
\label{dust_beta_fig}
\end{figure*}

However, the IRX-$\beta$ relation has been proven not to be tight for any kind of galaxies. \cite{Goldader2002} obtained that the SB relation does not reproduce the observed IRX-$\beta$ relation for local LIRGs and ULIRGs. These galaxies tend to be located above the SB curve: for a given UV continuum slope, local LIRGs and ULIRGs tend to have higher dust attenuation than that predicted by the SB relation \citep{Buat2005,Takeuchi2010}. For FIR-bright galaxies at $z \sim 0.3$ there is no a clear IRX-$\beta$ relation either, since they are located over a wide region of the IRX-$\beta$ diagram \citep{Buat2010,Oteo2012b}. \cite{Nordon2013} find that main sequence galaxies form a tight sequence in the IRX-$\beta$ diagram, which has a flatter slope than commonly used relation and favors a SMC-like UV extinction curve. Furthermore, they obtain that the location of galaxies in the IRX-$\beta$ diagram correlates with their location in the SFR-$M_*$ diagram. Additionally, \cite{Reddy2012} suggest that the location of galaxies in the IRX-$\beta$ diagram depends upon age; younger galaxies follow a relation that is stepper than the M99 and \cite{Calzetti2000} ones.

In this section, by analyzing the location of our PACS-detected galaxies at $z \sim 1$ in an IRX-$\beta$ diagram (see Figure \ref{dust_beta_fig}), we examine whether they follow a relation between dust attenuation and $\beta$ and, therefore, whether it is possible to recover their dust attenuation from their UV continuum slope in an accurate way. In Figure \ref{dust_beta_fig} our PACS-detected LBGs are represented with red dots and our PACS-detected UV-faint galaxies with green open triangles. The curves of M99 and B07 for local SB and SF galaxies, respectively, are also shown. We also represent the correction of T12 to the M99 law that takes into consideration the small aperture of \emph{IUE} that may have missed about half of the light \citep[see also][]{Overzier2011}. It can be seen that our PACS-detected LBGs are mostly distributed around the M99 relation and within the zone defined by the uncertainties in the IRX ratio of the M99 relation (orange shaded zone). This also applies for our PACS-detected UV-faint galaxies. The location of our PACS-detected galaxies is also compatible with the \cite{Buat2012} relation for PACS-detected UV-selected galaxies at $0.95 < z < 2.2$. \cite{Heinis2013} obtain a IRX-$\beta$ relation for galaxies at $z \sim 1.5$ that, for a given UV continuum slope, gives lower values of the dust attenuation than those of our PACS-detected LBGs. This is likely due to the higher redshifts of their sources and mainly to the stacking analysis employed in that work that allow to the authors to recover less dusty galaxies.

In Figure \ref{dust_beta_fig} we also include the points associated to the stacked PACS-160$\mu$m fluxes for our LBGs in the six bins of rest-frame UV luminosities, stellar mass, and age. The UV continuum slope in each bin is the median value of the galaxies included in each one. As expected from Figures \ref{mipsdetections} and \ref{properties_PACS}, the PACS-stacked LBGs (which are PACS-undetected) are bluer than the PACS-detected LBGs and, therefore, are located in a different zone of the IRX-$\beta$ diagram. In the two cases where a stacked detection is recovered, we obtain that the point follow the M99 relation, although the upper values in the cases where a stacked detections not recovered indicate that the actual dust attenuation is lower than that predicted by the M99 law, favoring the T12 relation. In the zone where the stacked points are located, the M99, B07, and T12 curves are very close to each other. Actually, the B07 law is contained within the uncertainty region associated to the M99 relation. Therefore, any of the laws might be used to recover the dust attenuation of those galaxies from their UV continuum slope.

\begin{figure}
\centering
\includegraphics[width=0.49\textwidth]{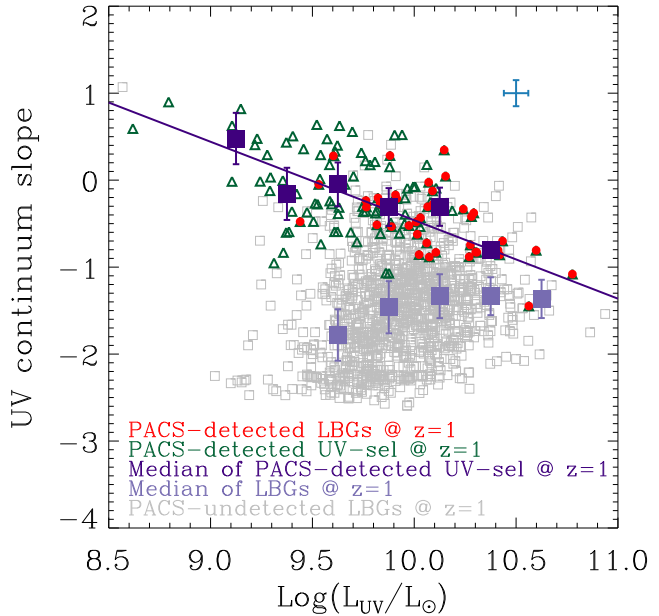}
\caption{Relation between the UV continuum slope and the rest-frame UV luminosity of our PACS-detected LBGs (red dots) and PACS-detected UV-faint galaxies (green open triangles). Filled dark purple squares represent the median value of the UV continuum slope of the PACS-detected galaxies in different bins of rest-frame UV luminosity, being the error bars the width of the distributions. The solid dark purple line is a linear fit to the dark purple filled squares. For comparison we also represent the PACS-undetected LBGs at $z \sim 1$ with grey open squares. Filled light purple squares represent the median value of the UV continuum slope of PACS-undetected galaxies in different bins of rest-frame UV luminosity, being the error bars the width of the distributions. The blue error bars indicate the typical uncertainties of the parameters shown in the horizontal and vertical axes.
              }
\label{lumi_beta}
\end{figure}

For a given value of the UV continuum slope, PACS-detected LBGs tend to have slightly lower values of dust attenuation than PACS-detected UV-faint galaxies. This is a direct consequence of the different selection criterion since LBGs are among the most UV-luminous galaxies and, therefore, have lower values of the IRX ratio. Furthermore, it can be seen in Figure \ref{dust_beta_fig} that PACS-detected LBGs tend to be located in a zone of the IRX-$\beta$ diagram corresponding to bluer values of the UV continuum slope than PACS-detected UV-faint galaxies. This way, we obtain that UV-brighter PACS-detected galaxies have bluer UV continuum slopes. At higher redshifts, \cite{Finkelstein2012} do not find significant correlation ($\lesssim 2\sigma$) between both parameters between $z \sim 4-7$. \cite{Dunlop2012} do not find significant trend between UV continuum slope and rest-frame UV luminosity between $z = 5-7$. However, \cite{Bouwens2012} report the existence of a well-defined rest-frame UV color-magnitude relation that becomes systematically bluer when the UV luminosity decreases. \cite{Bouwens2009} obtain a similar behavior between $z = 2.5-4$. \cite{Castellano2012} report that brighter LBGs at $z \sim 4$ are only slightly redder than fainter ones. Therefore, there seem to be some discrepancy in the results. What we have obtained for our PACS-detected LBGs and UV-faint galaxies does not apply to the whole population of SF galaxies at $z \sim 1$ since we have only considered the FIR-brightest galaxies. We further analyze the situation in Figure \ref{lumi_beta} including the PACS-undetected LBGs as well. As expected from Figure \ref{properties_PACS}, for each value of the rest-frame UV luminosity, PACS selects the reddest galaxies in the UV continuum. Furthermore, it can be seen that the trend found for PACS-detected and PACS-undetected galaxies are completely different as a consequence of their different FIR nature. Whereas for PACS-detected galaxies there is a clear reddening at lower luminosities, the median values of the UV continuum slope associated to the PACS-undetected galaxies seem to become bluer with decreasing luminosity. However, it should be remarked that the width of the distributions of the UV continuum slope for each bin of luminosity is high and, therefore, no correlation can be significantly constrained. Finally, it is also interesting to point out that the values of the UV continuum slope of our LBGs and UV-faint galaxies (both PACS-detected and PACS-undetected) are higher than those obtained in \cite{Finkelstein2012, Bouwens2009, Bouwens2012}, indicating an evolution of the UV continuum slope from $z \sim 7$ down to $z \sim 1$ \citep[see also][]{Oteo2013}.

\emph{Location of galaxies in the IRX-$\beta$ diagram as a function of redshift}: We compare now the location in the IRX-$\beta$ diagram of UV-selected PACS-detected galaxies as a function of redshift, focusing on our PACS-detected galaxies at $z \sim 1$ and $z \sim 0.3$. We use here the same sample of UV-selected galaxies at $z \sim 0.3$ than in Section \ref{bolometric_dust}. Note that this sample is limited to the same rest-frame UV luminosity than the UV-selected galaxies at $z \sim 1$ and, therefore, there is no bias in their UV selection. However, there is a bias in the IR selection since at $z \sim 0.3$ we can detect less IR luminous sources and this may have an effect in their dust properties. There is an additional difference between the two samples, which is the calculation of the UV continuum slope. As indicated in \cite{Oteo2013}, the UV continuum slope for the galaxies in our sample of UV-selected galaxies at $z \sim 1$ is obtained by fitting a power law to the the templates which fit the GALEX+ALHAMBRA photometry of each galaxy best. For the galaxies at $z \sim 0.3$, we do not have such a high-quality photometric coverage of the UV continuum and, as a consequence, their UV continuum slope is obtained from their photometry in FUV and NUV bands as in \cite{Kong2004}, \cite{Buat2010}, or \cite{Oteo2012b}, for example. Figure \ref{sum_coc} indicates that galaxies at $z \sim 0.3$ are located mostly around the SF relation. This represents a difference between $z \sim 1$ and $z \sim 0.3$ and implies that for a given value of the UV continuum slope, the dustiest galaxies at $z \sim 0.3$ are less attenuated than the dustiest UV-selected galaxies at $z \sim 1$. This tendency seem to continue to higher redshifts. According to \cite{Oteo2013_LBGsz3}, PACS-detected LBGs at $z \sim 3$ are more attenuated than PACS-detected LBGs at $z \sim 1$ for a given UV continuum slope. This higher dust attenuation for galaxies at higher redshifts might be due to an evolution of the IR luminosity of LBGs with redshift, at least in the IR-bright envelope. The dust attenuation recovered from the stacking analysis for the PACS-undetected LBGs at $z \sim 1$ is as low as the values found for galaxies at $z \sim 0.3$. However, they tend to be located in a zone of IRX-$\beta$ diagram associated to much bluer values of the UV continuum slope. This might indicate that, for the same dust attenuation, galaxies at higher redshifts tend to have bluer values of the UV continuum slope than galaxies at lower redshifts.

Joining all previous results, it turns out that the location of UV-selected galaxies in the IRX-$\beta$ depends upon both rest-frame UV luminosity, total IR luminosity, and redshift. Therefore, in general, dust attenuation for UV-selected galaxies at $z \sim 1$ cannot be recovered from the UV continuum slope with a single dust attenuation law. It should be noted that this result only applies for FIR-detected galaxies. For those FIR undetected, we can only give an approximation to their location in the IRX-$\beta$ diagram throughout stacking analysis and, therefore, the application of a single relation between UV continuum slope and dust attenuation is even riskier.

\begin{figure}
\centering
\includegraphics[width=0.49\textwidth]{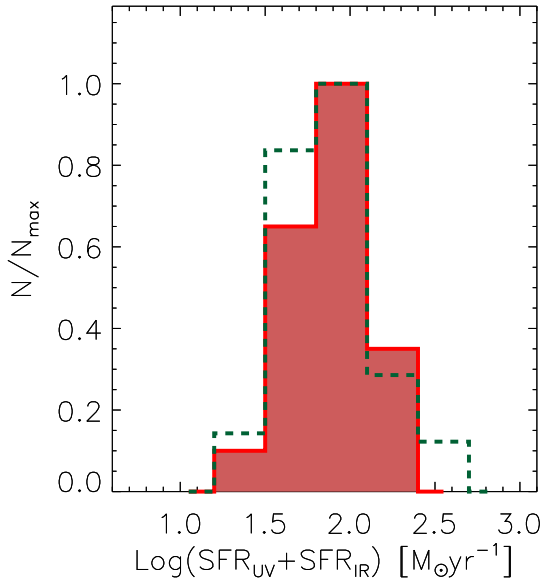}
\caption{Distribution of the total SFR as derived from the combination of UV and IR emissions for the galaxies studied in this work. PACS-detected LBGs and UV-selected galaxies are represented by red shaded and dashed green histograms. Histograms have been normalized to clarify the representation.
              }
\label{SFR}
\end{figure}

\section{Star formation rate}\label{starformation}

There are some estimators of the star-formation rate in galaxies, such as H$\alpha$ emission, UV emission, IR emission, or radio continuum. Maybe, the best determination of the total SFR in galaxies can be obtained by combining UV and FIR emissions, since they represent the measured UV light and the emission in the FIR of the light which is absorbed in the UV by dust, respectively. This way, the total SFR, SFR$_{\rm total}$, can be obtained by summing the star formation associated to the rest-frame UV light uncorrected by dust absorption, SFR$_{\rm UV,uncorrected}$, and the star formation associated to the emitted IR light, SFR$_{\rm IR}$. Each one can be obtained independently from the rest-frame UV and total IR luminosities by applying the \cite{Kennicutt1998} calibrations:

\begin{equation}\label{SFR_UV}
\textrm{SFR}_{\rm UV,uncorrected}[M_{\odot}\textrm{yr}^{-1}] = 1.4 \times 10^{-28}L_{1500}
\end{equation}

\begin{equation}\label{SFR_IR}
\textrm{SFR}_{\rm IR}[M_{\odot}\textrm{yr}^{-1}] = 4.5 \times 10^{-44}L_{\rm IR}
\end{equation}

\noindent where the IR luminosity is defined in the same way that in Section \ref{lumi_IR} and $L_{1500}$ is defined in a $L_\nu$ way. The total SFR can be obtained from:

\begin{equation}\label{SFR_total}
\textrm{SFR}_{\rm total} = \textrm{SFR}_{\rm UV, uncorrected} + \textrm{SFR}_{\rm IR}
\end{equation}

Figure \ref{SFR} represents the distribution of the total SFR for our sample of PACS-detected galaxies. The median values for our PACS-detected LBGs and UV-faint galaxies are $91 \pm 22 \, M_\odot$/yr and $73 \pm 39 \, M_\odot$/yr, respectively. The uncertainties correspond to the standard deviation of the distributions, as a measure of their width. Therefore, PACS-detected LBGs have higher values of the total SFR than UV-faint galaxies.


\begin{figure*}
\centering
\includegraphics[width=0.49\textwidth]{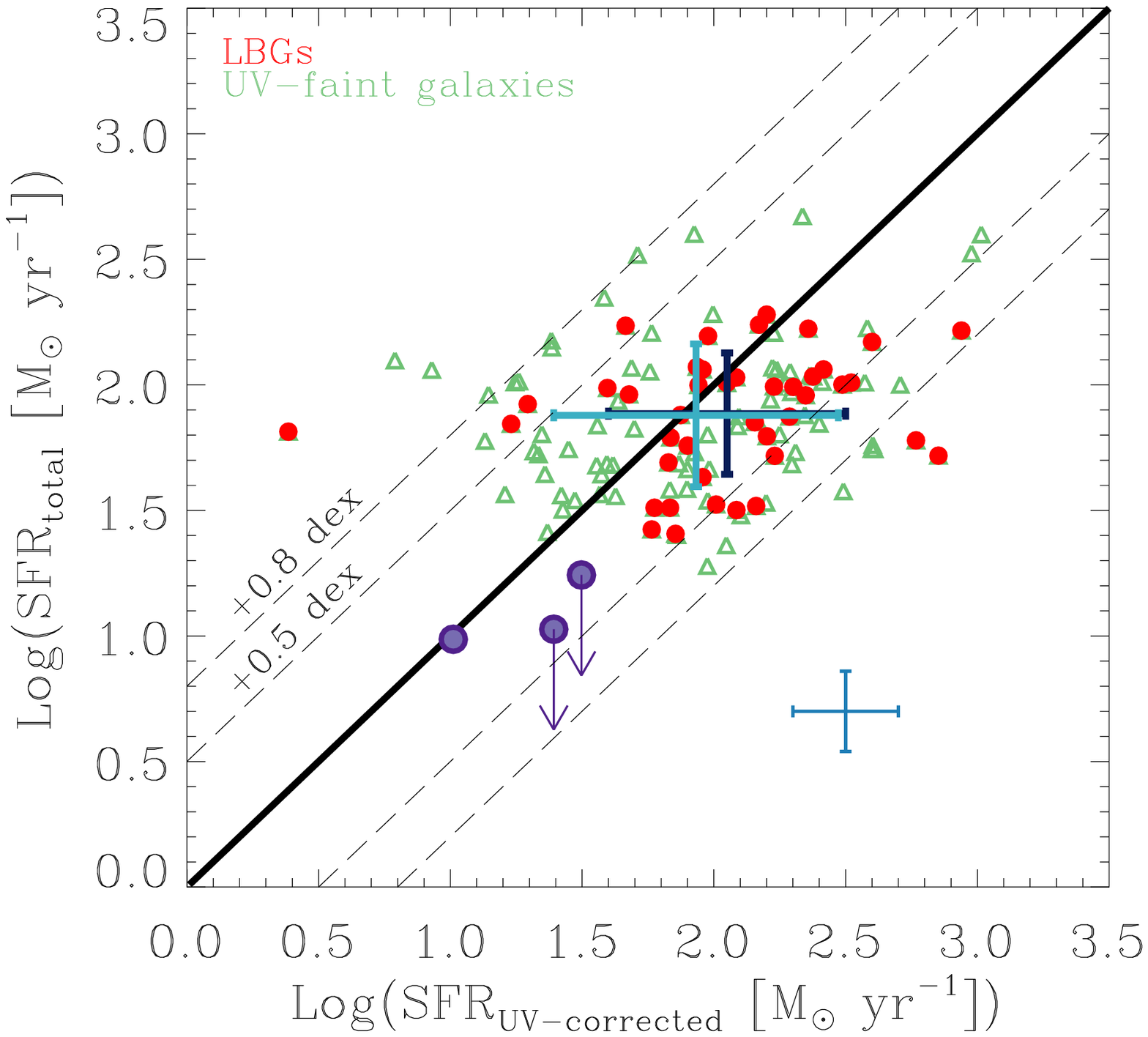}\\
\includegraphics[width=0.49\textwidth]{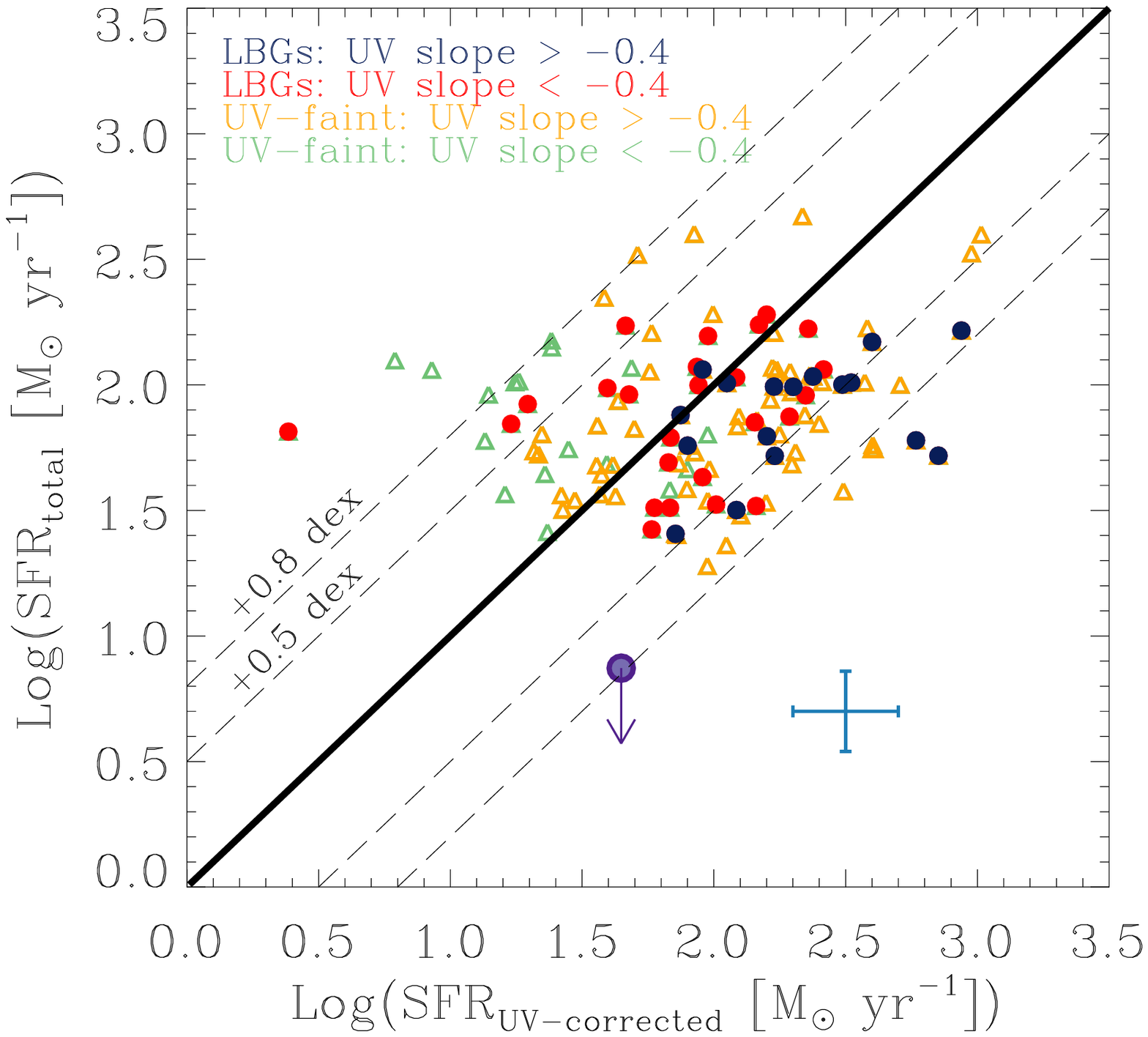}
\includegraphics[width=0.49\textwidth]{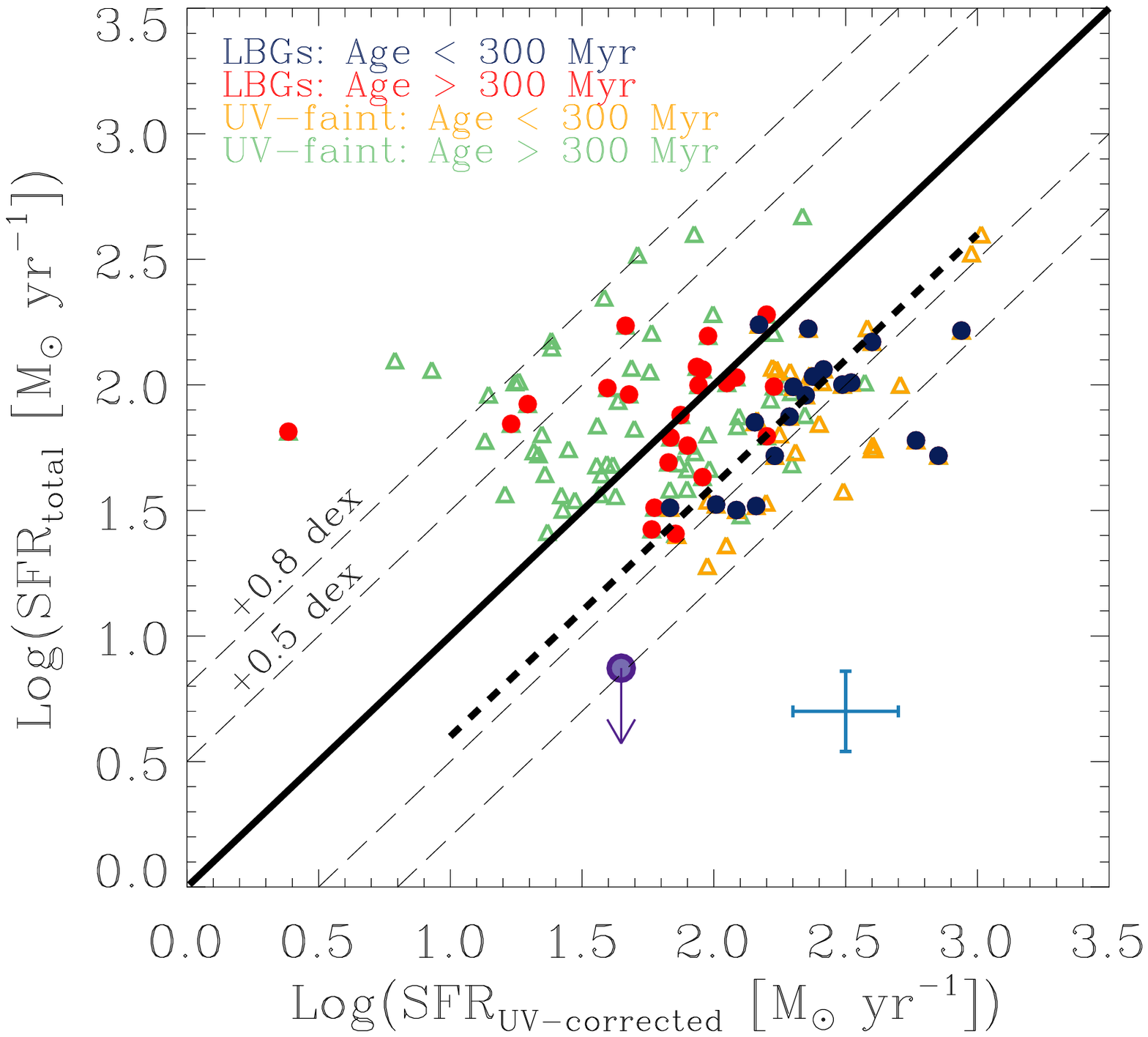}
\caption{\emph{Top}: Total SFR derived from the combination of UV and FIR measurements against the total SFR derived by correcting the rest-frame UV luminosity with the SED-derived $Es(B-V)$ value obtained from SED-fitting with BC03 templates associated to a constant star-formation rate and constant sub-solar metallicity $Z=0.4Z_\odot$. PACS-detected LBGs and UV-selected galaxies are represented by red filled dots and green open triangles, respectively. The solid line is the one-to-one relation. The median loci in the diagram of the PACS-detected LBGs and UV-selected galaxies are represented by the dark blue and light blue bars, respectively. The bars cross in the average value of each parameter for each kind of galaxy and the length of the bars is the standard deviation of the distribution of values. We include with purple filled dots the point corresponding to the stacking in the three bins of rest-frame UV luminosity. The arrows indicate upper limits for the two ranges where no stacked detections are recovered. The blue error bars indicate the typical uncertainties (taken from \citet{Oteo2013}) of the parameters shown in the horizontal and vertical axes. \emph{Bottom left}: The same than in the top panel but distinguishing points corresponding to LBGs and UV-faint galaxies with different values of the UV continuum slope, as indicated. Red and blue filled dots represent PACS-detected LBGs, while green and orange open triangles represent the general population of UV-selected galaxies. \emph{Bottom right}: Same than bottom left but we distinguishing in values of age, as indicated. The dashed straight line indicates a deviation of -0.6 dex with respect to the one-to-one relation (solid line) and it is represented to guide the eye about the overestimation of the SED-derived total SFR for young galaxies. We plot in the bottom right and left panels the upper value corresponding to stacked PACS-160$\mu$m flux of the galaxies younger than 100 Myr with a purple filled dot.
              }
\label{sfr_sfr_edad_beta}
\end{figure*}

\subsection{Dust-correction factors at $z \sim 1$}\label{dust_correction}


\begin{figure*}
\centering
\includegraphics[width=0.49\textwidth]{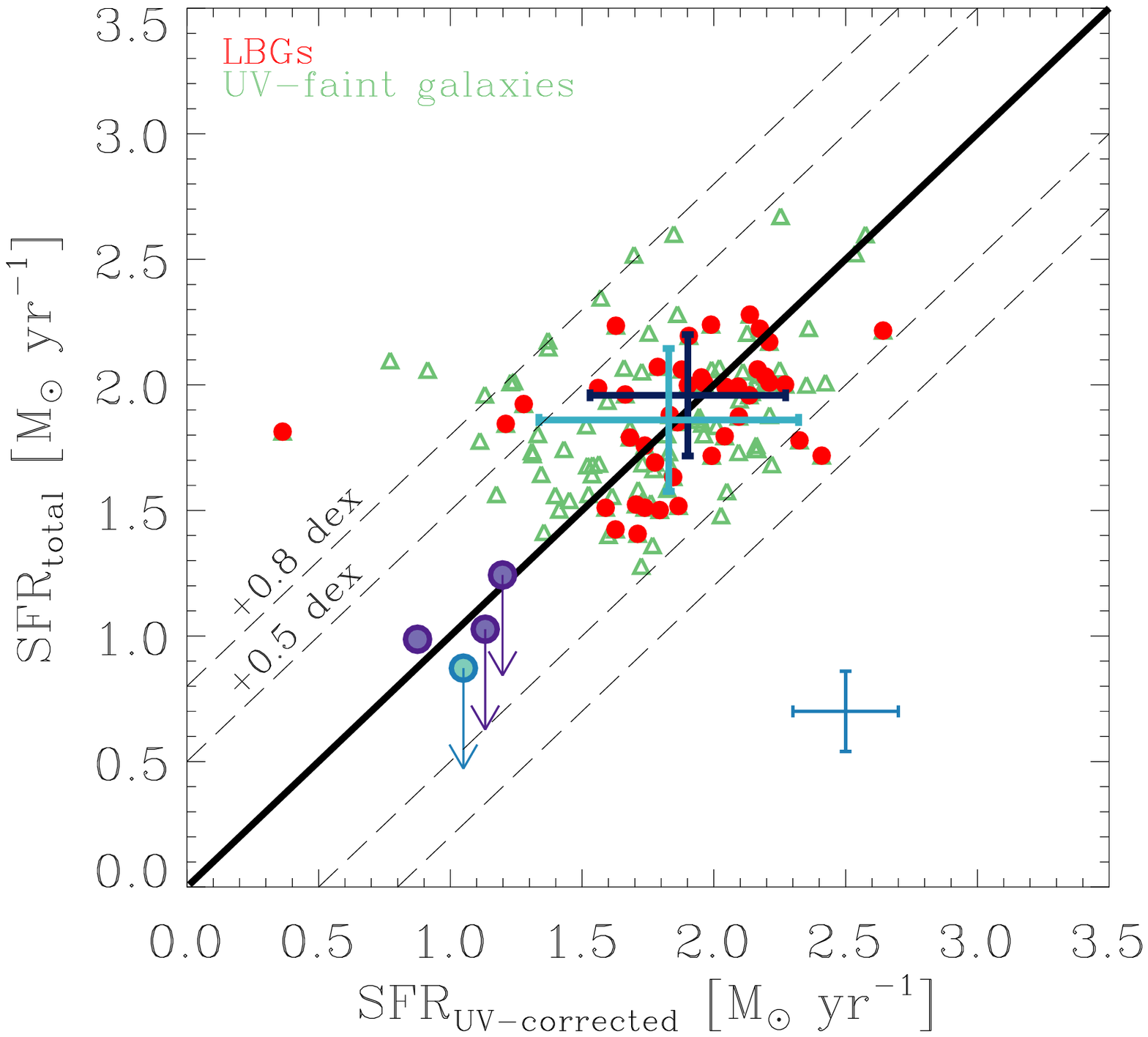}
\includegraphics[width=0.49\textwidth]{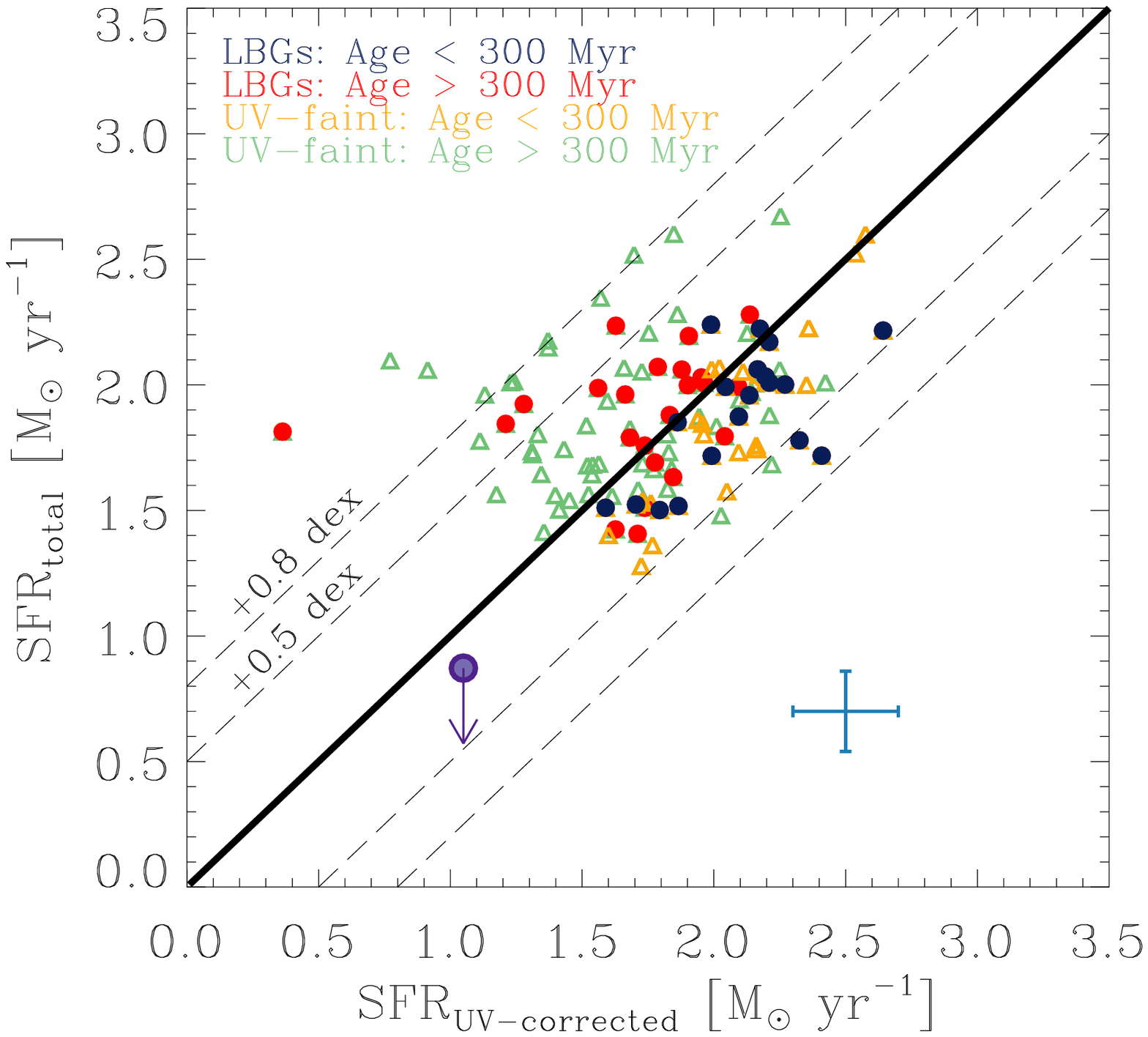}
\caption{\emph{Left}:Total SFR derived from the combination of UV and IR measurements against the total SFR obtained by correcting the rest-frame UV luminosity with the UV continuum slope and applying \citet{Meurer1999} relation. PACS-detected LBGs and UV-selected galaxies are represented with red filled dots and green open triangles, respectively. The median loci in the diagram of the PACS-detected LBGs and UV-selected galaxies are represented by the dark blue and light blue bars, respectively. The bars cross in the median value of each parameter for each kind of galaxy and the length of the bars is the standard deviation of the distribution of values. The solid line is the one-to-one relation. PACS-stacked LBGs in the bins of rest-frame UV luminosity are represented as filled purple dots and the PACS-stacked LBGs in the age bin are indicated with a filled blue dot. Arrows indicate upper values where no stacked detection is found. The blue error bars indicate the typical uncertainties (taken from \citet{Oteo2013}) of the parameters shown in the horizontal and vertical axes. \emph{Right}: The same than in the left panel but considering different IRX relations: \citet{Meurer1999} in green, \citet{Overzier2011} in orange, and \citet{Takeuchi2012} in purple. In this plot, and with the aim of clarifying the results, LBGs and UV-faint galaxies are considered as a whole population.
              }
\label{sfr_sfr}
\end{figure*}

In the upper panel of Figure \ref{sfr_sfr_edad_beta}, we represent the SED-derived dust-corrected total SFR against the total SFR obtained from the combination of UV and IR measurements (Equation \ref{SFR_total}) for our PACS-detected LBGs (red dots) and UV-selected galaxies (green triangles). We define the SED-derived total SFR as the one obtained by multiplying the rest-frame dust-uncorrected $SFR_{\rm UV,uncorrected}$ by the factor 10$^{0.4A_{1500}}$, where $A_{1500}$ is the dust attenuation in 1500\AA\ calculated from the SED-derived $E_s(B-V)$ assuming the \cite{Calzetti2000} law. We also represent the median loci of each kind of galaxy in the diagram. It can be seen that, although in average the SED-derived dust attenuation reproduces the UV+IR-derived total SFR with some underestimation, the dispersion around the one-to-one relation is relatively high. We also include in that panel the points associated to the stacking analysis in the three bins of rest-frame UV luminosity. It can be seen that there is a good agreement between the SED-derived and UV+IR-derived total SFRs, although the PACS-stacked points associated to the highest values of the rest-frame UV luminosity are more deviated from the one-to-one relation, suggesting again that the SED-derived $E_s(B-V)$ gives overestimated results of the total SFR for UV-bright galaxies. This is the opposite to what is found at $z \sim 3$, where the SED-derived $E_s(B-V)$ underestimate the total SFR for most PACS-detected LBGs. It should be remarked that the fact that the uncertainties in the SED-derived and UV+IR-derived total SFR are very similar does not mean, a priori, that the SED-derived is as good estimator of the total SFR as the UV+IR-derived. The uncertainties in the SED-derived total SFR only reflect the uncertainties in the SED-derived dust attenuation and rest-frame UV luminosity but do not reflect the capacity of the SED fitting procedure to reproduce the FIR emission of the galaxies. There some some physical reasons from that we expect that the UV+IR-derived total SFR is a good estimation of the real total SFR of a galaxy, but there is no physical reason to expect that it is possible to recover the total SFR for any kind of galaxy. This is only an expectation from the SED fitting procedure employed, but it is something that need to be checked.

In the lower panels of Figure \ref{sfr_sfr_edad_beta} we further analyze the relation between SED-derived and UV+IR-derived total SFRs by dividing the sample of PACS-detected galaxies as a function of the UV continuum slope and the SED-derived age of the galaxies. It can be seen that the PACS-detected LBGs and UV-faint galaxies which are further the one-to-one relation are those with younger stellar populations. The stacked point corresponding to the galaxies with age younger than 100 Myr is also in agreement with the previous tendency. However, we do not find significant correlation with the UV continuum slope. It should be noted that the tendency with age can be a consequence of the degeneracy between dust attenuation and stellar age in the SED-fitting procedures. The reddening of the SED of a galaxy can be caused either by the presence of old stellar population or by the presence of high values of dust attenuation. A galaxy with a red SED can be young and dusty but also old and less dusty. This way, the overestimation of the total SFR for young galaxies, produces by high values of the dust attenuation, can be due to the fact that the SED-fitting cannot distinguish between old and young and dust-free and old galaxies.

In the left panel of Figure \ref{sfr_sfr} we represent the SED-derived dust-corrected total SFR obtained from the dust attenuation derived from the values of the UV continuum slope of each galaxy by employing the M99 law. The accuracy of this method is directly related to the location of our PACS-detected galaxies in the IRX-$\beta$ diagram. When dust-correcting with this method, the SED-derived dust-corrected total SFR of PACS-detected LBGs and UV-selected galaxies is in very good agreement with the one derived from IR and UV measurements. The points associated to the PACS-stacked galaxies in the three bins of rest-frame UV luminosities are also very near to the one-to-one relation. Comparing the median loci of our PACS-detected galaxies in Figures \ref{sfr_sfr_edad_beta} and \ref{sfr_sfr} it can be seen that the dust attenuation derived from the UV continuum slope and the M99 relation gives much better estimations of the total SFR than that derived with the SED-derived $E_s(B-V)$. As it will be explained in Section \ref{radio_VLA} this differential behavior is also found when obtaining the total SFR with radio-VLA data instead of FIR ones. The right panel of Figure \ref{sfr_sfr} shows evidence that the dust correction factors obtained with the M99 law are also age-dependent. Younger galaxies than 300 Myr tend to be located below the one-to-one relation, where are older galaxies tend to be located above. This might be due to the dependence between the age and the location of galaxies in the IRX-$\beta$ diagram \citep[see also][]{Reddy2010,Reddy2012}.

\begin{figure}
\centering
\includegraphics[width=0.49\textwidth]{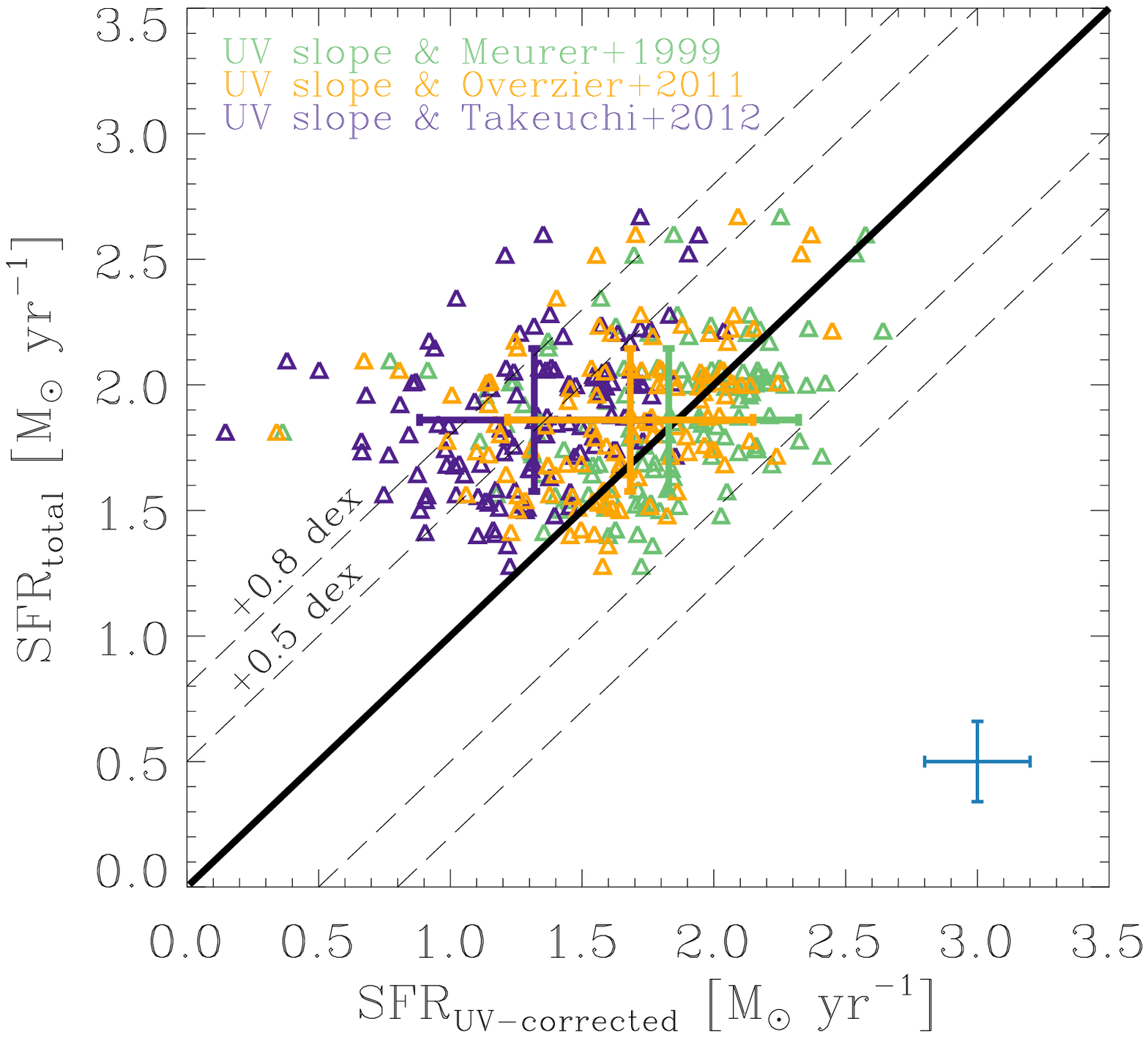}
\caption{Relation between the UV+IR-derived total SFR and the dust-corrected total SFR by using the UV continuum slope and the application of three different IRX-$\beta$ relations: Green triangles are associated to the \citet{Meurer1999} law, orange triangles to the \citet{Overzier2011}, and purple triangles to the \citet{Takeuchi2012} law. The median loci of the galaxies in the diagram for each of the three laws are represented with bars of the same color. The bars cross in the median value of the parameters presented in the x- and y-axis and the length of the bars is the standard deviation of the distribution of values. The solid line is the one-to-one relation and the dashed lines represent deviations of $\pm$ 0.5 and $\pm$0.8 dex with respect to the one-to-one relation. The blue bars represent the typical uncertainties in the derivation of the total SFR with each of the two methods.
              }
\label{sfr_sfr_Takeuchi}
\end{figure}

We compare in Figure \ref{sfr_sfr_Takeuchi} the dust correction factors obtained after the application of the M99 and the new relations of \cite{Takeuchi2012} (T12) and \cite{Overzier2011} (O11) for our PACS-detected galaxies. For clarity reasons, we consider in this plot our PACS-detected LBGs and UV-faint galaxies as a whole population of PACS-detected and UV-selected galaxies at $z \sim 1$. It can be seen that the IRX-$\beta$ relation that represent the best our PACS-detected galaxies is the M99 one. This is a direct consequence of the location of our galaxies in the IRX-$\beta$ diagram. However, our PACS-detected galaxies represent a low percentage of the whole population of SF galaxies at $z \sim 1$. Additionally, for each UV slope, PACS only detects the dustiest galaxies. Therefore, a IRX-$\beta$ relation that gives lower values of the dust attenuation for each UV continuum slope (such as the O11 and T12 ones) might be more appropriate to recover the dust attenuation of SF galaxies at $z \sim 1$ from their UV continuum slope.

\section{SFR vs stellar mass diagram}\label{sfr_mass_plane}

\begin{figure*}
\centering
\includegraphics[width=0.49\textwidth]{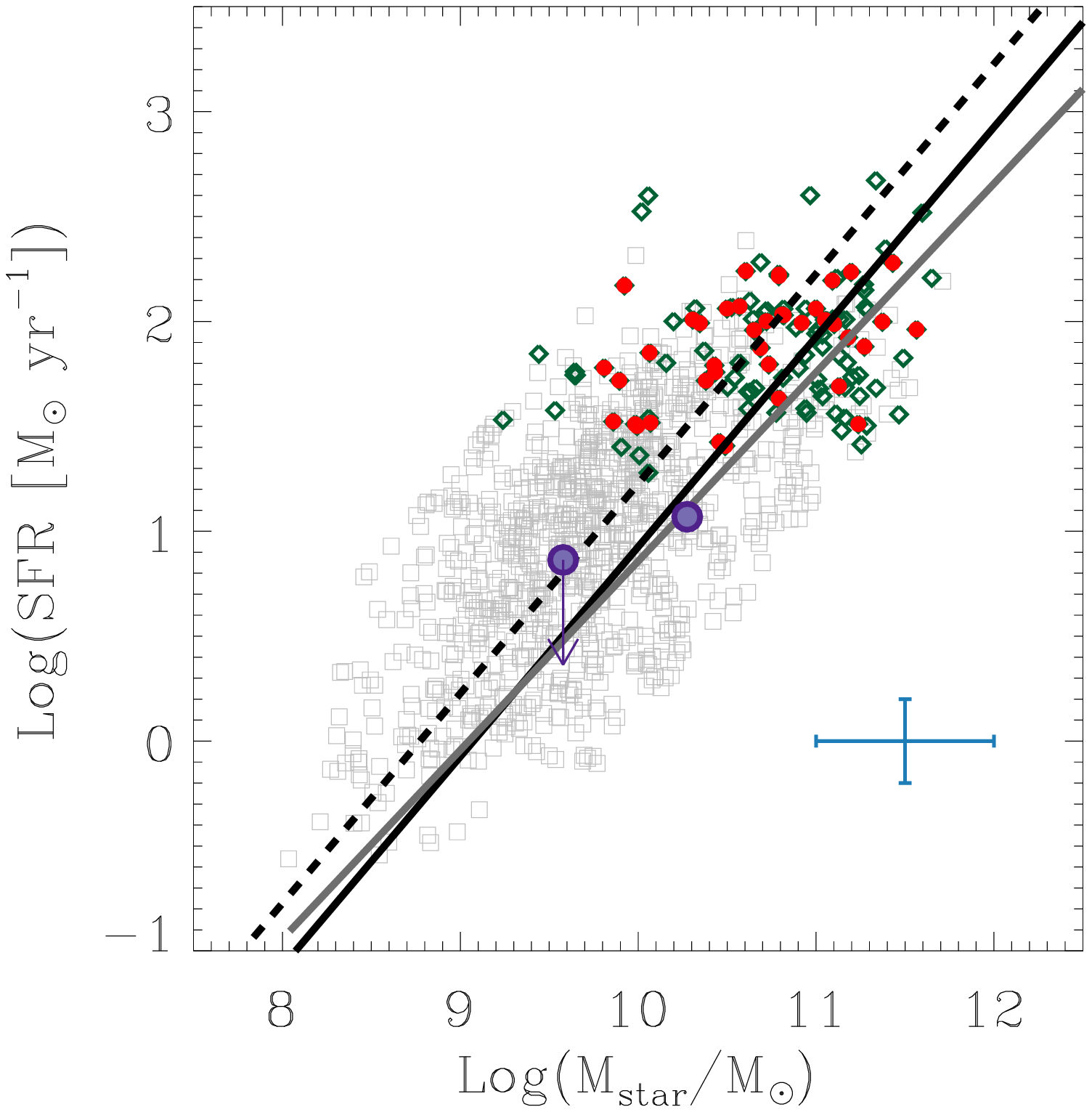}
\includegraphics[width=0.49\textwidth]{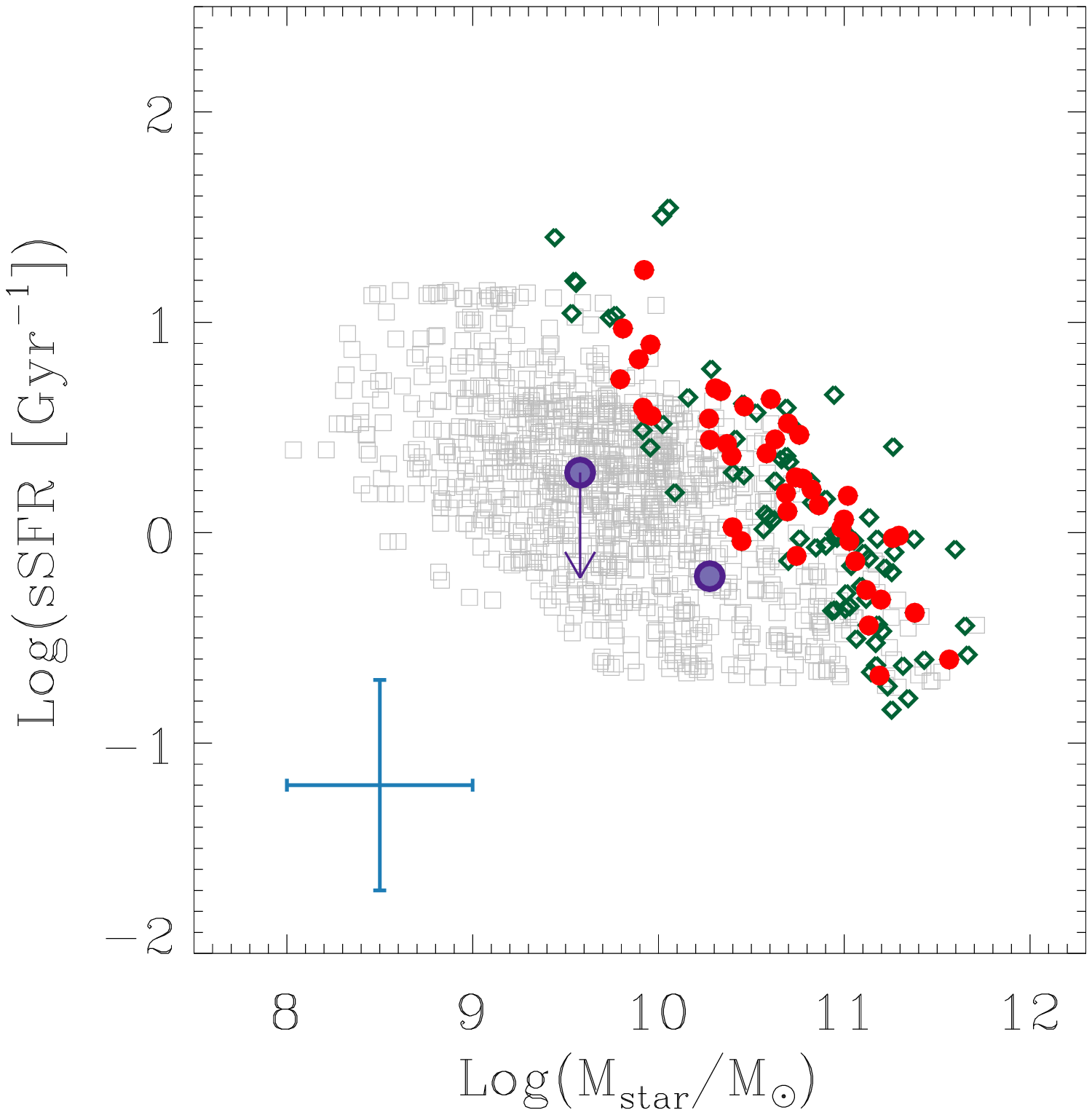}
\caption{\emph{Left:} Location of our PACS-detected galaxies in an SFR-mass diagram. PACS-detected LBGs and UV-selected galaxies are represented with red filled dots and green open diamonds, respectively. The solid black line is the main sequence (MS) of star forming galaxies at $z \sim 1$ taken from \citet{Elbaz2011} and the dashed curve is twice the values of that MS. The solid grey line is the MS at $z \sim 1$ of \citet{Elbaz2007}. The values of the dust-corrected total SFR adopted are those obtained from the UV continuum slope and the application of the \citet{Meurer1999} relation (see Section \ref{dust_correction}). The blue error bars indicate the typical uncertainties (taken from \citet{Oteo2013}) of the parameters shown in the horizontal and vertical axes. \emph{Right:} Specific star formation rate, $sSFR=SFR/M_*$ as a function of the stellar mass for our PACS-detected LBGs (red filled dots) and UV-selected galaxies (green open diamonds). Purple filled dots represent the results of the stacking in PACS-160$\mu$m in two bins of stellar mass. The arrow represent the upper limit of the total IR luminosity in the bin associated to the lowest stellar masses, where no stacked PACS-160$\mu$m detection is found. In both panels, for comparison, we also show the location in the SFR-mass and sSFR-mass diagrams of the PACS-undetected LBGs at $z \sim 1$ (grey open squares).  The blue error bars indicate the typical uncertainties (taken from \citet{Oteo2013}) of the parameters shown in the horizontal and vertical axes.
              }
\label{sfr_mass}
\end{figure*}

In \cite{Oteo2013} we study the location of the whole sample of GALEX-selected LBGs in a SFR vs stellar mass diagram \citep{Salim2007,Elbaz2007,Noeske2007,Daddi2007,Pannella2009,Rodighiero2010,Gonzalez2010,Karim2011,Elbaz2011,Salmi2012,Sawicki2012}. It was found that, according to the definition of \cite{Elbaz2011} of the main sequence (MS) of galaxies at $z \sim 1$, most GALEX-selected LBGs are located in the SB region, giving clues to their SF activity. Additionally, it was found that more massive galaxies have higher total SFR and lower sSFR. In that case, we used SED-derived values of the dust-corrected total SFR, but not those derived from direct IR measurements. Now, we go back to the analysis of the location of our UV-selected galaxies at $z \sim 1$ in the SFR-mass diagram but focusing on the PACS-detected ones and, thus, utilizing the values of the total SFR derived from the combination of UV and IR measurements (Equation \ref{SFR_total}).

Left panel of Figure \ref{sfr_mass} represents the $SFR-M_*$ for our PACS-detected LBGs and UV-faint galaxies. The SFRs which appear in the vertical axis of the plot are the total SFRs determined with the combination of UV and IR data (Equation \ref{SFR_total}) and the stellar masses are those obtained by fitting the GALEX+ALHAMBRA photometry with BC03 templates built by assuming a constant SFR \citep{Oteo2013}. Along with the data points, the SFR-mass relation for the MS of galaxies at $z \sim 1$ \citep{Elbaz2011} is also represented. \cite{Elbaz2011} assume for their definition of MS a slope of the $SFR$-mass relation equal to one and, therefore, that the $sSFR$ is mass independent at a fixed redshift. First, it can be seen that most PACS-detected galaxies at $z \sim 1$ are over the 'mean sequence' of galaxies at their redshift and, therefore, they are normal SF galaxies. Only the most massive galaxies are below the MS and the less massive are in the SB region. There is no significant relation between SFR$_{\rm total}$ and stellar mass. In \cite{Oteo2013} we reported that galaxies with higher masses tend to have higher star formation rates. The reason why we do not see such a relation here is because in the present work we only analyze PACS-detected galaxies and, therefore a subsample of IR-bright, dusty, and massive galaxies of the whole sample analysed in \cite{Oteo2013}. The range of IR luminosities of our galaxies does not allow to constrain a general relation between SFR$_{\rm total}$ and stellar mass. However, we do see that despite having a narrow range of values of SFR$_{\rm total}$ there is a large dispersion in the stellar mass values. This result is consistent to that found in \cite{Giovannoli2011} for their MIPS-24$\mu$m selected LIRGs at $ \sim 1$. They also found that galaxies with stellar populations younger than 300 Myr exhibit a larger dispersion in the stellar mass than galaxies whose SFR is constant over a longer period.

With the aim of gaining insights about the location of IR-faint galaxies in the $SFR-M_*$ diagram, we also plot the two points associated to the stacking results of the galaxies located in the two bins of stellar mass defined in Section \ref{stack}. It can be seen that the stacked point corresponding to the LBGs in the range of the highest stellar masses is right over the MS of galaxies. The upper limit of the total SFR for PACS-stacked LBGs in the range of the lowest values of stellar mass indicates that most of these galaxies might also be located over the MS. This indicates that most GALEX-selected LBGs, regardless their detection in the FIR, tend to be located over the main sequence of galaxies at their redshift.

We find (see right planel of Figure \ref{sfr_mass}) a trend between the sSFR and the stellar mass, which is similar than that found in \cite{Oteo2013} with UV, optical and near-IR data and in some other works at different redshifts \citep{Rodighiero2010,Feulner2005,Feulner2005b,Erb2006,Dunne2009,Damen2009}: galaxies with higher stellar masses tend to have lower sSFRs. This relation is characterized by its slope. In \cite{Oteo2013}, by fitting a linear relation in the form $\log{\left(sSFR [Gyr^{-1}] \right)} = a+b*\log{\left( M_* [M_\odot] \right)}$ to the whole population of GALEX-selected LBGs it was found $a=5.20\pm0.25$ and $b=-0.49\pm0.03$ when dust-correcting the rest-frame UV luminosity with the SED-derived $E_s(B-V)$ and $a=3.95\pm0.17$ and $b=-0.38\pm0.02$ when dust-correcting the rest-frame UV luminosity with the dust attenuation obtained from the UV continuum slope and the relation of M99. Carrying out the same linear fit to the PACS-detected LBGs studied in this work we find $a=10.1\pm0.8$ and $b=-0.95\pm0.07$. For PACS-detected LBGs the correlation between sSFR and stellar mass has much lower scatter and a higher slope than that for the whole population of LBGs due to the bias produced by the FIR selection. Actually, it can be seen in the right panel of Figure \ref{sfr_mass} that, for each stellar mass, PACS only sees the galaxies with the highest sSFRs. However, the sSFR for the most massive galaxies in the sample does not have values as high as the ones derived for the least massive galaxies. This agrees with the downsizing scenario where massive galaxies form their stars earlier and faster than lower massive ones. The absence of massive galaxies with $sSFR$ values as high as those for the least massive ones is not a bias due to the FIR selection criteria, since massive galaxies with high $sSFR$ should have been detected with PACS. We do not find any significant difference in the trend between LBGs UV-faint galaxies in the sSFR-mass diagram: all our PACS-detected galaxies have similar values of the sSFR.

To further analyze the correlation between sSFR and stellar mass for IR-fainter LBGs we also plot in the right plot of Figure \ref{sfr_mass} the two points associated to the stacked PACS-160$\mu$m fluxes for the LBGS in the two stellar mass ranges defined in Section \ref{stack}. In agreement with Figure \ref{properties_PACS}, the values of the $sSFR$ for PACS-detected and PACS-stacked galaxies are within a similar range. The stacked points are deviated from the linear relation found for PACS-detected galaxies towards lower values of the sSFR. This indicates that PACS detections segregate galaxies with the highest sSFR for a given value of stellar mass and that the relation between sSFR and stellar mass becomes much wider when introducing IR-faint galaxies.

\begin{figure*}
\centering
\includegraphics[width=0.8\textwidth]{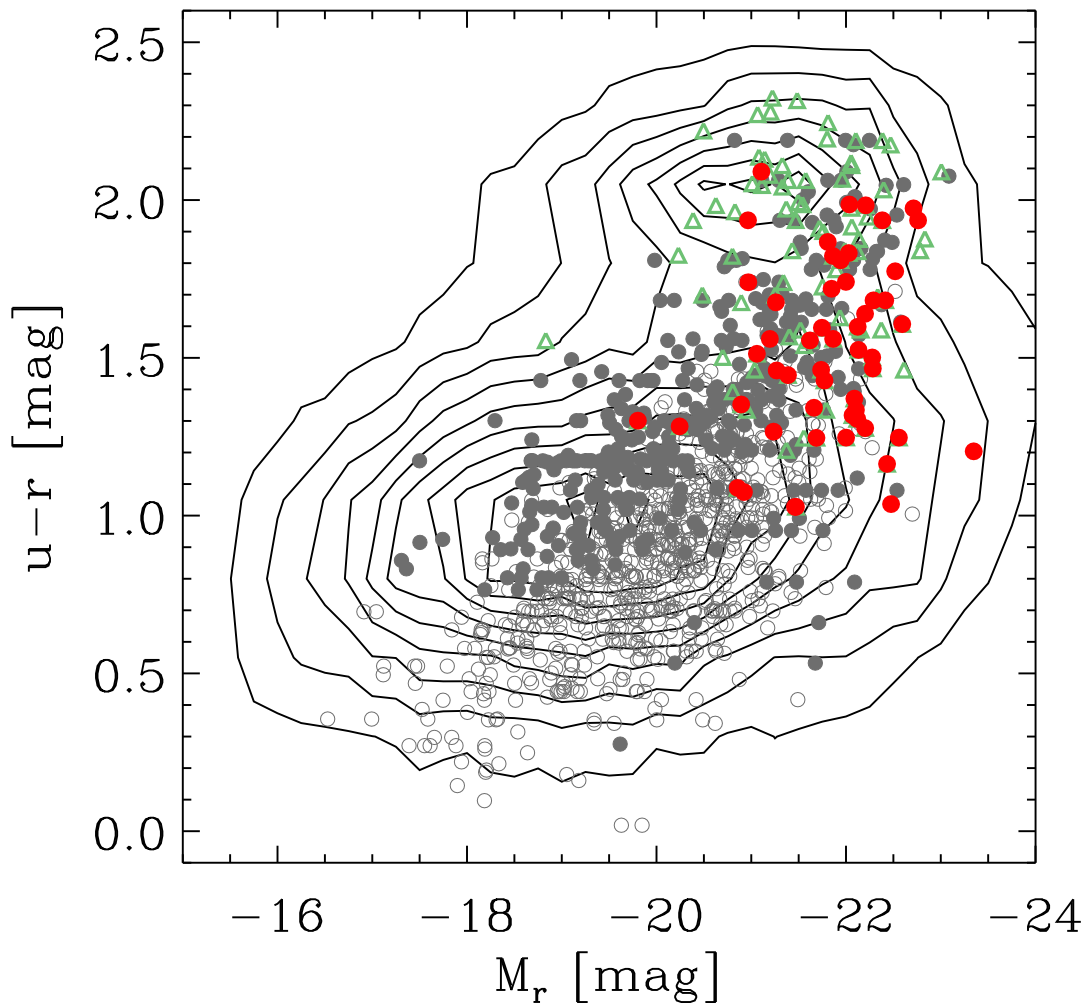}
\caption{Location of our PACS-detected galaxies in a color-magnitude diagram. The color $u-r$ is calculated in the rest-frame \citep[see][]{Oteo2013}. PACS-detected LBGs and UV-selected galaxies are represented with red filled dots and green open triangles, respectively. Grey filled dots represent PACS-undetected LBGs older than 1200 Myr and grey open dots represent PACS-undetected LBGs younger than 1200 Myr. Black contours represent the CMD of a general sample of galaxies $z \sim 1$ selected from the ALHAMBRA survey.
              }
\label{CMD_PACS}
\end{figure*}

\section{CMD of the PACS-detected sources}\label{CCMMDD}

In this section we study the location of our PACS-detected galaxies in a color-magnitude diagram (CMD). Such a diagram was already studied in \cite{Oteo2013} for the whole population of LBGs at $z \sim 1$ and it was found that most LBGs occupy the blue cloud of the galaxies at their redshift. Only the dustiest and/or oldest galaxies are distributed over the green valley or the red sequence. In Section \ref{PACS_counterparts} we saw that dustier sources are more likely detected in PACS. Therefore, it should be expected that PACS-detected galaxies are located mostly above the blue cloud. This is confirmed in Figure \ref{CMD_PACS}, where LBGs and UV-faint galaxies are represented in red filled dots and green open triangles, respectively. As in \cite{Oteo2013}, the $u-r$ are associated to the magnitudes in the $u$ and $r$ broad-band filters of the SDSS survey. The apparent $u$ and $r$ and absolute $r$ magnitudes are obtained by convolving the best-fitted template of each galaxy with the transmission of the $u$ and $r$ SDSS filters shifted in wavelength according the redshift of each source.

It can be seen that PACS-detected galaxies are mainly located over the green valley or the red sequence at their redshift. Additionally, there is a difference between LBGs and UV-faint galaxies in the sense that LBGs tend to have bluer optical colors than UV-faint galaxies as a consequence of their lower dust attenuation. UV-faint galaxies are more attenuated by dust and are mostly located near the red sequence despite being actively SF galaxies according to their total SFR.

\begin{figure*}
\centering
\includegraphics[width=0.10\textwidth]{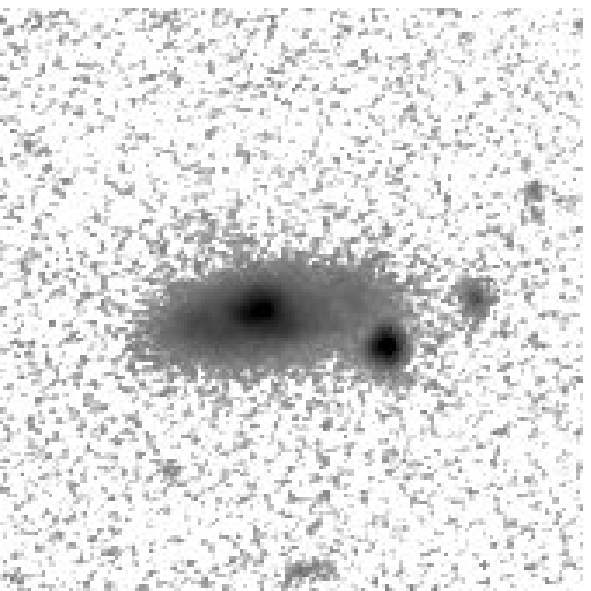}
\includegraphics[width=0.10\textwidth]{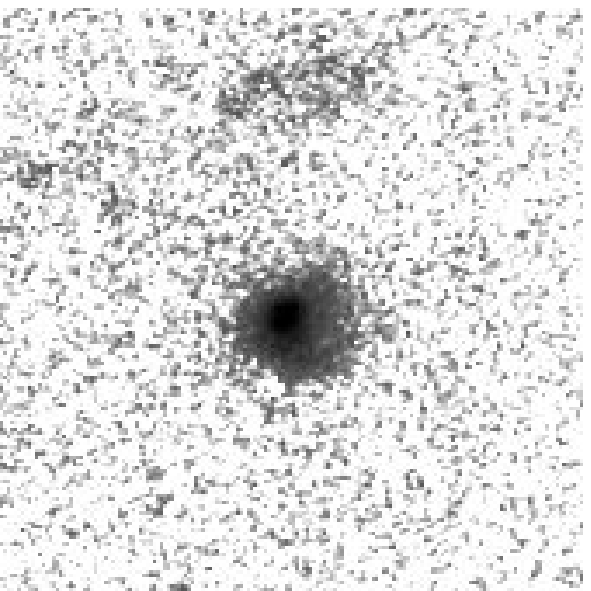}
\includegraphics[width=0.10\textwidth]{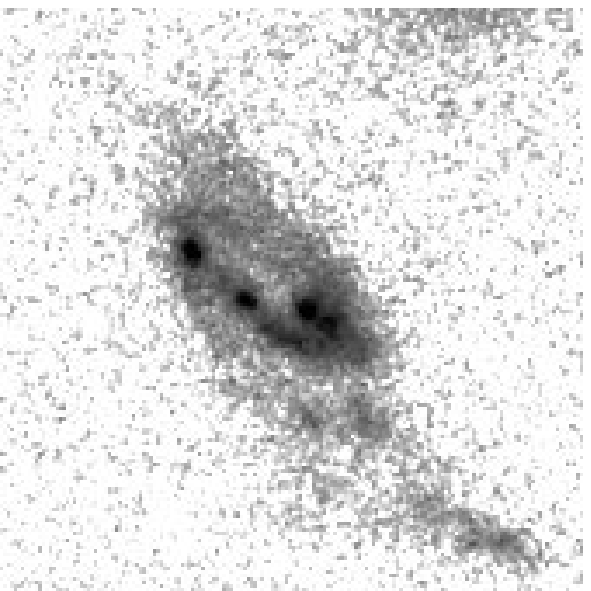}
\includegraphics[width=0.10\textwidth]{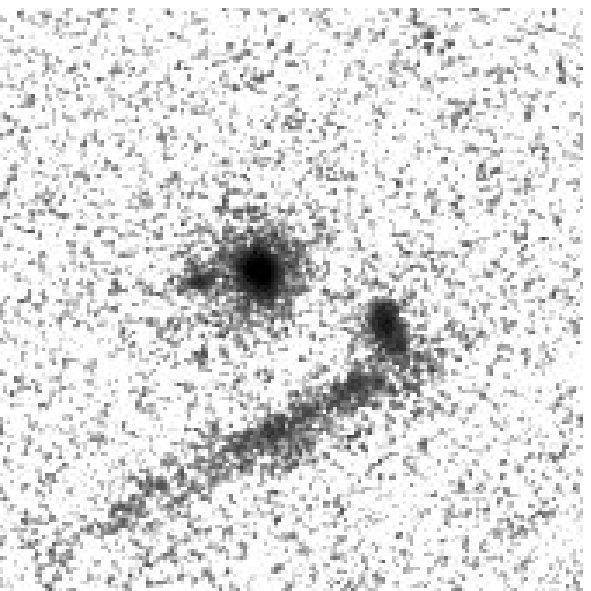}
\includegraphics[width=0.10\textwidth]{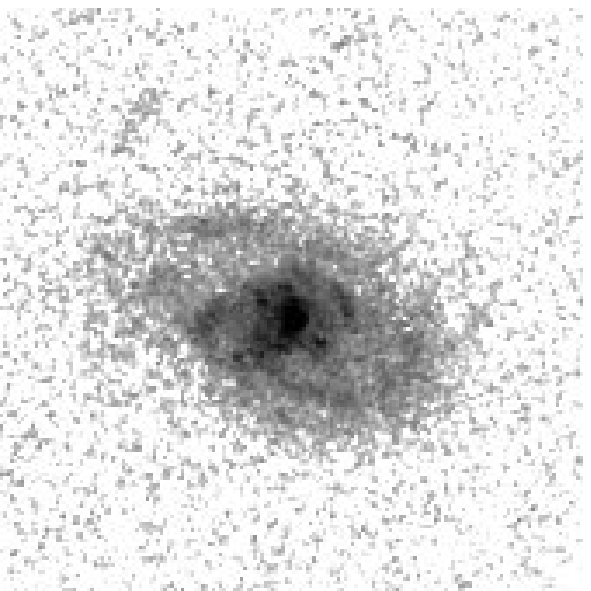}
\includegraphics[width=0.10\textwidth]{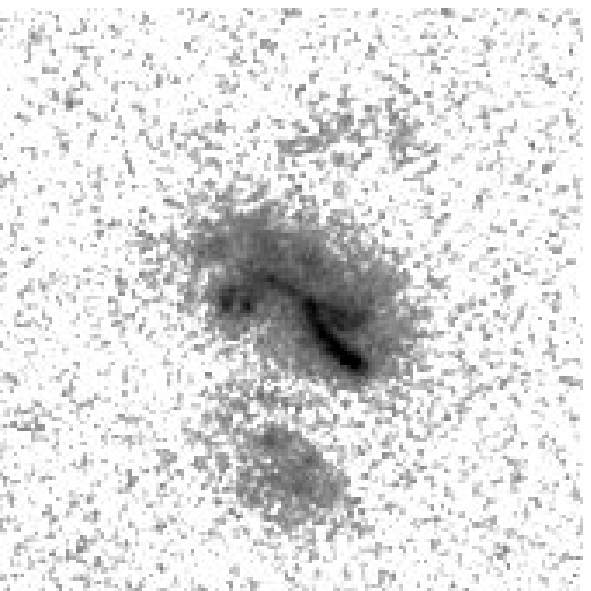}\\
\includegraphics[width=0.10\textwidth]{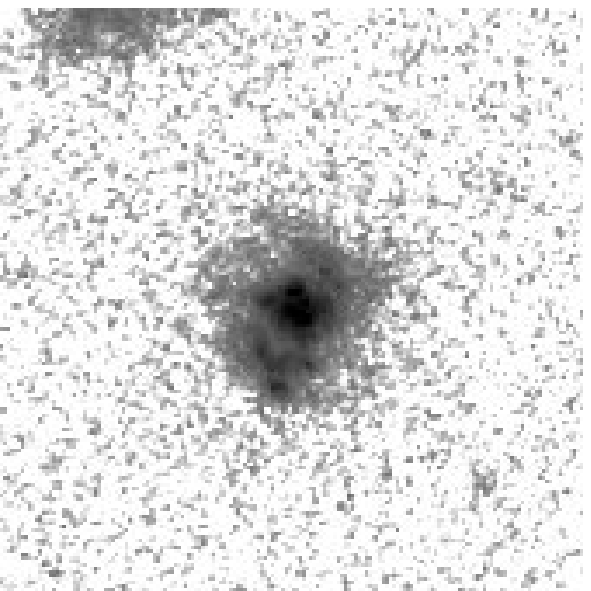}
\includegraphics[width=0.10\textwidth]{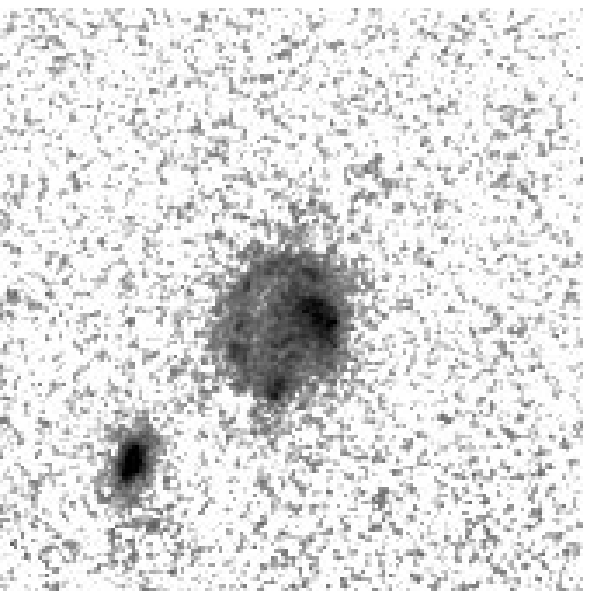}
\includegraphics[width=0.10\textwidth]{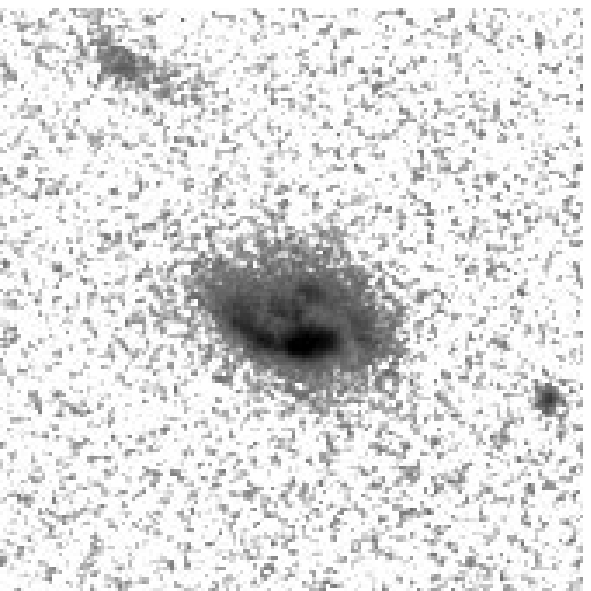}
\includegraphics[width=0.10\textwidth]{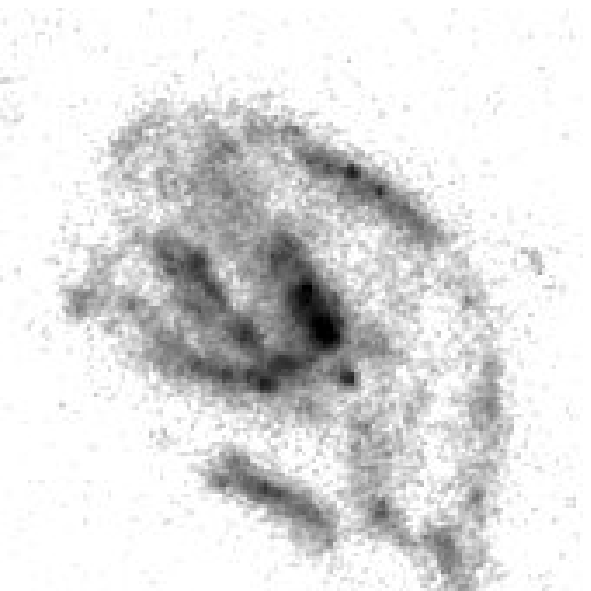}
\includegraphics[width=0.10\textwidth]{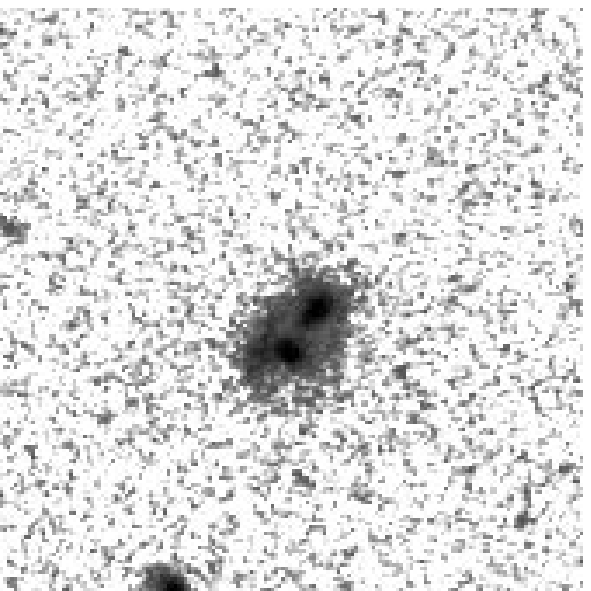}
\includegraphics[width=0.10\textwidth]{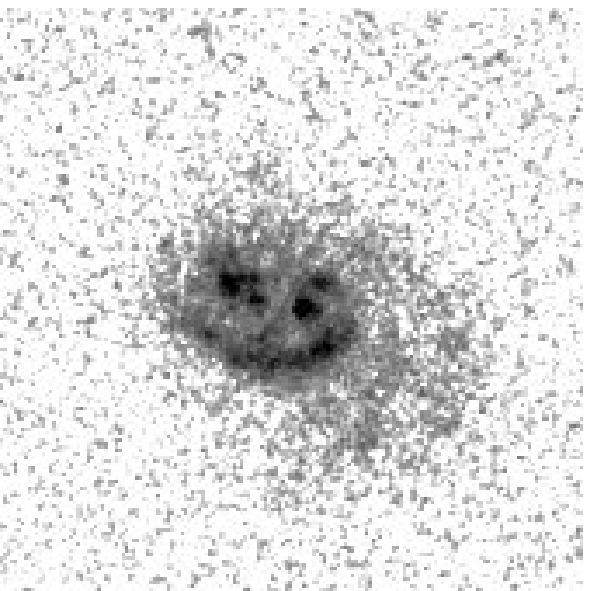}
\includegraphics[width=0.10\textwidth]{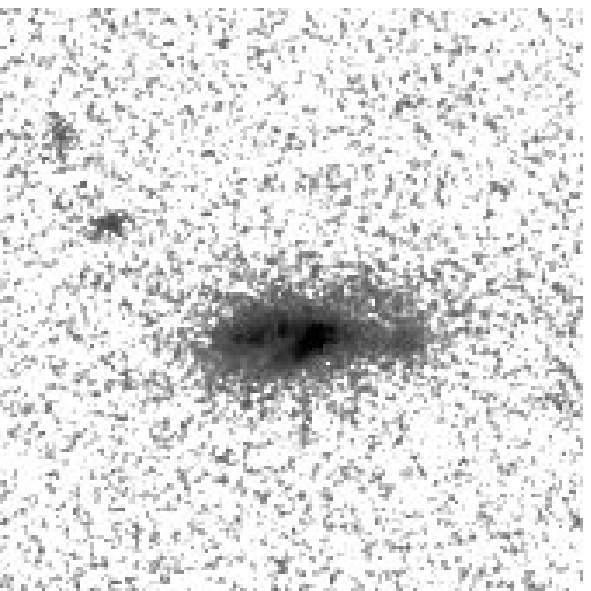}
\includegraphics[width=0.10\textwidth]{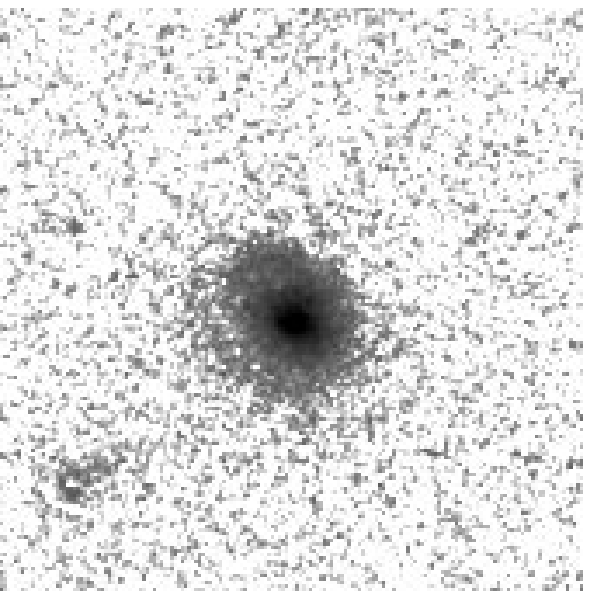}
\includegraphics[width=0.10\textwidth]{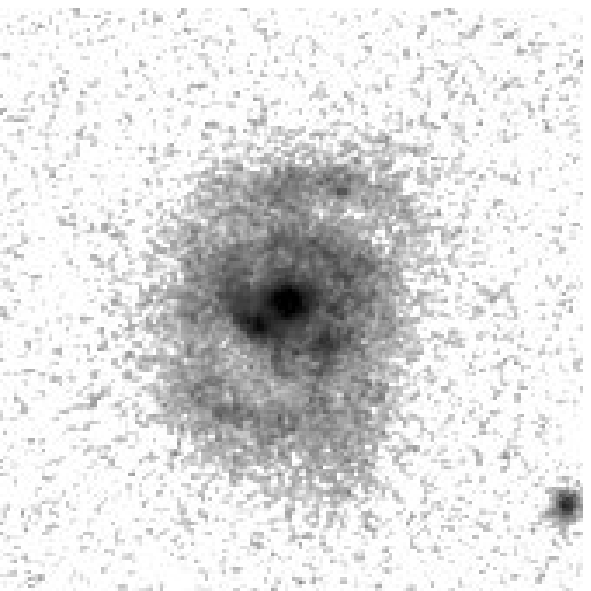}
\includegraphics[width=0.10\textwidth]{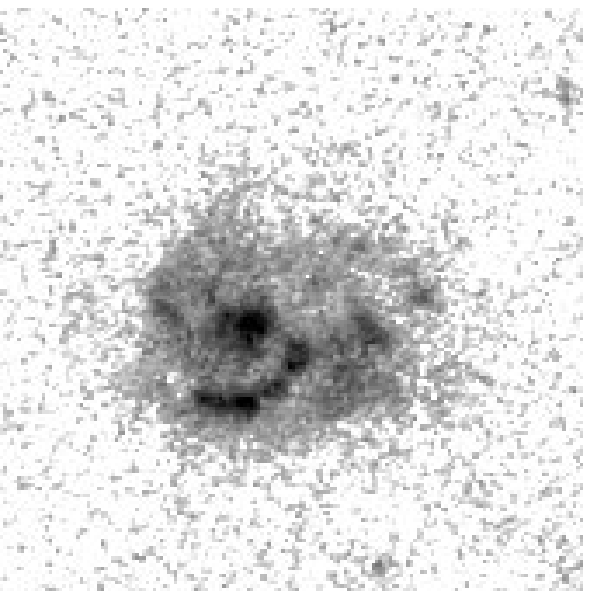}
\includegraphics[width=0.10\textwidth]{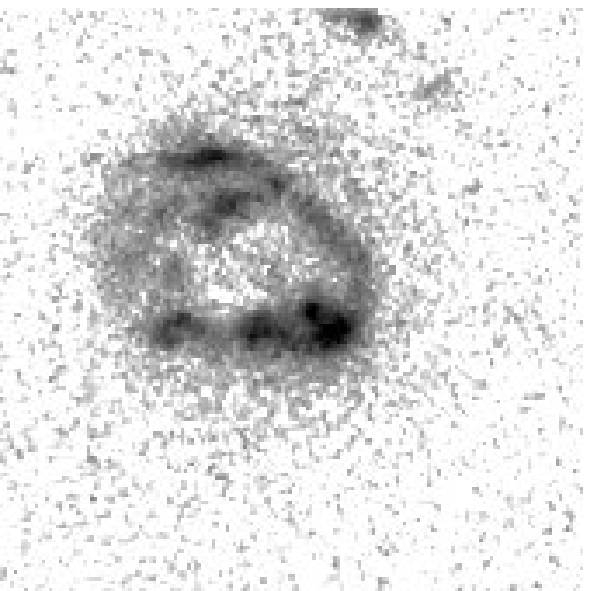}
\includegraphics[width=0.10\textwidth]{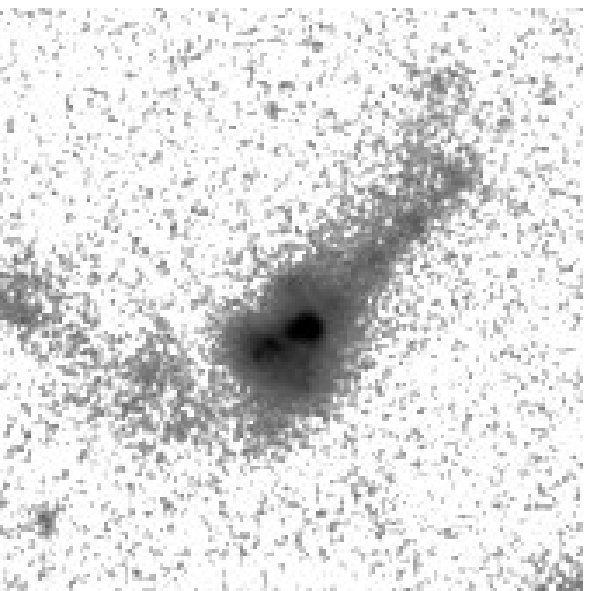}
\includegraphics[width=0.10\textwidth]{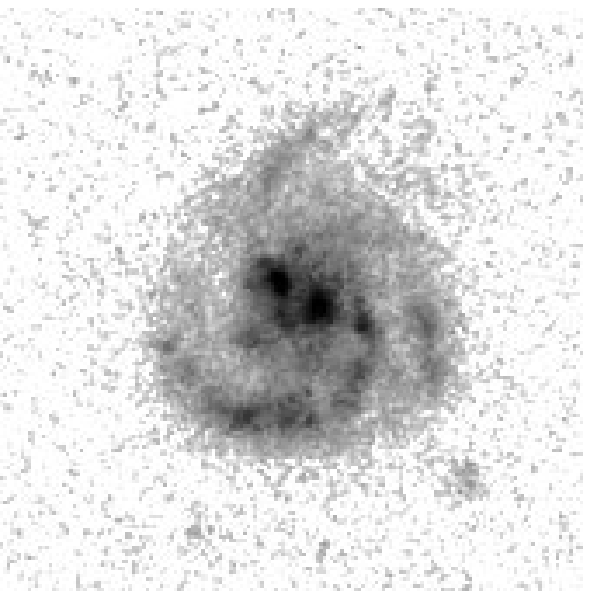}
\includegraphics[width=0.10\textwidth]{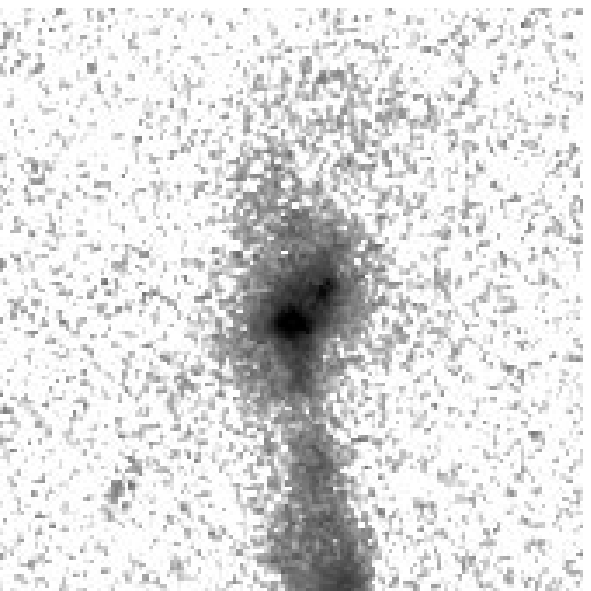}
\includegraphics[width=0.10\textwidth]{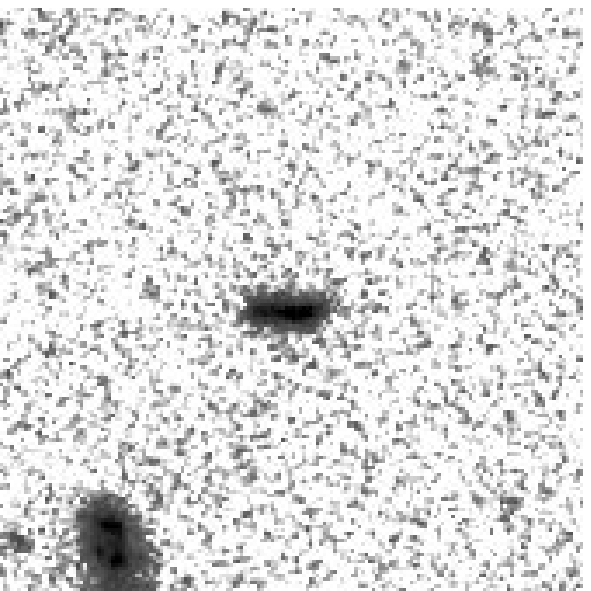}
\includegraphics[width=0.10\textwidth]{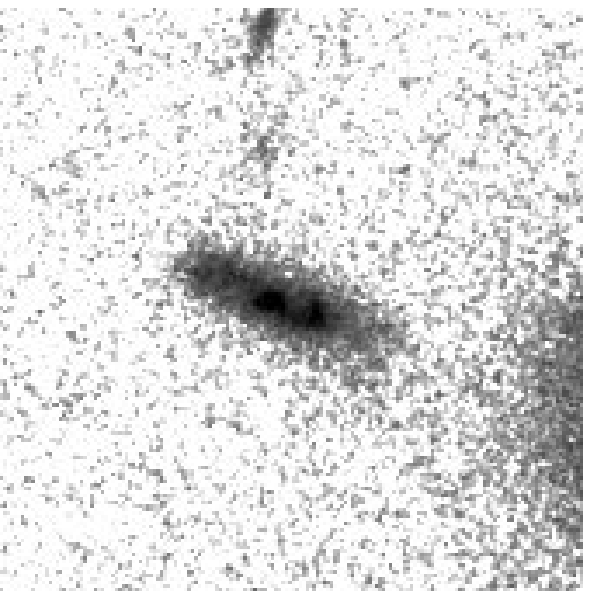}
\includegraphics[width=0.10\textwidth]{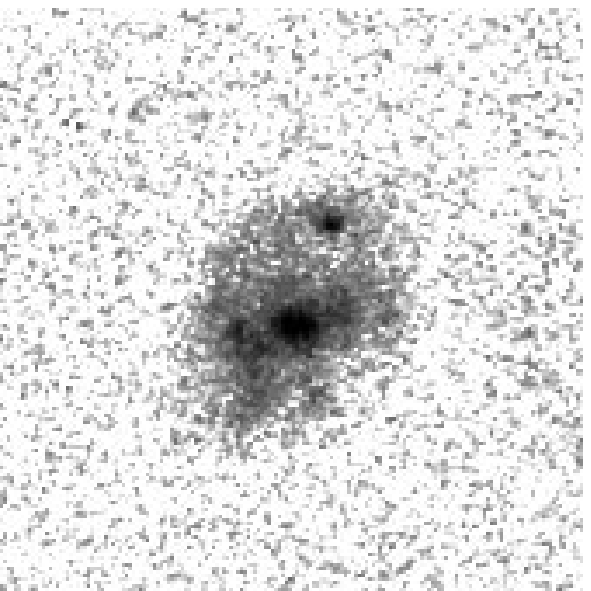}
\includegraphics[width=0.10\textwidth]{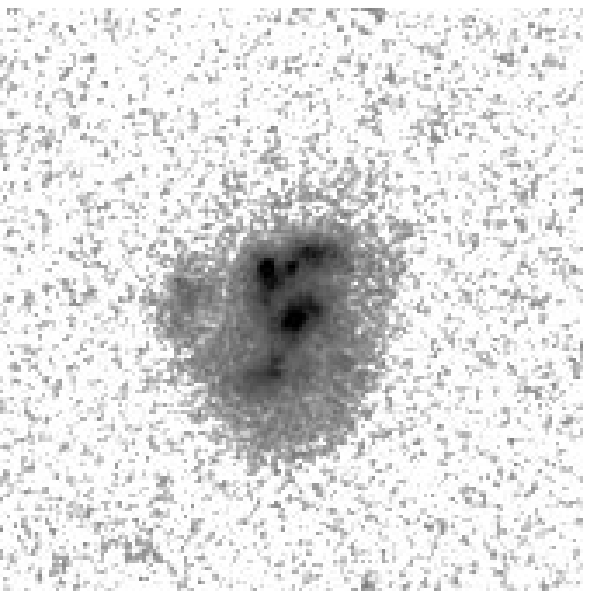}
\includegraphics[width=0.10\textwidth]{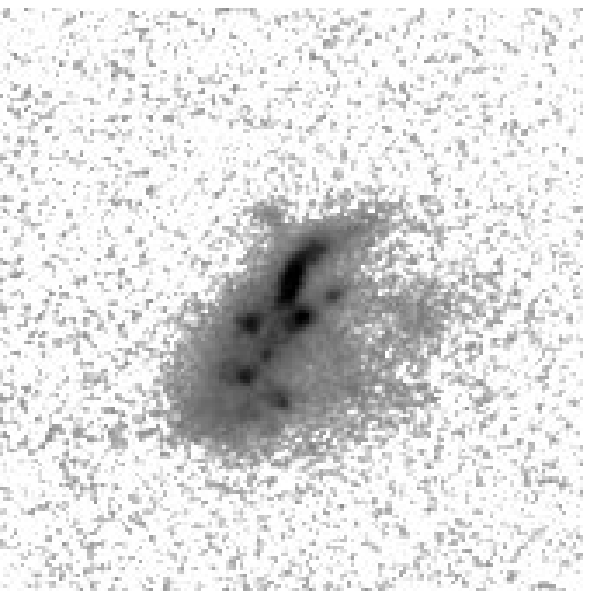}
\includegraphics[width=0.10\textwidth]{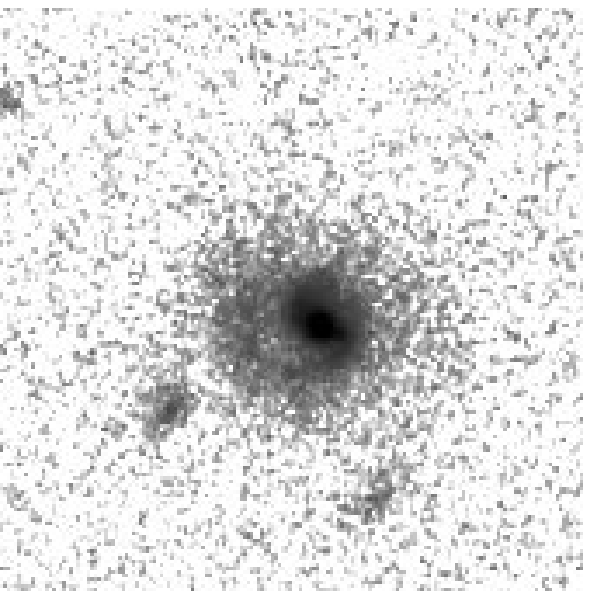}
\includegraphics[width=0.10\textwidth]{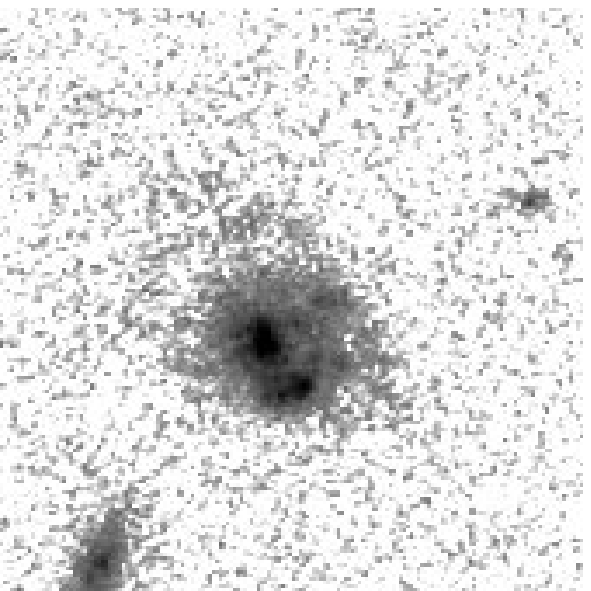}
\includegraphics[width=0.10\textwidth]{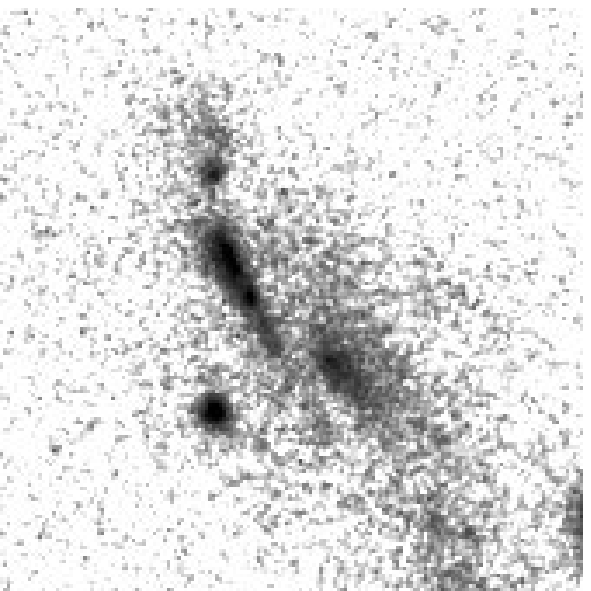}
\includegraphics[width=0.10\textwidth]{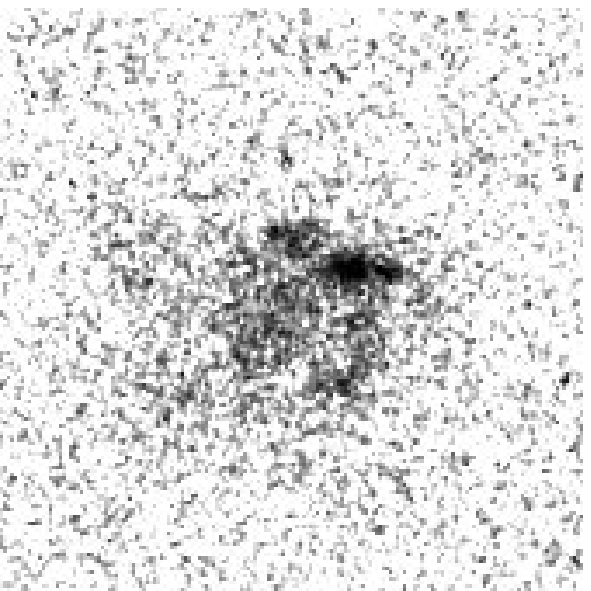}
\includegraphics[width=0.10\textwidth]{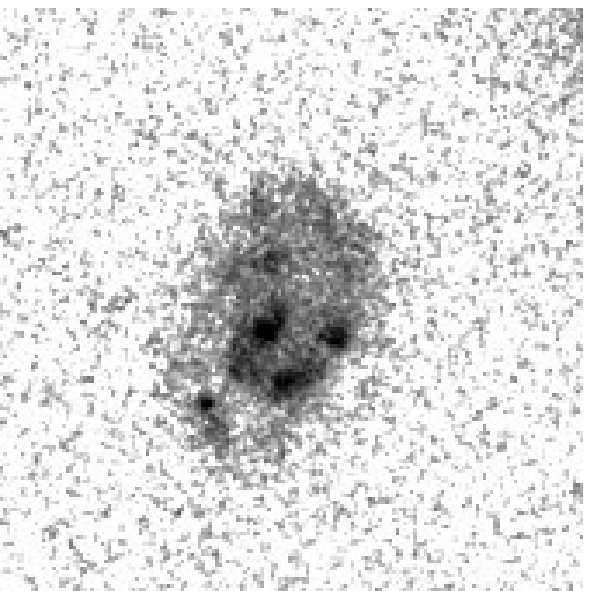}
\includegraphics[width=0.10\textwidth]{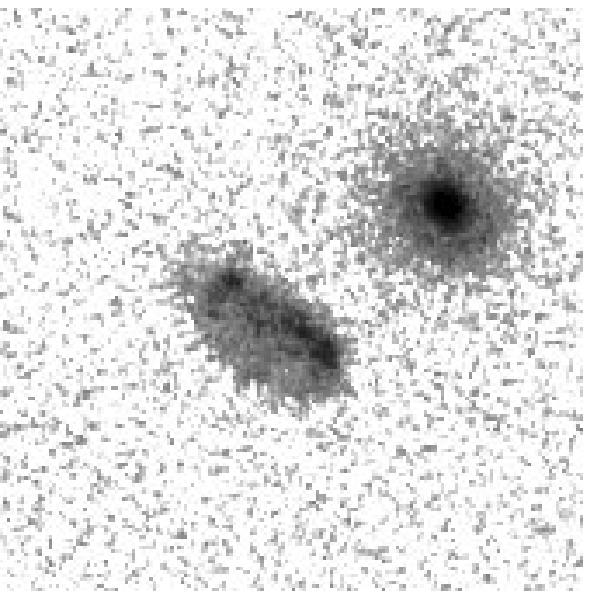}
\includegraphics[width=0.10\textwidth]{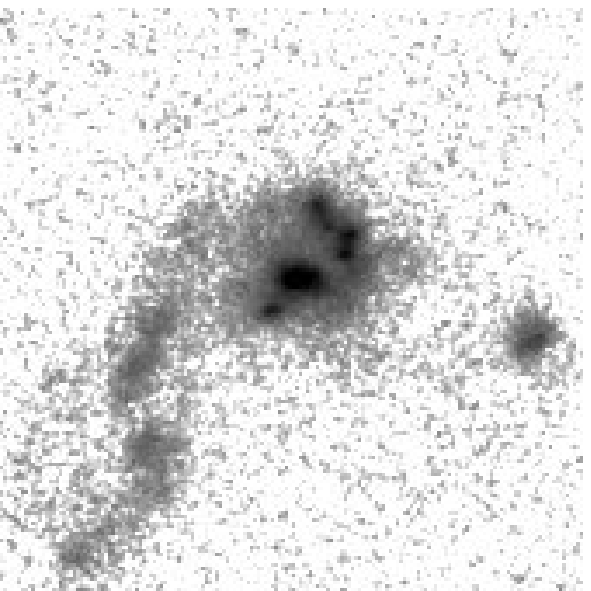}
\includegraphics[width=0.10\textwidth]{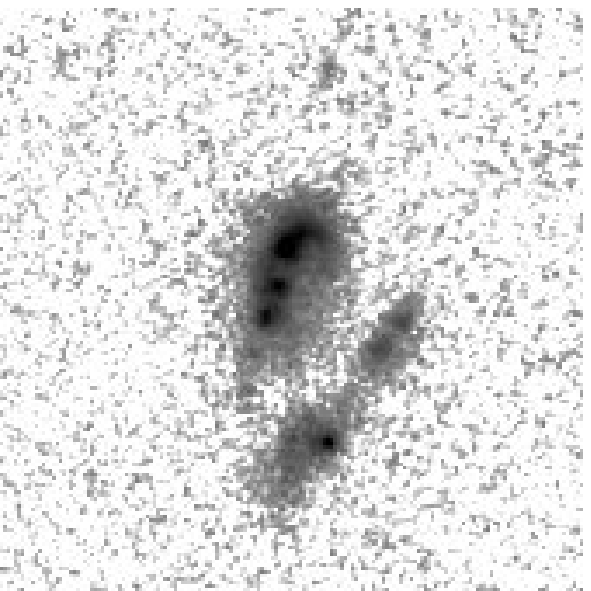}
\includegraphics[width=0.10\textwidth]{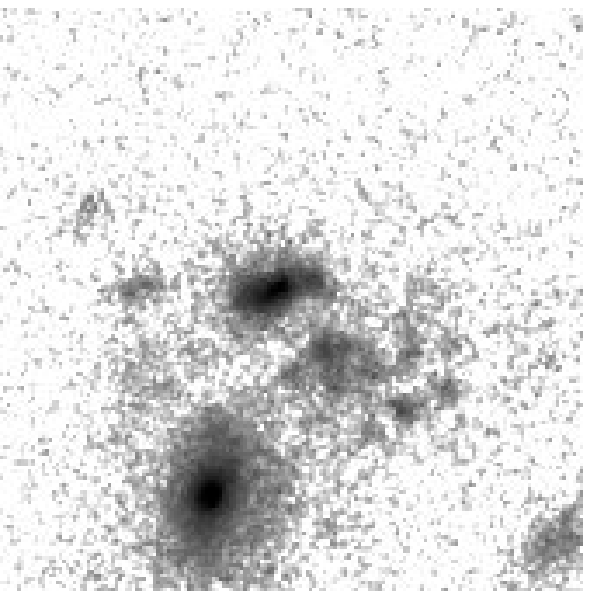}
\includegraphics[width=0.10\textwidth]{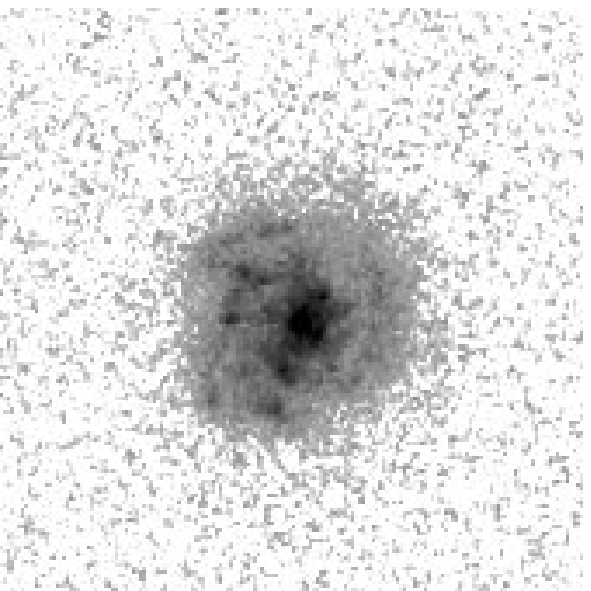}
\includegraphics[width=0.10\textwidth]{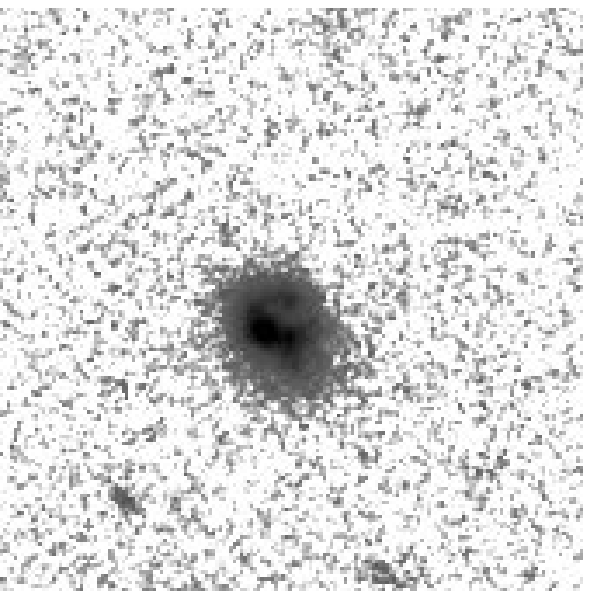}
\includegraphics[width=0.10\textwidth]{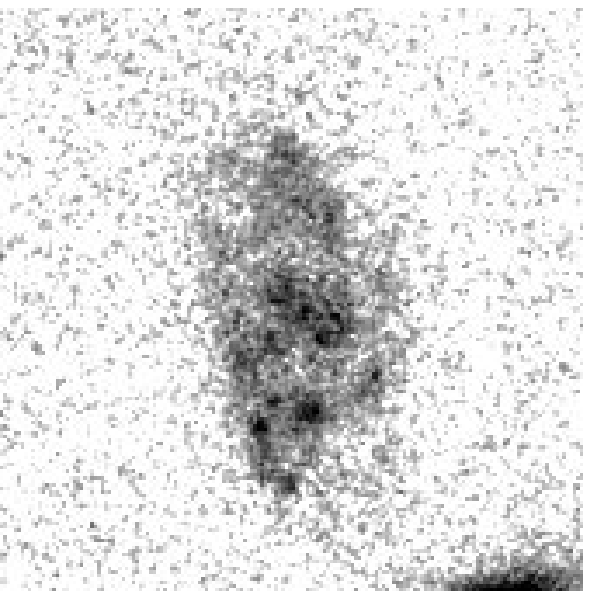}
\includegraphics[width=0.10\textwidth]{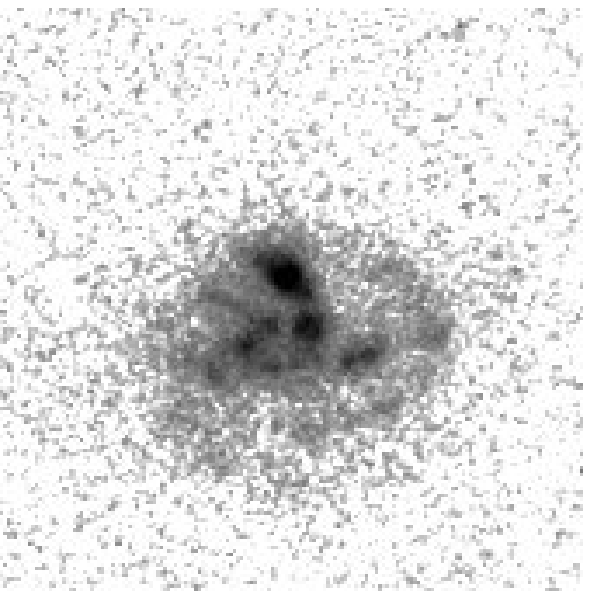}
\includegraphics[width=0.10\textwidth]{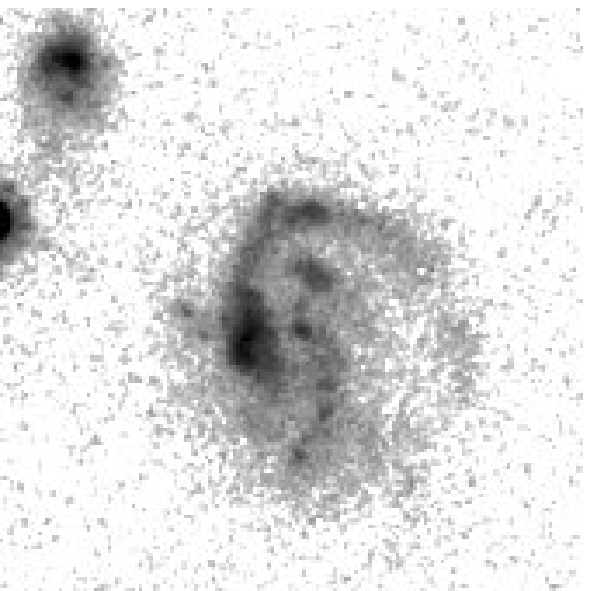}
\includegraphics[width=0.10\textwidth]{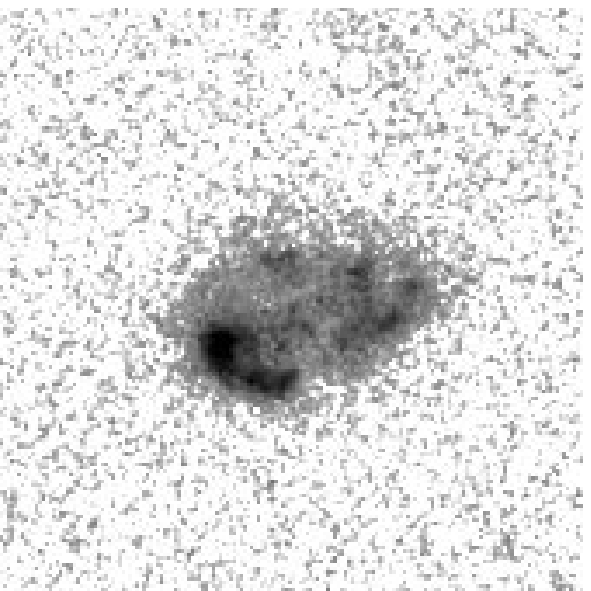}
\includegraphics[width=0.10\textwidth]{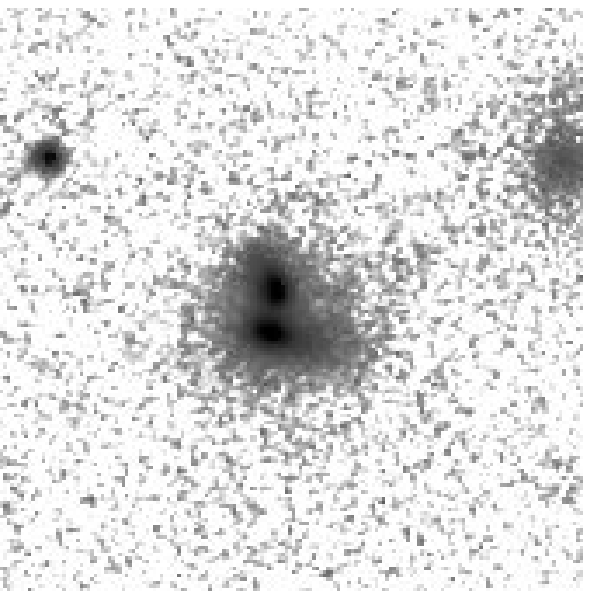}
\includegraphics[width=0.10\textwidth]{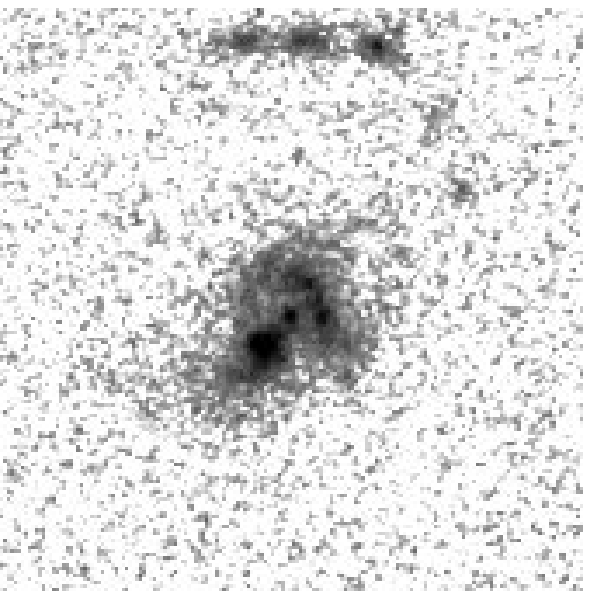}
\caption{I-band optical ACS cutouts of the 42 PACS-detected LBGs. The size of each image is 5''x5''. At $z \sim 1$, the plate scale is 8.008 kpc/" under the assumed cosmology and the ACS I-band samples the rest-frame U band approximately.
              }
\label{morpho_LBGs}
\end{figure*}

We also plot in Figure \ref{CMD_PACS} the location of the PACS-undetected LBGs in order to analyze the difference between PACS-detected and PACS-undetected sources regarding their location in a CMD. Following the same idea than in \cite{Oteo2013}, we divide the sample of PACS-undetected LBGs into \emph{old-LBGs}, those whose ages are older than 1200 Myr, and \emph{young-LBGs}, those whose ages are younger than 1200 Myr. They are represented by filled and open grey dots, respectively. It can be seen that PACS-detected LBGs are located over a similar zone than old-LBGs, which indicates that both old and/or dusty LBGs tend to depart from the blue cloud to the red sequence. This result is similar to that found in \cite{Oteo2013}. It should be noted that here, instead of dividing the sample depending on dust attenuation, we consider PACS-detected and PACS-undetected galaxies separately. Despite this, we still obtain the same result: PACS-detected and/or old LBGs tend to be located over the green valley or the red sequence rather than over the blue cloud. Those LBGs which are located over the blue cloud are only those young and less dusty.

Therefore, Figure \ref{CMD_PACS} shows a clear tendency in the CMD as a function of age and dust attenuation, from the less dusty and young galaxies located in the blue cloud to the UV-faint and dusty galaxies located over the red sequence. It also indicates that the properties of LBGs at $z \sim 1$ are diverse. The Lyman-break technique seems to select galaxies with different properties rather with specific characteristics.

\section{Morphology of PACS-detected LBGs}\label{morfo_ACS}

This section is devoted to the study of the morphological properties of our PACS-detected galaxies by carrying out both an analytical and visual procedure. To perform a precise study, high-spatial resolution ACS images are mandatory. We employ here the publicly available ACS I-band images of the COSMOS field. At $z \sim 1$ the ACS I-band samples the rest-frame U band approximately, with little variation according to the redshift distribution of the sources. All our PACS-detected galaxies have available ACS information.

On the visual side, we visually classify the PACS-detected galaxies into five groups, as it was done in \cite{Oteo2013}: disk-like galaxies, interacting/merging galaxies, compact galaxies, chain galaxies (CH) or clump clusters (CC), and irregular galaxies. We define disk galaxies as those having a clear disk structure, regardless the presence of spiral arms or a central bulge. For CH and CC we employ the definition given in \cite{Elmegreen2009}. Finally, we consider a galaxy as irregular if it cannot be classified within any of the other types. According to a visual classification of their ACS images (see Figure \ref{morpho_LBGs}) we obtain that most (68\%) PACS-detected LBGs have clear disk-like morphologies. Among them, 59\% have flocculent structures \cite[see for example][]{Elmegreen2009GEMS}. For the remaining galaxies, 13\% have signs of interaction, 7\% are irregular, 7\% are CC, and 5\% belong to the CH class. 

\begin{figure*}
\centering
\includegraphics[width=0.4\textwidth]{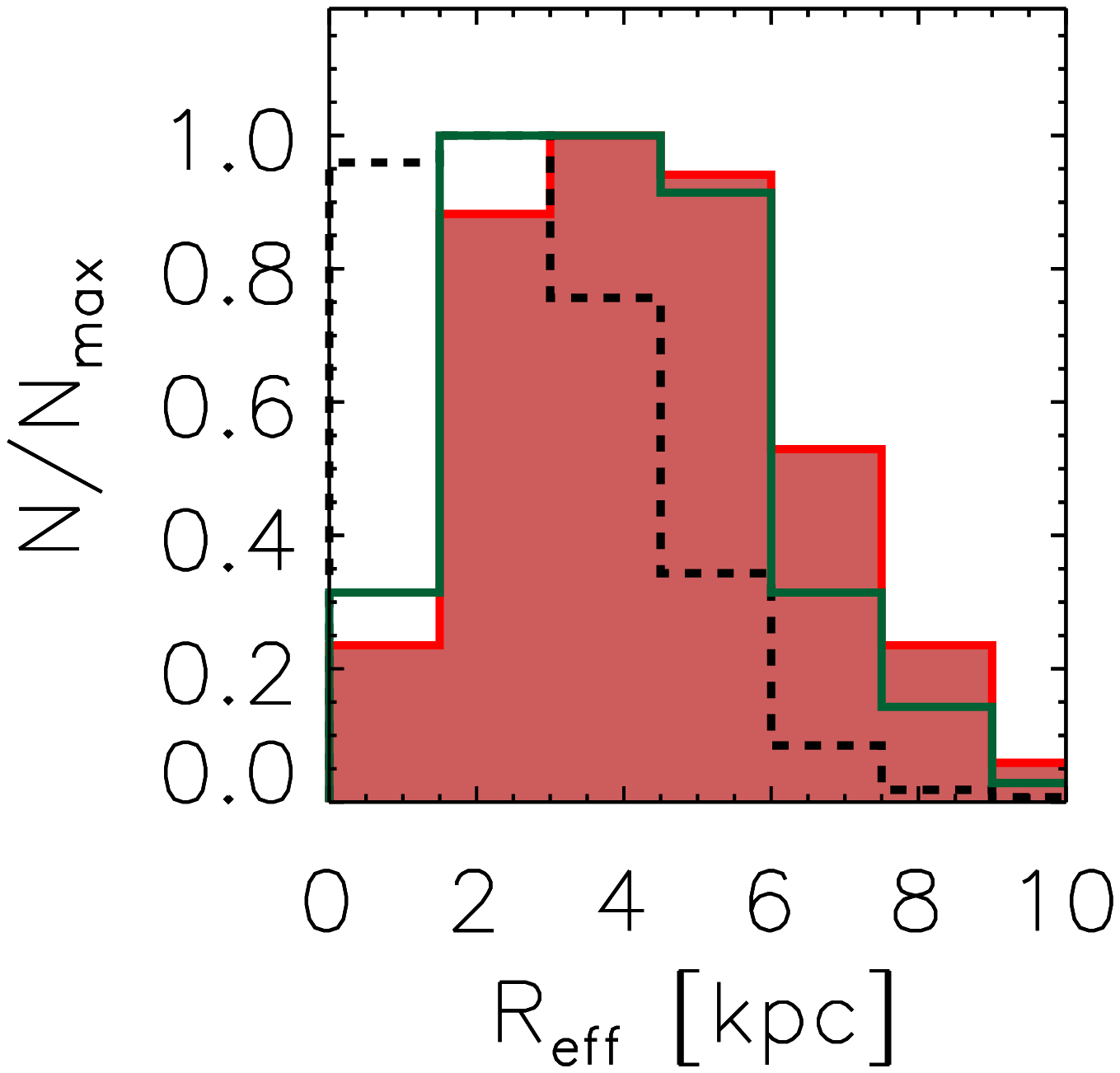}
\includegraphics[width=0.4\textwidth]{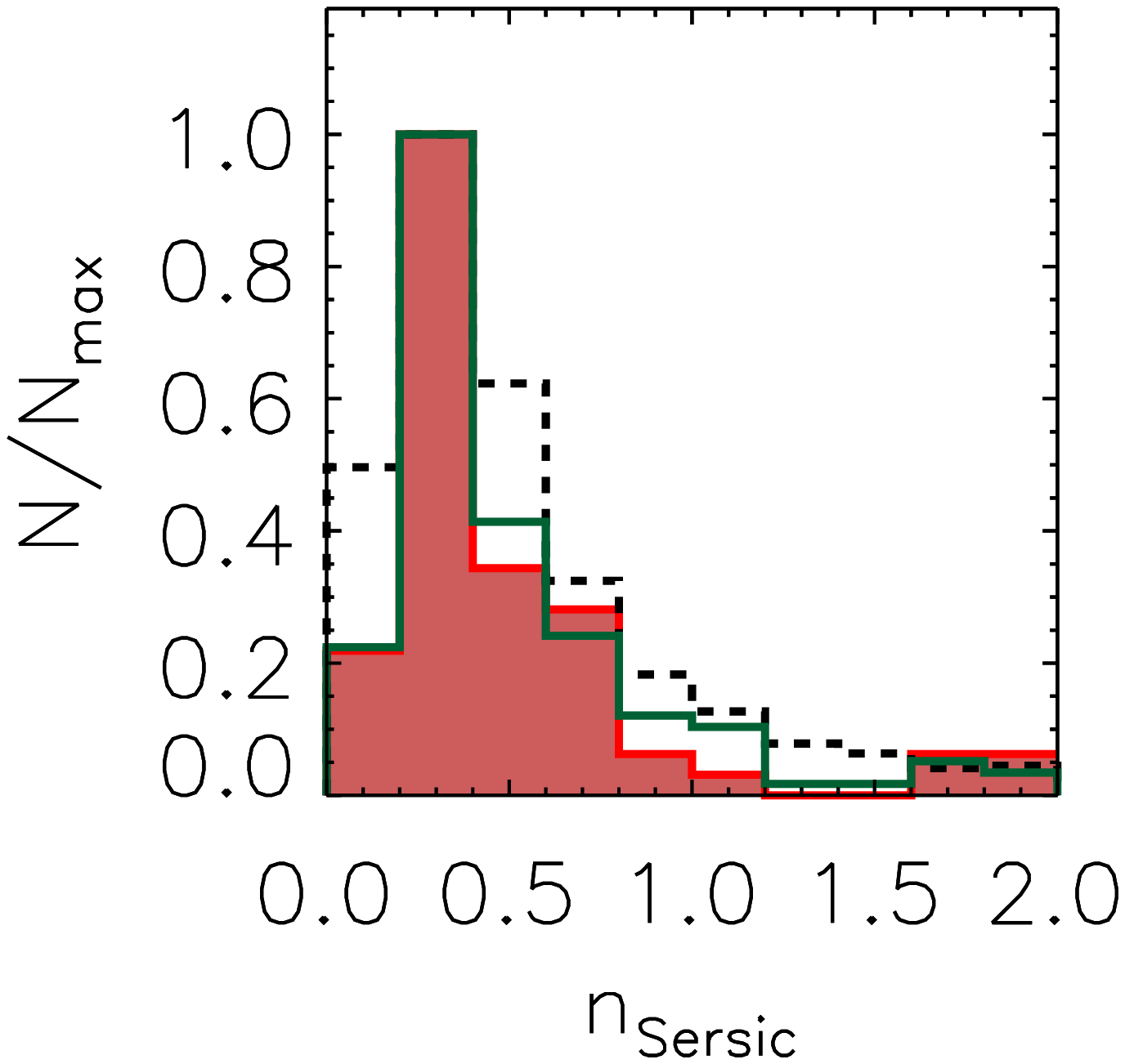}
\caption{Results of the analytical morphological study of our PACS-detected galaxies. \emph{Left}: Distribution of the physical sizes. \emph{Right}: Distribution of the Srsic indices. In both panels, PACS-detected LBGs and UV-selected galaxies, and PACS-undetected LBGs are represented with red shaded, green, and black dashed histograms, respectively. Histograms have been normalized to their maximum in order to clarify the representations.
              }
\label{reff_index}
\end{figure*}

On the analytical side, and with the aim of obtaining the physical sizes and examining the radial dependence of the light distribution of the PACS-detected galaxies, we  carry out fits to their radial light curves with \verb+GALFIT+ \citep{Peng2010}. In this step, we consider Sersic profiles \citep{Sersic1968}, which can be described as:

\begin{equation}
\Sigma (r) = \Sigma_e \exp{\left[ -\kappa \left(\left(\frac{r}{R_{\rm eff}}\right)^{1/n}-1\right) \right]}
\end{equation}

\noindent where $\Sigma_e$ is the pixel surface brightness at the effective radius $R_{\rm eff}$ and $n$ is the concentration parameter or Sersic index. The effective radius is the radius which encloses half the light of the galaxy. To make this definition true, the dependent variable $\kappa$ is coupled to $n$ \citep{Peng2010}. For each input galaxy, \verb+GALFIT+ provides the effective radius (in pixels) and the Sersic index. In order to convert the effective radius in pixels into the physical size in kpc we employ ACS pixel scale and the assumed cosmology for calculating the plate scale at the redshift of each galaxy. The left panel of Figure \ref{reff_index} represents the distributions of the effective radii of PACS-detected (red shaded histogram) and PACS-undetected LBGs (green histogram). PACS-detected LBGs are large galaxies with effective radii ranging from about 2 to 8 kpc, with a median value of 4 kpc. PACS-undetected LBGs have a median effective radius of $R_{\rm eff} = 2$ kpc, indicating that FIR detections tend to segregate LBGs with large sizes. There is no significant difference in the physical size between PACS-detected LBGs and PACS-detected UV-selected galaxies. Their Sersic indices of our PACS-detected LBGs and UV-selected galaxies, shown in the right panel of Figure \ref{reff_index}, vary mostly within $0 \lesssim n \lesssim 0.75$, in agreement with the visual classification that indicates that most PACS-detected LBGs are disk-like galaxies.


\section{Radio measurements}\label{radio_VLA}


Dust absorbs the UV radiation created from massive stars and re-radiates this light in the IR. When those massive stars explode as SNe they generate cosmic-ray electrons which lose their energy in radio wavelengths, mostly via synchrotron emission. Therefore, it is expected that the radio emission from galaxies is in part related to their emission in the IR. This way, radio measurements also provide a powerful tool to obtain SFR.

\begin{figure}
\centering
\includegraphics[width=0.49\textwidth]{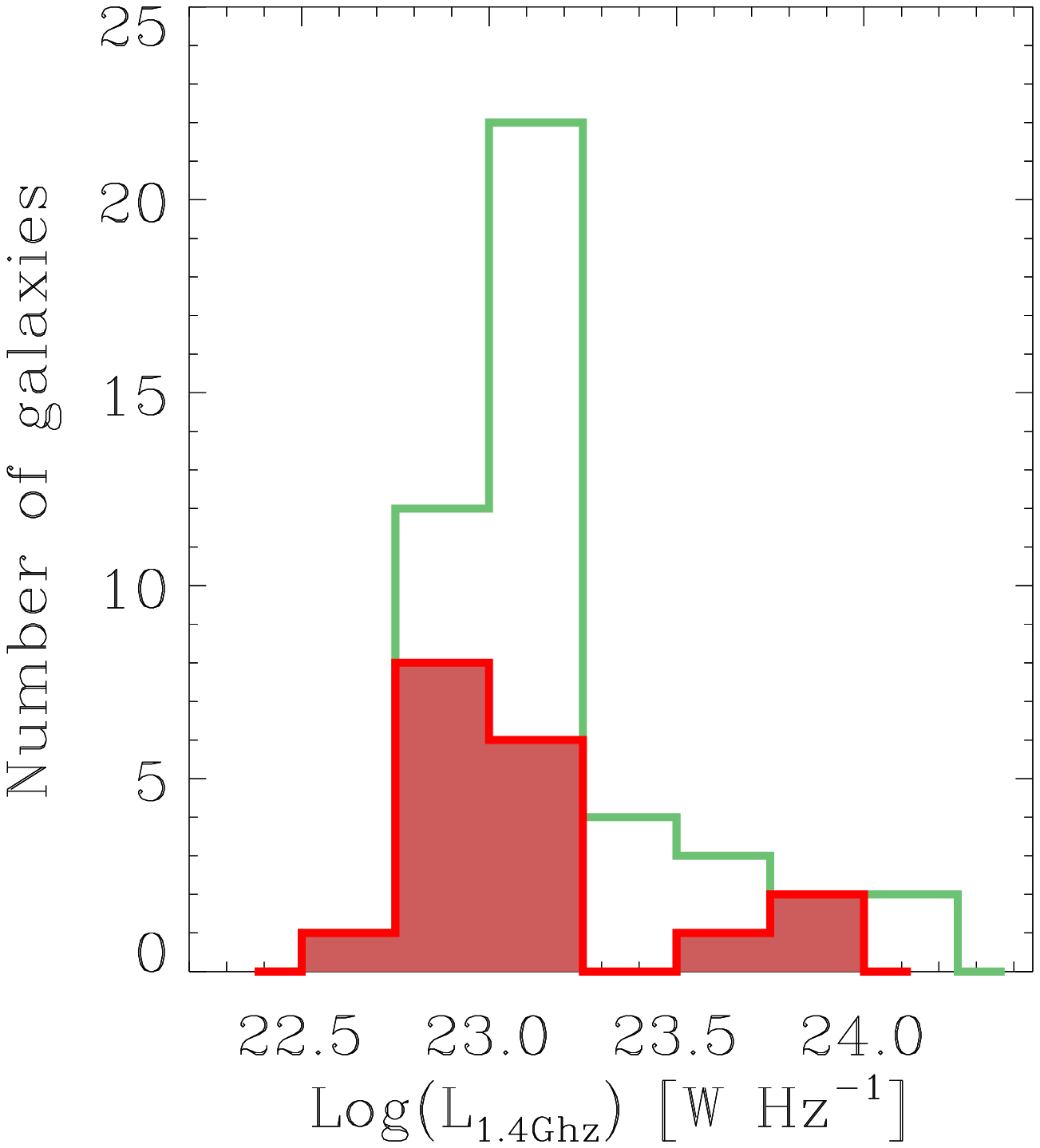}
\includegraphics[width=0.49\textwidth]{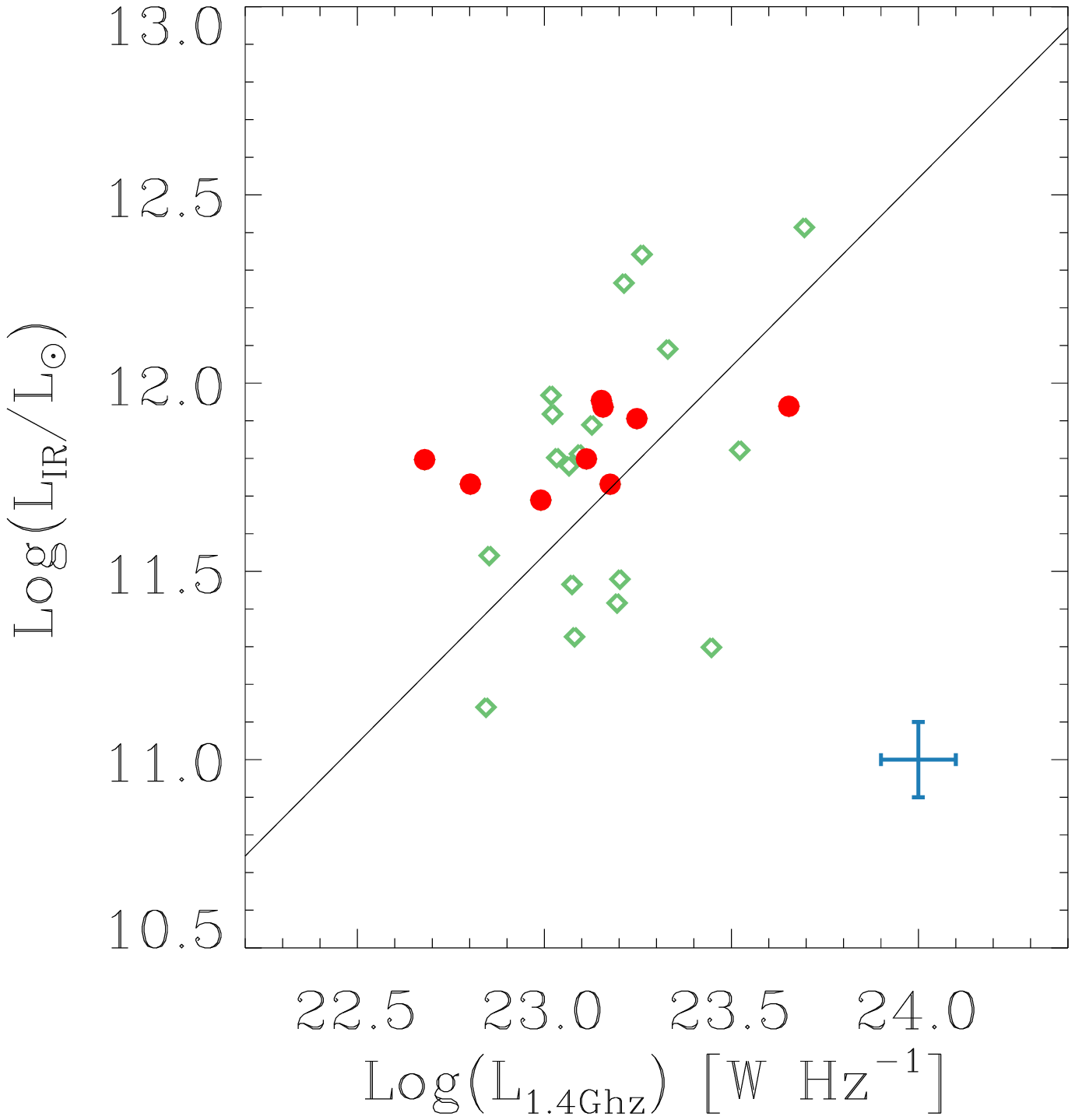}
\caption{\emph{Upper}: Distribution of the rest-frame 1.4GHz luminosities for our VLA-detected LBGs (red shaded histogram), and UV-selected galaxies (green histogram). \emph{Bottom}: Relation between the total IR and radio luminosities for our samples of PACS-detected LBGs (red filled dots) and UV-selected galaxies (green open diamonds). The straight line represents the relation of \citet{Condon1992}. The blue error bars represent the typical uncertainties of the parameters shown in the x and y axes.
              }
\label{lumi_radio}
\end{figure}

\begin{figure}
\centering
\includegraphics[width=0.49\textwidth]{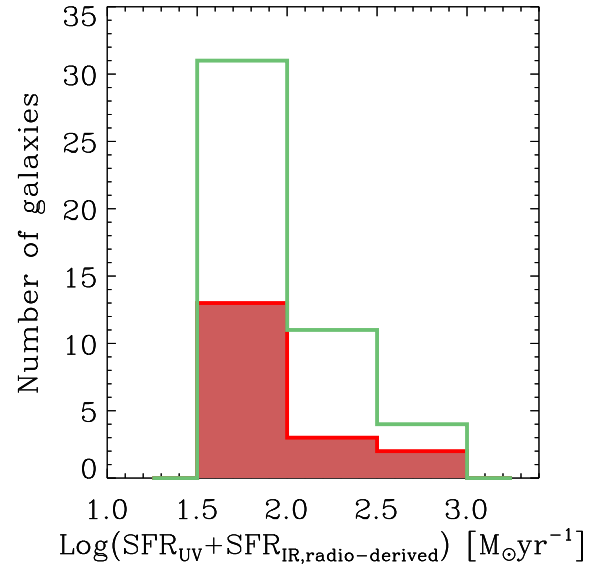}
\caption{Distribution of the total SFR as derived for the combination of UV and radio derived IR luminosities for the VLA-detected LBGs (red shaded histogram) and UV-selected galaxies (green histogram). Note that given that we assume that the IR-radio correlation is valid for both IR-detected and IR-undetected sources, in this plot are included all radio detected sources, regardless if they are detected or not in PACS.
              }
\label{sfr_radio}
\end{figure}

Following this idea, we analyze in this section the radio emission of the whole sample of LBGs, and UV-faint galaxies built in \cite{Oteo2013} in order to study a possible difference between them and to study the relation between IR and radio derived luminosities and SFRs. To this aim, we look for radio counterparts of the whole sample of LBGs and UV-faint galaxies selected in \cite{Oteo2013} by using data coming from VLA observations of the COSMOS field \citep{Schinnerer2010}. Using again a matching radius of 2'', we find that 18 LBGs and 46 UV-faint galaxies are detected at 1.4 GHz. These percentages of VLA detections are lower than those corresponding to PACS detections. In order to obtain the rest-frame 1.4 GHz luminosities, since our VLA-detected galaxies are located at $z \sim 1$, we must carry out a K-correction. To do that, we assume that the radio continuum behaves as $f_{radio}$$\sim\nu^{\alpha}$, where $\alpha$ is the radio spectral index, for which we consider a value of $\alpha = -0.8$ \citep{Magdis2010,Ivison2010_Herschel,Ivison2010_BLAST}. We plot in the upper panel of Figure \ref{lumi_radio} the distribution of rest-frame 1.4GHz luminosities for our VLA-detected LBGs, and UV-faint galaxies. The low number of detections in radio wavelength of both kinds of galaxies prevents us from distinguishing them according to their radio emission. Actually, both histograms are very similar. A KS-test yield a probability factor of 0.5, reinforcing the similarity between both distributions.


Obtaining radio derived SFR requires calculating the IR luminosities associated to the rest-frame 1.4GHz luminosities by  assuming a IR-radio correlation \cite[for example][]{Condon1992}. Then, we first compare in the bottom panel of Figure \ref{lumi_radio} the total IR luminosities and the rest-frame 1.4GHz luminosities for our VLA-detected and PACS-detected galaxies to check if the \cite{Condon1992} relation applies for our galaxies. Nine LBGs and 27 UV-faint galaxies are detected both in PACS and VLA. We also plot there the \cite{Condon1992} IR-radio correlation. It can be seen that PACS-detected UV-faint galaxies are around the \cite{Condon1992} relation and, therefore, for those galaxies, the IR-derived and radio-derived SFRs with the procedure outlined above would agree. PACS-detected LBGs tend to be located slightly above the \cite{Condon1992} relation. This could introduce some differences in the VLA-derived total IR luminosities with respect to those that would be obtained with PACS data. For those PACS-undetected and radio-detected galaxies, we can assume that the \cite{Condon1992} relation is still valid with the aim of calculating the IR luminosity associated to their radio emission. It should be noted that, according to the IR-radio correlation, the SFR$_{radio}$ is a tracer of the SFR$_{IR}$, but not of the SFR$_{total}$. Therefore, in order to calculate the total SFR from radio observations we also have to add the term associated the the UV emission (Equation \ref{SFR_total}). In Figure \ref{sfr_radio} we plot the distribution of the total SFR derived from the combination of UV and radio measurements for our VLA-detected galaxies. The median values are $67 \pm 65 \, M_\odot$/yr and $76 \pm 80 M_\odot$/yr for VLA-detected LBGs and UV-faint galaxies, respectively.  The uncertainties indicates the standard deviations of the distributions, as a measurement of their width. The total SFRs obtained in Section \ref{starformation} from the combination of UV and FIR measurements were $91 \pm 22 \, M_\odot$/yr and $73 \pm 39 M_\odot$/yr for PACS-detected LBGs and UV-faint galaxies, respectively. The distributions of radio-derived total SFR are much wider than those obtained from PACS measurements. This is due to the existence of galaxies with very high values of the radio-derived total SFR compared to the median of the distributions. Despite this, there is a very good agreement between the radio and FIR determinations.

\begin{figure*}
\centering
\includegraphics[width=0.49\textwidth]{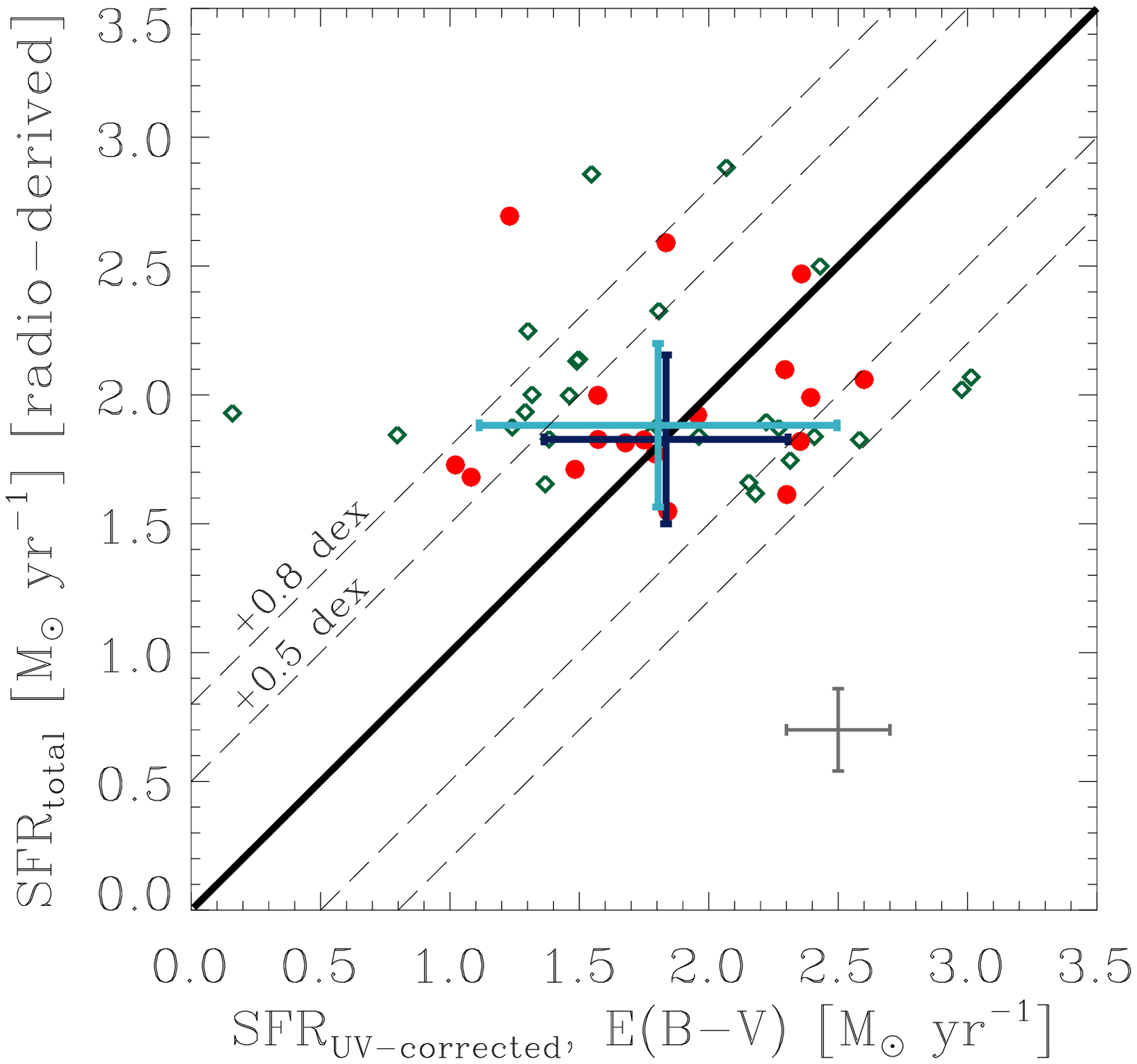}
\includegraphics[width=0.49\textwidth]{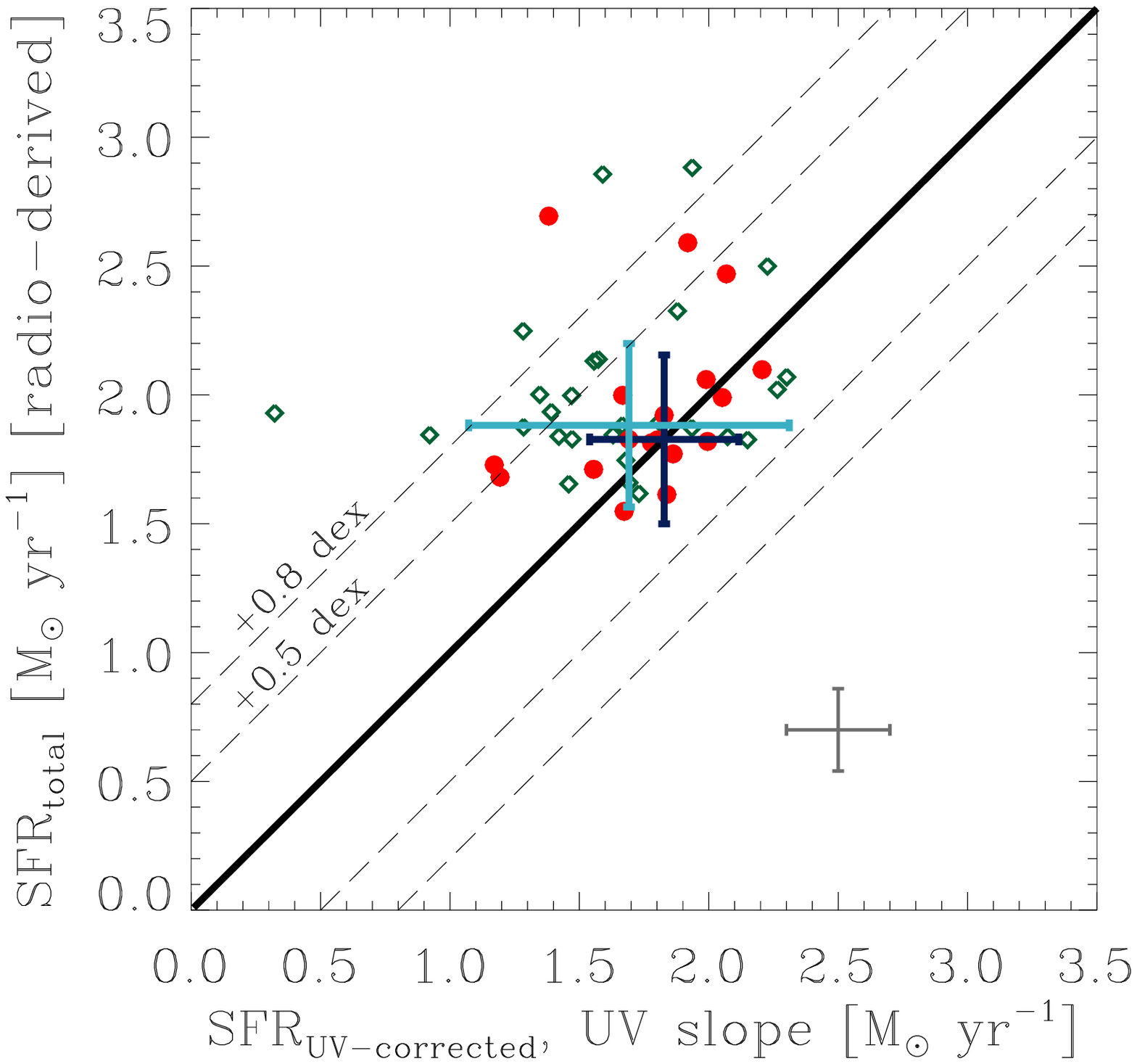}
\caption{\emph{Left}: Total SFR derived from the combination of UV and radio measurements against the total SFR derived by correcting the rest-frame UV luminosity with the $Es(B-V)$ value obtained from SED-fitting with BC03 templates associated to a constant star-formation rate and sub-solar metallicity $Z=0.4Z_\odot$. \emph{Right}: Total SFR derived from the combination of UV and radio measurements against the total SFR obtained by correcting the rest-frame UV luminosity with the UV continuum slope and applying \citet{Meurer1999} relation. In both panels, VLA-detected LBGs and UV-selected galaxies are represented with red open dots and green open diamonds, respectively. The median loci in the diagrams of the PACS-detected LBGs and UV-selected galaxies are represented by the dark blue and light blue bars, respectively. The bars cross in the median value of each parameter for each kind of galaxy and the length of the bars is the standard deviation of the distribution of values. The lines represent where both estimators would agree. The grey error bars represent the typical uncertainties of the parameters shown in the x and y axes.
              }
\label{SFR_radio_corrected}
\end{figure*}

Finally, we study in Figure \ref{SFR_radio_corrected} the relation between the SED-derived dust-corrected total SFR and the total SFR derived from the combination of UV and radio measurements. These are similar plots than those presented in Figures \ref{sfr_sfr_edad_beta} and \ref{sfr_sfr}. As in that case, we use the dust-corrected SED-derived total SFRs obtained from the best-fitted $E_s(B-V)$ value and from the dust attenuation obtained with UV continuum slope by applying the M99 relation. It can be seen that both procedures tend to give similar results, providing estimation of the total SFR within 0.8 dex for most of the galaxies. Furthermore, this spread around the one-to-one relation is similar to that found in Figures \ref{sfr_sfr_edad_beta} and \ref{sfr_sfr} when using PACS data to obtain the UV+IR-derived total SFR.


\section{Conclusions}\label{conclusions}

In this work the have analyzed the UV-to-FIR SED of a sample of LBGs and other UV-selected galaxies at $z \sim 1$ located in the COSMOS field. To this aim, we have used data coming from GALEX, ALHAMBRA survey, IRAC, MIPS, and PACS-100$\mu$m/160$\mu$m observations. The combination of all these measurements provides an unprecedented coverage of the SED of galaxies at the studied redshift. The sample of LBGs was selected via their $FUV-NUV$ color. Due to their selection criteria, LBGs are UV-bright galaxies. For comparison, we also retain a sample of UV-selected and PACS-detected galaxies fainter in the UV than LBGs but located at their same redshift range. We call these sources \emph{UV-faint galaxies}. The main conclusions of the work can be summarized as follows:

\begin{enumerate}

	\item We have found that, among a sample of 1225 UV-selected LBGs at $z \sim 1$, 42 of them are individually detected in either PACS-100$\mu$m or PACS-160$\mu$m under depths of 5.0 and 11.0 mJy in each band, respectively. These measurements, in combination with ancillary IRAC-8.0$\mu$m, and MIPS-24$\mu$m data provide a good sampling of the mid-IR/FIR SED of these galaxies. This allows an accurate determination of their total IR luminosity, dust attenuation, and total and  sSFR. PACS-detected LBGs tend to have higher rest-frame UV luminosities, dust attenuation, and stellar mass, and have redder UV continuum than PACS-undetected galaxies. There is no relation in age between PACS-detected and PACS-undetected LBGs.
	
	\item All PACS-detected LBGs have total IR luminosities $10^{11} \lesssim L_{\rm IR}/L_\odot \lesssim 10^{12}$ and, therefore, are all LIRGs. The lower limit range of total IR luminosities is a direct consequence of the depth of the PACS observations employed in this work. However, there are no PACS-detected LBGs in the ULIRG regime, $L_{\rm IR}/L_\odot \geq 10^{12}$, where the observations are complete. At $z \sim 3$, the PACS-detected LBGs found so far are all ULIRGs. This suggests that the FIR emission of galaxy is changing with cosmic time in the sense that at higher redshifts there is a population of extreme IR-bright, red, and dusty LBGs which is not found at intermediate redshifts.
	
	\item By using their UV and IR emissions and applying the \cite{Buat2005} calibration we find that PACS-detected LBGs have dust attenuation spanning $2 \lesssim A_{\rm 1200 \AA}{\rm [mag]} \lesssim 5$, with a median value of 4 mag. PACS-detected LBGs, due to their brightness in the UV as a consequence of their selection criteria, are less dusty than other UV-fainter galaxies at their same redshift. PACS-detected LBGs with higher rest-frame UV luminosities are less attenuated. Furthermore, the dust attenuation of PACS-detected LBGs increases with the total IR luminosity.
	
	\item According to the combination of their UV and IR emission, PACS-detected LBGs have a median total SFR of 90 $M_\odot {\rm yr}^{-1}$. The SED-derived dust attenuation obtained with BC03 templates associated to constant SFR tend to overestimate the total SFR obtained from the combination of UV and IR measurements for the youngest galaxies in the sample. This is likely due to the typical degeneracy between age and dust attenuation in the SED-fitting procedures with synthetic templates and highlight the importance of the usage of direct UV and IR measurements when obtaining the total SFR and dust attenuation in galaxies.
	
	\item The dust attenuation derived from rest-frame UV to NIR SED fitting overestimate the total SFR for most of our PACS-detected LBGs in a age-dependent way: the overestimation factor is higher for younger galaxies. This is likely due to the degeneracy in the SED-fitting procedures between dust attenuation, age, and SFH and highlights the importance of using direct FIR detections of galaxies to obtained accurate values of their IR luminosities, dust attenuation, and total SFR.
	
	\item By analyzing the location of our PACS-detected LBGs in a IRX-$\beta$ diagram we find that they are mostly around the \cite{Meurer1999} relation for local star-bursts, but above the corrections of \cite{Takeuchi2012} and \cite{Overzier2011} to the \cite{Meurer1999} law. This is also true for other UV-selected PACS-detected galaxies. Consequently, the dust-correction factors obtained with the \cite{Meurer1999} relation can be used to recover the dust attenuation of our PACS-detected galaxies. This does not have to be true for other less dusty and PACS-undetected galaxies, since PACS only detects the dustiest and IR-brightest galaxies at $z \sim 1$.
	
	\item In a CMD diagram, PACS-detected LBGs have red optical color compatible with them being mostly located over the green valley or the red sequence despite being active star-forming galaxies. UV-faint galaxies are also located over the green valley and the red sequence, although there is a larger percentage of galaxies populating the red sequence than in the sample of PACS-detected LBGs. PACS-undetected LBGs with ages larger than 1200 Myr tend to be also located over the green valley and the red sequence. This indicates that the properties of LBGs at $z \sim 1$ range are diverse, from blue, young, PACS-undetected LBGs to old or PACS-detected red LBGs.
	
	\item Morphologically, ACS I-band (sampling the rest-frame U band at $z \sim 1$) imaging indicate 68\% PACS-detected LBGs are disk-like galaxies, with 59\% of them having flocculent structures, and13\% have signs of interactions. There is a low percentage ($< 7\%$) of irregular, CH or CC galaxies. The median effective radius of the PACS-detected LBGs is 4 kpc. This value is higher than that for PACS-undetected LBGs, $R_{\rm eff} = 2$ kpc, indicating that FIR detections tend to segregate galaxies with larger sizes. The Sersic indices of PACS-detected are compatible with them being disk-galaxies.
	
	\item The radio measurements of our galaxies also provide accurate determinations of their total SFR. The IR/UV and radio/UV derived total SFRs are in agreement due to the validity of the \cite{Condon1992} relation for most of our galaxies at $z \sim 1$. The dust correction obtained from the UV continuum slope and the application of the \cite{Meurer1999} relation gives much better results for the total SFR than the dust correction carried out with the SED-derived $E_s(B-V)$.
	

\end{enumerate}

\section*{Acknowledgments}

The authors would like to thanks the referee for the motivating comments provided that have improved the presentations of our results. This research has been supported by the Spanish Ministerio de Econom\' ia y Competitividad (MINECO) under the grant AYA2011-29517-C03-01. AFS and VJM acknowledge support from the Spanish Ministry project AYA2010-22111-C03-02 and Generalitat Valenciana project PROMETEO-2008/132. Some/all of the data presented in this paper were obtained from the Multimission Archive at the Space Telescope Science Institute (MAST). STScI is operated by the Association of Universities for Research in Astronomy, Inc., under NASA contract NAS5-26555. Support for MAST for non-HST data is provided by the NASA Office of Space Science via grant NNX09AF08G and by other grants and contracts. Based on observations made with the European Southern Observatory telescopes obtained from the ESO/ST-ECF Science Archive Facility. Based on zCOSMOS observations carried out using the Very Large Telescope at the ESO Paranal Observatory under Programme ID: LP175.A-0839. {\it Herschel} is an ESA space observatory with science instruments provided by European-led Principal Investigator consortia and with important participation from NASA. The Herschel spacecraft was designed, built, tested, and launched under a contract to ESA managed by the Herschel/Planck Project team by an industrial consortium under the overall responsibility of the prime contractor Thales Alenia Space (Cannes), and including Astrium (Friedrichshafen) responsible for the payload module and for system testing at spacecraft level, Thales Alenia Space (Turin) responsible for the service module, and Astrium (Toulouse) responsible for the telescope, with in excess of a hundred subcontractors. PACS has been developed by a consortium of institutes led by MPE (Germany) and including
UVIE (Austria); KUL, CSL, IMEC (Belgium); CEA, OAMP (France); MPIA (Germany); IFSI, OAP/AOT, OAA/CAISMI,
LENS, SISSA (Italy); IAC (Spain). This development has been supported by the funding agencies BMVIT (Austria), ESA-
PRODEX (Belgium), CEA/CNES (France), DLR (Germany), ASI (Italy) and CICYT/MICINN (Spain). Funding for the SDSS and SDSS-II has been provided by the Alfred P. Sloan Foundation, the Participating Institutions, the National Science Foundation, the U.S. Department of Energy, the National Aeronautics and Space Administration, the Japanese Monbukagakusho, the Max Planck Society, and the Higher Education Funding Council for England. The SDSS Web Site is http://www.sdss.org/.

The SDSS is managed by the Astrophysical Research Consortium for the Participating Institutions. The Participating Institutions are the American Museum of Natural History, Astrophysical Institute Potsdam, University of Basel, University of Cambridge, Case Western Reserve University, University of Chicago, Drexel University, Fermilab, the Institute for Advanced Study, the Japan Participation Group, Johns Hopkins University, the Joint Institute for Nuclear Astrophysics, the Kavli Institute for Particle Astrophysics and Cosmology, the Korean Scientist Group, the Chinese Academy of Sciences (LAMOST), Los Alamos National Laboratory, the Max-Planck-Institute for Astronomy (MPIA), the Max-Planck-Institute for Astrophysics (MPA), New Mexico State University, Ohio State University, University of Pittsburgh, University of Portsmouth, Princeton University, the United States Naval Observatory, and the University of Washington. Financial support from the Spanish grant AYA2010-15169 and from the Junta de Andaluc\'ia through TIC-114 and the Excellence Project P08-TIC-03531 is acknowledged.

\bibliographystyle{mn2e}
\bibliography{ioteo_biblio}

\onecolumn

\appendix

%
%
%

\clearpage

\section{Dependence of the results on the assumed star-formation history}\label{other_SFH}

Throughout this work, we have used a set of BC03 templates which were built by using a SFH constant with time. In the elaboration of the BC03 templates, other temporal functions can be assumed for the SFH associated to the templates. One of the most used temporal variation is that when the SFH is exponentially declining with time. In this case, the SFH is characterized by the time-scale, $\tau_{\rm SFH}$. This is an extra parameter to fit in the SED-fitting procedure in addition to age and dust attenuation (assuming that we only consider one value of metallicity). Three questions arises now: Do the SED-fitting derived values of age and dust attenuation for our PACS-detected galaxies change with the assumed SFH? Once we assume an exponentially declining SFH, do the SED-fitting derived values of age and dust attenuation for our PACS-detected galaxies change with different assumed values of the time-scale $\tau_{\rm SFH}$? and finally, and most important, in the case that there is a dependence of age and dust attenuation on $\tau_{\rm SFH}$, can an SED-fitting with BC03 templates to the observed UV-to-near-IR photometry distinguish which is the right value of the star-formation time scale and, therefore, give the right values of age and dust attenuation?

In this appendix we analyze the possible differences in the SED-derived parameters when using different temporal dependences of the SFH with different values of the time-scale $\tau_{\rm SFH}$. To this aim, we build another set of BC03 templates assuming time declining SFHs with values of the time-scale $\tau_{\rm SFH} = [0.0001, 0.001, 0.01, 0.1, 1.0, 2.0, 5.0, 10.0, 50.0]$ Gyr. It should be noted that a constant SFH is a exponentially declining SFH when the star-formation time-scale tends to infinity.

\begin{figure*}
\centering
\includegraphics[width=0.49\textwidth]{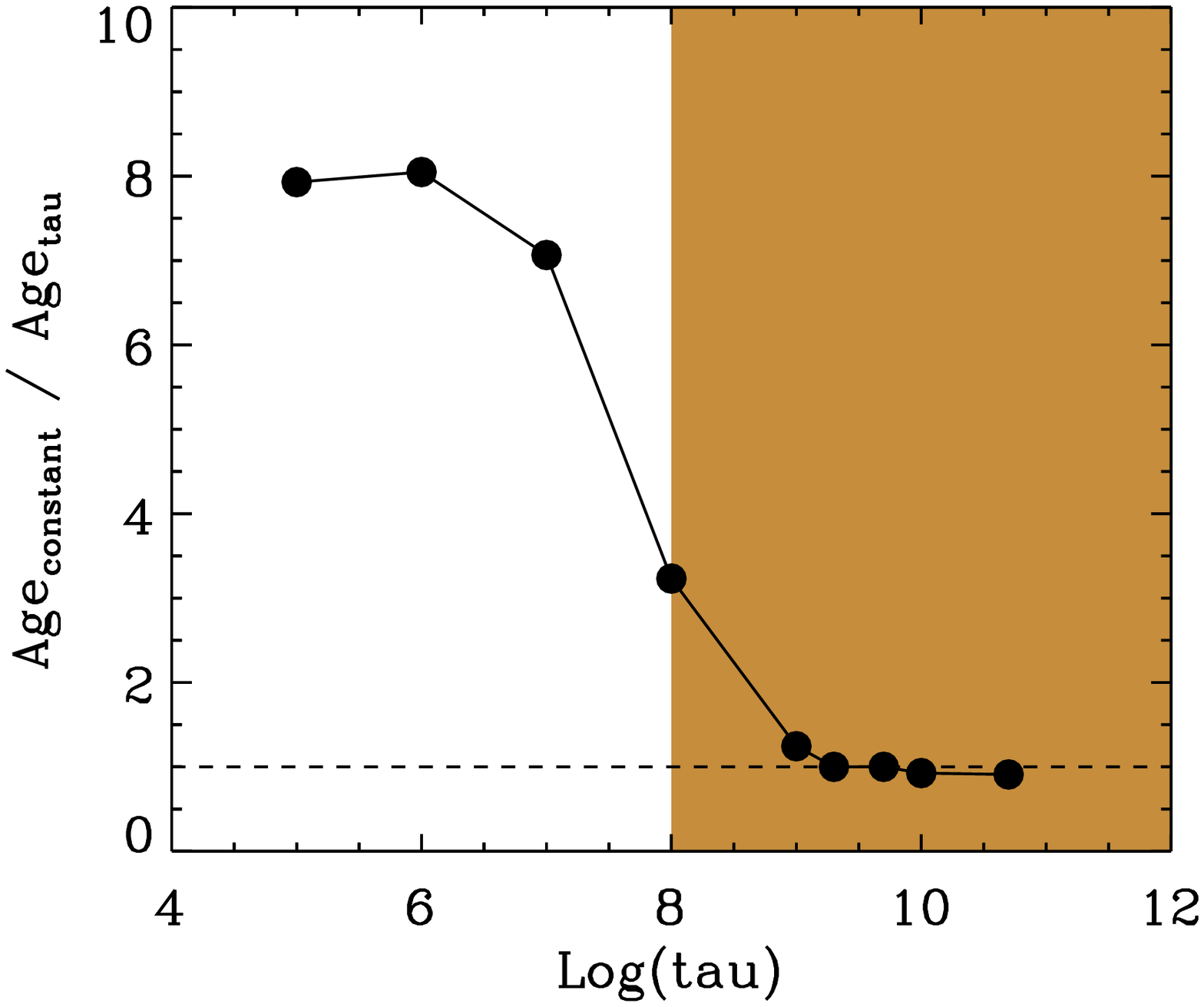}
\includegraphics[width=0.49\textwidth]{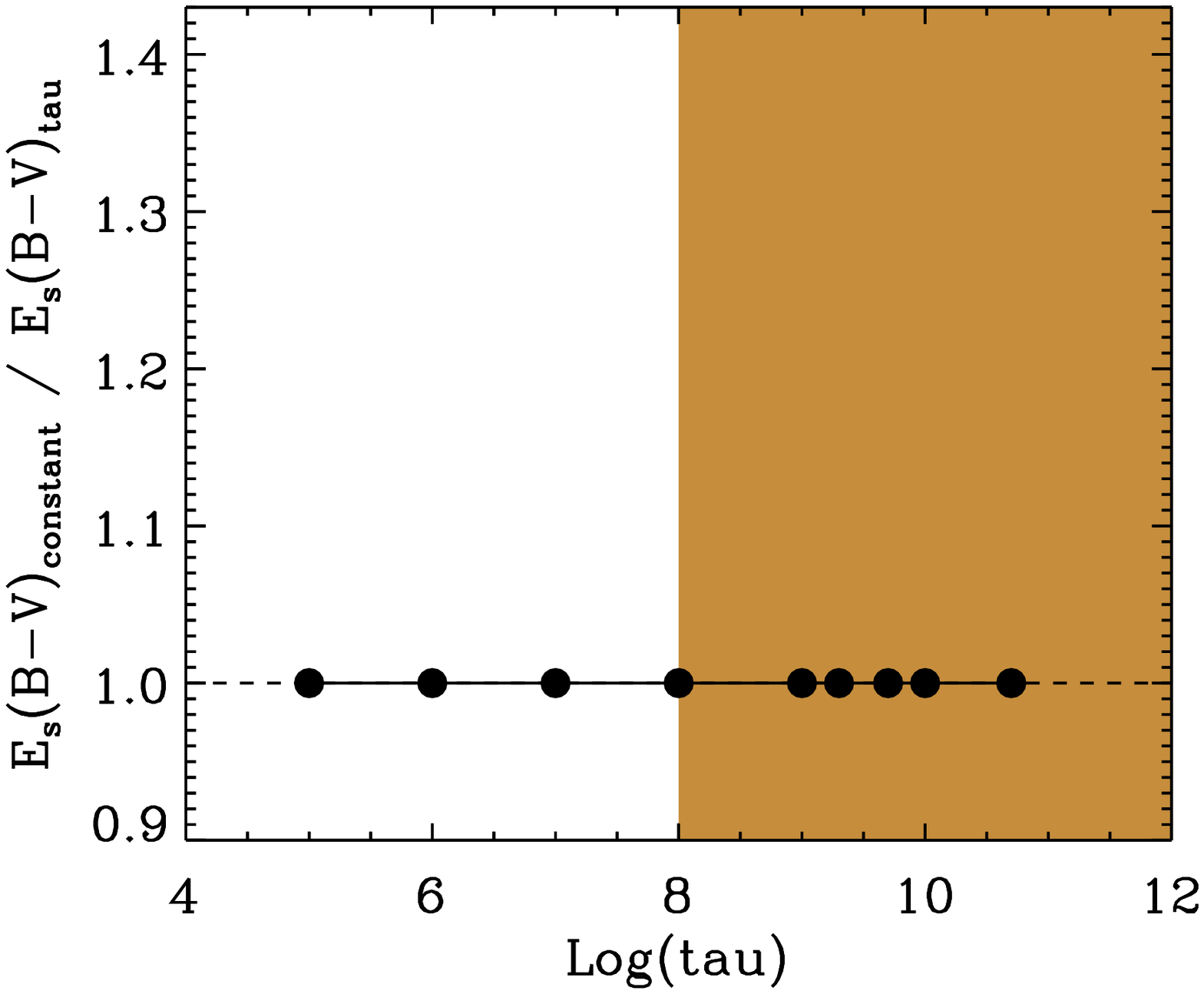}
\includegraphics[width=0.49\textwidth]{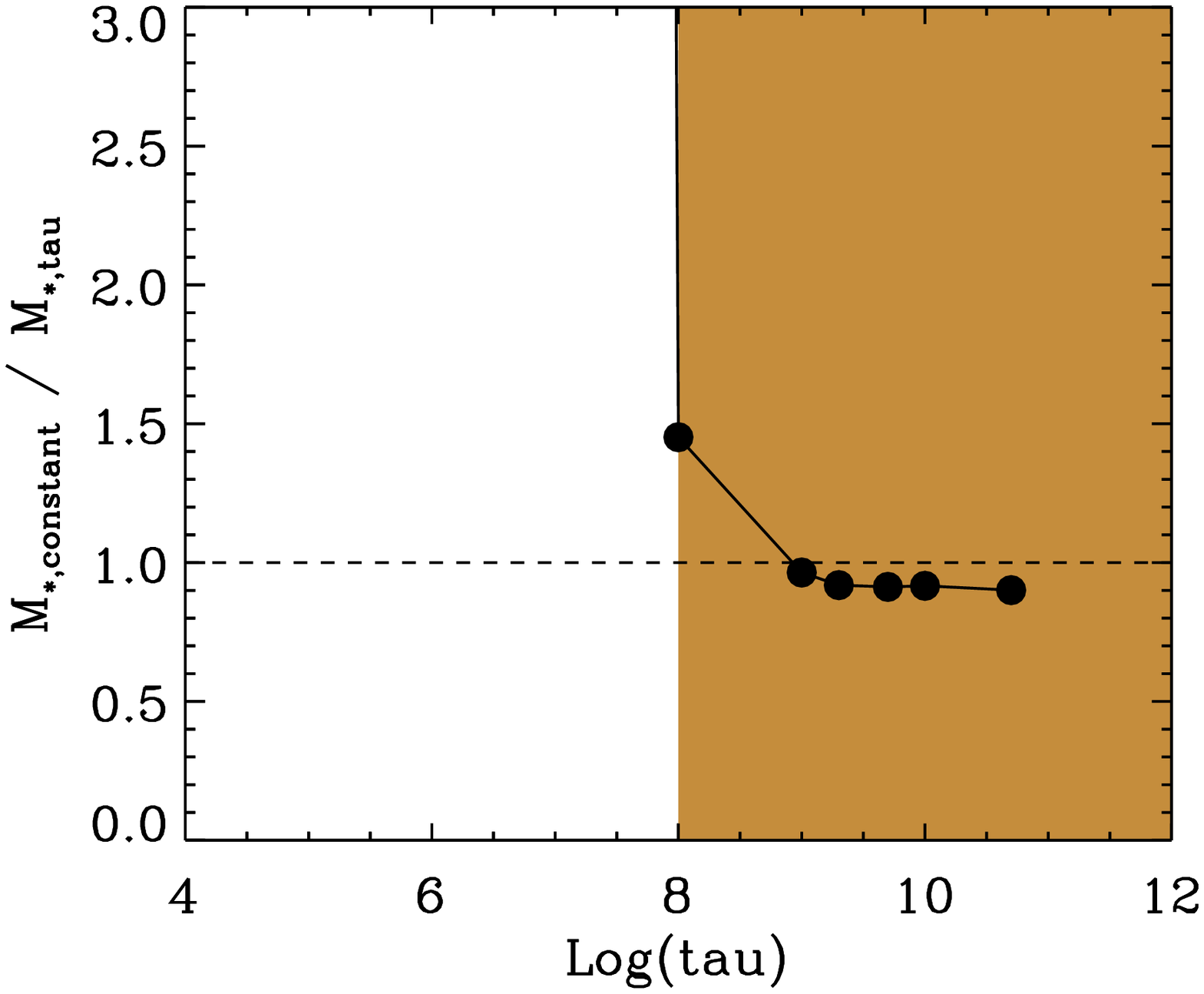}
\includegraphics[width=0.49\textwidth]{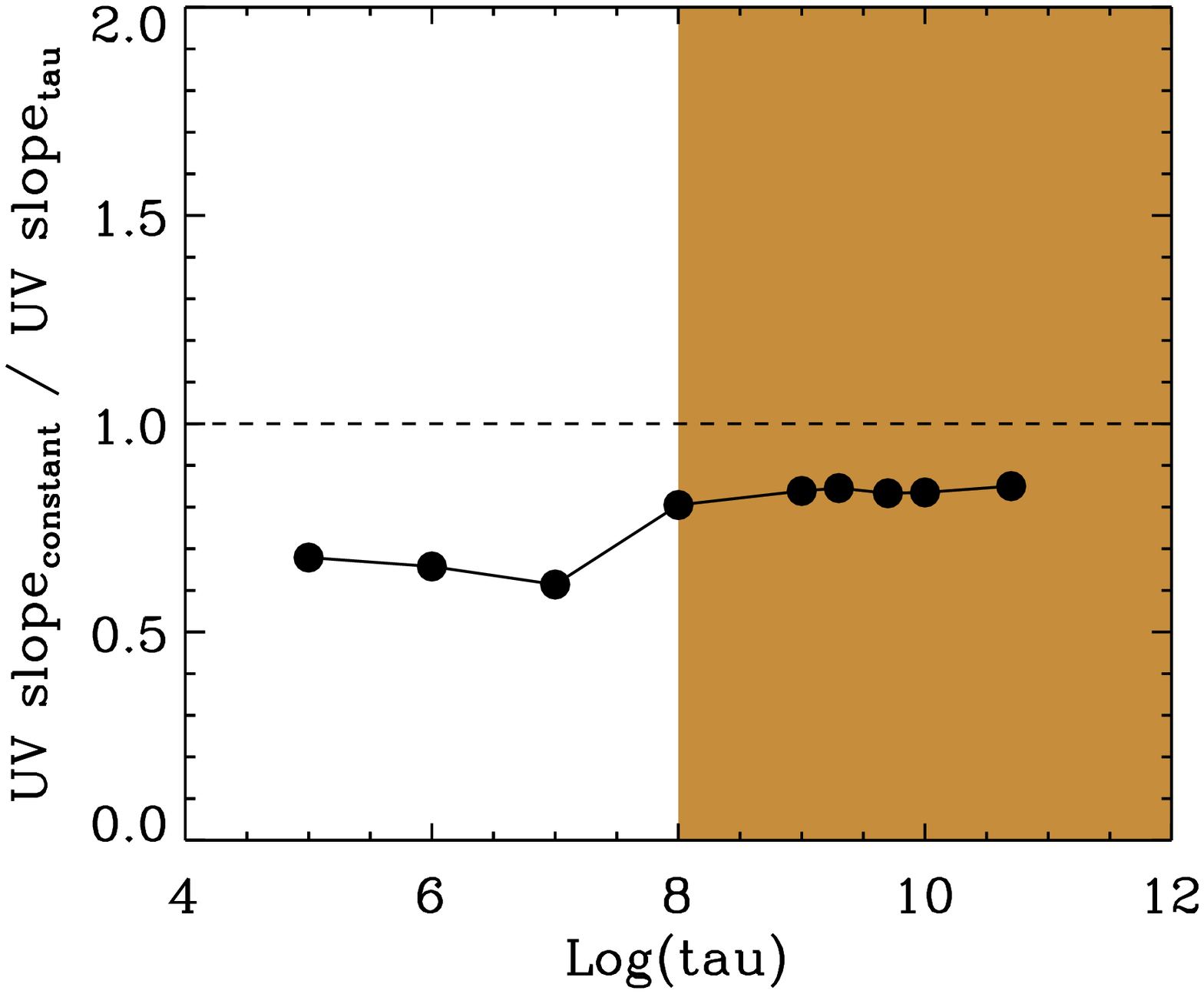}
\caption{Ratios between the SED-derived age (\emph{upper left}), dust attenuation (\emph{upper right}), stellar mass (\emph{bottom left}), and UV continuum slope \emph{bottom right} when assuming a constant SFH and SFHs associated to different values of the time-scale. In this plot, all our PACS-detected sources are included as a whole sample of UV-selected FIR-detected galaxies at $z \sim 1$. These ratios are represented as a function of the logarithm of the star-formation time-scale expressed in Gyr. Horizontal dashed lines represent where the SED-derived values associated to different SFHs would agree. Shaded zones are those corresponding to realistic values of the star-formation time-scale for our PACS-detected galaxies at $z \sim 1$.
              }
\label{tau}
\end{figure*}

In order to analyze if the SED-fitting derived results depend on the assumed SFH, we show in Figure \ref{tau} the relation between the SED-derived age, dust attenuation, stellar mass, and UV continuum slope for our PACS-detected galaxies when assuming a constant SFH and temporal exponentially declining SFHs associated to different star-formation time scales. Each point in the plots represents the median value for all the PACS-detected galaxies. It can be seen that age and stellar mass are the most dependent parameters on the assumption of the SFH. Both properties are lower with respect the value associated to a constant star-formation rate when the value of $\tau_{\rm SFH}$ decreases. The median values of the dust attenuation do not strongly depend on the assumed star-formation rate and regarding the UV continuum slope, it tends to change to lower values when the $\tau_{\rm SFR}$ is low. In this analysis, we have considered a wide range of values for the star-formation time scale in order to give a general vision of what happens to the SED-fittting results when the the SFH is modified. However, among the values considered, only those $\tau_{\rm SFH} > 0.1$ could be considered as realistic cases for our galaxies at the studied redshifts. Lower values of the star-formation time-scale would led to a passively evolving galaxy in which a instant burst took place in the onset of the star-formation. This is not probably the most realistic picture for UV-selected FIR detected galaxies, which are actively star-forming galaxies according to their selection in the rest-frame UV and their detection in the FIR. The regions associated to realistic values for our PACS-detected galaxies are those shaded in Figure \ref{tau}. It can be seen that considering only star-formation time-scales in the realistic zone, the SED-derived ages when considering declining SFH are lower than in the constant SFR case in a maximum factor of about 3. Dust attenuation is the same in all the cases. The SED-derived stellar masses are slightly higher when considering a declining SFH for values of the star-formation time-scale higher than about 1Gyr, and are higher in a maximum factor of 1.5 for $\tau_{\rm SFH} < 1 {\rm Gyr}$. Regarding the UV continuum slopes, they are lower in the case of declining SFH in a maximum factor of 0.8.

\begin{figure*}
\centering
\includegraphics[width=0.30\textwidth]{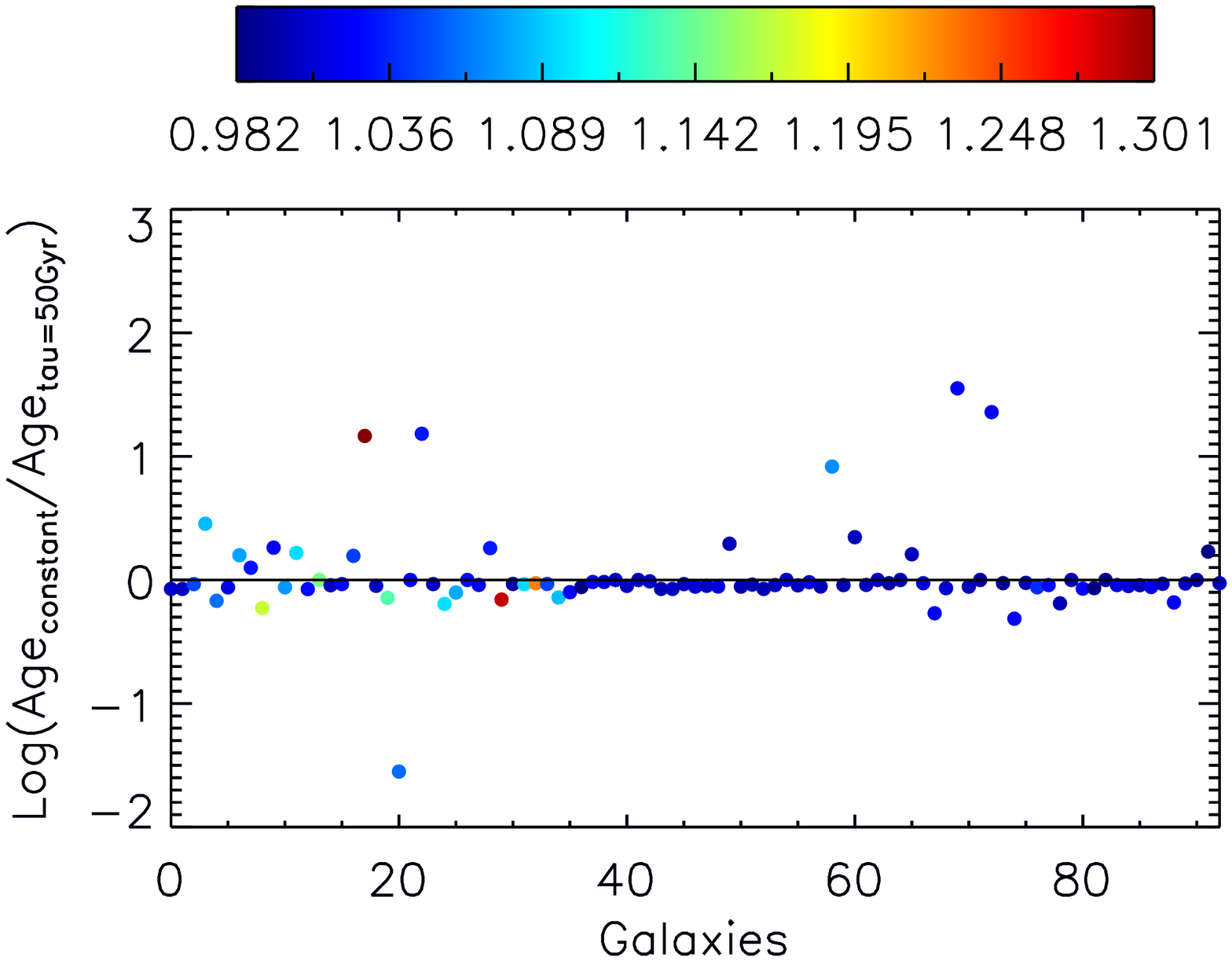}
\includegraphics[width=0.30\textwidth]{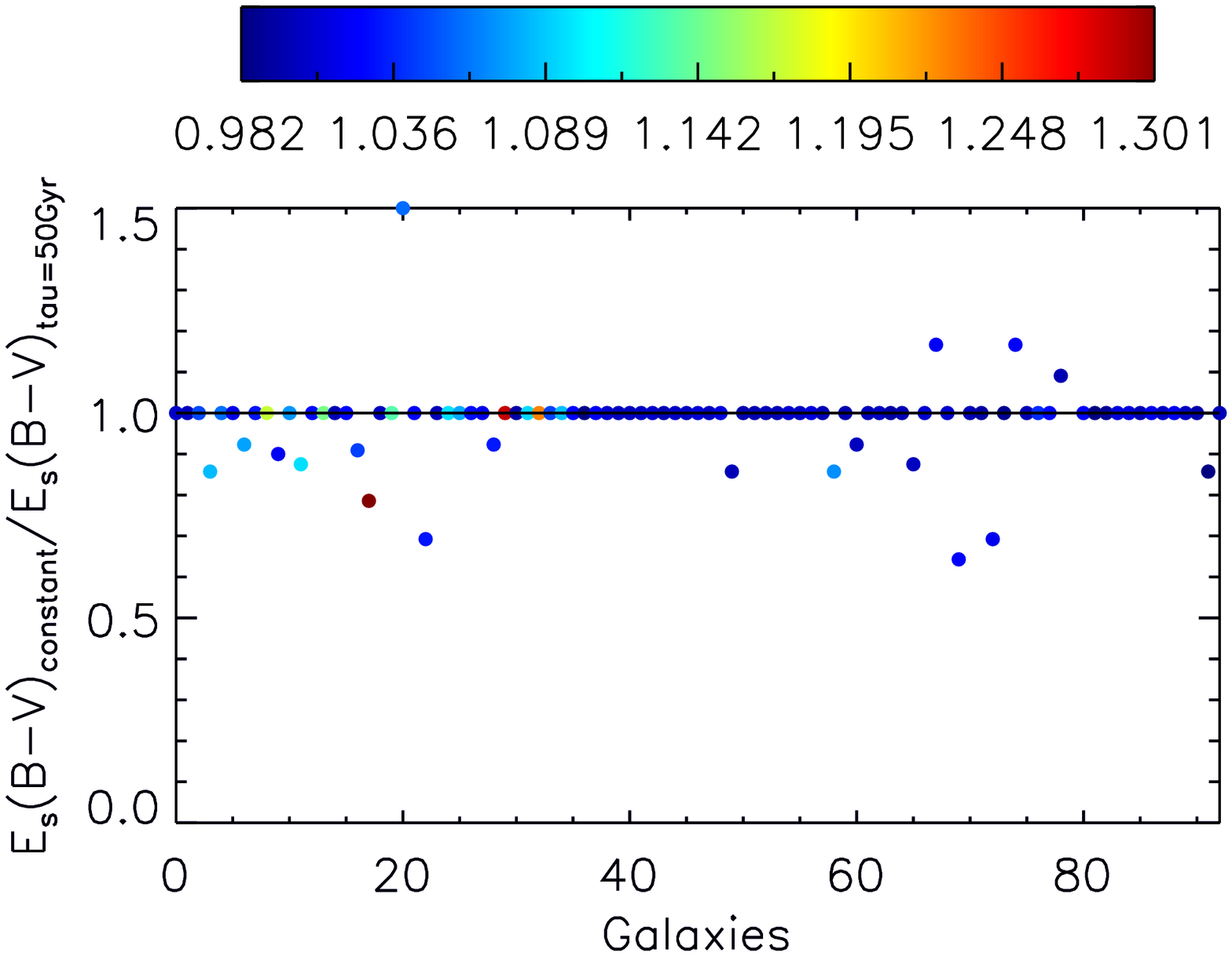}

\includegraphics[width=0.30\textwidth]{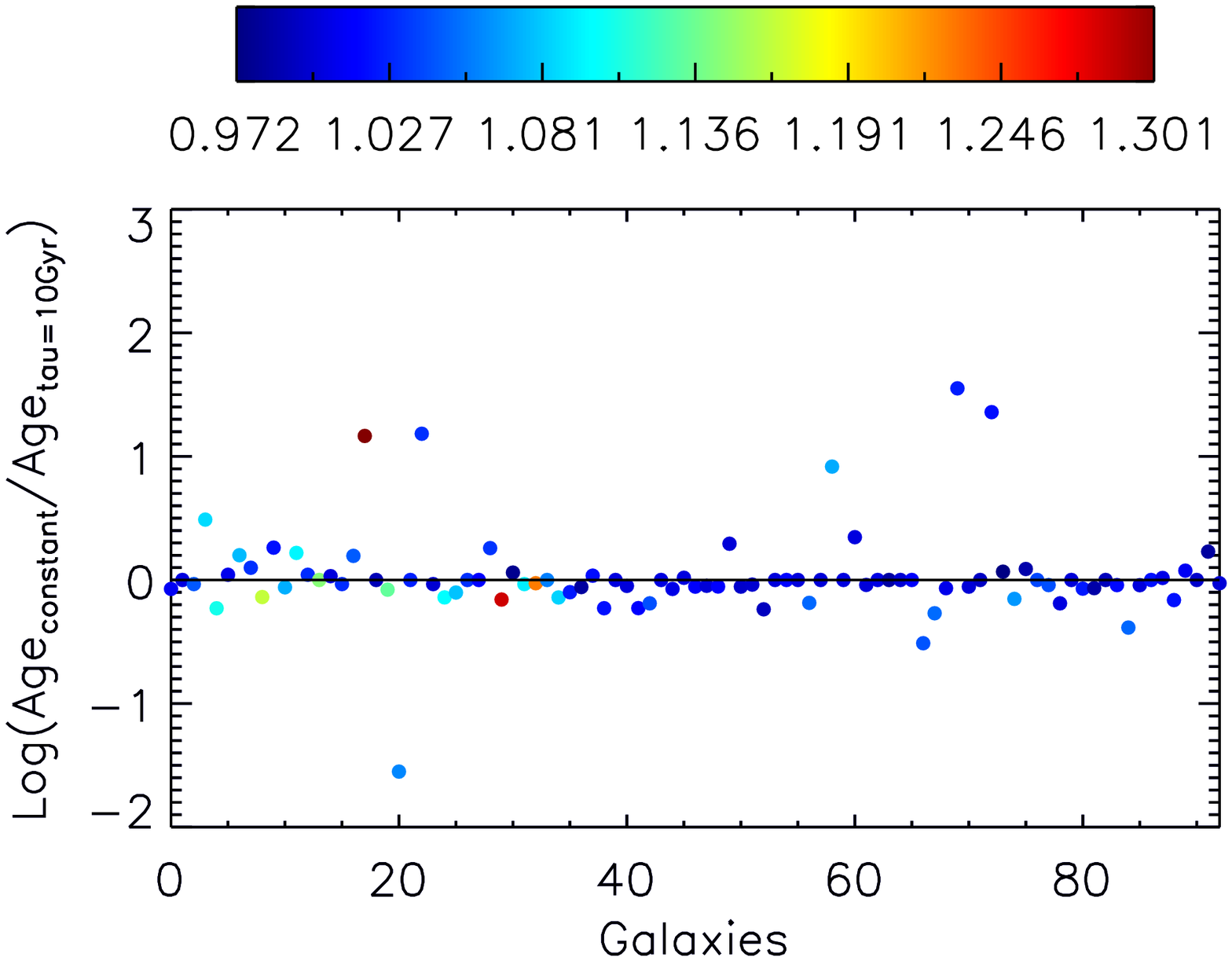}
\includegraphics[width=0.30\textwidth]{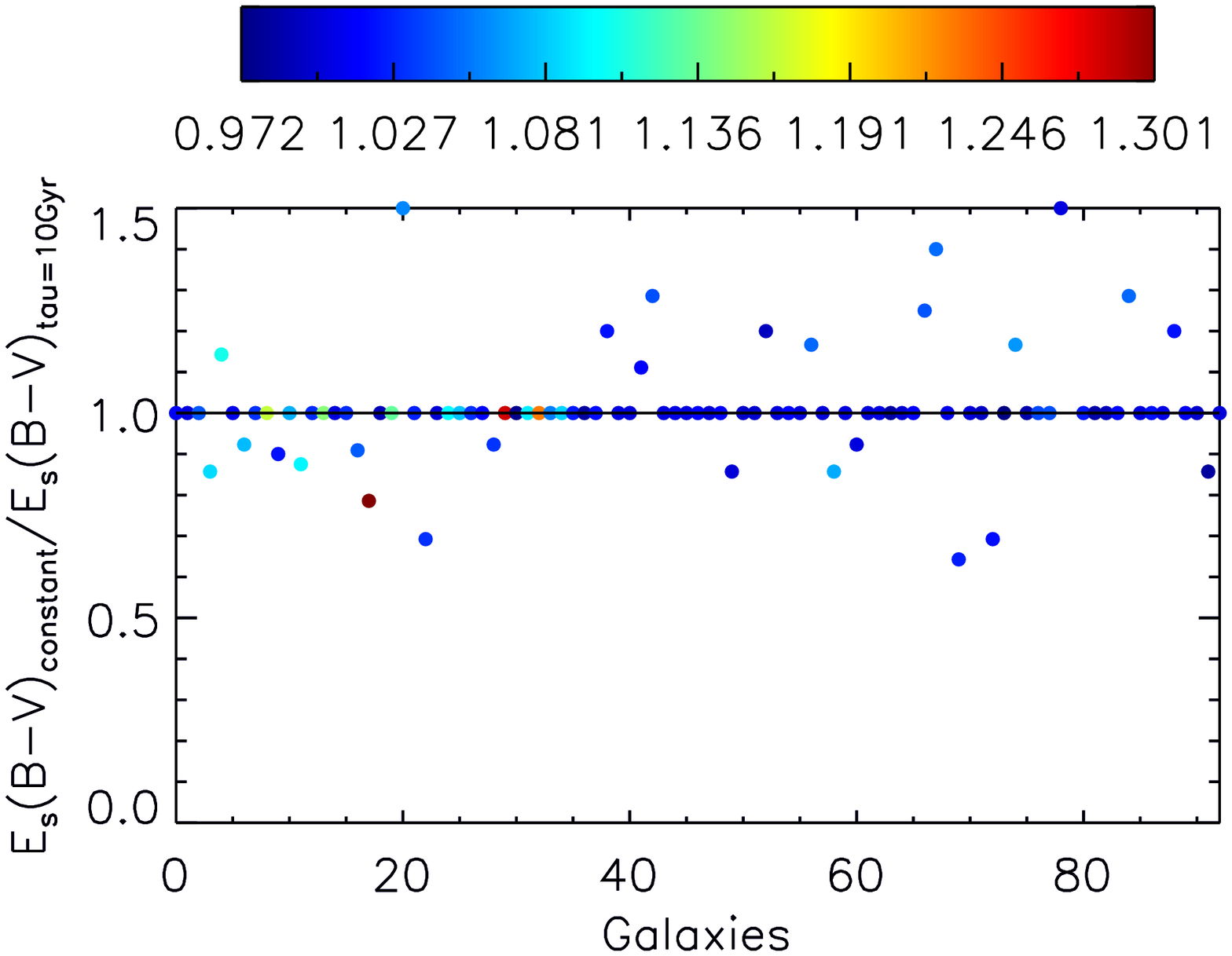}

\includegraphics[width=0.30\textwidth]{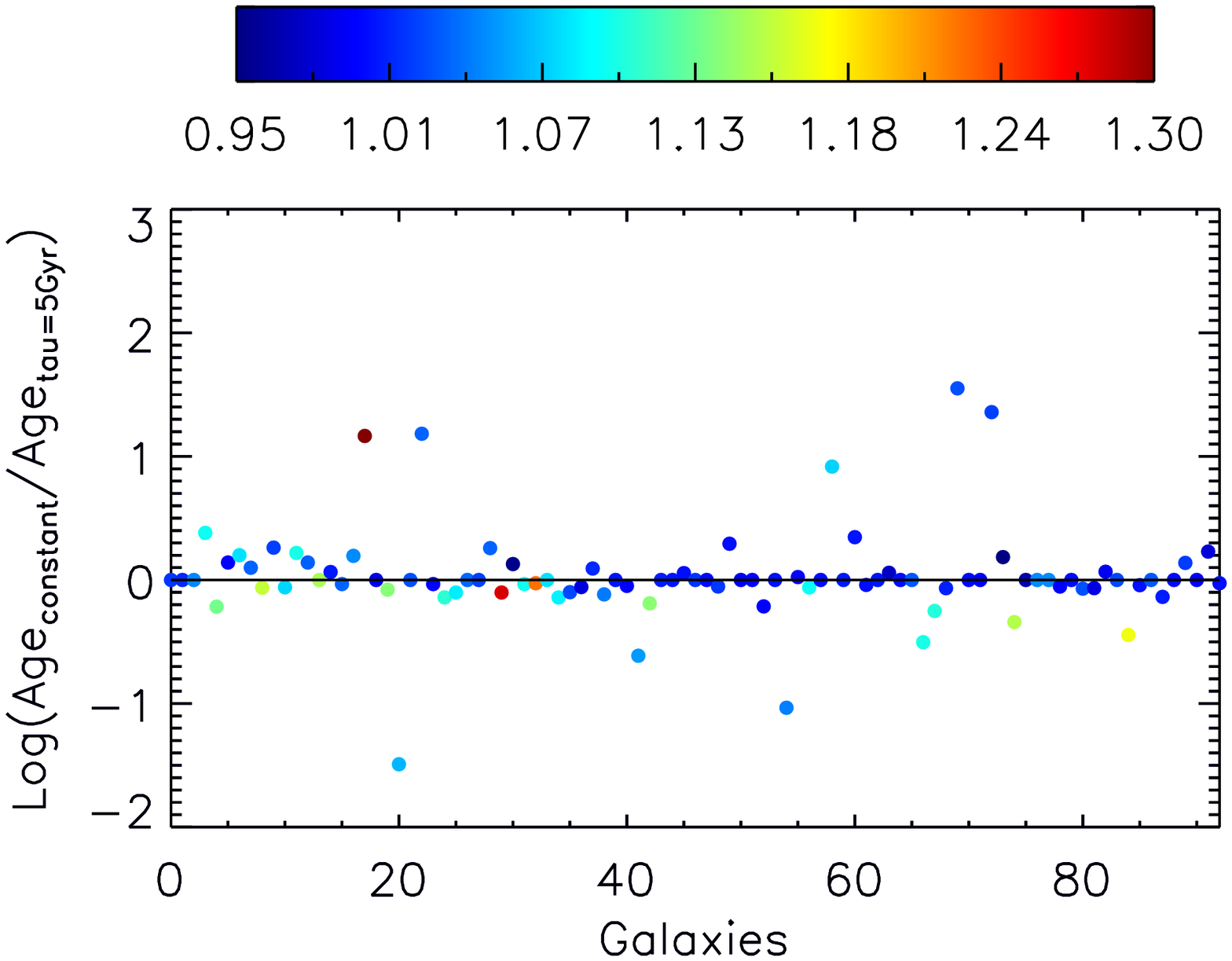}
\includegraphics[width=0.30\textwidth]{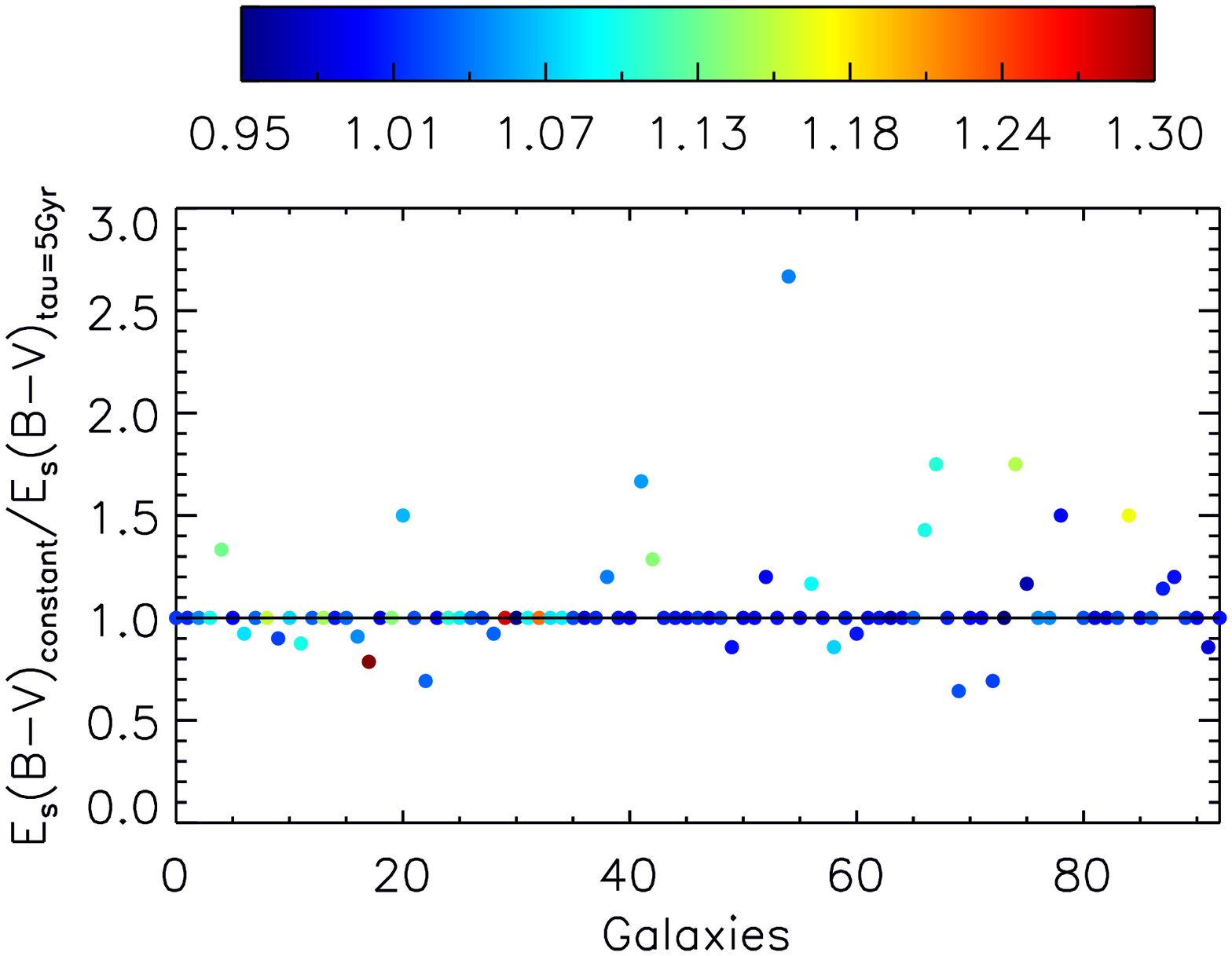}

\includegraphics[width=0.30\textwidth]{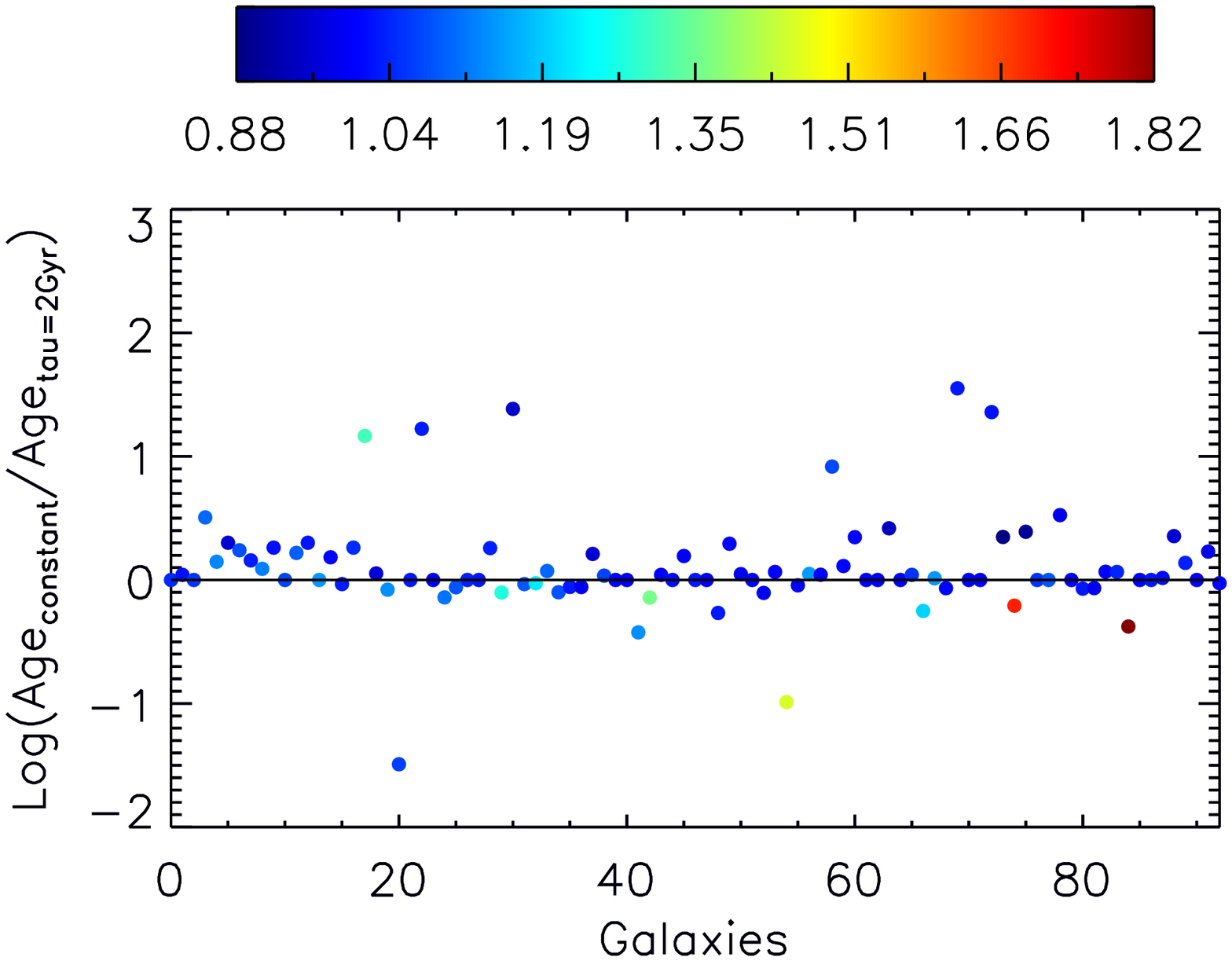}
\includegraphics[width=0.30\textwidth]{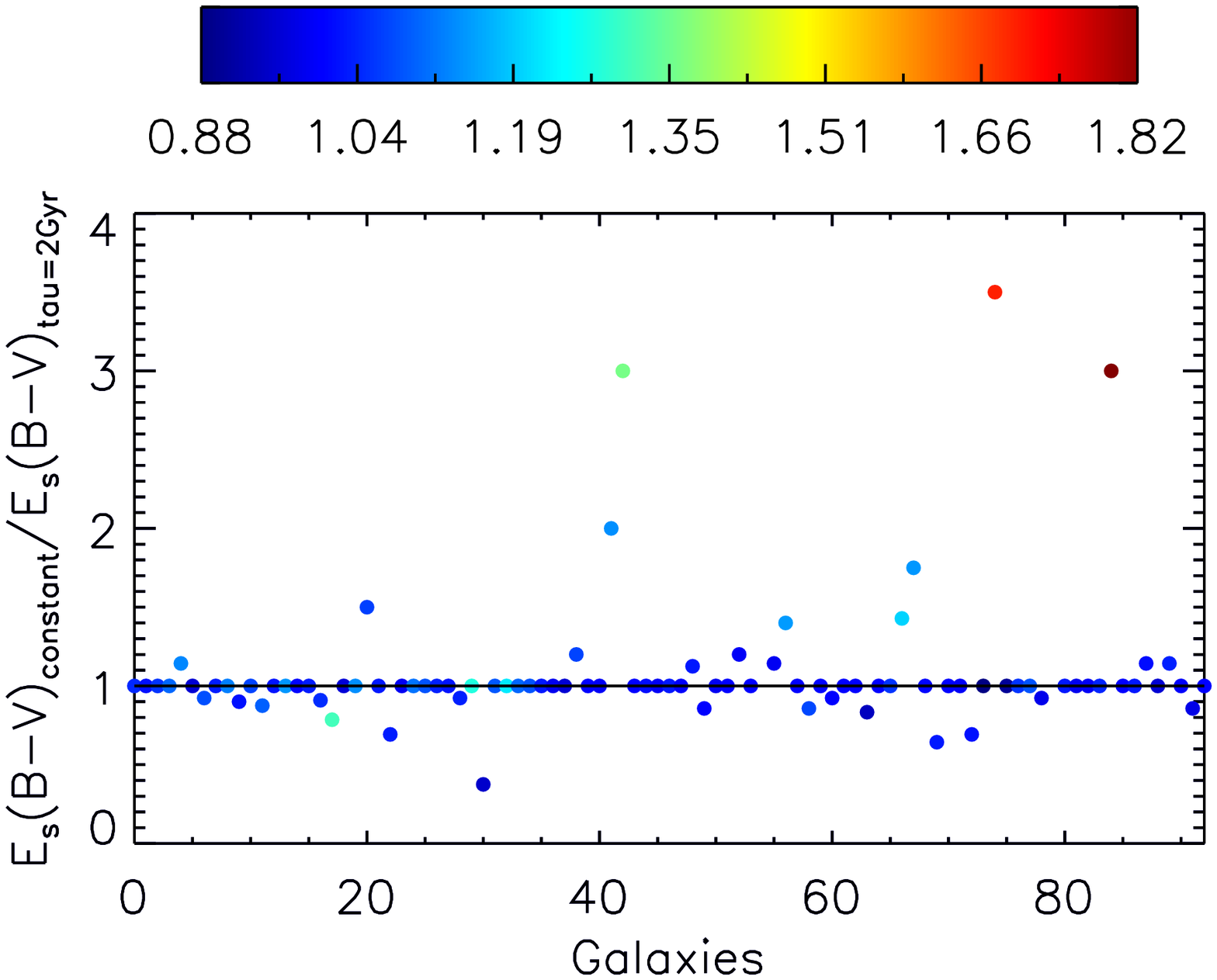}

\includegraphics[width=0.30\textwidth]{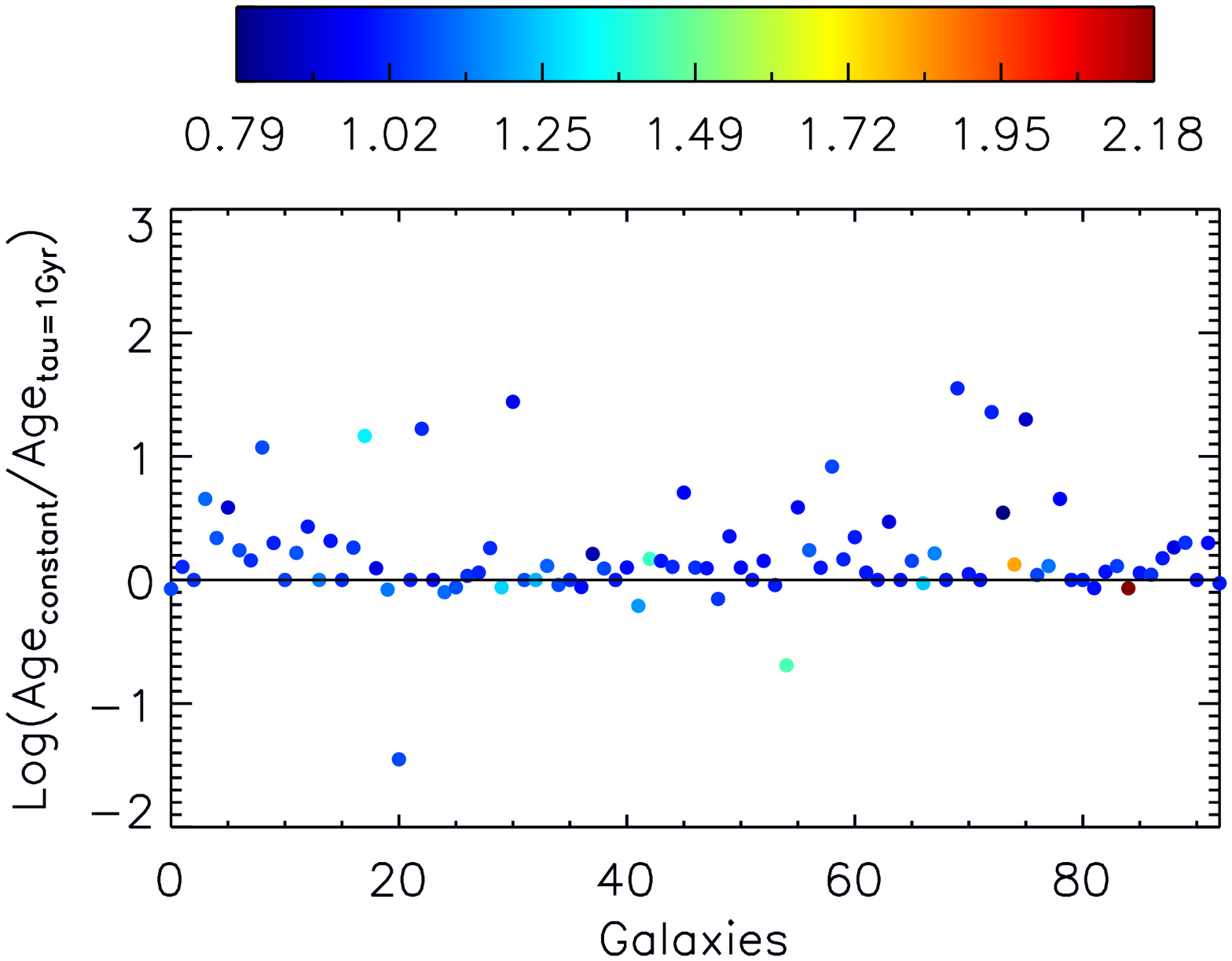}
\includegraphics[width=0.30\textwidth]{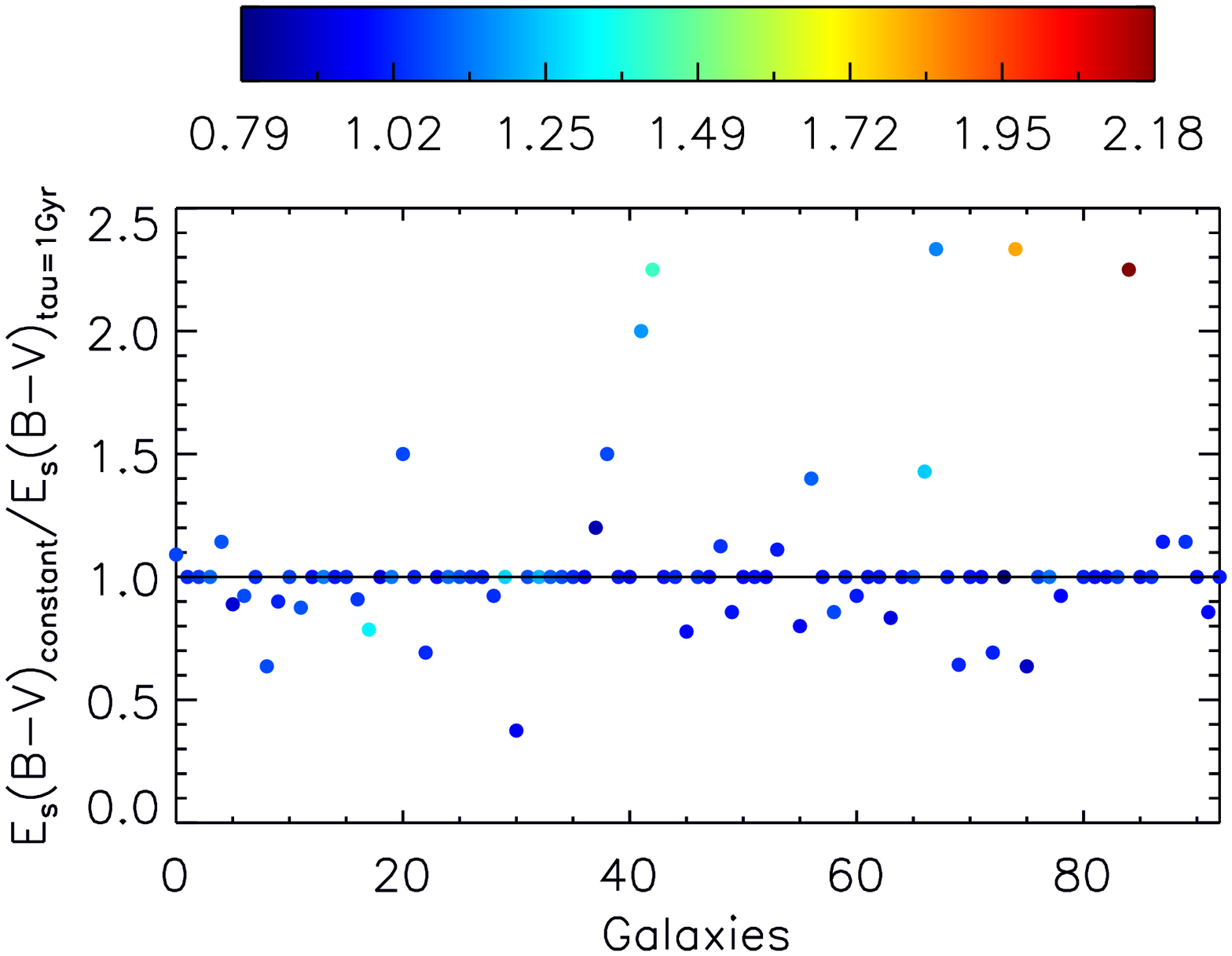}

\caption{\emph{Left column}: Relation between the ages derived with an SED-fitting procedure when considering a constant SFH and an exponentially declining SFH for different values of the SFH time-scale. \emph{Right column}: Relation between the dust attenuation derived with an SED-fitting procedure when considering a constant SFH and an exponentially declining SFH for different values of the SFH time-scale. Each point in the plots represents one PACS-detected galaxy. Each row is related to a specific value of the star-formation time-scale, $\tau_{\rm SFH}$, as indicated in each vertical axis. The SFR time-scales considered are $\tau_{SFH}$=50, 10, 5, 2, and 1 Gyr in the first, second, third, fourth, fifth and sixth rows, respectively. The color of each point is related to the ratio of the $\chi^2$ values between the SED-fittings carried out by considering a constant and an exponentially declining SFH: $\chi^2_{\rm constant}/\chi^2_{\tau_{\rm SFH}}$. The color code for the values of $\chi^2_{\rm constant}/\chi^2_{\tau_{\rm SFH}}$ are indicated in the color-bars. In all the plots, the horizontal line represent where both quantities represented in each vertical axis would agree.
              }
\label{tau_color}
\end{figure*}

\begin{figure*}
\centering
\includegraphics[width=0.30\textwidth]{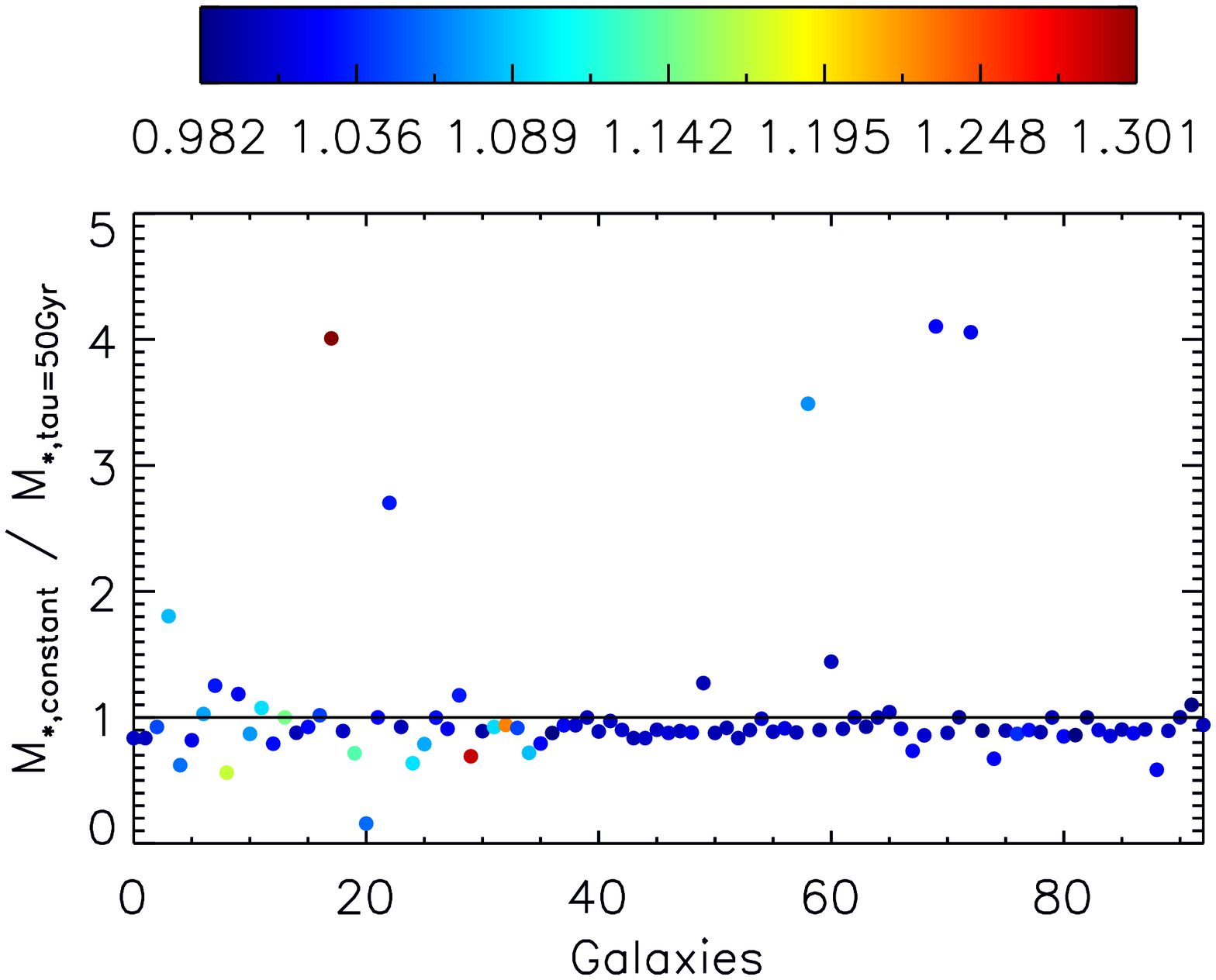}
\includegraphics[width=0.30\textwidth]{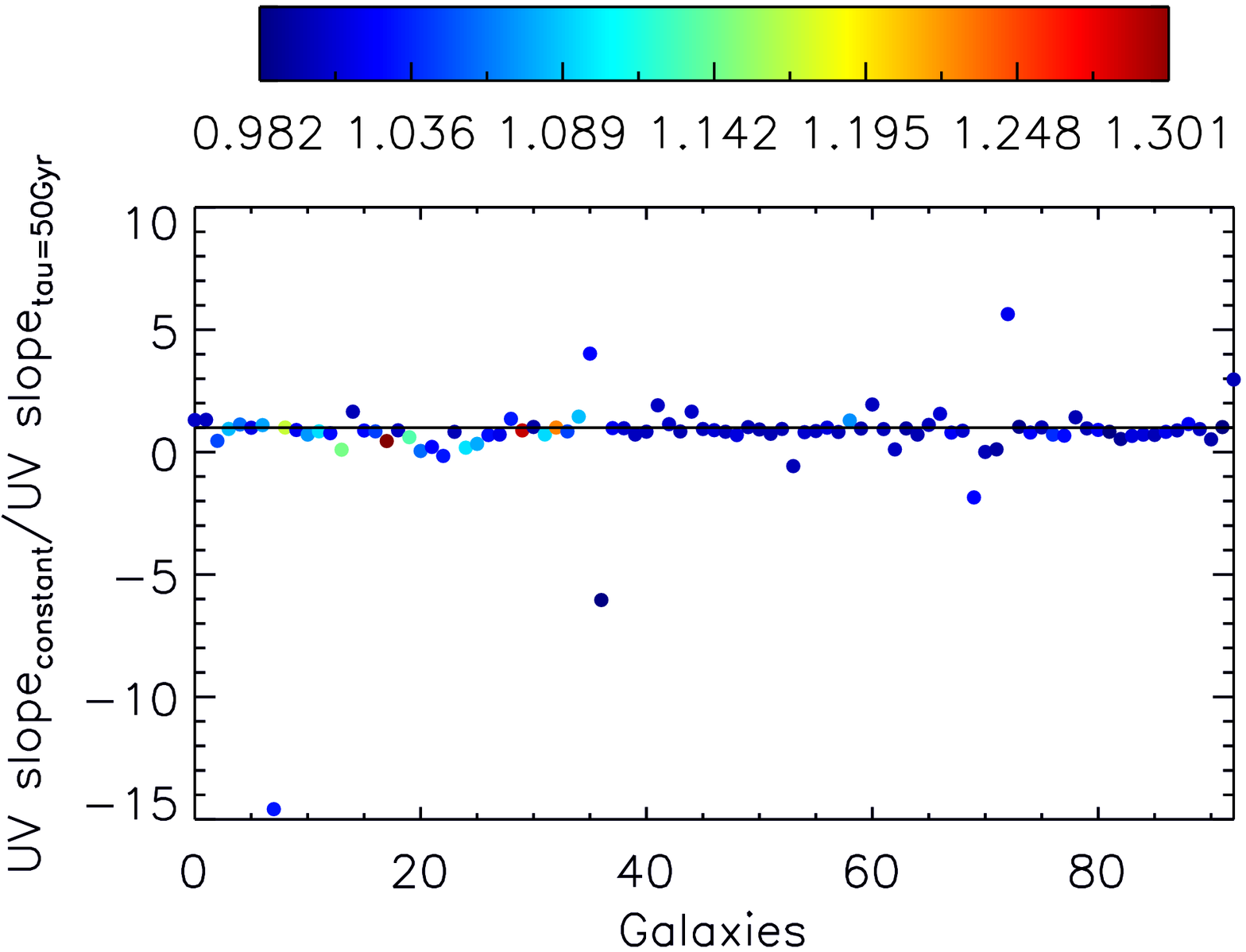}

\includegraphics[width=0.30\textwidth]{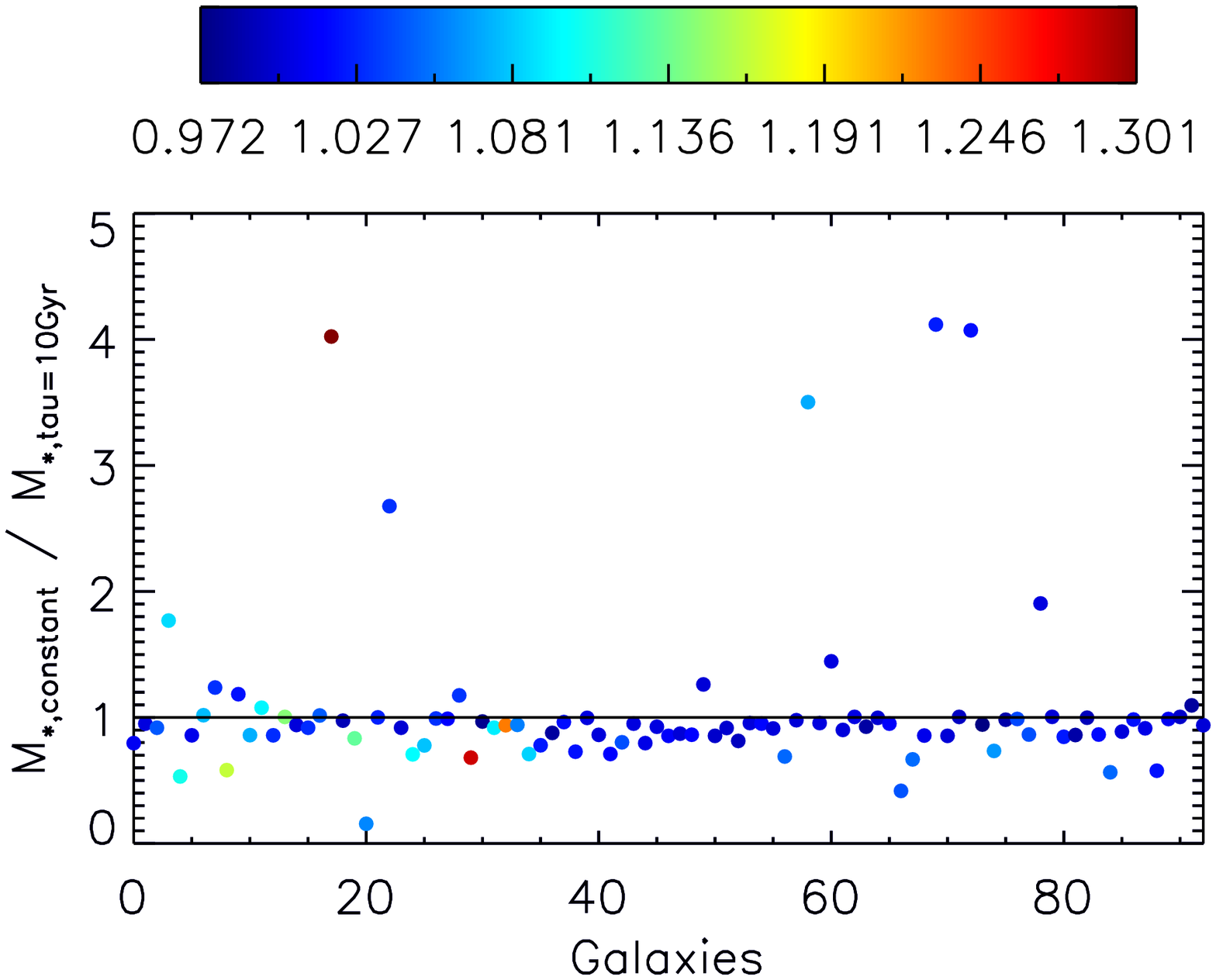}
\includegraphics[width=0.30\textwidth]{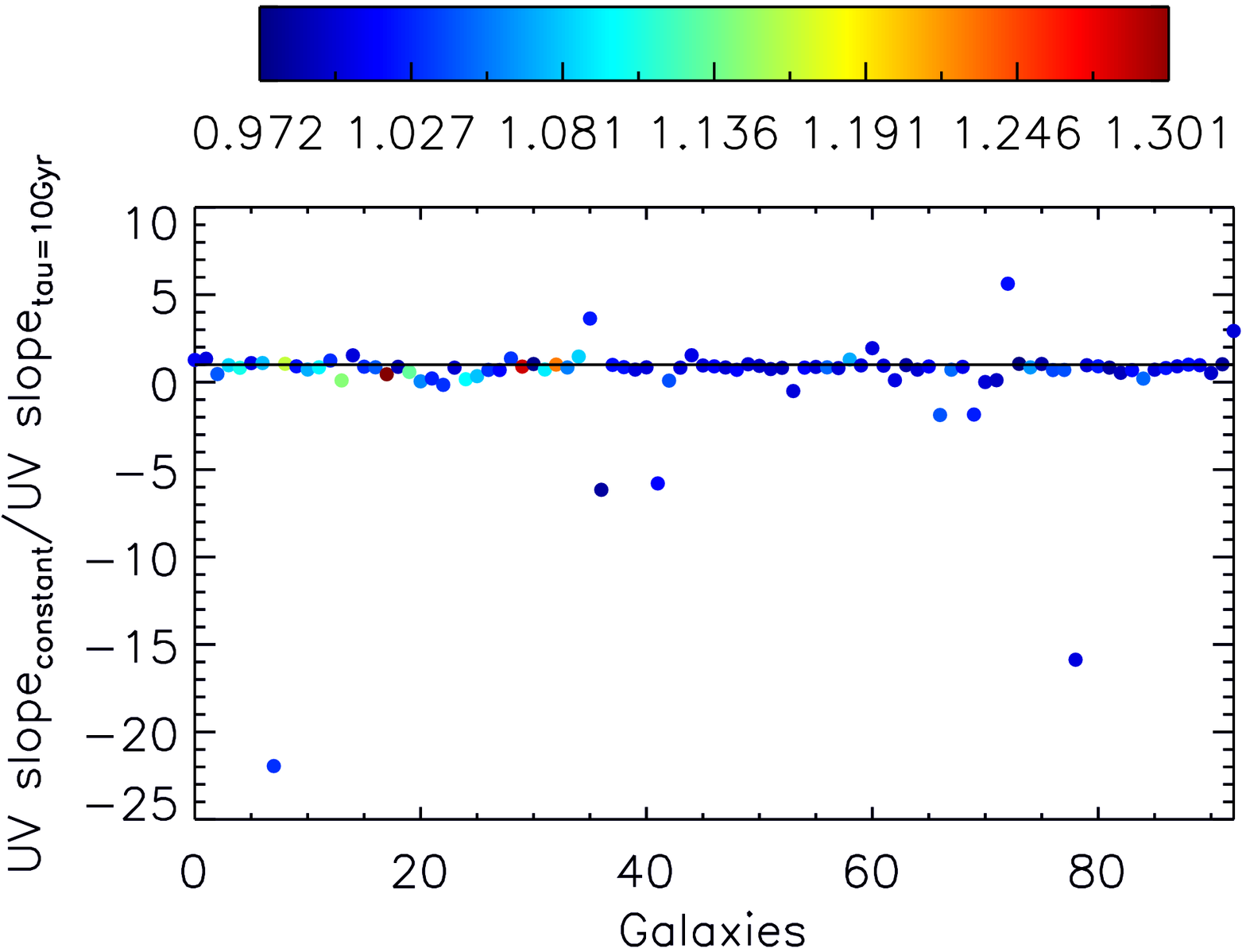}

\includegraphics[width=0.30\textwidth]{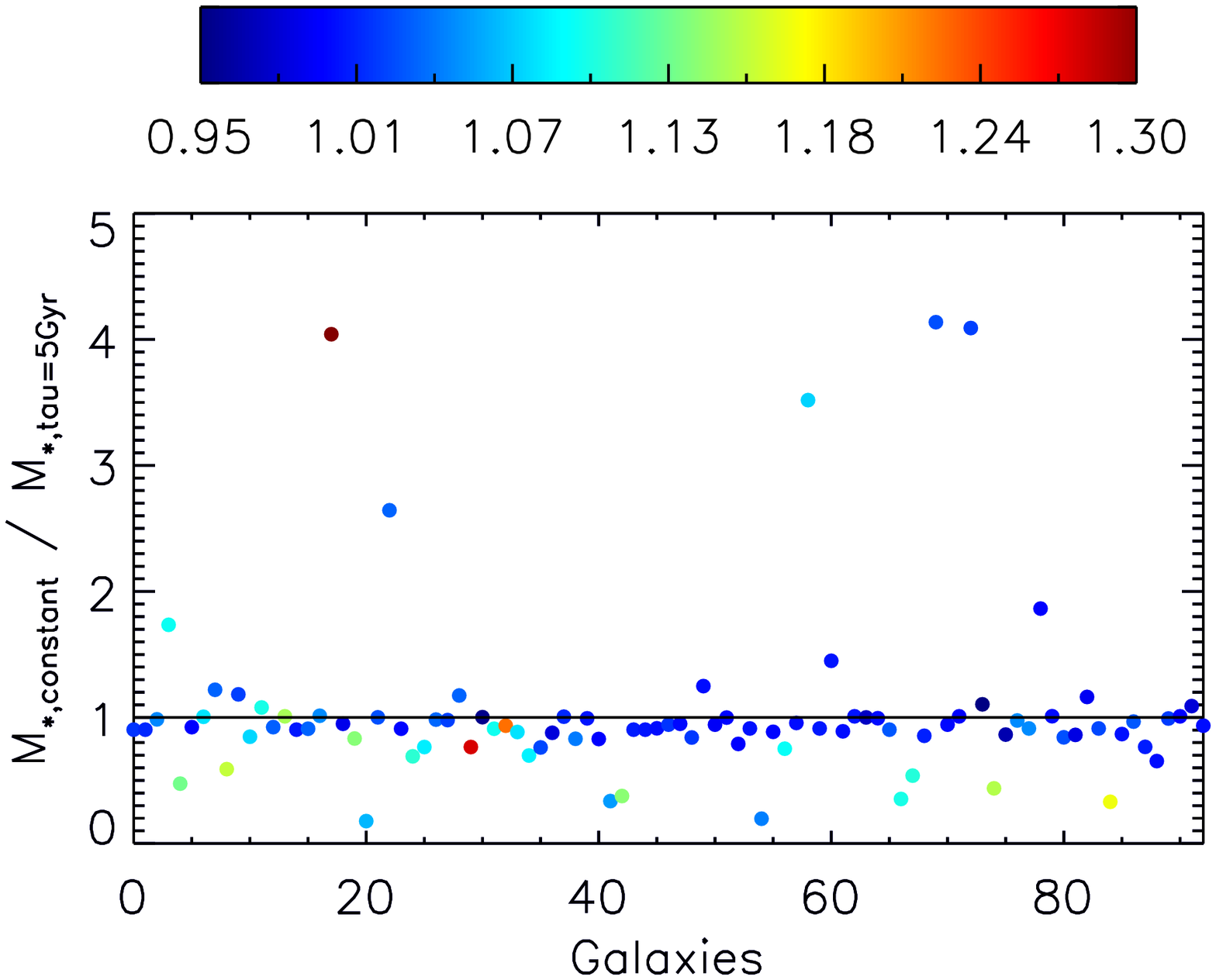}
\includegraphics[width=0.30\textwidth]{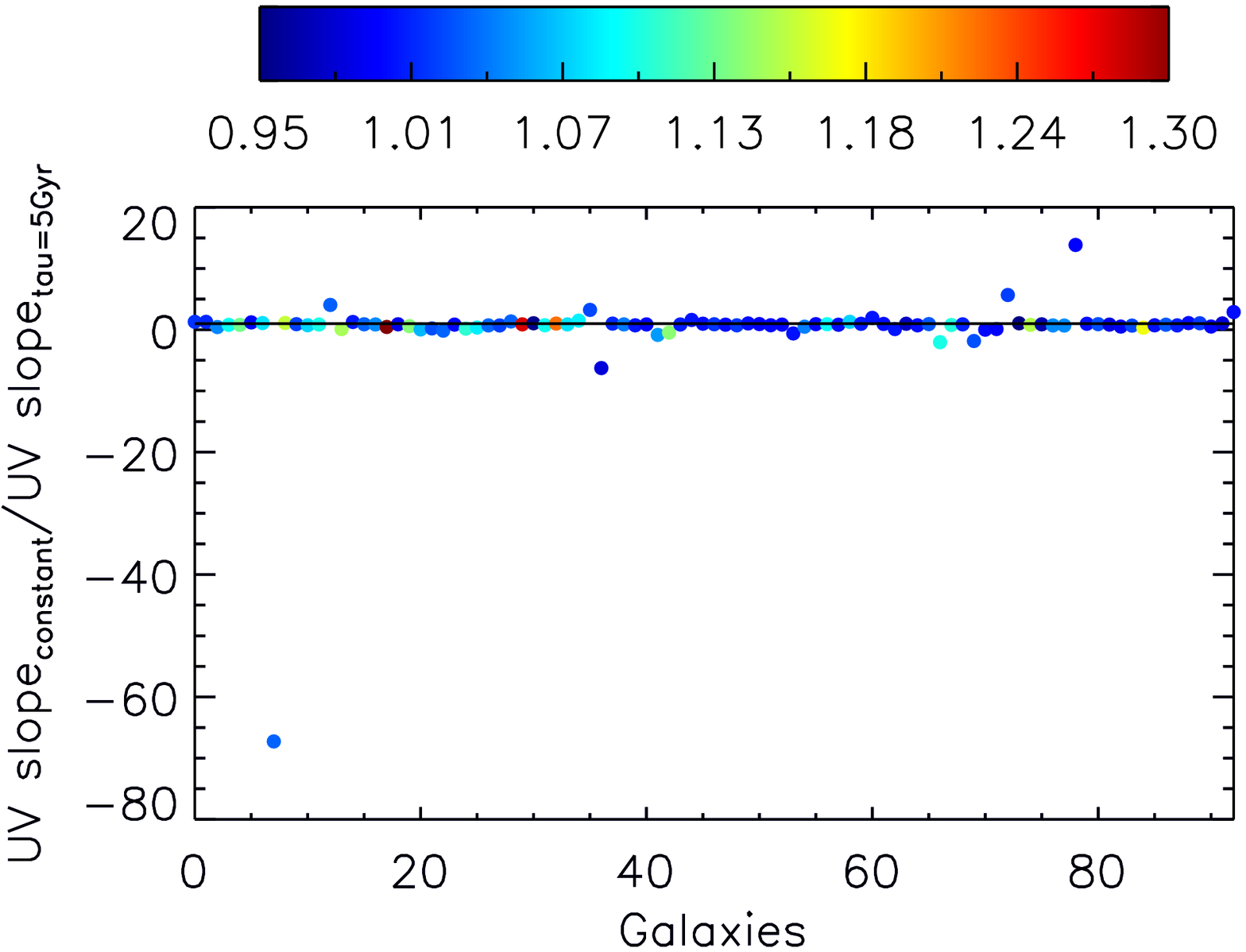}

\includegraphics[width=0.30\textwidth]{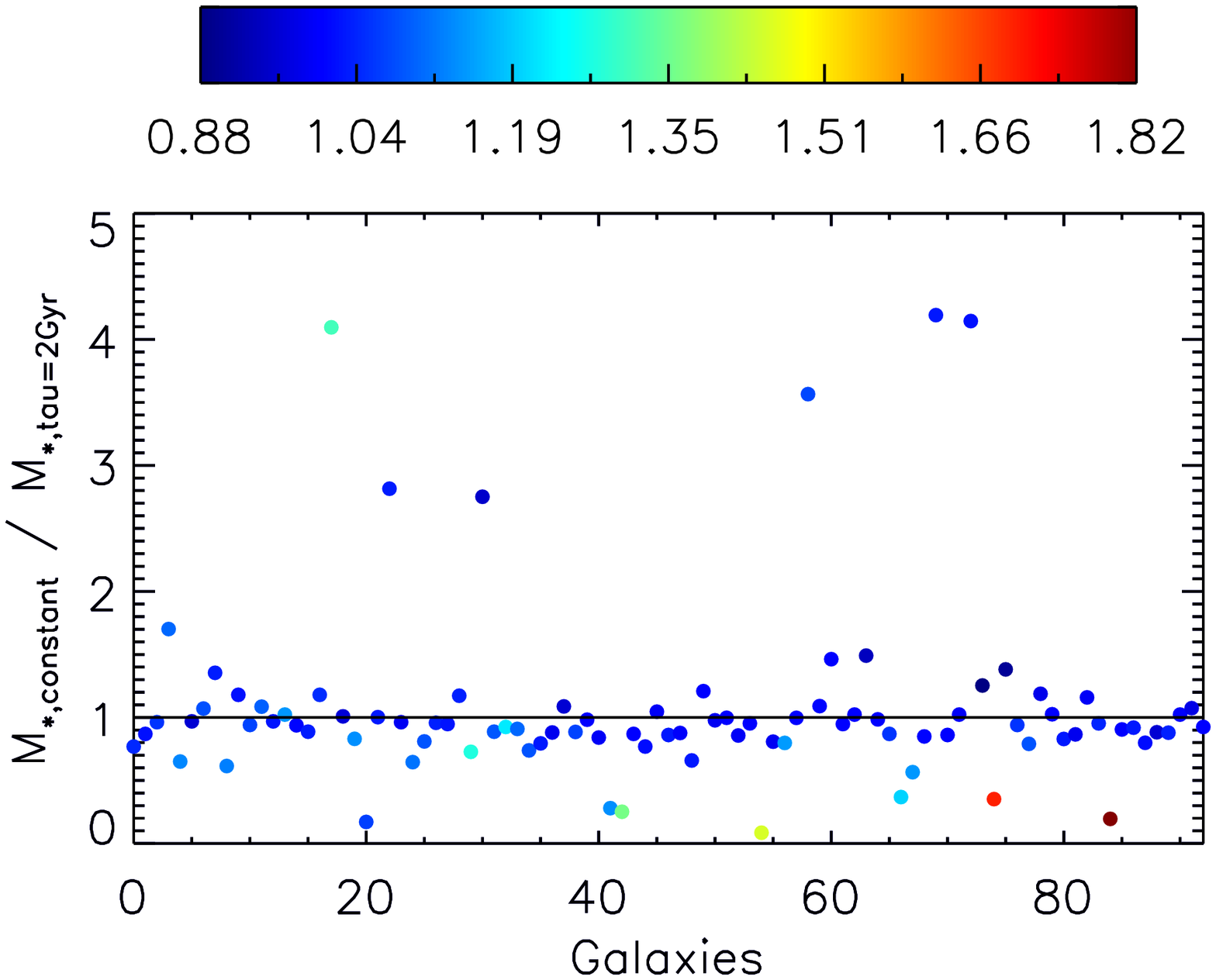}
\includegraphics[width=0.30\textwidth]{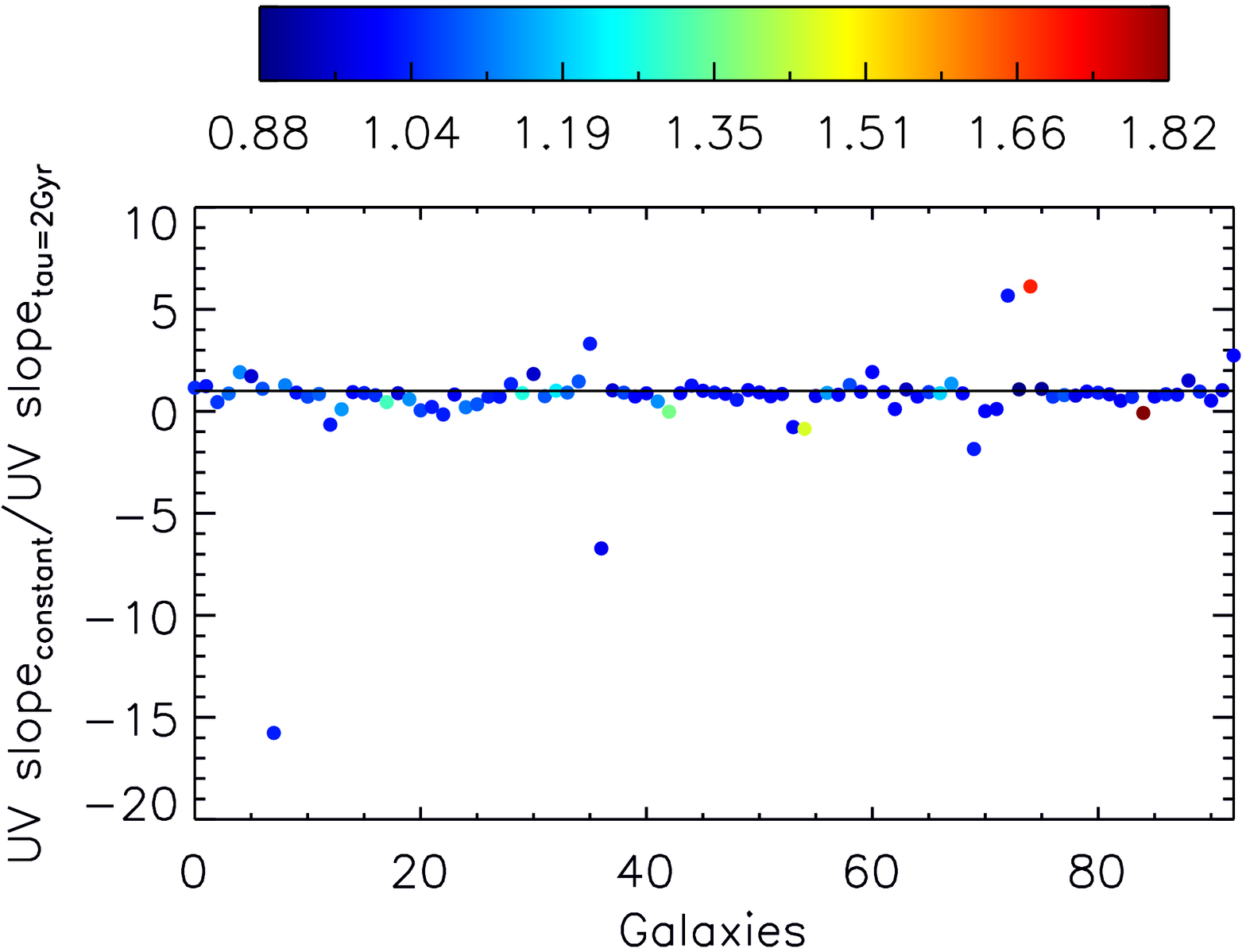}

\includegraphics[width=0.30\textwidth]{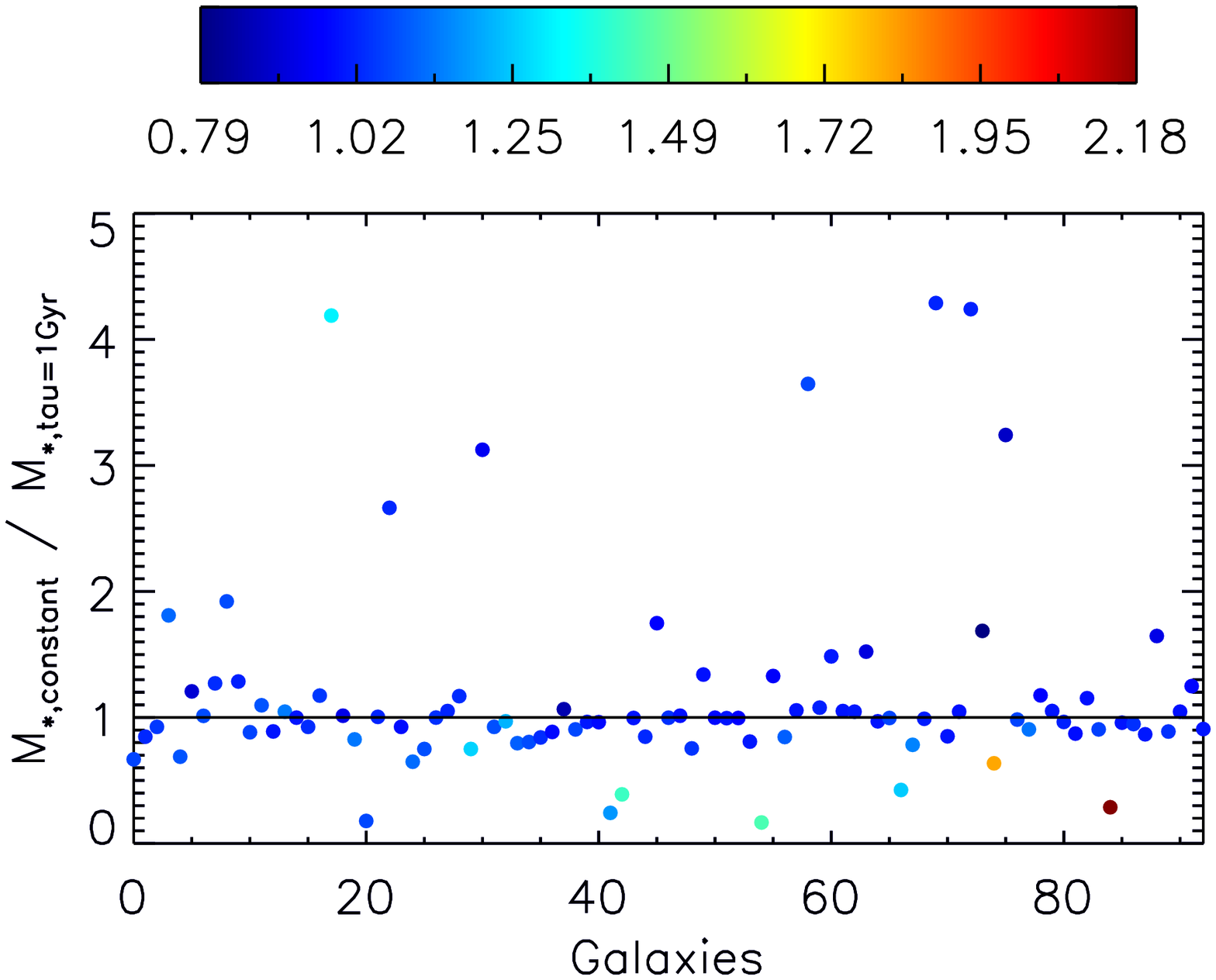}
\includegraphics[width=0.30\textwidth]{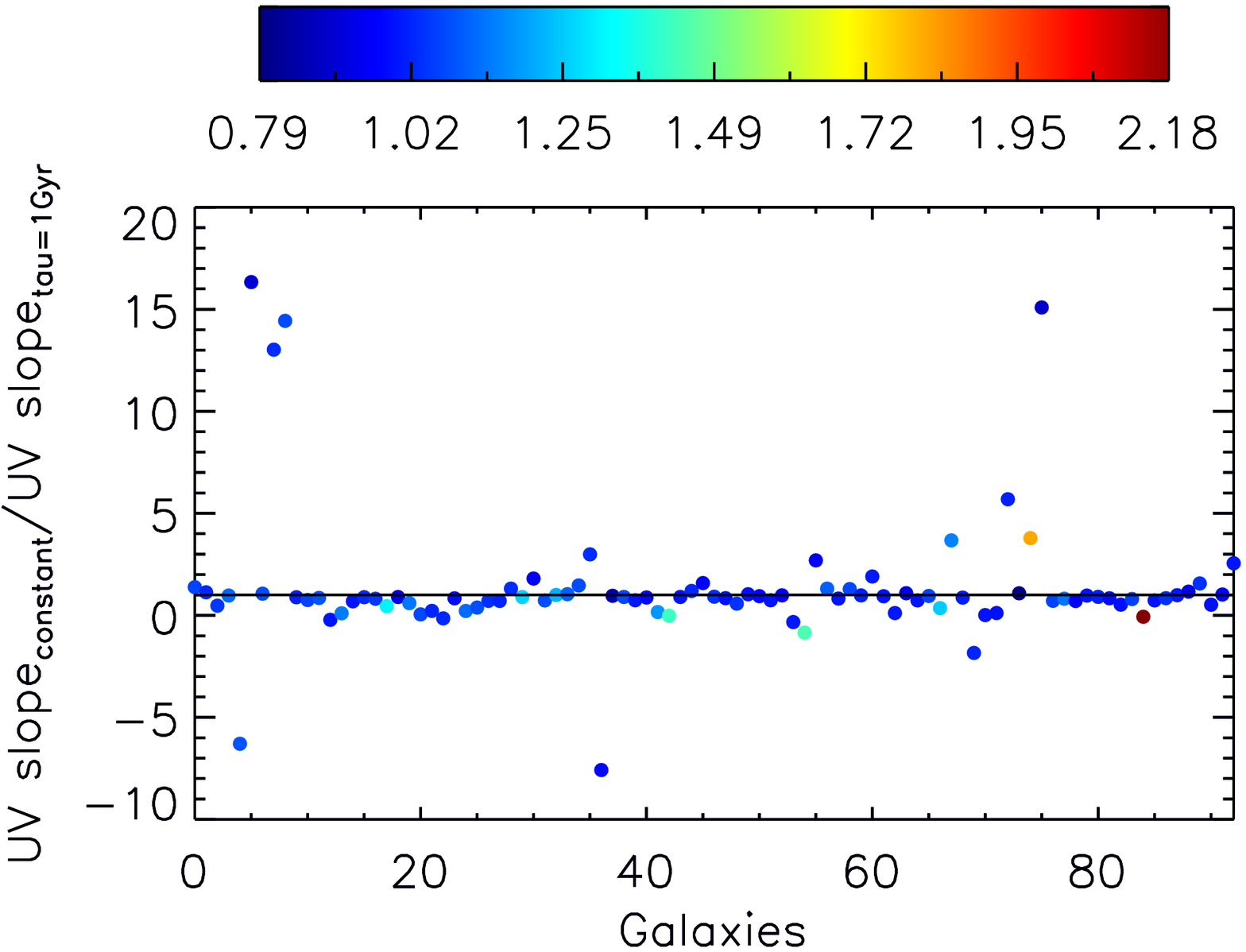}

\caption{\emph{Left column}: Relation between the stellar mass derived with an SED-fitting procedure when considering a constant SFH and an exponentially declining SFH for different values of the SFH time-scale. \emph{Right column}: Relation between the UV continuum slope derived with an SED-fitting procedure when considering a constant SFH and an exponentially declining SFH for different values of the SFH time-scale. Each point in the plots represents one PACS-detected galaxy. Each row is related to a specific value of the star-formation time-scale, $\tau_{\rm SFH}$, as indicated in each vertical axis. The SFR time-scales considered are $\tau_{SFH}$=50, 10, 5, 2, and 1 Gyr in the first, second, third, fourth, fifth and sixth rows, respectively. The color of each point is related to the ratio of the $\chi^2$ values between the SED-fittings carried out by considering a constant and an exponentially declining SFH: $\chi^2_{\rm constant}/\chi^2_{\tau_{\rm SFH}}$. The color code for the values of $\chi^2_{\rm constant}/\chi^2_{\tau_{\rm SFH}}$ are indicated in the color-bars. In all the plots, the horizontal line represent where both quantities represented in each vertical axis would agree.
              }
\label{tau_color_2}
\end{figure*}

We have just seen that the SED-fitting derived physical properties of our galaxies slightly change with the assumption of the SFH. Now we address the question if we can distinguish between different kinds of SFH with an SED fitting procedure with the good photometric coverage that the combination of GALEX and ALHAMBRA provides. To this aim, we have to study if the $\chi^2$ of the SED fits change significantly with the assumed SFH. If the $\chi^2$ value associated to an SED-fit obtained, for example, with a constant SFH is much lower than the $\chi^2$ associated to any other SED-fit obtained with a exponentially decaying SFH we could ensure that the right SFH is that constant with time, and therefore, establish that the age, dust attenuation, stellar mass, and UV continuum slope are those associated to the template of constant SFH. In Figures \ref{tau_color} and \ref{tau_color_2} we present the ratio between the SED-derived age, dust attenuation, stellar mass, and UV continuum slope when considering BC03 templates associated to a constant star-formation rate and exponentially decaying SFHs. This ratio is represented as a function of an ID number which represents each PACS-detected galaxy. The color of each point is related to $\chi^2_{\rm constant}/\chi^2_{\tau_{\rm SFH}}$, the ratio of $\chi^2$ values of the SED-fits carried out with BC03 templates associated to constant and exponentially declining SFH, respectively. Each plot in Figure \ref{tau_color} is linked to a specific value of the star-formation time-scale. It can be seen that the $\chi^2_{\rm constant}/\chi^2_{\tau_{\rm SFH}}$ ratios are close to unity for most of the galaxies and for most values of $\tau_{\rm SFH}$. Therefore, we conclude that by employing SED-fittings with BC03 templates we cannot distinguish between different types of SFHs and, therefore, we have a lack of knowledge of the specific SFH of each galaxy. Joining this result to the fact that different values of the assumed star-formation time-scale tend to give different values of some SED-derived physical properties, we conclude that even when using the exceptional photometric coverage that the combination of GALEX and ALHAMBRA data provides, we cannot obtain completely accurate values of the SED-derived properties of our PACS-detected galaxies. For example, in the case of the SED-derived age, as it tends to decrease with respect to the constant star-formation rate case when the star-formation time-scale decreases, the derived values of the age derived with BC03 templates associated to a constant star-formation rate could be considered as an upper limit of the real age of the galaxy. This way, if we obtain that a galaxy is young, it would be also young when considering other kinds of SFHs.

We wonder now what are the implications of the degeneration of the assumed SFH on the figures analyzed in this work. It only affects to the analysis involving the age, stellar mass, and the UV continuum slope since, as we have just seen, the SED-derived dust attenuation does not strongly change with the assumed SFH. Therefore, we have to pay special attention to the location of the galaxies in the star-formation vs stellar mass diagram (Figure \ref{sfr_mass}) and in the IRX-$\beta$ diagram (Figure \ref{dust_beta_fig}). As commented in Section \ref{sfr_mass_plane}, the total SFR shown in the vertical axis are derived from the combination of UV and IR emission, where the stellar masses are those SED-derived. Therefore, the x-axis values are those depending on the assumed SFH. Since the stellar mass derived with BC03 templates associated to exponentially decaying SFH are higher with respect to the value associated to a constant star-formation rate it is expected that we would have obtained that the galaxies are closer to the MS. In any case, the differences found in the SED-derived parameters when assuming different SFH scenarios are within their typical uncertainties and, therefore, the choice of a fixed SFH does not strongly affect our results.

\clearpage

\bsp

\label{lastpage}

\end{document}